\documentclass[monophys,vecphys,envcountchap,natbib,openany]{svmono}

\usepackage{makeidx}         
\usepackage{graphicx}        
\usepackage{multicol}        
\usepackage[bottom]{footmisc}
\usepackage{amsmath}
\usepackage{amssymb}
\usepackage{bm}
\usepackage{amsbsy}
\usepackage{bbm}
\usepackage{nicefrac}
\usepackage{mathrsfs}
\usepackage{dcolumn}         
\usepackage{txfonts}
\usepackage{comment}
\usepackage{exscale}         
\usepackage{theorem}         
\usepackage{txfonts}
\usepackage{wrapfig}
\usepackage{feynmf}

\usepackage{psfrag}

\psfrag{alpha}{$\alpha$}
\psfrag{gap}{$\Delta$}
\psfrag{lambda}{$\lambda$}
\psfrag{Lamb}{$\Lambda$}
\psfrag{L_crit}{$\Lambda_c$}
\psfrag{alphgap}{$\alpha\,,\,\,\Delta$}
\psfrag{Lc_alph_gap}{$\Lambda_{c}\,,\,\,\alpha\,,\,\,\Delta$}

\psfrag{pp_etas}{$\eta_{f},\eta_{b}$}
\psfrag{p_e_b}{$\eta_{b}$}
\psfrag{p_e_f}{$\eta_{f}$}
\psfrag{Z_b}{$Z_{b}$}
\psfrag{Z_f}{$Z_{f}$}
\psfrag{pp_g}{$\tilde{g}$}
\psfrag{pp_mass}{$\tilde{m}^{2}_{b}$}
\psfrag{pp_u}{$\tilde{u}$}
\psfrag{log_chi}{$\log\left[\xi^{-2}\right]$}
\psfrag{log_mb}{$\log\left[m^{2}_{b}-m^{2}_{b,\text{crit}}\right]$}

\psfrag{rhodivten}{$0.1\,\tilde{\rho}$}
\psfrag{rhodivfive}{$0.2\,\tilde{\rho}$}

\psfrag{rhoinv}{$1/\tilde{\rho}$}
\psfrag{etaASig}{$\eta_{A_{\sg}}$}
\psfrag{etaA}{$\eta_{A}$}
\psfrag{eta}{$\eta$}
\psfrag{ZSig}{$\tilde{Z}_{\sg}$}
\psfrag{ZPi}{$\tilde{Z}_{\pi}$}

\psfrag{t_rho}{$\tilde{\rho}$}
\psfrag{T_c}{$T_{c}$}
\psfrag{d_del}{$\delta-\delta_{\text{c}}$}
\psfrag{chi_inv}{$\chi^{-1}$}

\psfrag{m_bosonic}{$m^{2}_{b},\,\,m^{2}_{\sigma}$}
\psfrag{Z_factors}{$Z_{\pi},\,\,Z_{\sigma}$}
\psfrag{g_factors}{$g_{\pi},\,\,g_{\sigma}$}

\psfrag{tyldu}{$\tilde{u}$}
\psfrag{tyldel}{$\tilde{\delta}$}
\psfrag{delta}{$\delta$}
\psfrag{logdelt}{$\log (\delta-\delta_c)$}
\psfrag{logphi}{$\log (\phi_0)$}
\psfrag{dgdc}{$\frac{\delta_G-\delta_c}{\delta_c}$}
\psfrag{Temp}{$T$}
\psfrag{etaom}{$\eta_\omega$}

\def\bi{{\mathbf i}}
\def\bj{{\mathbf j}}
\def\bk{{\mathbf k}}

\def\bq{{\mathbf q}}
\def\bA{{\mathbf A}}
\def\bG{{\mathbf G}}
\def\bR{{\mathbf R}}
\def\b0{{\mathbf 0}}
\def\bGam{{\mathbf\Gamma}}
\def\bSg{{\mathbf\Sigma}}

\def\cR{{\cal R}}

\def\bra{\langle}
\def\ket{\rangle}
\def\up{\uparrow}
\def\down{\downarrow}

\def\alf{\alpha}
\def\eps{\epsilon}
\def\gam{\gamma}
\def\Gam{\Gamma}
\def\lam{\lambda}
\def\Lam{\Lambda}
\def\om{\omega}

\def\sg{\sigma}

\def\psib{\bar\psi}
\def\Psib{\bar\Psi}
\def\Phib{\bar\Phi}

\def\dag{\dagger}
\def\pdag{\phantom\dagger}

\makeindex   

\begin{document}
\setlength{\unitlength}{1mm}
\thispagestyle{empty}

\author{P. Strack}
\title{
\begin{center}
Renormalization group theory for fermions and order parameter fluctuations 
in interacting Fermi systems 
\end{center}
}

\subtitle{\vspace{20mm}
\begin{center}
\large{Von der Fakult\"at Mathematik und Physik der Universit\"at
Stuttgart\\
zur Erlangung der W\"urde eines Doktors der Naturwissenschaften\\
(Dr. rer. nat.) genehmigte Abhandlung}\\
%
\vspace*{20mm}
vorgelegt von\\
\textbf{Philipp Strack}\\
aus Frankfurt am Main\\[20mm]
Hauptberichter: Prof. Dr. Walter Metzner\\
Mitberichter: Prof. Dr. Alejandro Muramatsu\\[15mm]
%
%
Max-Planck-Institut f\"ur Festk\"orperforschung, Stuttgart,
2009
\end{center}}

\maketitle
%
%
\frontmatter
\cleardoublepage
\thispagestyle{empty}

\acknowledgments

Walter Metzner's supervision privileged my scientific
efforts: timely availability for questions, significant
calculational help, and extremely flexible work arrangements left
hardly any external factors to blame. Though not always easy to
digest, his sober and matter-of-factly Wittgensteinian 
approach to penetrate problems has sharpened my wit during the last three years 
for what I am deeply thankful. Alejandro Muramatsu is thanked for co-refereeing this thesis.

\bigskip

Pawel Jakubczyk, So Takei, Sebastian Diehl, and Johannes Bauer are thanked wholeheartedly for contributing important stimuli and corrections during the course of this work. During initial stages of my PhD time, Sabine Andergassen, Tilman
Enss, and Carsten Honerkamp were always available for questions.
Roland Gersch and Julius Reiss have additionally provided me with
code examples and many useful hints on programming. Further
interactions with Inga Fischer, Andrey Katanin, Manfred Salmhofer, 
and Roland Zeyher are gratefully acknowledged.

\bigskip

Christof Wetterich is thanked for providing the
opportunity to interact frequently with his group at the Institute
for Theoretical Physics in Heidelberg. Intense discussions with 
Jan Pawlowski, Holger Gies, and Hans Christian Krahl have shaped some ideas of this thesis. Jens
Braun, Stefan Fl\"orchinger, Jens M\"uller, Georg Robbers, and
Michael Scherer are also acknowledged for useful conversations.

Gil Lonzarich is gratefully acknowledged for being an inspirational 
host during the summer 2007 in
Cambridge, UK. The Quantum Matter group at Cavendish Laboratory and
especially Stephen Rowley, Leszek Spalek, Montu Saxena are thanked
for insightful exchanges.

\bigskip

Anne Gerrit Knepel makes my life better in any
respect; Rolf Dieter Strack's
engineering skills made our apartment more livable and freed
up valuable time; Irmgard and
Georg Walter Strack's liquidity injections eased costs associated
with frequent travel and coexisting apartments; 
Elisabeth Strack's and Uwe Gs\"anger's 
policies insured me safely; 
Eva Maria and Helmut Knepel enabled a most luxurious lifestyle for our two cats Mia and Momo
during much of this PhD time. Thank you.

%
%
\newpage
\mbox{}
\newpage

{\abs

\vspace{-2mm}

The physics of interacting Fermi systems is extremely sensitive to
the energy scale. Of particular interest is
the low energy regime where correlation induced collective behavior
emerges. The theory of interacting Fermi systems is confronted with the occurence
of very different phenomena along a continuum of
scales calling for methods capable of computing
physical observables as a function of energy scale.

In this thesis, we perform a comprehensive renormalization group
analysis of two- and three-dimensional Fermi systems at low and zero
temperature. We examine systems with spontaneous symmetry-breaking
and quantum critical behavior by deriving and solving flow equations
within the functional renormalization group framework.

We extend the Hertz-Millis theory of quantum phase transitions in
itinerant fermion systems to phases with discrete and continuous
symmetry-breaking, and to quantum critical points where the zero
temperature theory is associated with a non-Gaussian fixed point.
The order parameter is implemented by a bosonic Hubbard-Stratonovich
field, which --for continuous symmetry-breaking-- splits into two
components corresponding to longitudinal and transversal Goldstone fluctuations. We compute the finite temperature phase boundary near the quantum critical point explicitly including non-Gaussian fluctuations.

We then set up a coupled fermion-boson renormalization group theory
that captures the mutual interplay of gapless fermions with massless
order parameter fluctuations when approaching a quantum critical
point. As a first application, we compute the complete set of
quantum critical exponents at the semimetal-to-superfluid quantum
phase transition of attractively interacting Dirac fermions in two dimensions. 
Both, the order parameter propagator and the fermion propagator become non-analytic 
functions of momenta destroying the Fermi liquid behavior.

We finally compute the effects of quantum fluctuations in the
superfluid ground state of an attractively interacting Fermi system,
employing the attractive Hubbard model as a prototype.
The flow equations capture the influence of longitudinal and
Goldstone order parameter fluctuations on non-universal quantities
such as the fermionic gap and the fermion-boson vertex, as well as
the exact universal infrared asymptotics present in every fermionic
superfluid.

%
\tableofcontents
%
\mainmatter
%
%
%
\chapter[Introduction]{Introduction}
\label{chap:intro}




This thesis is concerned with low- and zero-temperature
properties of correlated fermion systems in two and three
dimensions. Interaction effects accompanied by quantum and thermal
fluctuations play an important role and produce fascinating
phenomena such as, for example, new scaling laws in the vicinity of
quantum critical points and exotic forms of superfluidity in cold atomic gases.

\section{Experiments}

Before presenting the outline of this thesis, we refer to two recent hallmark experiments
which are related to computations performed in the present work.

\subsection{Phase boundary close to a quantum critical point}

First, we discuss the recent measurement of the finite temperature
phase boundary of the antiferromagnetically ordered phase in
BaCuSi$_{2}$O$_{6}$ performed in Stanford (Sebastian 2006). In this
compound, the line of finite temperature phase transitions remains
continuous down to the lowest temperatures and ends at a
quantum critical point (QCP). In the vicinity of the QCP, the shape
of the phase boundary becomes universal and follows the power-law:
$T_{c}\sim\left(\delta-\delta_{c}\right)^{\psi}$, where $\delta$ is
the non-thermal control parameter --in this case the external
magnetic field. The so-called shift exponent $\psi$ is
independent of microscopic details and depends only on
dimensionality ($d$) and the additional dimensionality incurred from
quantum fluctuations ($z$):
\begin{eqnarray}
\psi=\frac{z}{d+z-2}\;,
\label{eq:intro_psi}
\end{eqnarray}
as first derived by Millis (1993). The relevant excitations of the material measured
by the Stanford group (Sebastian 2006) are spin-dimers which can be described
by an interacting Bose gas undergoing Bose-Einstein condensation
at the critical temperature.
The original spin degrees of freedom
exhibit antiferromagnetic order when the (collective) spin-dimer gas condenses.
In such cases, the dynamical exponent is $z=2$.

\begin{figure} 
 \begin{center}
\includegraphics*[width=125mm,angle=0]{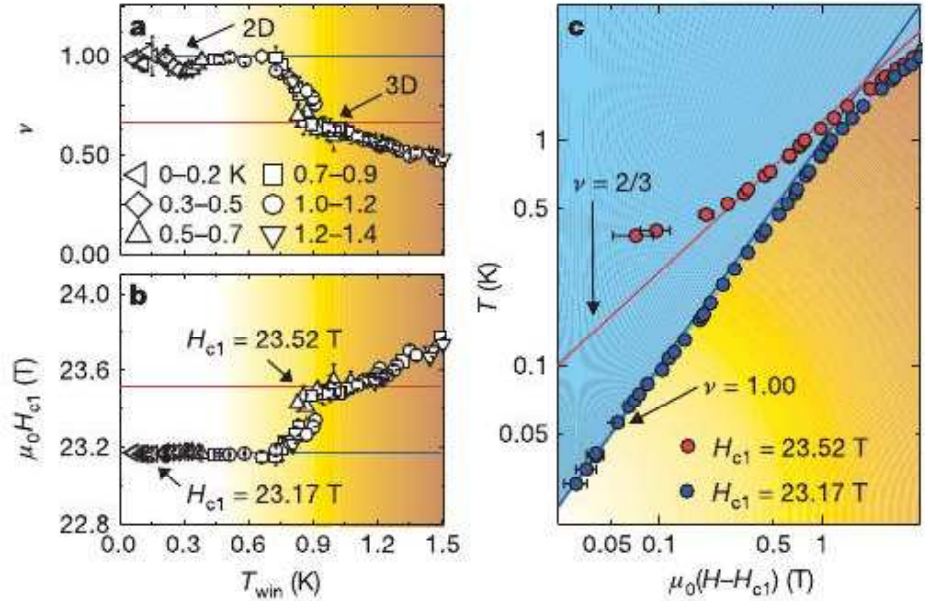}
\end{center}
\caption{\textit{Phase boundary of the antiferromagnetic phase in BaCuSi$_{2}$O$_{6}$ close to the QCP (Sebastian 2006). (a) Values of the shift exponent (here $\nu$) 
obtained from fitting experimental points on the phase boundary. The data 
approach $\nu=2/3$ in the intermediate regime, and there is a distinct 
crossover toward $\nu=1$ before the QCP is reached. 
(b): 
Estimates of the location of the QCP, $H_{c1}$, obtained along with $\nu$ during the fit. 
(c): Best fits to the phase boundary in the intermediate and 
low temperature regime on a logarithmic scale.}}
\label{fig:psi_experiment}
\end{figure}

In Fig. \ref{fig:psi_experiment} (c), data from torque magnetometry 
for the phase boundary is exhibited,
clearly in agreement with Eq. (\ref{eq:intro_psi}). The area coloured in blue is the paramagnetic
region and the yellow area corresponds to the antiferromagnetically ordered phase. 
Upon lowering the temperature when approaching the QCP, there is a crossover 
from three-dimensional scaling to two-dimensional scaling. This has been argued to 
be a consequence of effective decoupling of the $2d$-layers of the material at 
very low temperatures and is specific to the geometrically frustrated lattice structure 
of BaCuSi$_{2}$O$_{6}$. 



\bigskip

In chapters \ref{chap:bosonicqcp_discrete} and \ref{chap:bosonicqcp_goldstone}, we
compute zero- and finite-temperature properties of correlated systems close to a QCP. 
The interplay of thermal and quantum fluctuations makes this 
an interesting subject for theoretical studies. 
Among other things, we introduce a new way to compute the shift exponent coming
from the symmetry-broken phase with and without Goldstone modes by the use
of modern renormalization group equations which 
continuously connect the quantum fluctuation dominated regime directly at the QCP to the 
regime dominated by non-Gaussian classical fluctuations further up the $T_{c}$-line.

\subsection{Phase diagram of an attractive two-component Fermi gas}

As the second experiment, we make reference to the seminal measurement
of the phase diagram for a
spin-polarized Fermi gas performed
by the Ketterle group at MIT (Shin 2008). In this experiment, a different
amount of spin-up and spin-down components of $^{6}$Li are loaded in an optical trap.
By applying an external magnetic field 
close to a Feshbach resonance the population imbalanced gas is
tuned to the resonantly (attractively) interacting regime, forms bosonic spin-singlets, and condenses
to the superfluid state upon lowering the temperature. The non-thermal control parameter
here is the population imbalance or polarization $\sigma=\left(n_{\uparrow}-n_{\downarrow}\right)/\left(n_{\uparrow}+n_{\downarrow}\right)$,  
\begin{figure}[b]
\includegraphics*[width=75mm,angle=0]{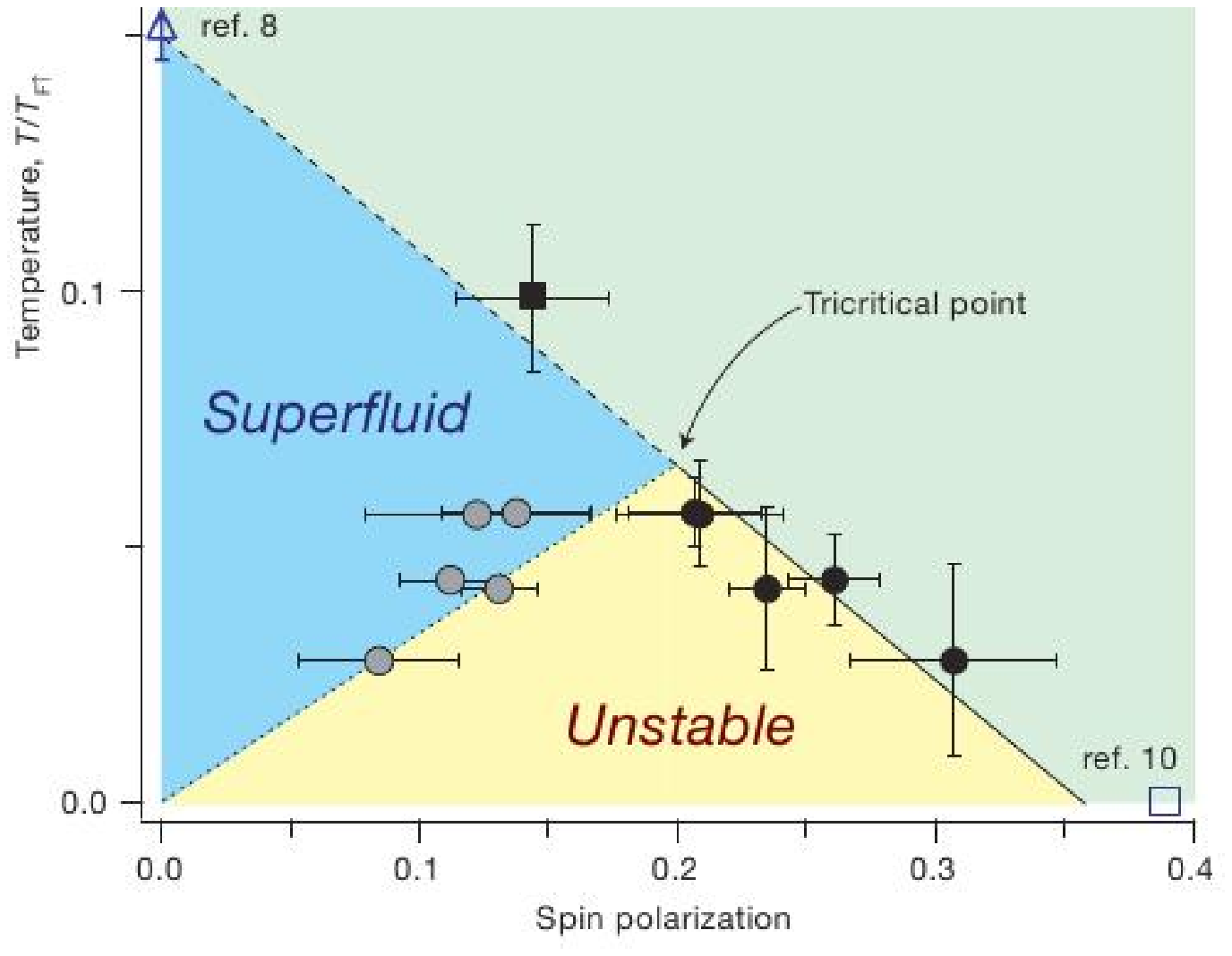}
\hspace*{2mm}
\includegraphics*[width=57mm,angle=0]{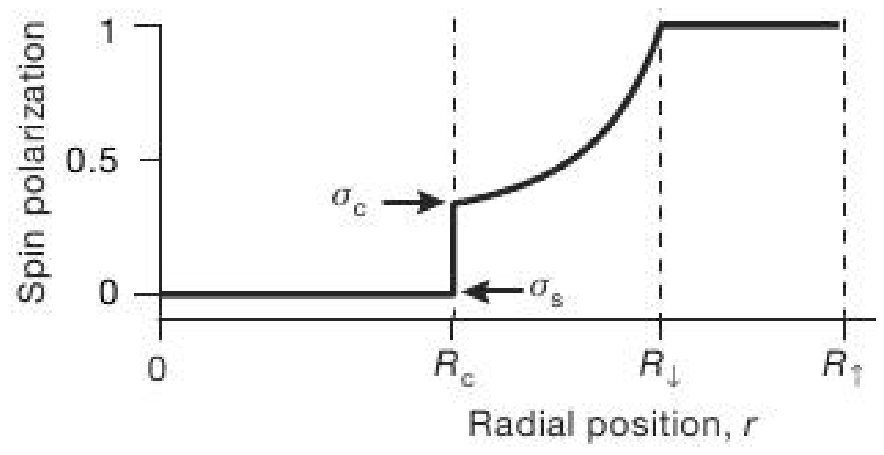}
\caption{\textit{Left: The $\sigma-T$-phase diagram for a homogeneous 
spin-polarized Fermi gas with resonant interactions (Shin 2008). 
The critical polarizations $\sigma_{c}$ (black solid circles and square) 
and $\sigma_{s}$ (grey solid circles) are displayed along the local $T/T_{F\uparrow}$, 
with $T_{F\uparrow}$ the Fermi temperature of the majority component 
in the spin-up state, at the phase boundary. The yellow area ($\sigma_{s}<\sigma<\sigma_{c}$) 
represents a region with phase separation until 
$\sigma_{s}=\sigma_{c}$ at the 
tricritial point. The triangle with ref. 8 marks $T_{c}$ for the unpolarized Fermi gas at unitarity from Burovski et. al (2006) and the 
square with ref. 10 marks the position 
of the first order quantum phase transition from Lobo et. al. (2006).
Right: At low $T$, the sample has a three-layer 
structure (Shin 2008): the core region ($0\leq r <R_{c}$) of a 
fully paired superfluid with $n_{\uparrow}=n_{\downarrow}$; 
the intermediate region ($R_{c}\leq r <R_{\downarrow}$) of a 
partially polarized normal gas; and the outer region $(R_{\downarrow}<
r<R_{\uparrow})$ of a fully polarized normal gas.
}}
\label{fig:ketterle}
\end{figure}
where $\uparrow$ and $\downarrow$ refer to the two spin components with densities 
$n_{\uparrow,\downarrow}$. Increasing $\sigma$ deprives
the attractively interacting fermions of their respective opposite-spin pairing partner thus
inhibiting the superfluid pairing beyond a critical polarization $\sigma_{c}(T)$.

As shown in the left plot of Fig. \ref{fig:ketterle}, an increasing polarization indeed 
suppresses the critical temperature and
destabilizes the superfluid phase at low temperatures. In the inner region of the trap, 
a superfluid core forms and the spin polarization shows a discontinuity 
at the boundary of the superfluid core $r=R_{c}$, a signature 
of the phase separation of a superfluid and a normal gas (Bedaque 2003). 
The critical polarization $\sigma_{c}=\text{lim}_{r\rightarrow R_{c}^{+}}\sigma(r)$ 
represents the minimum spin polarization for a stable normal gas; $\sigma_{s}=\text{lim}_{r\rightarrow R_{c}^{-}}\sigma(r)$ represents the 
maximum spin polarization for a stable superfluid gas as illustrated 
in on the right plot in Fig. \ref{fig:ketterle}. At $\sigma_{c0}\approx 0.36$, in the bottom right 
corner of the left plot in Fig. \ref{fig:ketterle}, the Clogston limit of 
superfluidity which is a first order quantum phase transition 
was experimentally verified. 

One way to achieve a \emph{continuous} quantum phase transition in
attractive fermion systems involves cold atomic gases loaded in a
honeycomb optical lattice (see Zhao 2006 and references therein) and
possibly graphene (Uchoa 2007, Castro Neto 2008). The attractive
Hubbard model on the honeycomb lattice exhibits, at half-filling, a
quantum critical point between a semimetal with massless Dirac
fermions and an s-wave superconductor.

\bigskip

Fermionic systems displaying superfluidity form a 
cornerstone of this thesis. 
In chapter \ref{chap:fermibosetoy}, we investigate
attractively interacting Dirac fermions in the vicinity of a
continuous semimetal-to-superfluid quantum phase transition and in 
chapter \ref{chap:fermionsuperfluids}, we consider the 
superfluid ground state of the attractive Hubbard model at quarter-filling.

\section[Thesis outline]{Thesis outline}
\label{sec:outline}

This thesis is structured in two parts. In the first part, the
methodological framework is presented. The second part contains
various applications of these methods and concepts. In detail, the
chapters in Part \ref{part:one} have the following contents:

\begin{itemize}

 \item In chapter \ref{chap:theo_concepts}, we introduce the fundamental
concepts necessary for an understanding of the applications in the
second part. First, the standard model of interacting Fermi systems,
Landau's Fermi liquid theory, and instabilities thereof are phrased
in the language of the renormalization group (RG). The notion of
spontaneous symmetry-breaking with a focus on Goldstone bosons is
introduced. These massless bosonic excitations arise when a
continuous symmetry is broken, as is the case for example in
magnetic or superfluid systems. Then, we --after having presented
basic features of phase transitions in general-- expose the
conventional RG approach to quantum critical systems referred to as
the Hertz-Millis theory. In transiting to chapter
\ref{chap:functional} we recapitulate the underlying ideas of the
Wilsonian RG approach to quantum field theory.

\bigskip

\item In chapter \ref{chap:functional}, the functional integral approach
to many-particle systems containing fermionic and bosonic fields is
laid out. At the heart of the chapter is the derivation of the
functional RG framework in its one-particle irreducible representation geared toward 
systems with spontaneously broken symmetry.

\end{itemize}

\noindent
The salient points of Part \ref{part:one} are summarized in chapter
\ref{chap:summary_part1}. The applications presented in Part
\ref{part:apple} are organized as follows.

\begin{itemize}

 \item In chapter \ref{chap:bosonicqcp_discrete}, we extend the
Hertz-Millis theory for quantum criticality to phases with broken
discrete symmetry. Using the Hertz action, we compute the RG flow of
the bosonic propagator and the effective potential with flow
equations derived from the functional RG framework. Different
dynamical exponents are distinguished and the shape of the phase
boundary at finite temperature is computed for each. Large parts of
this chapter have been published previously in Phys. Rev. B {\bf 77}, 195120
(2008) and the results for the Quantum Ising model are published 
in Phys. Rev. B {\bf 80}, 085108 (2009). 

\bigskip

\item In chapter \ref{chap:bosonicqcp_goldstone}, we analyze
quantum critical systems with Goldstone modes. We employ an
exclusively bosonic effective action with two propagators: one
associated with the transverse (Goldstone) dispersion and the
other stemming from longitudinal fluctuations. The interactions
between longitudinal and Goldstone fluctuations lead to strong
renormalizations of the longitudinal propagator away from and at
the phase boundary where both modes become degenerate. 
We compute RG flows to determine the phase boundary at
finite temperatures in three dimensions. 

\bigskip

\item In chapter \ref{chap:fermibosetoy}, the main inconsistency of the 
conventional Hertz-Millis approach --that massless fermions are integrated out 
in one sweep-- is cured by explicitly including
fermions into a coupled RG flow for (Dirac)
fermions and order parameter fluctuations. This way, we assess 
the mutual interplay of both types of fluctuations for the semimetal-to-superfluid quantum
phase transition in two dimensions. At the QCP, we find that
both the order parameter propagator and the single-particle
propagator are non-analytic functions of frequency and momenta
signalling the breakdown of the Fermi liquid. An improved 
version of this chapter is published in Phys. Rev. B {\bf 81}, 125103 (2010).

\bigskip

\item In chapter \ref{chap:fermionsuperfluids}, the low energy
behavior of the
attractive Hubbard model at quarter-filling as a prototype for systems with a superfluid ground state is analyzed. Various
non-universal quantities such as the fermionic gap are computed.
At the same time, the universal infrared behavior correctly
emerges from the flow of the coupled fermion-boson action in
agreement with the exact behavior of an interacting Bose gas. 
This chapter has been published in Phys. Rev. B {\bf 78}, 014522 (2008).

\end{itemize}

\noindent
Part \ref{part:summary} contains the conclusions of this thesis,
criticism, and a description of future research projects. The numerical procedure developed to solve the flow equations is explained in Appendix 
\ref{app:num_proc}.

\cleardoublepage \thispagestyle{empty}
%
%
%
%
\part{Theoretical Framework}
\label{part:one}
\chapter[Underlying concepts]{Underlying concepts}
\label{chap:theo_concepts}

\section{Fermi liquid instabilities}
\label{sec:fermi_lickit}

This section investigates conditions under which an interacting Fermi system might
not be described well anymore by its
original degrees of freedom, namely fermionic (quasi-) particles.
To lay a basis, we first describe the standard model of interacting fermions: the Landau Fermi liquid theory (for references and review see Nozieres 1964). This theory is formulated entirely
in terms of fermionic quasi-particles --bare fermions dressed by interaction effects, hence the prefix quasi.
The validity of the Landau Fermi liquid theory rests upon fulfillment of the following conditions:
\begin{itemize}

 \item the existence of a well-defined Fermi surface, that is, a $\left(d-1\right)$-dimensional hypersurface
       in momentum space, to the vicinity of which the low-energy fermionic excitations are restricted to,

 \item the existence of quasi-particles with a one-to-one correspondence to the non-interacting particles
       by adiabatically turning on the interaction.

\end{itemize}
In the functional integral terminology of the renormalization group, the Fermi liquid is said
to be described by the \emph{fixed point}
%
%
or \emph{scale-invariant} action:
\begin{eqnarray}
S\left[\bar{\psi},\psi\right]=
\int^{<\Lam}_{k}\bar{\psi}_{k}\left(i k_{0} - v_{\mathbf{k}_{\text{F}}}k_{r}\right)\psi_{k}
-\frac{1}{2}
\int_{k,k',q}^{<\Lam}
f_{\mathbf{k}_{\text{F}},\mathbf{k}'_{\text{F}}}
 \bar{\psi}_{k-q/2} \bar{\psi}_{k'+q/2}
 \psi_{k'-q/2}\psi_{k+q/2}\;,\nonumber\\
\label{eq:fermi_liquid_action}
\end{eqnarray}
where
$\int_{k}=T\sum_{k_{0}}\int\frac{d^{d}k}{\left(2\pi\right)^{d}}$
comprises (Matsubara) frequency summation and momentum integration
restricted to a shell of width $2\Lambda$ around the Fermi surface,
$\bar{\psi}$, $\psi$ are renormalized Grassmann fields rescaled by
the fermionic $Z_{\mathbf{k}_{\text{F}}}$-factor, $k_{r}$ is the
momentum deviation in radial direction from the Fermi surface,
$v_{\mathbf{k}_{\text{F}}}$ is the Fermi velocity projected on the
Fermi surface, and
$f_{\mathbf{k}_{\text{F}},\mathbf{k}'_{\text{F}}}$ the Landau
function obtained as the (non-commutative) limit
lim$_{q_{0}\rightarrow 0}$ lim$_{\mathbf{q}\rightarrow 0}$ with
$\mathbf{q}/q_{0}\rightarrow 0$ of the two-particle
vertex (for a review see Metzner 1998). Here and in the following,
we employ the four-vector notation
$k=\left(k_{0},\mathbf{k}\right)$. Considering only processes with 
zero-momentum transfer by 
taking the limit $q\rightarrow0$, the Hartree
mean-field theory of this action leads to the energy functional
phenomenologically postulated by Landau
\begin{eqnarray}
\delta E\left[\delta n\right]=\int_{\mathbf{k}} v_{\mathbf{k}_{\text{F}}}k_{r}
\delta n_{\mathbf{k}} + \frac{1}{2}\int_{\mathbf{k},\mathbf{k}'}
f_{\mathbf{k}_{\text{F}},\mathbf{k}'_{\text{F}}}
 \delta n_{\mathbf{k}_{\text{F}}}\delta n_{\mathbf{k}_{\text{F}}'}\;,
\label{eq:landau}
\end{eqnarray}
with $\delta n_{\mathbf{k}}$ the quasi-particle distribution function.
In addition to constituting the standard model of interacting fermions, 
the merit of Landau's theory lies in its predictive power for 
systems with spherical Fermi surfaces, the epitome being liquid $^{3}$He. 
Upon expanding the Landau function $f_{\mathbf{k}_{\text{F}},\mathbf{k}'_{\text{F}}}$
in Legendre polynomials, one can --with knowledge of only the first few Legendre coefficients--
compute a larger number of physical observables such as, e.g., the specific heat or
compressibility (Negele 1987).

An important trademark of the Fermi liquid is the renormalized single-particle
propagator determined by the quadratic part of the fixed point action as
\begin{eqnarray}
G_{f}(k)=\frac{1}{ik_{0}- v_{\mathbf{k}_{\text{F}}}k_{r}}\;,
\label{eq:landau_prop}
\end{eqnarray}
which describes \emph{stable} quasi-particle excitations with a
velocity $v_{\mathbf{k}_{\text{F}}}$. Note that the
interaction-induced reduction in quasi-particle weight by
$Z_{\mathbf{k}_{\text{F}}}$ has been absorbed into renormalized
field variables. Standard higher-order corrections to the
fermion self-energy of Landau's theory are expected to be quadratic in 
frequency and temperature (Chubukov 2003). 
Fresh research by Millis, Chubukov and
others, however, revealed --already for the generic weakly and
locally interacting Fermi gas-- non-analytic corrections to the
Fermi liquid behavior of the susceptibilities and the
specific heat (Chitov 2001, Chubukov 2003, Chubukov 2005).

\bigskip

Indeed, much of the more recent research in correlated fermion systems has been devoted to systems
where the Fermi liquid paradigm breaks down and other theoretical descriptions
must be invoked (Stewart 2001). Often, tendencies to destabilize the Fermi liquid are already visible in the bare bosonic response functions of the particle-particle (pp) channel:
\begin{fmffile}{20081028_5}
\vspace*{-5mm}
\begin{eqnarray}
\Pi_{\text{pp}}(q_{0},\mathbf{q})&=&-\int_{k}
   G_{f0}\left(k+q\right)
   G_{f0}\left(-k\right)
\propto
\parbox{20mm}{\unitlength=1mm\fmfframe(2,2)(1,1){
\begin{fmfgraph*}(20,10)\fmfpen{thin}
 \fmfleft{l1}
 \fmfright{r1}
 \fmfpolyn{empty,tension=0.4}{G}{3}
 \fmfpolyn{empty,tension=0.4}{K}{3}
  \fmf{dbl_wiggly}{l1,G1}
 \fmf{fermion,tension=0.3,right=0.6}{G2,K3}
 \fmf{fermion,tension=0.3,left=0.6}{G3,K2}
 \fmf{dbl_wiggly}{K1,r1}
 \end{fmfgraph*}
}}\;,
\label{eq:ph_bubble}
\end{eqnarray}
and the particle-hole (ph) channel,
\begin{eqnarray}
\Pi_{\text{ph}}(q_{0},\mathbf{q})&=&-\int_{k}
   G_{f0}\left(k+q\right)
   G_{f0}\left(k\right)
\propto
\parbox{20mm}{\unitlength=1mm\fmfframe(2,2)(1,1){
\begin{fmfgraph*}(20,10)\fmfpen{thin}
 \fmfleft{l1}
 \fmfright{r1}
 \fmfpolyn{empty,tension=0.4}{G}{3}
 \fmfpolyn{empty,tension=0.4}{K}{3}
  \fmf{dbl_wiggly}{l1,G1}
 \fmf{fermion,tension=0.3,left=0.6}{K3,G2}
 \fmf{fermion,tension=0.3,left=0.6}{G3,K2}
 \fmf{dbl_wiggly}{K1,r1}
 \end{fmfgraph*}
}}
\,\,\,\,\,\,,
\label{eq:ph_bubble}
\end{eqnarray}
\end{fmffile}
\hspace*{-1mm}where $G_{f0}^{-1}=
ik_{0}-\xi_{\mathbf{k}}$ is the bare propagator with $\xi_{\mathbf{k}}$ the
single-particle energy relative to the Fermi surface. The 
magnitudes of $\Pi_{\text{ph}}(q_{0},\mathbf{q})$ and
$\Pi_{\text{pp}}(q_{0},\mathbf{q})$ in conjunction with the sign and momentum 
structure of the interaction may point at the low-energy fate of
the system.
If, in a lattice structure for example, the interaction is local and repulsive and max$\left\{\Pi_{\text{pp}}(q_{0},\mathbf{q}),\Pi_{\text{ph}}(q_{0},\mathbf{q})\right\}$ 
occurs at $\Pi_{\text{ph}}(0,\mathbf{Q}=2\mathbf{k}_{F})$, the system may form a charge- or spin-density wave eventually leading to charge or magnetic order. On the other hand, 
strong particle-hole fluctuations in some cases enhance pairing phenomena in the particle-particle channel via the Kohn-Luttinger effect. Therefore, the interplay of both channels 
has to be assessed to make a statement about the true low-energy fate of the system. 
Moreover, possible infrared singularities in one or 
sometimes even multiple scattering channels invalidate 
perturbative approaches. For certain geometries, such
as a nested Fermi surface or when the Fermi surface intersects the van-Hove points,
$\Pi_{\text{ph}}$ exhibits a logarithmic divergence for small energies. 
$\Pi_{\text{pp}}$ also diverges logarithmically as a function of energy for an inflection-symmetric Fermi surface in \emph{any} dimension (for a review see Abrikosov 1975).

A situation of competing Fermi liquid instabilities where both, $\Pi_{\text{ph}}$
and $\Pi_{\text{pp}}$ are important, occurs for example in
the repulsive Hubbard model and can be analyzed within
a fermionic functional RG treatment which takes into account both, the particle-particle
and particle-hole channel (and their mutual interplay)
on equal footing (Zanchi 2000, Halboth 2000, Honerkamp 2001). In these RG flows, it is the summed up
\emph{non-universal} contributions to the flow that determine which of
the several \emph{universal} fixed points (Fermi liquid, antiferromagnetic, superconducting, etc.)
the system scales towards.

\bigskip

A drawback of the hitherto employed fermionic functional RG is its failure to penetrate phases
with broken-symmetry as the vertex functions grow large at a critical scale thus invalidating
the truncation devised for weakly to intermediately coupled systems. If one attempts to go
beyond this scale, the vertices eventually diverge and the flow has to be stopped. To continue the flow
into phases with broken symmetry is clearly desirable because many systems
have symmetry-broken ground states and many experiments are able to verify or falsify quantitative predictions from calculations in the symmetry-broken phase such as, for instance, the magnitude and symmetry properties
of the superconducting gap in the cuprates.

At present, two remedial approaches have been developed.
In the first approach, a small initial
anomalous self-energy for one or multiple channels is added to the bare action and as such offered to the flow (Salmhofer 2004, Gersch 2005, 2006, 2008). If now the symmetry-breaking occurs in one of the offered channels, this
initial anomalous self-energy regularizes the vertices around the critical scale and the flow can be continued into the symmetry-broken phase. 
This method requires no decoupling of the interaction.

In the second approach, the fermionic interaction is decoupled with a suitable
Hubbard-Stratonovich field and one subsequently deals with a coupled
fermion-boson theory (Baier 2004, Strack 2008). This approach yields easy access to the phenomena
associated with spontaneous symmetry-breaking alluded to in section \ref{sec:ssb} and we employ it in chapter \ref{chap:fermionsuperfluids} of this thesis in the context of fermionic superfluids.

\section{Spontaneous symmetry-breaking}
\label{sec:ssb}

Spontaneous symmetry-breaking occurs if the energetically lowest state of a
system described by a quantum or statistical field theory exhibits different symmetry properties than
the original Lagrangian. The importance of spontaneous symmetry-breaking for physics
can hardly be overestimated as has been recognized by the Nobel committee in the last 50 years with a
series of Nobel prizes. At the time of writing,
Nambu has been awarded the Nobel prize ``for the discovery of the mechanism of
spontaneously broken symmetry in subatomic physics'' (Nambu 2008).
Quite appropriately, we will in parts employ Nambu's formalism in chapter
\ref{chap:fermionsuperfluids} of this thesis.

In elementary particle physics as well as condensed matter physics, a stunning range of observed phenomena finds an explanation with the help of symmetry considerations (Negele 1987, Peskin 1996).
Perhaps most fundamentally, the Anderson-Higgs mechanism (Negele 1987) bridges the gap between
quantum excitations of the vacuum in high-energy physics and statistical many-particle systems at low temperatures. In the former context, this mechanism is the generally accepted
\begin{wrapfigure}{r}{0.5\textwidth}
  \vspace{-9mm}
  \begin{center}
\includegraphics*[width=40mm,angle=0]{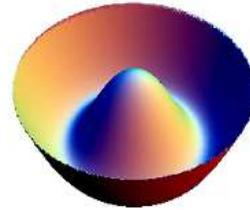}
\end{center}
\vspace{-5mm} \caption{\textit{Potential for spontaneous breaking of
continuous $O(N)$-symmetry, drawn for the case $N=2$. Fluctuations
along the trough in the potential correspond to the Goldstone
excitations (Peskin 1996)}.} \label{fig:mexhat} \vspace{-4mm}
\end{wrapfigure}
scenario for dynamical mass generation of fundamental particles such as 
gauge bosons: by (spontaneous) condensation of the Higgs field the order parameter thereof generates a mass term for particles covariantly coupled to the Higgs field. In the latter context of statistical physics, it was Anderson who first proposed this mechanism in 1958 to explain the Meissner effect in superconductors (Negele 1987): the photon (gauge boson) mediating the electromagnetic interaction becomes massive from the superconducting condensate (Higgs field).

\bigskip

Goldstone bosons emerge when spontaneously breaking a continuous
symmetry. In the low-energy sector of massless Quantum
Chromodynamics, for example, the formation of mesonic two-quark
bound states spontaneously breaks the chiral symmetry of the
original Lagrangian. As a result, the emergent pions can be
understood as the Goldstone bosons of massless QCD (Peskin 1996).
Many striking properties of low-temperature condensed matter systems
such as magnets, superfluid Helium, and superconducting materials
are attributed to Goldstone bosons. In superfluid systems, Goldstone
bosons are an indicator of long-range phase coherence. In this case,
the order parameter contains two degrees of freedom, one associated
with the amplitude and the other with the phase of the
order parameter. Above the critical temperature, the Hamiltonian is
invariant under the rotational symmetry group $O(2)$. Below the
critical temperature, the potential attains the form shown in Fig.
\ref{fig:mexhat} and the energetically lowest state is
\emph{degenerate} with the respect to phase transformations (i.e.
location in the trough). Generalized to an $N$-component
order parameter, the original $O(N)$-symmetry is hidden leaving only
the subgroup $O(N-1)$, which rotates the Goldstone modes among
themselves (Peskin 1996). Note that the form of the potential must
respect the full $O(N)$-symmetry even below the critical
temperature, but the field evaluated at its \emph{minimum} breaks
and reduces the symmetry. Theoretically, Goldstone modes are 
difficult to cope with in perturbation theory as the
massless propagators cause severe infrared singularities in low
enough dimensions necessitating a renormalization group (RG)
treatment.

\bigskip

In this thesis, the physics of Goldstone modes plays a central role.
In chapter \ref{chap:bosonicqcp_goldstone}, we consider a
quantum critical point adjacent to an ordered phase with Goldstone
bosons. In magnetic systems, for instance, the original $O(3)$
symmetry is broken down to $O(2)$ and the Goldstone bosons are
collective, gapless fluctuations in the direction of the electron
spins with respect to which the ground state is degenerate. For
antiferromagnetically ordered spins, these excitations form spin
waves.

In chapter \ref{chap:fermionsuperfluids}, we compute the RG flow of
various observables in fermionic superfluids. Here, the Goldstone
bosons correspond to fluctuations in the phase of the superfluid
order parameter and the original $O(2)$ symmetry is broken spontaneously.

\section{Quantum criticality}
\label{sec:quantum_crit}

In certain cases, the departure from the Fermi liquid
occurs in the vicinity of a quantum critical point (QCP) where spontaneous symmetry-breaking 
occurs at zero temperature.
This quantum criticality induced breakdown of the Fermi liquid has attracted enormous attention
in recent years (Stewart 2001, Loehneysen 2007).
This section recollects the most popular approach to quantum critical fermion systems, 
the Hertz-Millis theory (Hertz 1976, Millis 1993) and its shortcomings, after having recapitulated
similarities and differences between classical and quantum phase transitions.

\bigskip

The amount of thermal agitation in a multi-particle system can severly alter its macroscopic properties although it is described by the same particles, the same interactions,
and the same Hamiltonian (Goldenfeld 1992). The point separating the two distinct --for example para- versus ferromagnetic-- phases by tuning the temperature to $T_{c}$, is referred to as the critical point if the correlation length among the constituent particles diverges. At $T_{c}$, thermal fluctuations are so strong that the system becomes \emph{self-similar} on
all scales, i.e., it becomes \emph{scale-invariant}. The free energy develops non-analyticities which can be
subsumed under a set of critical exponents two of which are independent related to the existence of two relevant scaling fields, namely temperature and the external symmetry-breaking field (Goldenfeld 1992). It now turns out
that --rather remarkably-- lavish reservoirs of physical systems can be described by
the \emph{same} set of critical exponents distinguished only by their most basic symmetry properties and spatial dimensionality into universality classes.

\bigskip

\begin{wrapfigure}{r}{0.5\textwidth}
  \vspace{-15mm}
  \begin{center}
\hspace*{-3mm}
\includegraphics*[width=73mm,angle=-90]{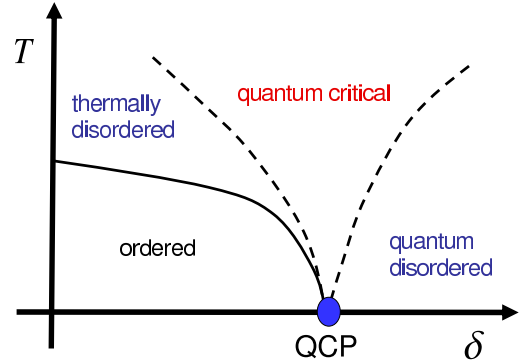}
\end{center}
\vspace{-15mm}
\caption{\textit{Generic phase diagram in the
vicinity of a continuous quantum phase transition.}}
\label{fig:intro_qcp}
\vspace{2mm}
\end{wrapfigure}

In recent times, one has come to envisage systems where continuous phase transitions
occur at zero temperature. Here, a non-thermal control parameter drives 
the phase transition and zero-point quantum fluctuations trigger 
quantum critical scaling of various quantities often described by 
different exponents than their classical counterparts (Sachdev 1999, Loehneysen 2007).
Such a quantum critical point lies between neighboring ground states --the
generic situation is exhibited in Fig. \ref{fig:intro_qcp}.
A peculiarity of quantum phase transitions
is their increased effective dimensionality incurred by quantum fluctuations in the direction
of imaginary time (Hertz 1976). In the functional integral formalism, the frequency term
in the quadratic part of the action, $\phi\left(|\Omega|^{2/z}+\mathbf{q}^{2}\right)\phi$ for $z=1,2$, and $\phi\left(|\Omega|/|\mathbf{q}|^{z-2}+\mathbf{q}^{2}\right)\phi$ for $z\geq3$, 
with $\phi$ being the bosonic order parameter field, indicates that characteristic frequencies scale as
$\Omega\sim |\mathbf{q}|^{z}$. When approaching the transition at $T=0$, both the order parameter correlation length $\xi$ and
correlation time $\xi_{\tau}$ diverge as function of control parameter:
\begin{eqnarray}
\xi\sim|\delta-\delta_{c}|^{-\nu}\,\,,\;\;\xi_{\tau}\sim\xi^{z}\;.
\end{eqnarray}
At finite temperature, however, the $(d+z)$-dimensional quantum system
has finite length in the time direction, $L_{\tau}\sim 1/T$.
In Fig. \ref{fig:intro_qcp}, we observe
that the existence of a QCP shapes large portions of the phase diagram. The black line of
finite-temperature classical phase transitions separates the thermally disordered phase
from the ordered phase and terminates at the QCP. Its shape is determined by the shift-exponent
as alluded to in chapter \ref{chap:intro}. A completely different regime is the high-temperature
regime above the QCP where $\xi_{\tau}\gg L_{\tau}$, the critical singularity is cut off
by the finite temperature, and the boundaries are the crossover
lines $T\sim|\delta-\delta_{c}|^{\nu z}$. Strikingly, in this quantum critical regime
one can measure power-law dependencies of physical observables up to rather high temperature
with exponents \emph{not} assuming the expected Fermi liquid values (Loehneysen 2007).

For a computation of these non-Fermi liquid power-laws in itinerant fermion systems,
Millis (1993) has set up RG equations capitalizing
on the fact that the zero-temperature fixed point for the order parameter is Gaussian as long as $d+z>4$.
The procedure is the following. The microscopic four-fermion interaction is decoupled with a Hubbard-Stratonovich field corresponding to the incipient order. The fermions are integrated out from
the functional integral in one sweep, but leaving their mark on the subsequent,
usually polynomial, expansion in powers of the order parameter field.
One then integrates out order parameter fluctuations on a Gaussian level (Millis 1993).
Already this results in a rich phase diagram with some materials fulfilling 
the Hertz-Millis exponents but others not (Stewart 2001, Sebastian 2006, Loehneysen 2007). For certain cases, such as the SU(2)-symmetric antiferromagnetic channel, this procedure is ill-defined as the coefficients are singular functions of frequency and momentum (Chubukov 2003,
Belitz 2005, Rech 2006, Loehneysen 2007). Experimental discrepancies and the questionable procedure to integrate out gapless fermions \emph{before} considering order parameter fluctuations suggest that there is room for improving the Hertz-Millis theory.

\bigskip

In this thesis, we go beyond the Hertz-Millis approach in three ways. First,
we show how to account for discrete symmetry-breaking and non-Gaussian classical
fluctuations by use of truncated functional RG equations in chapter \ref{chap:bosonicqcp_discrete}.
Second, possible effects of Goldstone modes and their fluctuations on the quantum phase transition when breaking a continuous symmetry are analyzed in chapter \ref{chap:bosonicqcp_goldstone}.
Third, we present a coupled fermion-boson RG that is able to assess the interplay of gapless fermions
with massless order parameter fluctuations in chapter \ref{chap:fermibosetoy}.

\section{The renormalization group}
\label{sec:rg_basics}

The perturbative expansion of most quantum field theories is plagued by unbounded
expressions. Either the integrands are singular for small momenta and
the phase space volume in low dimensions does not shrink fast enough to tame the infrared
divergence or the phase space volume blows up for
large momenta leading to ultraviolet divergences in higher-dimensional continuum theories.
%
These divergences are either of \emph{physical} origin signalling
for instance the proximity to a phase transition with truly infinite observables, or of \emph{technical}
origin as result of an inefficient organization of the perturbation
theory with all physical observables being in fact finite.

In the 1950's and 60's, the obnoxious presence of these singularities in calculations
proved a pertinacious problem to solve and severely
hindered progress in statistical and quantum field theory. Then, in the beginning of the 1970's, Wilson --at the top of a scientist iceberg--
%
%
reformulated the functional integral in terms of
differential equations along the continuous flow parameter $\Lambda$.
Typically, but not always, $\Lam$ is associated with the characteristic
\emph{kinetic} energy scale implemented by a cutoff for the propagator. This
cutoff carves out the singularity from the spectrum and excludes modes with
$|$energy$|$ $<\Lam$ from the functional integral.
This way, the singularity is approached
peu-\`a-peu only at the end of the functional integration when $\Lam\rightarrow 0$ and the
renormalized theory is delivered.
The solution of these so-called RG or flow equations
enabled, among countless
other groundbreaking discoveries, a controlled computation of critical exponents
for bosonic field theories (for a review see Wilson 1974).
The extension of these ideas to non-Abelian gauge theories involving fermions and bosons
led to the discovery of asymptotic freedom (Politzer 1973, Gross 1973).

\bigskip

Purely fermionic systems in one dimension were considered with
RG methods even before Wilson's Nobel prize in 1982 by Solyom (1979). 
Originally taken up by mathematical physicists (see references in Salmhofer 1999), the RG was 
subsequently applied to interacting Fermi systems also in
higher dimensions (for a review see Shankar 1994, Metzner 1998, and Salmhofer 1999).
The existence of a Fermi surface greatly enriches the analysis as the fermion propagator
becomes unbounded on a $\left(d-1\right)$-dimensional hypersurface to be compared with the much simpler
situation for bosons where the propagator becomes large only at one point (the origin) of momentum space.
Rigorous proofs involving phase space arguments and the curvature of the Fermi-surface
have established the RG as a mathematically well defined starting point for practical
computations. Therein, the low energy projection
of the vertices onto the Fermi surface involves \emph{flowing functions} to parametrize
\emph{transversal} momentum dependences parallel to the Fermi surface which
remain relevant until the very end of the flow (Zanchi 2000, Halboth 2000, Honerkamp 2001).
For this reason, the Fermi surface is partitioned into patches and mapped onto a grid in momentum space. The vertex function is then parametrized by several thousand variables and the
flow equations are solved numerically.

\bigskip

Flowing functions can be dealt with conveniently within modern functional renormalization group
formulations (Polchinski 1984, Wetterich 1993, Salmhofer 2001, for a review see Berges 2002, Metzner 2005).
Three formally exact flow equations have been devised: (i) the Polchinski (1984) scheme
for the generating functional of connected amputated vertex functions ($\mathcal{V}$) utilized by Zanchi (2000), (ii) the Wick-ordered scheme (Salmhofer 1999) where
the vertex functions are the expansion coefficients when expanding $\mathcal{V}$ in Wick-ordered
polynomials of the source-field utilized by Halboth (2000),
and (iii) the \emph{one-particle irreducible} (1PI) scheme
(Wetterich 1993, Salmhofer 2001) for the generating functional of the 1PI-vertex functions utilized
by Honerkamp (2001) and in a variety of contexts in Berges (2002).

\bigskip

In the next chapter, we thoroughly present the most popular, the
1PI-scheme of the functional RG. A particular strength
of this scheme is the natural inclusion of self-energy corrections as all
internal lines are fully dressed. Further, coupled fermion-boson
theories and spontaneous symmetry-breaking can be treated easily within this scheme.
We will apply this scheme for the computations in Part \ref{part:apple} of this thesis.

\cleardoublepage
\thispagestyle{empty}

\chapter[Functional renormalization group]{Functional renormalization group}
\label{chap:functional}

In this chapter, we introduce the main computational tools employed in this thesis.
First, we set up a unifying framework for fermionic and bosonic models by defining
superfields whose fermionic and bosonic components are distinguished by a
statistics index.
Second, we define the one-particle irreducible generating
functional, the effective action. Third, the functional flow equation for the
effective action is derived. Finally, we explain how to devise truncations with
non-zero expectation values of field variables to account for spontaneous
symmetry-breaking.

\section{Functional integral for quantum many-particle systems}
We consider interacting Fermi systems described by the \emph{bare} action,
\begin{eqnarray}
\Gamma_{0}[\bar{\psi},\psi]=-\int_{K}\bar{\psi}_{K} G^{-1}_{f0}(K)\psi_{K}+
\int_{K,K',Q} V^{f}_{K,K',Q}\,\bar{\psi}_{K} \bar{\psi}_{K'+Q} \psi_{K'} \psi_{K+Q} \;,
\label{eq:four_fermi}
\end{eqnarray}
where $\bar{\psi}$ and $\psi$ are Grassmann-valued fields, the index
$(K=(k_{0},\mathbf{k},\sigma)$ and its corresponding integration
$\int_{K}$ comprises frequency, momentum and internal states such
as, for example, the spin projection, and $G_{f0}^{-1}(K)= ik_{0}
-\xi_{\mathbf{k}}$ is the bare inverse Green function. The Fermi
surface is defined as the $\left(d-1\right)$-dimensional manifold
$\xi_{\mathbf{k}}=0$. $V^{f}_{K,K',Q}$ denotes an arbitrary and in
general momentum- and spin-dependent many-body interaction of the
lowest non-trivial (quartic) order
\footnote{\textit{ Models of the form Eq. (\ref{eq:four_fermi}) have
attracted attention for decades: in one-dimensional systems, various
phases such as the Luttinger liquid (Solyom 1979, Voit 1994) occur;
in the context of gauge theories, in vacuum and with more elaborate
internal degrees of freedom, Eq. (\ref{eq:four_fermi}) may describe
the Thirring or Gross-Neveu Model (Hands 1993); on the lattice in
two and higher dimensions, Eq. (\ref{eq:four_fermi}) includes the 
attractive or repulsive (depending on the sign of the in that 
case momentum-independent interaction)
Hubbard model often employed in contemporary condensed matter
physics (Micnas 1990, Baeriswyl 1995).}}.
In many circumstances, correlations induce fermion-pairing phenomena and one finds
the description in terms of the order parameter useful.
The simplest bosonic action is the $\phi^{4}$-model
\begin{eqnarray}
\Gamma_{0}[\phi]=-\int_{Q}\phi^{\ast}_{Q} G_{b0}^{-1}(Q) \phi_{Q}+
\int_{K,K',Q} V^{b}_{K,K',Q}\,\phi^{\ast}_{K-Q} \phi^{\ast}_{K'+Q} \phi_{K'} \phi_{K} \;,
\label{eq:phi4}
\end{eqnarray}
where $\phi$ is here a complex-valued field but may in general
contain $N$ components referred to as $O(N)$-models (Amit 1995).
$G_{b0}^{-1}(q)=-\left(q_{0}^{2} + \omega_{\mathbf{q}} + m_{b}^{2}\right)$ is
the bosonic Green's function with $\omega_{\mathbf{q}}$ being the
dispersion, usually but not always quadratic in momenta, and
$m_{b}^{2}=\xi^{-2}$ the mass or inverse correlation length. 
The frequency term is also not always quadratic; for the
non-relativistic Bose gas one has a complex linear term $\sim i
q_{0}$ and for certain quantum-critical systems Landau-damping
entails $|q_{0}|$ or even $|q_{0}|/\mathbf{q}$ (Hertz 1976, Millis
1993).

\bigskip

Finally, models with both, fermionic and bosonic fields, may be the subject of analysis. In addition
to the kinetic terms with $G^{-1}_{b0}(Q)$ and  $G^{-1}_{f0}(K)$, to lowest order we then have to consider the
fermion-boson vertex
\begin{eqnarray}
\Gamma_{0}[\phi,\bar{\psi},\psi]= \int_{K,Q} g_{K,Q}\,\left(\phi^{\ast}_{Q}
\psi_{-K+Q} \psi_{K}+\phi_{Q}
\bar{\psi}_{K+Q} \bar{\psi}_{-K}\right)\;.
\label{eq:yukawa}
\end{eqnarray}
\subsection{Superfield formulation for fermionic and bosonic models}
\label{subsec:super}
To develop a unifying framework, we combine fermionic and bosonic fields in
a superfield $\mathcal{S}$, where fermions
and bosons are distinguished by a statistics index $s= b,f$, that is,
\begin{eqnarray}
\mathcal{S}_b = \Phi\;,\;\;\mathcal{S}_f = \Psi\,\,.
\end{eqnarray}
To account for matrix propagators necessary to describe
phases with broken symmetry such as, for example, superfluidity, we here use the fermionic Nambu fields
\begin{equation}
 \Psi_k = \left( \begin{array}{c}
 \psi_{k\up} \\ \psib_{-k\down}
 \end{array} \right) \; , \quad
 \Psib_k = \left( \psib_{k\up}, \psi_{-k\down} \right)
\end{equation}
and bosonic Nambu fields
\begin{equation}
 \Phi_q = \left( \begin{array}{c}
 \phi_q \\ \phi^*_{-q}
 \end{array} \right) \; , \quad
 \Phib_q =
 \left( \phi^*_q, \phi_{-q} \right) \; .
\end{equation}
The fermionic and bosonic matrix propagators are then given by
$\bG_f(k) = - \bra \Psi_k \Psib_k \ket$ and
$\bG_b(q) = - \bra \Phi_q \Phib_q \ket \,$, respectively.
The superpropagator $\bG(q) = - \bra \mathcal{S}_q \bar{\mathcal{S}}_q \ket$
is diagonal in the statistics index. 

Note that the specific choice of the fermionic and bosonic Nambu fields depends on the
physical situation under investigation and can be adjusted straightforwardly.

\subsection{One-particle irreducible generating functional}
\label{sec:rgschemes}
Functional integration of the bare action containing fermionic and bosonic fields, $\Gamma_{0}[\mathcal{S},\bar{\mathcal{S}}]$, and subsequent functional differentiation
with respect to source fields $\mathcal{S}'$ and
$\bar{\mathcal{S}}'$ linearly coupled to $\mathcal{S}$ and
$\bar{\mathcal{S}}$ yields the connected m-particle Green functions (Negele 1987)
\begin{eqnarray}
G_{m}\left(K_{1}',...,K_{m}';K_{1},...,K_{m}\right)&=&
-\bra\mathcal{S}_{K_{1}'}...\mathcal{S}_{K_{m}'}
 \bar{\mathcal{S}}_{K_{m}}  ...\bar{\mathcal{S}}_{K_{1}} \ket_{c}\nonumber\\
&=&
\frac{\partial^{m}}{\partial \mathcal{S}'_{K_{1}} ... \partial \mathcal{S}'_{K_{m}}}
\frac{\partial^{m}}{\partial \bar{\mathcal{S}}'_{K_{m}} ... \partial \bar{\mathcal{S}}'_{K_{1}}}
G[\mathcal{S}',\bar{\mathcal{S}}']\Bigg|_{\mathcal{S}'=\bar{\mathcal{S}}'=0}\;,
\end{eqnarray}
where $\bra ...\ket_{c}$ denotes the connected average and the generating functional
is obtained from the logarithm of the partition function
\begin{eqnarray}
 G[\mathcal{S}',\bar{\mathcal{S}}'] = - \log \int D\mathcal{S} D\bar{\mathcal{S}} \,
 e^{-\Gam_0[\mathcal{S},\bar{\mathcal{S}}]+
 (\mathcal{S}',\bar{\mathcal{S}}) + (\mathcal{S},\bar{\mathcal{S}}')} \;,
\end{eqnarray}
where the bracket $(.,.)$ is a shorthand notation for the inner product for superfields.
The Legendre transform of $G[\mathcal{S}',\bar{\mathcal{S}}']$ is the effective action
\begin{eqnarray}
\Gamma[\mathcal{S},\bar{\mathcal{S}}]=\mathcal{L}G[\mathcal{S}',\bar{\mathcal{S}}']
=G[\mathcal{S}',\bar{\mathcal{S}}']-
 (\mathcal{S}',\bar{\mathcal{S}}) - (\mathcal{S},\bar{\mathcal{S}}')\;,
\label{eq:eff_action}
\end{eqnarray}
where the conjugated field variables are
\begin{eqnarray}
\mathcal{S}=\frac{\partial G[\mathcal{S}',\bar{\mathcal{S}}']}{\partial\bar{\mathcal{S}}'}\;,
\;\;\;
\bar{\mathcal{S}}=\frac{\partial G[\mathcal{S}',\bar{\mathcal{S}}']}{\partial\mathcal{S}'}\;.
\label{eq:conjugated}
\end{eqnarray}
$\Gamma[\mathcal{S},\bar{\mathcal{S}}]$
generates the one-particle-irreducible (1PI) Green functions. Topologically,
all diagrams are generated where cutting \emph{one} \emph{line}, irrespective if fermionic or bosonic, does not
separate the diagram into two disconnected parts (Negele 1987, Amit 2005).

\section{Flow equations}
\label{sec:flow_eq}

In this section, we derive the exact flow equation (Wetterich 1993, Salmhofer 2001, Berges 2002,
Metzner 2005, Enss 2005) for the effective action,
that is, the 1PI generating functional, adapted to superfields.

\subsection{Exact flow equation}
\label{subsec:exact_flow}

The exact flow equation is a reformulation of the functional
integral as a functional differential equation. It describes the
evolution of the effective action as a function of a flow parameter
$\Lam$, usually a cutoff. The cutoff can be implemented by adding
the regulator term quadratic in the fields
\begin{equation}
 \cR^{\Lam} = (\bar{\mathcal{S}},\bR^{\Lam}\mathcal{S})=
 \frac{1}{2} \int_q \Phib_q \, \bR_b^{\Lam}(q) \, \Phi_q +
 \int_k \Psib_k \, \bR_f^{\Lam}(k) \, \Psi_k
\label{eq:cutoff}
\end{equation}
to the bare action. The purpose of the cutoff is twofold. First, it
regularizes fermionic and bosonic infrared divergences when dealing
with, for example, a massless Goldstone boson or zero-temperature
fermions whose propagator becomes unbounded in the vicinity of the
Fermi surface. Second, $\bR^{\Lam}$ incorporates the continuous flow
parameter along which the functional integration is performed: as a
function of decreasing cutoff, $\Gam^{\Lam}$ interpolates smoothly
between the bare action $\Gam_0$ for $\Lam = \infty$ and the full
effective action $\Gam$, recovered in the limit $\Lam \to 0$.

To \emph{regularize} and \emph{interpolate}, the matrix elements of $\bR^{\Lam}$ have
to fulfill the following properties:
\begin{eqnarray}
&(i)&\;\; R^{\Lam}(Q)\geq 0\nonumber\\
&(ii)&\;\; \text{lim}_{\Lambda\rightarrow 0}\,R^{\Lam}(Q)= 0\nonumber\\
&(iii)&\;\; \text{lim}_{\Lambda\rightarrow \infty}\,R^{\Lam}(Q)=\infty\;.
\end{eqnarray}

Integrating $e^{-\Gam_0 - \cR^{\Lam}}$ in the presence of source
fields coupling linearly to $\mathcal{S}$ and $\bar{\mathcal{S}}$ yields the
cutoff-dependent generating functional for connected Green
functions
\begin{equation}
 G^{\Lam}[\mathcal{S}',\bar{\mathcal{S}}'] = - \log \int D\mathcal{S} D\bar{\mathcal{S}} \,
 e^{-\Gam_0[\mathcal{S},\bar{\mathcal{S}}] - \cR^{\Lam}[\mathcal{S},\bar{\mathcal{S}}] +
 (\mathcal{S}',\bar{\mathcal{S}}) + (\mathcal{S},\bar{\mathcal{S}}')} \; .
\end{equation}
The cutoff-dependent effective action $\Gam^{\Lam}$ is defined as
\begin{equation}
 \Gam^{\Lam}[\mathcal{S},\bar{\mathcal{S}}] =
 {\cal L} G^{\Lam}[\mathcal{S}',\bar{\mathcal{S}}'] - \cR^{\Lam}[\mathcal{S},\bar{\mathcal{S}}] \; ,
\label{eq:gamma_cutoff}
\end{equation}
where ${\cal L} G^{\Lam}$ is the Legendre transform of $G^{\Lam}$ as specified without a cutoff in
Eq. (\ref{eq:eff_action}).

Executing a scale-derivative
on both sides of Eq. (\ref{eq:gamma_cutoff}) and using Eq. (\ref{eq:cutoff}), we obtain
\begin{eqnarray}
\partial_{\Lam}\Gam^{\Lam}[\mathcal{S},\bar{\mathcal{S}}]=
\partial_{\Lam} G^{\Lam}[\mathcal{S}',\bar{\mathcal{S}}']-
 (\bar{\mathcal{S}},\dot{\bR}^{\Lam}\mathcal{S})\;,
\label{eq:Gam_deriv}
\end{eqnarray}
where $\dot{\bR}^{\Lam} = \partial_{\Lam} \bR^{\Lam}$ and we further compute
\begin{eqnarray}
\partial_{\Lam} G^{\Lam}[\mathcal{S}',\bar{\mathcal{S}}']&=&
-e^{G^{\Lam}[\mathcal{S}',\bar{\mathcal{S}}']}\partial_{\Lam}
e^{-G^{\Lam}[\mathcal{S}',\bar{\mathcal{S}}']}\nonumber\\
&=&
e^{G^{\Lam}[\mathcal{S}',\bar{\mathcal{S}}']}
\int D\mathcal{S} D\bar{\mathcal{S}} \,
(\bar{\mathcal{S}},\dot{\bR}^{\Lam}\mathcal{S})\,
e^{-\Gam_0[\mathcal{S},\bar{\mathcal{S}}] - \cR^{\Lam}[\mathcal{S},\bar{\mathcal{S}}] +
 (\mathcal{S}',\bar{\mathcal{S}}) + (\mathcal{S},\bar{\mathcal{S}}')}\nonumber\\
&=&
e^{G^{\Lam}[\mathcal{S}',\bar{\mathcal{S}}']}
(\partial_{\mathcal{S}'},\dot{\bR}^{\Lam}\partial_{\bar{\mathcal{S}}'})\,
e^{-G^{\Lam}[\mathcal{S}',\bar{\mathcal{S}}']}\nonumber\\[2mm]
&=&
\left(
\frac{\partial G^{\Lam}[\mathcal{S}',\bar{\mathcal{S}}']} {\partial \mathcal{S}'},\dot{\bR}^{\Lam}
\frac{\partial G^{\Lam}[\mathcal{S}',\bar{\mathcal{S}}']}{\partial \bar{\mathcal{S}}'}
\right)
+
\text{Str}\,
\dot{\bR}^{\Lam}\frac{\partial^{2} G^{\Lam}[\mathcal{S}',\bar{\mathcal{S}}']}
{\partial \mathcal{S}'\partial \bar{\mathcal{S}}'}\nonumber\\[2mm]
&=&
\left(
\bar{\mathcal{S}},\dot{\bR}^{\Lam}
\mathcal{S}
\right)
+
\text{Str}\,
\dot{\bR}^{\Lam}\left(\frac{\partial^{2}\left( \Gamma^{\Lam}[\mathcal{S},\bar{\mathcal{S}}]
+ \mathcal{R}^{\Lam}[\mathcal{S},\bar{\mathcal{S}}]\right)}
{\partial \mathcal{S}\partial \bar{\mathcal{S}}}\right)^{-1}\;,
\label{eq:G_deriv}
\end{eqnarray}
where the supertrace $\rm Str$ traces over all indices with a plus
sign for bosons and a minus sign for fermions. For the first term in the last
line, we have used Eq. (\ref{eq:conjugated}) and for the second term in the last line
we have used the relation $\partial^{2} G/\partial \mathcal{S}' \partial \bar{\mathcal{S}}'
= \left(\partial^{2} \Gamma/\partial \mathcal{S} \partial \bar{\mathcal{S}}\right)^{-1}$ augmented
with the cutoff term from Eq. (\ref{eq:gamma_cutoff}). Denoting
\begin{equation}
 \bGam^{(2) \,\Lam}[\mathcal{S},\bar{\mathcal{S}}] =
 \frac{\partial^2 \Gam^{\Lam}[\mathcal{S},\bar{\mathcal{S}}]}
 {\partial\mathcal{S} \partial\bar{\mathcal{S}}} \;,
\end{equation}
and assembling Eqs. (\ref{eq:G_deriv}, \ref{eq:Gam_deriv}),
we obtain the exact flow equation for the effective action
\begin{equation}
 \frac{d}{d\Lam} \Gam^{\Lam}[\mathcal{S},\bar{\mathcal{S}}] =
 {\rm Str} \, \frac{\dot{\bR}^{\Lam}}
 {\bGam^{(2) \, \Lam}[\mathcal{S},\bar{\mathcal{S}}] + \bR^{\Lam}} \;.
\label{eq:exact_flow_eqn}
\end{equation}
Note that the definitions of $\Gam^{\Lam}$ vary slightly in the
literature. In particular, $\Gam^{\Lam}$ is frequently defined
as the Legendre transform of $G^{\Lam}$ without subtracting
the regulator term $\cR^{\Lam}$, which leads to a simple
additional term in the flow equation.

To expand the functional flow equation (\ref{eq:exact_flow_eqn}) in powers of the
fields, we write the Hessian of $\Gam^{\Lam}$ as
\begin{equation}
 \bGam^{(2) \,\Lam}[\mathcal{S},\bar{\mathcal{S}}] =
 - (\bG^{\Lam})^{-1} + \tilde\bGam^{(2) \,\Lam}[\mathcal{S},\bar{\mathcal{S}}] \; ,
\end{equation}
where $\tilde\bGam^{(2) \,\Lam}[\mathcal{S},\bar{\mathcal{S}}]$
comes from terms which are at least quadratic in the fields.
Defining $\bG_R^{\Lam} = [(\bG^{\Lam})^{-1} - \bR^{\Lam}]^{-1}$ and
expanding in powers of $\tilde\bGam^{(2) \,\Lam}$ yields
\begin{eqnarray}
 \frac{d}{d\Lam} \Gam^{\Lam} &=&
 - {\rm Str}\, (\dot{\bR}^{\Lam} \bG_R^{\Lam})
 - {\rm Str} \left[
 {\bG'_R}^{\!\Lam} \big( \tilde\bGam^{(2) \,\Lam} +
 \tilde\bGam^{(2) \,\Lam} \bG_R^{\Lam} \tilde\bGam^{(2) \,\Lam} + \dots
 \big) \right] \; , \quad
\label{eq:flow_expansion}
\end{eqnarray}
where
\begin{equation}
 {\bG'_R}^{\!\Lam} =
 \bG_R^{\Lam} \dot{\bR}^{\Lam} \bG_R^{\Lam} \; .
 \label{eq:single_scale}
\end{equation}
For a sharp cutoff, Eq. (\ref{eq:single_scale}) denotes the
so-called single-scale propagator (Salmhofer 2001, Metzner 2005) as
it then has support only for energies $=\Lam$ due to a
$\delta$-function. In general, ${\bG'_R}^{\!\Lam}$ is smoothened and
may thus have support on more energies, typically in an energy
window around energies $\approx\Lam$.

Expanding both sides of Eq.~(\ref{eq:flow_expansion}) in powers of
the fields and comparing coefficients yields an (infinite) hierarchy
of flow equations for the vertex functions. The zero-point vertex,
$\gamma^{(0)\Lam}$, see also the first line of Fig.
\ref{fig:ssb_hierarchy} equals the effective potential with the
analytic flow equation from the first term in Eq.
(\ref{eq:flow_expansion}):
\begin{eqnarray}
 \frac{d}{d\Lam} U^{\Lam} &=&
 - {\rm Str}\, (\dot{\bR}^{\Lam} \bG_R^{\Lam})\;.
\label{eq:eff_pot}
\end{eqnarray}
Flow equations for thermodynamic properties of the system such as
the specific heat or the entropy follow from derivatives with
respect to the temperature of Eq. (\ref{eq:eff_pot}).

\subsection{Spontaneous symmetry-breaking}
\label{subsec:ssb_rg}

The low energy regimes of the models described in Eqs.
(\ref{eq:four_fermi}, \ref{eq:phi4}) are often dominated by phases
with broken symmetry, that is, states where interaction effects
spontaneously break one, or multiple, of the symmetries present in
the original bare action. The physical properties in such a phase
are vastly different from the phase where the original symmetry is
preserved.

In terms of our superfields, the symmetry-breaking is associated
with a non-vanishing expectation value of one of the bosonic
components of $\mathcal{S}$, referred to as the order parameter.
While in the symmetric phase the flow equations are expanded around
$\phi=0$, in the symmetry-broken regime we expand around a generally
non-zero value of the $q=0$ component of the bosonic field, that is,
we expand $\Gam^{\Lam}[\psi,\psib,\alf^{\Lam}+\phi]$ in powers of
$\psi,\psib,\phi,\phi^*$, where $\alf^{\Lam}$ may be non-zero
(see also Delamotte 2004, Sch\"utz 2006).

We now illustrate how this expansion works. For this purpose,
we restrict ourselves to the case of a real-valued scalar field and
consider only local terms of the effective action. 
The ansatz for the bosonic components of the left-hand-side of
Eq.~(\ref{eq:flow_expansion}) is
\begin{eqnarray}
\Gamma^{\Lam}[\phi]=\int\sum^{\infty}_{n=0}\frac{1}{n!}\gamma^{(n)\,\Lam}
\left(\phi-\alf^{\Lam}\right)^{n}\;,
\label{eq:ssb_expansion}
\end{eqnarray}
where the integration symbol denotes the standard position space integration. 
Performing a scale-derivative on both
sides of Eq. (\ref{eq:ssb_expansion}) yields
\begin{eqnarray}
\partial_{\Lambda}\Gamma^{\Lam}[\phi]=
\int\sum^{\infty}_{n=0}
\frac{1}{n!}\dot{\gamma}^{(n)\,\Lam}
\left(\phi-\alf^{\Lam}\right)^{n}
-
\frac{\dot{\alpha}^{\Lam}}{(n-1)!}\gamma^{(n)\,\Lam}
\left(\phi-\alf^{\Lam}\right)^{n-1}\;,
\label{eq:alpha_dot}
\end{eqnarray}
Comparing powers of $\left(\phi-\alf^{\Lam}\right)$ between Eq.
(\ref{eq:alpha_dot}) and the right-hand-side of Eq.
(\ref{eq:flow_expansion}) yields the following prescription for the
flow equation for the n-point vertex
\begin{eqnarray}
\dot{\gamma}^{(n)\,\Lam}&=&\dot{\mathbf{R}}^{\Lam}\partial_{\mathbf{R}}
\left(\text{all 1-loop 1PI diagrams
generated by $\mathbf{G}^{\Lam}_{R}$ with n external legs}\right)\nonumber\\
&& +
\dot{\alpha}^{\Lam}\gamma^{(n+1)\,\Lam}\;.
\label{eq:recipe}
\end{eqnarray}
The expansion point $\alf^{\Lam}$ will be chosen such that the
bosonic 1-point function $\gam^{(1)\Lam}=0$ for all
$\Lam$:\\[-3mm]
\begin{fmffile}{20080906_1point}
\begin{eqnarray}
\partial_{\Lambda}\gam^{(1)\Lam}=
\hspace{-3mm}
\parbox{15mm}{\unitlength=1mm\fmfframe(2,2)(1,1){
\begin{fmfgraph*}(15,30)\fmfpen{thin} 
\fmfleft{i1}\fmfright{o1}
\fmftop{t1}
\fmfpolyn{full,tension=0.2}{T}{3}
\fmfpolyn{phantom,tension=100.}{B}{3}
\fmf{phantom,straight,tension=1.2}{i1,B1}
\fmf{phantom,straight,tension=1.2}{o1,B2}
\fmf{dashes,tension=0.6}{B3,T3}
\fmf{dashes,tension=0.2,right=0.8}{T1,t1}\fmf{dashes,tension=0.2,left=0.8}{T2,t1}
\end{fmfgraph*}
}}
+...+
\dot{\alpha}^{\Lam}\gamma^{(2)\,\Lam}:=0\;,
\;\;\;
\Rightarrow
\;\;\;
\partial_{\Lambda}\alpha^{\Lam}=\frac{-1}{\gamma^{(2)\,\Lam}}
\Big(
\hspace{-3mm}
\parbox{15mm}{\unitlength=1mm\fmfframe(2,2)(1,1){
\begin{fmfgraph*}(15,30)\fmfpen{thin} 
\fmfleft{i1}\fmfright{o1}
\fmftop{t1}
\fmfpolyn{full,tension=0.2}{T}{3}
\fmfpolyn{phantom,tension=100.}{B}{3}
\fmf{phantom,straight,tension=1.2}{i1,B1}
\fmf{phantom,straight,tension=1.2}{o1,B2}
\fmf{dashes,tension=0.6}{B3,T3}
\fmf{dashes,tension=0.2,right=0.8}{T1,t1}\fmf{dashes,tension=0.2,left=0.8}{T2,t1}
\end{fmfgraph*}
}}
+...
\Big)\;.\nonumber\\[-15mm]
\label{eq:alpha_dot_graph}
\end{eqnarray}
\end{fmffile}
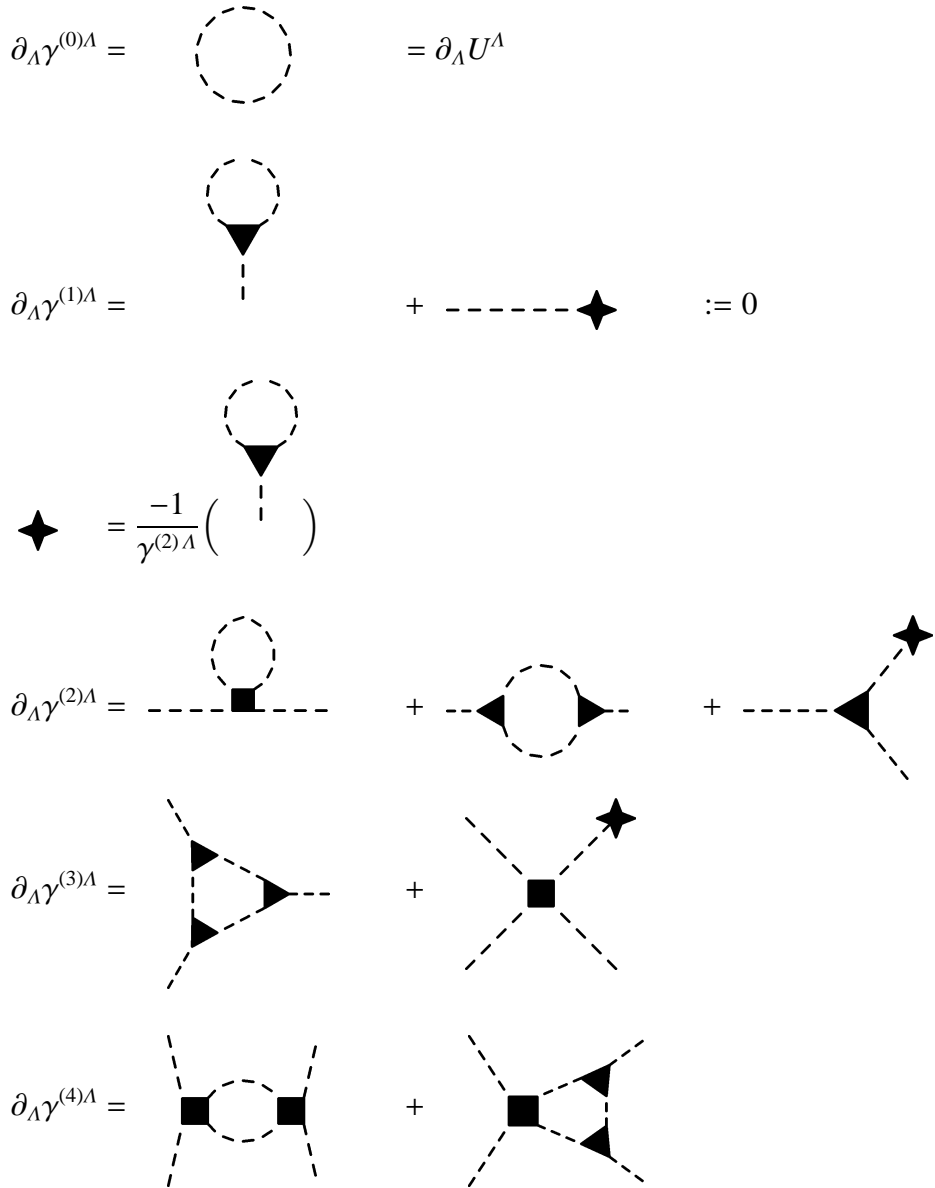
\begin{figure}
\vspace{-15mm}
\begin{fmffile}{20080907_hierarch_61}
\begin{eqnarray}
\partial_{\Lam}\gamma^{(0)\Lam} &=&
\parbox{35mm}{\unitlength=1mm\fmfframe(2,2)(1,1){
\begin{fmfgraph*}(25,25)\fmfpen{thin} 
 \fmfi{dashes}{fullcircle scaled .5w shifted (.5w,.5h)}
\end{fmfgraph*}
}}=\partial_{\Lam} U^{\Lam}
\nonumber\\[-3mm]
\partial_{\Lambda}\gam^{(1)\Lam}&=&
\parbox{35mm}{\unitlength=1mm\fmfframe(2,2)(1,1){
\begin{fmfgraph*}(25,40)\fmfpen{thin}
\fmfleft{i1}\fmfright{o1}
\fmftop{t1}
\fmfpolyn{full,tension=0.2}{T}{3}
\fmfpolyn{phantom,tension=100.}{B}{3}
\fmf{phantom,straight,tension=1.2}{i1,B1}
\fmf{phantom,straight,tension=1.2}{o1,B2}
\fmf{dashes,tension=0.6}{B3,T3}
\fmf{dashes,tension=0.2,right=0.8}{T1,t1}
\fmf{dashes,tension=0.2,left=0.8}{T2,t1}
\end{fmfgraph*}
}}
+
\parbox{35mm}{\unitlength=1mm\fmfframe(2,2)(1,1){
\begin{fmfgraph*}(20,25)\fmfpen{thin}
\fmfleft{l1}
\fmfright{r1}
\fmf{dashes}{l1,r1}
\fmfv{decor.shape=tetragram,decor.filled=1}{r1}
\end{fmfgraph*}
}}
:=0\nonumber\\[-15mm]
\parbox{10mm}{\unitlength=1mm\fmfframe(2,2)(1,1){
\begin{fmfgraph*}(5,10)
\fmfpen{thin}
\fmfleft{v}
\fmfv{decor.shape=tetragram,decor.filled=1}{v}
\end{fmfgraph*}
}}
&=&\frac{-1}{\gamma^{(2)\,\Lam}} \Big( \hspace{-9mm}
\parbox{35mm}{\unitlength=1mm\fmfframe(2,2)(1,1){
\begin{fmfgraph*}(25,40)\fmfpen{thin} 
\fmfleft{i1}\fmfright{o1}
\fmftop{t1}
\fmfpolyn{full,tension=0.2}{T}{3}
\fmfpolyn{phantom,tension=100.}{B}{3}
\fmf{phantom,straight,tension=1.2}{i1,B1}
\fmf{phantom,straight,tension=1.2}{o1,B2}
\fmf{dashes,tension=0.6}{B3,T3}
\fmf{dashes,tension=0.2,right=0.8}{T1,t1}
\fmf{dashes,tension=0.2,left=0.8}{T2,t1}
\end{fmfgraph*}
}}
\hspace{-15mm}
\Big)\;\nonumber\\[-13mm]\nonumber
\partial_{\Lam}\gamma^{(2)\Lam} &=&
\parbox{35mm}{\unitlength=1mm\fmfframe(2,2)(1,1){
\begin{fmfgraph*}(25,25)\fmfpen{thin}
 \fmfleft{l1}
 \fmfright{r1}
 \fmftop{v1}
 \fmfpolyn{full,tension=1}{G}{4}
 \fmf{dashes,straight}{l1,G4}
 \fmf{dashes,straight}{G1,r1}
 \fmffreeze
\fmf{dashes,tension=0.1,right=0.7}{G2,v1}
\fmf{dashes,tension=0.1,right=0.7}{v1,G3}
\end{fmfgraph*}
}}
+
\parbox{35mm}{\unitlength=1mm\fmfframe(2,2)(1,1){
\begin{fmfgraph*}(25,25)\fmfpen{thin}
 \fmfleft{l1}
 \fmfright{r1}
 \fmfpolyn{full,tension=0.3}{G}{3}
 \fmfpolyn{full,tension=0.3}{K}{3}
  \fmf{dashes}{l1,G1}
 \fmf{dashes,tension=0.2,right=0.8}{G2,K3}
 \fmf{dashes,tension=0.2,right=0.8}{K2,G3}
 \fmf{dashes}{K1,r1}
 \end{fmfgraph*}
}}
+
\parbox{35mm}{\unitlength=1mm\fmfframe(2,2)(1,1){
\begin{fmfgraph*}(25,20)\fmfpen{thin}
\fmfleft{l1}
\fmfrightn{r}{2}
\fmf{dashes}{l1,G1}
\fmfpolyn{full,tension=1.2}{G}{3}
\fmf{dashes}{G2,r1}
\fmf{dashes}{G3,r2}
\fmfv{decor.shape=tetragram,decor.filled=1}{r2}
\end{fmfgraph*}
}}
\nonumber\\[-5mm]
\partial_{\Lam}\gamma^{(3)\Lam} &=&
\parbox{35mm}{\unitlength=1mm\fmfframe(2,2)(1,1){
\begin{fmfgraph*}(25,25)
\fmfpen{thin} 
\fmfleftn{l}{2}\fmfright{r}
\fmfrpolyn{full,tension=0.65}{T}{3}
\fmfrpolyn{full,tension=0.65}{B}{3}
\fmfpolyn{full, tension=0.65}{R}{3}
\fmf{dashes}{l1,T2}\fmf{dashes}{l2,B2}
\fmf{dashes,tension=.6}{T3,B1}
\fmf{dashes,left=0.,tension=.5}{R3,T1}
\fmf{dashes,right=0.,tension=.5}{R2,B3}
\fmf{dashes}{R1,r}
\end{fmfgraph*}
}}
+
\parbox{35mm}{\unitlength=1mm\fmfframe(2,2)(1,1){
\begin{fmfgraph*}(25,20)
\fmfcmd{%
vardef cross_bar (expr p, len, ang) =
(( -len/2,0)--(len/2,0))
rotated (ang + angle direction length(p)/2 of p)
shifted point length(p)/2 of p
enddef;
style_def crossed expr p =
cdraw p;
ccutdraw cross_bar (p,5mm, 45);
ccutdraw cross_bar (p,5mm,-45)
enddef;}
\fmfpen{thin}
\fmfleftn{l}{2}
\fmfrightn{r}{2}
\fmf{dashes}{l2,G1}
\fmf{dashes}{l1,G2}
\fmfpolyn{full,tension=2.5}{G}{4}
\fmf{dashes}{r1,G3}
\fmf{dashes}{r2,G4}
\fmfv{decor.shape=tetragram,decor.filled=1}{r2}
 \end{fmfgraph*}
}}\nonumber\\[2mm]
\partial_{\Lam}\gamma^{(4)\Lam} &=&
\parbox{35mm}{\unitlength=1mm\fmfframe(2,2)(1,1){
\begin{fmfgraph*}(25,20)
\fmfpen{thin}
\fmfleftn{l}{2}\fmfrightn{r}{2}
\fmfrpolyn{full,tension=0.9}{G}{4}
\fmfpolyn{full,tension=0.9}{K}{4}
\fmf{dashes}{l1,G1}\fmf{dashes}{l2,G2}
\fmf{dashes}{K1,r1}\fmf{dashes}{K2,r2}
\fmf{dashes,left=.5,tension=.2}{G3,K3}
\fmf{dashes,right=.5,tension=.2}{G4,K4}
\end{fmfgraph*}
}}
+
\parbox{35mm}{\unitlength=1mm\fmfframe(2,2)(1,1){
\begin{fmfgraph*}(30,20)
\fmfpen{thin}
\fmfleftn{l}{2}\fmfrightn{r}{2}
\fmfrpolyn{full,tension=1.4}{K}{4}
\fmfpolyn{full,tension=0.6}{L}{3}
\fmfpolyn{full,tension=0.6}{M}{3}
\fmf{dashes}{l1,K1}\fmf{dashes}{l2,K2}
\fmf{dashes}{L1,r1}\fmf{dashes}{M2,r2}
\fmf{dashes,left=0.,tension=.9}{K4,L3}
\fmf{dashes,left=0.,tension=.4}{L2,M1}
\fmf{dashes,right=0.,tension=.9}{M3,K3}
\end{fmfgraph*}
}}
\nonumber\\[-5mm]\nonumber
\end{eqnarray}
\end{fmffile}
\caption{Truncated hierarchy of 1PI vertex functions up to quartic order in the fields
for phases with spontaneously broken symmetry. The cross denotes the flow
of the order parameter $\partial_{\Lam}\alf^{\Lam}$. Usually, the presence of vertices
with an odd number of legs is peculiar to phases with broken symmetry. It is important
to note that for the general case of mixed Fermi-Bose theories, many more diagrams with
normal and anomalous propagators as well as normal and anomalous vertices appear, see Chapter \ref{chap:fermionsuperfluids}.}
\label{fig:ssb_hierarchy}
\end{figure}
In this way tadpole contributions are absorbed into the flow of
$\alf$. The vanishing of the bosonic one-point function is
equivalent to the condition that the minimum of the effective
potential coincides with its first field derivative for all $\Lam$:
$U'[\phi]|_{\phi=\alf}=0$ (Berges 2002).

In a general polynomial truncation of Eq.(\ref{eq:ssb_expansion}) up
to quartic order in the bosonic fields, the hierarchy for the
n-point vertices can be represented in terms of Feynman diagrams,
see Fig. (\ref{fig:ssb_hierarchy}).

One aspect of the practicability
of the 1PI-formulation of the functional RG lies in its intimate kinship
to conventional perturbation theory (Negele 1987).
Eq. (\ref{eq:recipe}) contains
precisely the diagrammatic corrections that would show up in a 1PI-expansion of perturbation theory.
Technically, a computation of the flow equations therefore involves the determination of prefactors and signs of the various Feynman diagrams. One way to achieve this is to compare coefficients in
Eq. (\ref{eq:flow_expansion}). However, in symmetry-broken phases, the Nambu structure of the fields
leads to at least $4\times4$-matrices for $\bG_R^{\Lam}$ and $\tilde\bGam^{(2) \,\Lam}$,
thus complicating the trace operation on matrix products
with several of these matrices, and necessitating systematic
use of software packages, for example Mathematica.

In this thesis, we determine the
prefactors and signs by use of Wick's theorem and Feynman rules directly 
(see chapter 2.3, Negele 1987; chapter 4.4, Peskin 1995; chapter 6, Weinberg 2005). 

\chapter[Summary Part I]{Summary Part I}
\label{chap:summary_part1}

\vspace*{-5mm}

In chapter \ref{chap:theo_concepts}, some scenarios going beyond the
conventional Fermi liquid description for interacting Fermi systems
were discussed. After recapitulating the Fermi liquid
fixed point in the functional integral formalism in section
\ref{sec:fermi_lickit}, prominent competing and conspiring ordering
tendencies of two-dimensional lattice fermions were introduced. We
have argued that the fermionic functional RG is a useful tool to
treat these cases but for computations within symmetry-broken phases
a description in terms of the collective bosonic degrees of freedom
seems more practicable. In section \ref{sec:ssb}, the mechanism of
spontaneous symmetry-breaking is introduced in detail. We zoomed in
on the phenomenology of Goldstone bosons which are emitted in
interacting Fermi systems with broken continuous symmetry for
example with magnetic or superfluid order. Critical points that
separate phases with different symmetry properties were brought into
play in section \ref{sec:quantum_crit}. The standard phase diagram
in the vicinity of quantum critical point was exhibited in Fig.
\ref{fig:intro_qcp} and the conventional Hertz-Millis approach to
theoretically describe quantum critical behavior was outlined
briefly. Essential renormalization group ideas and 
references are summarized in section \ref{sec:rg_basics}.

\bigskip

In chapter \ref{chap:functional}, the functional RG framework was
presented as a promising tool to transform the before-mentioned
theoretical concepts into real computations for systems containing
both fermionic and bosonic degrees of freedom. We defined the effective action,
that is, the functional which generates the one-particle irreducible vertex
functions for superfields in section \ref{sec:rgschemes}. The exact flow equation
for the effective action is derived in section \ref{subsec:exact_flow}.
The central result of this chapter is contained in subsection \ref{subsec:ssb_rg}: the
prescription how to include scale-dependent expectation values into
the hierarchy of vertex functions to account for spontaneous
symmetry-breaking is pictorially shown in Fig. \ref{fig:ssb_hierarchy}.

\bigskip

In the following Part \ref{part:apple}, we merge the
theoretical concepts with the functional RG framework and apply it in
four cases. The chapters are sequenced in order of increasing
truncation complexity, consecutively build on and extend the previous one, but may
yet be read independently.

\cleardoublepage \thispagestyle{empty}
%

\part{Applications}
\label{part:apple}

\chapter[Hertz-Millis theory with discrete symmetry-breaking]
{Hertz-Millis theory with discrete symmetry-breaking}
\label{chap:bosonicqcp_discrete}

\section{Introduction}

Quantum phase transitions in itinerant electron systems continue to
ignite considerable interest (Sachdev 1999, Belitz 2005, Loehneysen
2007). In many
physical situations, a line of finite temperature second order phase
transitions in the phase diagram terminates at a quantum critical
point at $T=0$. In such cases quantum fluctuations influence the
system also at finite temperatures altering physical quantities such
as the shape of the phase boundary. Consequently, in a complete
description of the system at finite $T$, quantum and thermal
fluctuations have to be accounted for simultaneously.

\bigskip

The conventional renormalization group (RG) approach to quantum
criticality in itinerant electron systems (Hertz 1976, Millis 1993), see 
also section \ref{sec:quantum_crit}, relies on the assumption that it is sensible to integrate out
fermionic degrees of freedom from the functional integral
representation of the partition function and then to expand the
resulting effective action in powers of the order parameter alone.
This approach has been questioned for magnetic phase transitions
associated with spontaneous breaking of continuous spin rotation
invariance, since integrating the fermions leads to singular
interactions of the order parameter field (Belitz 2005, Loehneysen
2007).

\bigskip

In this chapter, we focus on quantum phase transitions to phases
with broken {\em discrete} symmetry. We analyze quantum and
classical fluctuations in the {\em symmetry-broken} phase with an
Ising-like order parameter near a quantum critical point. Our
calculations are based on a set of coupled flow equations obtained
by approximating the exact flow equations of the one-particle
irreducible version of the functional RG. Quantum and classical
(thermal) fluctuations are treated on equal footing. The functional
RG has been applied extensively to classical critical phenomena
(Berges 2002), where it provides a unified description of
$O(N)$-symmetric scalar models, including two-dimensional systems.
The classical Ising universality class has been analyzed in Refs.
9-11. In our approach, we
can compute the RG flow in any region of the phase diagram,
including the region governed by non-Gaussian critical fluctuations.
This allows comparison between the true transition line and the
Ginzburg line, which so far has been used as an estimate of the
former (Loehneysen 2007, Millis 1993). Specifically, we capture the
strong-coupling behavior emergent in the vicinity of the transition
line as well as the correct classical fixed point for $T_c$,
including the anomalous dimension of the order parameter field.

\bigskip

Our results may be applied to Pomeranchuk transitions (Metzner 2003, Dell'Anna 2006, W\"olfle 2007)
which spontaneously break the discrete point group lattice symmetry,
and to the Quantum Ising model (Sachdev 1999). An example of a fermionic model displaying a genuine quantum critical point already on
mean-field level is the so-called f-model, where forward-scattering processes
tend to deform the Fermi surface leading to Pomeranchuk transitions (Yamase 2005, Dell'Anna 2006).
The control parameter here is the density/chemical potential.

\bigskip

In Section \ref{sec:hertz} we introduce
Hertz's (1976) action and the effective action 
for the Quantum Ising model, adapted to the
symmetry-broken phase, which serves as a starting point for the
subsequent analysis. In Section \ref{sec:method} we describe the functional RG
method and its application in the present context, and subsequently
derive the RG flow equations. In Section \ref{sec:zero_temp} we present a solution of
the theory in the case $T=0$. Section \ref{sec:finite} contains numerical results
for the finite $T$ phase diagram in the region with broken symmetry.
Different cases are discussed, distinguished by the dimensionality
$d$ and the dynamical exponent $z$. For Pomeranchuk transitions $z=3$ and for 
the Quantum Ising model $z=1$. In particular, we compare the
$T_c$ line with the Ginzburg line, thus providing an estimate of the
critical region size. In Section \ref{sec:bosonic_qcp_conclusion} we summarize and discuss the
results.

\section{Bosonic action}
\label{sec:hertz}

The starting point of the standard RG approach to quantum critical
phenomena in itinerant electron systems is the Hertz action (Hertz
1976, Millis 1993). It can be derived from a microscopic Hamiltonian
by applying a Hubbard-Stratonovich transformation to the
path-integral representation of the partition function and
subsequently integrating out the fermionic degrees of freedom. The
resulting action is then expanded in powers of the order parameter
field, usually to quartic order.

The validity of this expansion is
dubious in several physically interesting cases, in particular for
magnetic transitions with SU(2)-symmetry, since the integration over
gapless fermionic modes can lead to singular effective interactions
of the order parameter field, which may invalidate the conventional
power counting (Belitz 2005, Loehneysen 2006, Abanov 2004). Such
complications do probably not affect transitions in the charge
channel and magnetic transitions with Ising symmetry. There are
several indications that singularities cancel in that case, namely
the cancellation of singularities in effective interactions upon
symmetrization of fermion loops (Neumayr 1998, Kopper 2001), and the
cancellation of non-analyticities in susceptibilities (Rech 2006).

We therefore rely on the usual expansion of the
action to quartic order in the order parameter field and also on the
conventional parametrization of momentum and frequency dependences,
which leads to the Hertz action (Hertz 1976, Millis 1993)
\begin{eqnarray}
 S_{\text{H}}[\phi] =
 \frac{T}{2} \sum_{\omega_n} \int\frac{d^dp}{(2\pi)^d} \,
 \phi_p \left( \frac{|\omega_{n}|}{|\mathbf{p}|^{z-2}}
 + \mathbf{p}^{2} \right) \phi_{-p} + U[\phi] \; .
 \label{eq:lagrangian}
\end{eqnarray}
Here $\phi$ is the scalar order parameter field and $\phi_p$ with $p
= (\mathbf{p},\omega_n)$ its momentum representation; $\omega_n =
2\pi n T$ with integer $n$ denotes the (bosonic) Matsubara
frequencies. Momentum and energy units are chosen such that the
prefactors in front of $\frac{|\omega_{n}|}{|\mathbf{p}|^{z-2}}$ and
$\mathbf{p}^{2}$ are equal to unity. The Hertz action is regularized in the ultraviolet by
restricting momenta to $|\mathbf{p}| \leq \Lambda_0$.
The value of the dynamical
exponent is restricted to $z \geq 2$. The case $z=3$ for Pomeranchuk transitions 
is of our main interest. Formally, the results obtained
with $S_{\text{H}}[\phi]$ as our starting point may be applied to systems with arbitrary $z\geq 2$.
\footnote{\textit{
A comprehensive discussion of the origin of the $\omega_n$ dependence
of the action is given by Millis (1993).
Eq.~(\ref{eq:lagrangian}) is valid in the limit
$\frac{|\omega_{n}|}{|\mathbf{p}|^{z-2} } \ll 1$ which is relevant
here since the dominant fluctuations occur only in this regime}.
}

Other interesting cases include spin systems such as the Quantum Ising model
for which $z=1$ and the associated continuum action reads (Sachdev 1999)
\begin{eqnarray}
 S_{\text{QI}}[\phi] =
 \frac{T}{2} \sum_{\omega_n} \int\frac{d^dp}{(2\pi)^d} \,
 \phi_p \left( \omega^{2}_{n} + \mathbf{p}^{2} \right) \phi_{-p} + U[\phi] \; ,
 \label{eq:quantum_ising}
\end{eqnarray}
where the frequency term enters with the same power as the momentum term.

In the symmetric phase the potential $U[\phi]$ is minimal at
$\phi=0$, and is usually parametrized by a positive quadratic and a
positive quartic term (Hertz 1976, Millis 1993). Since we approach
the quantum critical point from the symmetry-broken region of the
phase diagram, we assume a potential $U[\phi]$ with a minimum at a
non-zero order parameter $\phi_0$:
\begin{eqnarray}
 U[\phi] &=& \frac{u}{4!} \int_0^{1/T} d\tau \int d^d x
 \left( \phi^{2} - \phi_{0}^{2} \right)^{2} \nonumber\\
 &=& \int_0^{1/T} \!\!\! d\tau \int \! d^d x
 \left[u\,\frac{\phi'^{4}}{4!}+\sqrt{3 \, u\, \delta} \,
 \frac{\phi'^{3}}{3!} + \delta \, \frac{\phi'^{2}}{2!}\right] \, ,
\label{eq:ef_potential}
\end{eqnarray}
where $\phi$ and $\phi'$ are functions of $x$ and $\tau$
with $\phi= \phi_{0}+\phi'$.
The parameter
\begin{eqnarray}
\delta = \frac{u\,\phi_{0}^{2}}{3}=\frac{2u\,\rho_{0}}{3}\;,
\end{eqnarray}
controls the distance from criticality. Approaching the phase
boundary in the $(\delta, T)$-plane from the symmetry-broken phase
gives rise to the three-point vertex $\sqrt{3\, u\, \delta}$, which
generates an anomalous dimension of the order parameter field
already at one-loop level.

\bigskip

Although formally correct as a result of integrating out the
fermions, the Hertz action (\ref{eq:lagrangian}) is not a good starting point
for symmetry-broken phases with a fermionic gap,
such as charge density wave phases or antiferromagnets.
A fermionic gap leads to a suppression of the dynamical
term (linear in frequency) in the action, since it suppresses
low energy particle-hole excitations.
If not treated by a suitable resummation in the beginning,
this effect is hidden in high orders of perturbation theory
(Rosch 2001).
We do not deal with this complication in the present chapter as our 
main interest lies in systems with Pomeranchuk deformed Fermi surfaces where the particle-hole continuum remains gapless in the phase 
with broken symmetry. However, our results for the transition temperature should not
be affected by a gap in the symmetry-broken phase, since it
vanishes continuously at $T_c$.

\section{Method}
\label{sec:method}

To analyze the quantum field theory defined by $S[\phi]$ we compute
the flow of the effective action $\Gamma^{\Lambda}[\phi]$ with 
approximate flow equations derived from an exact functional
RG flow equation (Wetterich 1993, Berges 2002, Delamotte 2004, Gies
2006). The exact flow equation and the vertex expansion including 
symmetry-breaking was derived and discussed in detail in chapter 
\ref{chap:functional}. The effective action $\Gamma^{\Lambda}[\phi]$ is the
generating functional for one-particle irreducible vertex functions
in presence of an infrared cutoff $\Lambda$. The latter is
implemented by adding a regulator term of the form $\int \frac{1}{2}
\phi R^{\Lambda} \phi$ to the bare action. The effective action
interpolates smoothly between the bare action $S[\phi]$ for large
$\Lambda$ and the full effective action $\Gamma[\phi]$ in the limit
$\Lambda \to 0$ (cutoff removed). Its flow is given by the exact
functional equation (Wetterich 1993)
\begin{eqnarray}
\frac{d}{d \Lambda}\Gamma^{\Lambda}\left[\phi\right]=
\frac{1}{2}\text{Tr}\frac{\dot{R}^{\Lambda}}{\Gamma^{(2)}\left[\phi\right]
+ R^{\Lambda}}\,\,,
\label{eq:flow_eqn}
\end{eqnarray}
where $\dot{R}^{\Lambda}=\partial_{\Lambda}R^{\Lambda}$, and
$\Gamma^{(2)}\left[\phi\right] = \delta^{2}\Gamma^{\Lambda}[\phi]/
 \delta \phi^{2}$.
In momentum representation ($\phi_p$),
the trace sums over momenta and frequencies:
$\text{Tr} = T \sum_{\omega_{n}}
 \int \frac{d^{d} p}{\left(2\pi\right)^{d}}$.
For the regulator function $R^{\Lambda}(\mathbf{p})$ we choose the
optimized Litim cutoff (Litim 2001)
\begin{equation}
 R^{\Lambda}(\mathbf{p}) =
 Z_{\mathbf{p}} \left( \Lambda^{2}-\mathbf{p}^{2}\right)
 \theta\left(\Lambda^{2}-\mathbf{p}^{2} \right) \; ,
\label{Litim_fun}
\end{equation}
where $Z_{\mathbf{p}}$ is a renormalization factor (see below).
This regulator function $R^{\Lambda}(\textbf{p})$ replaces
$Z_{\mathbf{p}}\mathbf{p}^2$ with $Z_{\mathbf{p}} \Lambda^2$ for
$|\mathbf{p}| < \Lambda$. Following (Berges 2002) we neglect
$\dot{Z}_{\mathbf{p}}$ in $\dot{R}^{\Lambda}(\mathbf{p})$, such that
$\dot{R}^{\Lambda}(\mathbf{p}) =
 2 Z_{\mathbf{p}} \Lambda \Theta(\Lambda^2 - \mathbf{p}^2) \,$. 
The scale-derivative of the Z-factor leads to additional terms on the right-hand-side 
of the flow equations proportional to the anomalous 
dimension $\eta$ defined in below in Eq. (\ref{eq:eta}). 
These terms are neglected here as they are of higher order 
in the vertices. 

\subsection{Truncation}
\label{subsec:discrete_trunc}

The RG flow of the local potential of the form Eq. (\ref{eq:ef_potential}) will be
followed by cutoff-dependent parameters $u$ and $\phi_0$ (or, alternatively, $\delta$).

The quadratic part of the effective action for the Hertz action
contains two renormalization factors, one for the frequency and one
for the momentum dependence:
\begin{equation}
 \Gamma_{\phi\phi,\text{H}}=
 \int_{p}\frac{1}{2}\phi_{-p}
 \left(Z_{\omega} \frac{|\omega_{n}|}{|\mathbf{p}|^{z-2}} +
 Z_{\mathbf{p}}\mathbf{p}^2 + \delta\right)\phi_{p}\; ,
\label{eq:prop_hertz}
\end{equation}
where  $\int_{p}$ comprises frequency integration (Matsubara
summation) and momentum integration at zero (finite) temperature.
As
it will turn out below, the frequency renormalization for the Hertz
action does not play a role, in contrast to the Quantum Ising
model, where we have
\begin{eqnarray}
 \Gamma_{\phi\phi,\text{QI}}=
 \int_{p}\frac{1}{2}\phi_{-p}\left(
 Z_{\omega} \omega^{2}_{n}+ Z_{\mathbf{p}}\mathbf{p}^2 + \delta\right)
 \phi_{p}\; .
\label{eq:prop_QI}
\end{eqnarray}
The inverse propagator exhibits relativistic invariance and
$Z_{\omega}$ is equally important to $Z_{\mathbf{p}}$ at zero
temperature.

The Green function for the Hertz action endowed with the regulator
reads accordingly,
\begin{eqnarray}
G_{R}\left(p\right)=-\langle \phi_{-p}\phi_{p} \rangle=\frac{-1}
{Z_{\omega} \frac{|\omega_{n}|}{|\mathbf{p}|^{z-2}} +
 Z_{\mathbf{p}}\mathbf{p}^2 + \delta + R^{\Lam}\left(\mathbf{p}\right)}\;,
 \label{eq:green}
\end{eqnarray}
and similarly for the Quantum Ising model with the frequency term
replacement
$Z_{\omega}\frac{|\omega_n|}{|\textbf{p}|^{z-2}}\rightarrow Z_{\omega}
\omega^{2}_n$. In all expressions with the frequency term of
the Hertz action, Eq. (\ref{eq:prop_hertz}), the corresponding terms
for the Quantum Ising model are obtained by the just described
replacement of frequency terms.

\bigskip

We emphasize that we have used a relatively simple parametrization
of the effective action $\Gamma^{\Lambda}[\phi]$. In particular, we
kept only the dominant terms in the derivative expansion (Berges
2002) and neglected the field dependence of the $Z$-factors.
Furthermore, the simple parametrization of the effective potential
Eq.~(\ref{eq:ef_potential}) allowed us to substitute the partial
differential equation (\ref{ef_pot_flow}) governing the flow of
$U[\phi]$ by the two ordinary differential equations
(\ref{eq:delta_delta}) and (\ref{eq:delta_u}). The latter
approximation is equivalent to neglecting all higher order vertices
generated during the flow.  More sophisticated truncations have been
applied, among others, in the context of the classical Ising
universality class, where they lead to improved results for the
critical exponents (Ballhausen 2004).

\subsection{Flow equations}
\label{subsec:discrete_flow_eqns}

We derive our flow equations keeping the notation general so as to
account for zero and finite temperature. Evaluating
Eq.~(\ref{eq:flow_eqn}) for a momentum-independent field $\phi$
yields the flow of the effective potential $U[\phi]$
\begin{equation}
 \partial_\Lambda U[\phi] =
 \frac{1}{2} \, \text{Tr} \, \frac{\dot{R}^{\Lambda}(\mathbf{p})}
 {Z_{\omega}\frac{|\omega_n|}{|\textbf{p}|^{z-2}} +
 Z_{\textbf{p}}\textbf{p}^2 + R^\Lambda(\mathbf{p}) + U''[\phi]}\; ,
\label{ef_pot_flow}
\end{equation}
from which we derive the flows of the parameters $\phi_0$ and $u$,
following the procedure in (Berges 2002).
Viewing $U$ as a function of $\rho = \frac{1}{2} \phi^2$ and using
$U'[\rho_0] = 0$, we can write
\begin{eqnarray}
0 = \frac{d}{d\Lambda} U'[\rho_0] =
 \partial_\Lambda U'[\rho_0] +
 U''[\rho_0] \, \partial_\Lambda\rho_0\;.
\end{eqnarray}
Inserting $\partial_\Lambda U'[\rho_0]$ as obtained by differentiating
Eq.~(\ref{ef_pot_flow}) with respect to $\rho$ at $\rho = \rho_0$,
and using $U''[\rho_0] = \frac{1}{3} u$,
one obtains the flow equation for $\rho_0$
\begin{equation}
 \partial_\Lambda \rho_0 =
 \frac{3}{2} \, \text{Tr} \, \frac{\dot{R}^{\Lambda}(\mathbf{p})}
 {\left[Z_{\omega}\frac{|\omega_n|}{|\textbf{p}|^{z-2}} +
 Z_{\textbf{p}}\textbf{p}^2 + R^\Lambda(\mathbf{p}) +
 \frac{2}{3} u \rho_0 \right]^2} \; .
\label{eq:rho_flow}
\end{equation}
The flow of $u$ is obtained by differentiating Eq.~(\ref{ef_pot_flow})
twice with respect to $\rho$:
\begin{equation}
 \partial_\Lambda u =
 3 u^2 \, \text{Tr} \, \frac{\dot{R}^{\Lambda}(\mathbf{p})}
 {\left[Z_{\omega}\frac{|\omega_n|}{|\textbf{p}|^{z-2}} +
 Z_{\textbf{p}}\textbf{p}^2 + R^\Lambda(\mathbf{p}) +
 \frac{2}{3} u \rho_0 \right]^3} \; .
\label{eq:u_flow}
\end{equation}
Inserting the above flow equations into the $\Lambda$-derivative
of $\delta = \frac{2}{3} u \rho_0$, we obtain the flow of
$\delta$. To complete the system of flow equations one still needs to derive
the evolution of $Z_{\textbf{p}}$ and $Z_{\omega}$, which
parametrize the momentum and frequency dependence of the propagator.
Taking the second functional derivative of Eq.~(\ref{eq:flow_eqn}),
we obtain the flow equation for the propagator
\begin{equation}
 \partial_\Lambda G_{R}^{-1}(p) =
 3u \delta \, \text{Tr} \left[ \dot{R}^{\Lambda}\,
 G_{R}^2(q) \, G_{R}(q+p) \right],
\label{Gamma2flow}
\end{equation}
where $G_{R}(q)$ is given by Eq.~(\ref{eq:green}). Here we skipped
the contribution from the tadpole diagram, since it involves no
dependence on momentum and frequency, and therefore does not
contribute to the flow of the $Z$-factors. The momentum
renormalization factor is then given by
\begin{equation}
 Z_{\textbf{p}} = \left.
 \frac{1}{2d} \, \partial^{2}_{\textbf{p}}
 \left[G_{R}^{-1}(\textbf{p},\omega_n=0) \right]
 \right|_{\textbf{p} = 0} \; ,
\label{eq:Zp}
\end{equation}
where $\partial^{2}_{\textbf{p}}$ is the Laplace operator evaluated
at constant cutoff function, that is, $\partial^{2}_{\textbf{p}}$ does not
act on $R^{\Lambda}(\mathbf{p})$. We now turn to the renormalization 
of $Z_{\omega}$. 
\begin{wrapfigure}{r}{0.5\textwidth}
  \vspace{2mm}
\begin{fmffile}{20070802_loops_6}
\begin{eqnarray}
\delta &:&
\parbox{35mm}{\unitlength=1mm\fmfframe(2,2)(1,1){
\begin{fmfgraph*}(20,25)\fmfpen{thin}
 \fmfleft{l1}
 \fmfright{r1}
 \fmftop{v1}
 \fmfpolyn{full,tension=0.6}{G}{4}
 \fmf{dbl_wiggly,straight}{l1,G4}
 \fmf{dbl_wiggly,straight}{G1,r1}
 \fmffreeze
\fmf{dbl_wiggly,tension=0.1,right=0.7}{G2,v1}
\fmf{dbl_wiggly,tension=0.1,right=0.7}{v1,G3}
\end{fmfgraph*}
}}\hspace{-10mm}
+
\parbox{35mm}{\unitlength=1mm\fmfframe(2,2)(1,1){
\begin{fmfgraph*}(25,23)\fmfpen{thin}
 \fmfleft{l1}
 \fmfright{r1}
 \fmfpolyn{full,tension=0.3}{G}{3}
 \fmfpolyn{full,tension=0.3}{K}{3}
  \fmf{dbl_wiggly}{l1,G1}
  \fmf{dbl_wiggly,tension=0.2,right=0.8}{G2,K3}
 \fmf{dbl_wiggly,tension=0.2,right=0.8}{K2,G3}
 \fmf{dbl_wiggly}{K1,r1}
 \end{fmfgraph*}
}}\hspace{-5mm}\nonumber\\[-3mm]
u&:&
\parbox{30mm}{\unitlength=1mm\fmfframe(2,2)(1,1){
\begin{fmfgraph*}(25,17)
\fmfpen{thin}
\fmfleftn{l}{2}\fmfrightn{r}{2}
\fmfrpolyn{full,tension=0.9}{G}{4}
\fmfpolyn{full,tension=0.9}{K}{4}
\fmf{dbl_wiggly}{l1,G1}\fmf{dbl_wiggly}{l2,G2}
\fmf{dbl_wiggly}{K1,r1}\fmf{dbl_wiggly}{K2,r2}
\fmf{dbl_wiggly,left=.5,tension=.2}{G3,K3}
\fmf{dbl_wiggly,right=.5,tension=.2}{G4,K4}
\end{fmfgraph*}
}}
\nonumber\\
Z_{\omega},\;Z_{\mathbf{p}}&:&
\parbox{30mm}{\unitlength=1mm\fmfframe(2,2)(1,1){
\begin{fmfgraph*}(25,23)\fmfpen{thin}
 \fmfleft{l1}
 \fmfright{r1}
 \fmfpolyn{full,tension=0.3}{G}{3}
 \fmfpolyn{full,tension=0.3}{K}{3}
  \fmf{dbl_wiggly}{l1,G1}
  \fmf{dbl_wiggly,tension=0.2,right=0.8}{G2,K3}
 \fmf{dbl_wiggly,tension=0.2,right=0.8}{K2,G3}
 \fmf{dbl_wiggly}{K1,r1}
 \end{fmfgraph*}
}}\nonumber\\[-10mm]\nonumber
\end{eqnarray}
\end{fmffile}
\caption{\textit{Feynman diagrams representing the contributions to
the flow equations (\ref{eq:rho_flow}, \ref{eq:u_flow},
\ref{Gamma2flow})}.} \label{fig:fRG_flow} \vspace{0mm}
\end{wrapfigure}
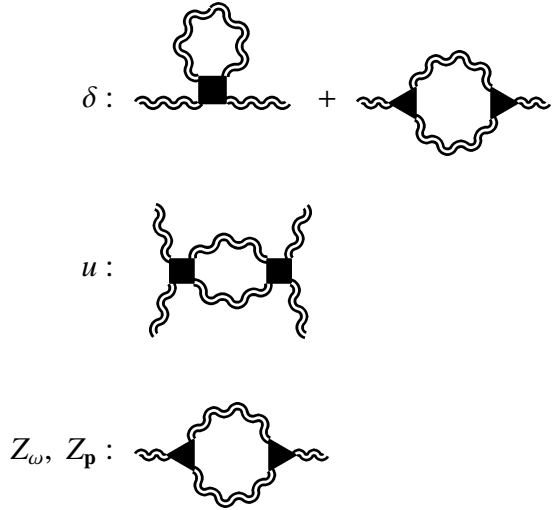
We first
consider the case $z=2$, where at finite temperatures $Z_{\omega}$
can be related to the propagator by
\begin{eqnarray}
 Z_{\omega}&=& \frac{1}{2\pi T} \Big[
 G_{R}^{-1}(\textbf{p}=0,2\pi T)\nonumber\\
&&-  G_{R}^{-1}(\textbf{p}=0,0) \Big]\; .
\end{eqnarray}
But we shall later show that $Z_{\omega}$ renormalizes only very weakly 
at finite temperatures for $z=2$ and therefore even more weakly for 
Pomeranchuk transitions where $z=3$.
At $T=0$, the derivatives become continuous and for the Quantum Ising model 
$z=1$, $Z_{\omega}$ renormalizes as strongly as $Z_{\mathbf{p}}$. In the
following, we will often employ the anomalous momentum scaling
exponent defined as:
\begin{equation}
 \eta = - \frac{d\log Z_{\mathbf{p}}}{d\log\Lambda} \; ,
\label{eq:eta}
\end{equation}
and similarly for $\eta_{\omega}$. 
The contributions to the flow of $\delta$, $u$, $Z_{\omega}$ and
$Z_{\mathbf{p}}$ are illustrated in terms of Feynman diagrams in
Fig.~\ref{fig:fRG_flow}.

\section{Zero-temperature solution at the quantum critical point}
\label{sec:zero_temp}

In this section we present a solution of the flow equations
(\ref{eq:rho_flow}, \ref{eq:u_flow}, \ref{Gamma2flow}) at zero
temperature. For $z\geq 2$, we can linearize the flow equations
around the Gaussian fixed in $d=2$ and we provide an analytic
expression for the value of the control parameter $\delta_0$
corresponding to the quantum critical point.

For $z=1$, the two-dimensional QCP is described by a non-Gaussian
fixed point and we resort to a numerical solution of our equations.

\subsection{$z\geq 2$}
\label{subsec:zero_hertz}

Convenient handling of the flow equations is achieved by introducing
the variables
\begin{eqnarray}
 \tilde v &=&  \frac{A_d}{4\pi Z_{\mathbf{p}} Z_{\omega} \, \Lambda^{4-(d+z)}} \, u
 \nonumber\\
 \tilde \delta & = & \frac{\delta}{Z_{\mathbf{p}}\Lambda^{2}}\;.
\end{eqnarray}
For the Hertz action, the resulting flow equations are
\begin{eqnarray}
 \frac{d\tilde\delta}{d\log\Lambda} &=&
 (\eta-2)\tilde{\delta} +
 \frac{12\tilde{\delta} \tilde v}{(1+\tilde\delta)^2}\frac{1}{d+z-2} +
 \frac{8 \tilde v}{1+\tilde\delta}\frac{1}{d+z-2} \; , \nonumber \\[2mm]
 \frac{d\tilde v}{d\log\Lambda} &=& (d+z-4+2\eta) \tilde v +
 \frac{12 \tilde v^2}{(1+\tilde\delta)^2} \frac{1}{d+z-2} \; ,
 \nonumber\\[2mm]
 \eta &=&
 \frac{6}{d}\frac{\tilde{\delta} \tilde v}{(1+\tilde\delta)^4}
 \left[\frac{4}{3}(1+\tilde\delta) \right.
 - \left. \frac{(d-2)(z-2)}{3(d+z-4)}(1 + \tilde\delta)^2 -
 \frac{4}{d+z}\right] \; .
\label{eq:beta_f_T0}
\end{eqnarray}
The flow equations (\ref{eq:beta_f_T0}) reveal that $\tilde v$ is an
irrelevant variable for $d+z > 4$ as pointed out earlier by Hertz
(Hertz 1976). In this case Eqs.~(\ref{eq:beta_f_T0}) have a stable
Gaussian fixed point in $\tilde v = 0$, $\tilde\delta = 0$, $\eta =
0$. Linearizing the flow equations around the fixed point, one
obtains the solution
\begin{eqnarray}
 \tilde\delta(\Lambda) &=& \left[ \tilde\delta +
 \frac{8 \tilde v}{(d+z-2)^2}
 \left(\left(\Lambda/\Lambda_0\right)^{d+z-2} - 1 \right) \right]
 \left(\Lambda/\Lambda_0\right)^{-2} \; , \nonumber \\[2mm]
 \tilde v(\Lambda) &=&
 \left(\Lambda/\Lambda_0\right)^{d+z-4} \tilde v \; , \nonumber \\[2mm]
 \eta(\Lambda) &=& 0 \; ,
\label{gdl2}
\end{eqnarray}
where $\tilde\delta$ and $\tilde v$ on the right hand sides are the initial
values of the parameters at $\Lambda=\Lambda_0$.
In the marginal case $d+z=4$ one finds logarithmic convergence of
$\tilde v$ and $\tilde{\delta}$ to zero.
Expressing the order parameter $\phi_0$ in terms of $\tilde\delta$
and $\tilde v$, substituting the above solution, and taking the limit
$\Lambda \to 0$ yields
\begin{equation}
 \phi_0 \propto \sqrt{\delta - \delta_0} \; ,
\label{gdl}
\end{equation}
where
\begin{equation}
 \delta_0 = \frac{2A_d}{\pi} \,
 \frac{\Lambda_0^{d+z-2}}{(d+z-2)^2} \, u
\end{equation}
is the quantum critical point's coordinate. From Eq.~(\ref{gdl}) we read
off the value of the exponent $\beta = \frac{1}{2}$, consistent with
mean-field theory. From Eq.~(\ref{gdl2}) one also straightforwardly
evaluates the correlation length $\xi$, using
$\xi^{-2} = \lim_{\Lambda\to 0} \delta(\Lambda) =
\lim_{\Lambda \to 0} Z_{\textbf{p}} \Lambda^2 \tilde{\delta}$,
which yields
\begin{equation}
 \xi = (\delta - \delta_0)^{-1/2} \; ,
\end{equation}
as expected within mean field theory.

We have recovered the well-known fact that quantum phase transitions
have properties similar to their classical counterparts in effective
dimensionality $\mathcal{D} = d+z$ (Sachdev 1999). In the case
studied here $\mathcal{D} \geq 4 $ and one obtains mean-field
behavior governed by a Gaussian fixed point.

\subsection{$z=1$}
\label{subsec:zero_QI}

We now deal with the case where $\mathcal{D}=d+z$ is smaller than
four and thus even at zero temperature the QCP falls into a
non-Gaussian universality class for $1<d<3$. Although this case is
covered by the quantum-classical mapping and the zero-temperature
quantum theory matches that of the classical theory in $d+1$, it is
worthfile to inspect the zero-temperature flow
equations in hindsight to a subsequent comparison with the 
finite-temperature equations in the following chapter.

We put $Z_{\omega}=Z_{\mathbf{q}}$ as there is nothing that breaks
the relativistic invariance except for our regulator function which
cuts off momenta but leaves the frequencies untouched. 
In addition to $\eta$ defined in Eq.
(\ref{eq:eta}), we here employ the rescaled variables
\begin{eqnarray}
\tilde{\rho}&=&\frac{\rho d Z_{\mathbf{p}}}{2 \Lam^{d-1}K_{d}}\nonumber\\
\tilde{u}&=&\frac{2u K_{d}}{d Z_{\mathbf{p}}^{2}\Lam^{3-d}}\;.
\end{eqnarray}
The frequency integrations can be performed analytically yielding
the flow equations
\begin{eqnarray}
 \frac{d\tilde\rho}{d\log\Lambda} &=&
 \left(1-d-\eta\right)\tilde{\rho} +
 \frac{3}{2}
\frac{1}{4\left(1+\frac{2 \tilde{u}\tilde{\rho}}{3}\right)^{3/2}}\nonumber\\
\frac{d\tilde u}{d\log\Lambda} &=&
\left(d-3+2\eta\right)\tilde{u}
+
3\tilde{u}^{2}
\left(
\frac{3}{16\left(1+\frac{2 \tilde{u}\tilde{\rho}}{3}\right)^{5/2}}
\right)\;,
\label{eq:zero_betas_QI}
\end{eqnarray}
and for the anomalous dimension:
\begin{eqnarray}
\eta=2\tilde{u}^{2}\tilde{\rho}
\frac{135\sqrt{3}\left(3+12 d +
8\left(d+2\right)\tilde{u}\tilde{\rho}\right)}
{128\left(d+2\right)\left(3+2\tilde{u}\tilde{\rho}\right)^{9/2}}\;.
\label{eq:zero_eta_QI}
\end{eqnarray}
We solve these equations numerically at the fixed point for $1<d<3$. 
The anomalous dimension comes out as $\eta=0.126$ for $d=2$. This 
can be compared with the best known estimates for the three-dimensional classical 
$O(1)$-model, where $\eta\approx0.04$ from seven-loop perturbation theory 
and $\epsilon$-expansions (Berges 2002). It is a general feature of 
functional RG computations in simple truncations that the anomalous dimension 
is overestimated. Improving the truncation by including a field-dependent 
effective potential and higher-order terms in the derivative expansion leads 
to accurate critical exponents (Berges 2002).

A thorough discussion and flows versus the cutoff-parameter follow 
in subsection \ref{subsec:z_1}.

\section{Finite temperatures}
\label{sec:finite}

Temperature provides for an additional scale in our problem that
significantly enriches the analysis. There are now \emph{two}
relevant parameters, the first being the control parameter and the
second the temperature. In the $(\delta,T)$-plane there is now line
of fixed points ending at the QCP $(\delta_{\text{crit}},0)$.
Notably, for \emph{any}, however small $T$, the critical behavior at
$\Lam\rightarrow 0$ is classical. Yet, we shall see below that the 
shape of the $T_{c}$-line is dominated by quantum
fluctuations.

\bigskip

At finite temperatures, the continuous frequency integrations are
substituted by discrete Matsubara sums. These sums in the flow
equations (\ref{eq:rho_flow}) and (\ref{eq:u_flow}) can be expressed
in terms of polygamma functions $\Psi_n(x)$, defined recursively by
$\Psi_{n+1}(x) = \Psi_n'(x)$ for $n=0,1,2,\dots$, and $\Psi_0(x) =
\Gamma'(x)/\Gamma(x)$, where $\Gamma(x)$ is the gamma function. From
the Weierstrass representation, $\Gamma(x)^{-1} =
 x e^{\gamma x}\prod_{n=1}^{\infty}(1+\frac{x}{n})e^{-x/n}$,
where $\gamma$ is the Euler constant, one can derive the relation
\begin{equation}
 \sum_{n=-\infty}^{\infty} \frac{1}{(|n|+x)^2} =
 \frac{1}{x^2} + 2 \Psi_1(x+1) \; .
\label{freqsum}
\end{equation}
Taking derivatives with respect to $x$ yields expressions for sums
involving higher negative powers of $(|n|+x)$ in terms of polygamma
functions of higher order. The $d$-dimensional momentum integrals on
the right hand side of the flow equations can be reduced to
one-dimensional integrals, since the integrands depend only on the
modulus of $\mathbf{p}$.

Explicit dependences on $\Lambda$, $Z$-factors, and lengthy
numerical prefactors in the flow equations can be eliminated by
using the following rescaled variables:
\begin{eqnarray}
 \tilde p &=& |\mathbf{p}|/\Lambda \; , \\
 \tilde T &=&
 \frac{2\pi Z_{\omega}}{Z_{\mathbf{p}}\Lambda^z} \, T \; , \\
 \tilde\delta &=& \frac{\delta}{Z_{\mathbf{p}}\Lambda^2} \; , \\
 \tilde u &=&
 \frac{A_d T}{2 Z_{\mathbf{p}}^2 \Lambda^{4-d}} \, u \; ,
\end{eqnarray}
with $A_d = (2\pi)^{-d} S_{d-1}$, where $S_{d-1} =
2\pi^{d/2}/\Gamma(d/2)$ is the area of the $(d\!-\!1)$-dimensional
unit sphere.

\subsection{$z=3$}
\label{subsec:z_3}

For the case $z=3$, the flow equations for $\tilde\delta$ and $\tilde u$ are 
obtained as
\begin{eqnarray}
 \frac{d\tilde\delta}{d\log\Lambda} =
 (\eta - 2) \, \tilde\delta
 &+&
 4 \tilde{u} \, \Bigg[
 \frac{1}{d} \frac{1}{(1+\tilde{\delta})^2} +
 \frac{1}{\tilde T^2} \int_0^1 \! d \tilde p \,
 \tilde p^{d+2z-5} \Psi_1(x) \Bigg]  \nonumber \\
 &+& \!
 12 \tilde u \tilde\delta \, \Bigg[
 \frac{1}{d} \frac{1}{(1+\tilde\delta)^3} -
 \frac{1}{\tilde T^3} \int_0^1 \! d\tilde p \,
 \tilde p^{d+3z-7} \Psi_2(x) \Bigg] \; , \hskip 7mm
\label{eq:delta_delta}
\end{eqnarray}
\begin{eqnarray}
 \frac{d\tilde u}{d\log\Lambda}
  =
 (d-4+2\eta) \, \tilde{u}
 &+&
 12 \tilde u^2 \Bigg[
 \frac{1}{d} \frac{1}{(1+\tilde{\delta})^3} -
 \frac{1}{\tilde T^3} \int_0^1 \! d\tilde p \,
 \tilde p^{d+3z-7} \Psi_2(x) \Bigg] \; , \hskip 7mm
\label{eq:delta_u}
\end{eqnarray}
where $x = 1 + \tilde T^{-1} (1 + \tilde\delta) \tilde p^{z-2}$.
Inserting Eqs.~(\ref{Gamma2flow}) and (\ref{eq:Zp}) into
Eq.~(\ref{eq:eta}), and performing the frequency sum, one obtains 
for the anomalous dimension
\begin{eqnarray}
 \eta &=&
 \frac{6}{d}\frac{\tilde{u}\tilde{\delta}}{(1+\tilde{\delta})^5}
 \Big[2(1+\tilde{\delta})-\frac{8}{d+2}\Big] -
 \frac{\tilde{u}\tilde{\delta}}{d} {\tilde T}^{-5}
 \int_0^1 d{\tilde p} {\tilde p}^{d+3z-13}
\nonumber\\
&&
 \Bigg[-6 {\tilde p}^4 (z-2)(d+z-4) \tilde{T}^2
 \Psi_2(x) - 2{\tilde p}^{2+z} \Big(2(8-d)
\nonumber\\
&&
 (1 + \tilde{\delta} - {\tilde p}^2) +
 [(d-14)(1+ \tilde\delta) + 8{\tilde p}^2]z + 3(1+\tilde\delta) z^2 \Big)
\nonumber\\
&&
 \tilde{T} \Psi_3(x) - {\tilde p}^{2z}
 [2({\tilde p}^2 - 1) + \tilde\delta(z-2)+z]^2 \Psi_4(x) \Bigg] \; .
\label{eq:etas}
\end{eqnarray}
Taking the logarithmic derivative with respect to $\Lambda$,
inserting Eq.~(\ref{Gamma2flow}), and performing the trace 
yields for $\eta_{\omega}$ for the case $z=2$
\begin{equation}
 \eta_{\omega} =
 \frac{12}{d} \,\tilde{u} \tilde\delta \tilde{T}^{-1}
 \left[\frac{\tilde{T} - (1+\tilde\delta)}{\tilde{T} (1+\tilde\delta)^3} +
 \tilde{T}^{-3} \Psi_2 \left(1 + \frac{1+\tilde\delta}{\tilde{T}} \right)
 \right] \; .
\end{equation}
From the numerical solution of the flow equation we observe that
$\eta_\omega$ is small at all scales and vanishes for $\Lambda \to
0$. For example, in two dimensions $\eta_\omega$ varies between
$-0.033$ and $0.005$ for $u=1$ at $T=e^{-4}$ and has practically no
influence on the phase diagram. In three dimensions the values are
at least one order of magnitude lower. For $z>2$ one expects an even
smaller $\eta_{\omega}$, since a larger $z$ reduces the strength of
fluctuations near the quantum critical point. Therefore we set
$\eta_\omega = 0$ and $Z_{\omega}=1$ from now on. The scaling
variable $\tilde{T}$ then obeys the flow equation
\begin{equation}
 \frac{d \tilde T}{d\log\Lambda} =
 (\eta - z) \tilde T \; .
\end{equation}
In the flow equations
(\ref{eq:delta_delta},\ref{eq:delta_u},\ref{eq:etas}) one identifies
the classical mean-field (involving only one power of
$\tilde{\delta}$ or $\tilde{u}$), classical non-Gaussian and quantum
(involving $\tilde T$) terms. The quantum contributions vanish as
the infrared cutoff tends to zero at constant non-zero temperature
($\tilde{T}^{-1} \ll 1$). On the other hand the quantum terms
dominate the high energy part of the flow, where $\tilde{T}^{-1} \gg
1$. Our  framework allows for a continuous connection of these two
regimes of the flow. The scale $\Lambda_{cl}$ below which quantum
fluctuations become irrelevant depends on temperature. It vanishes
at $T=0$. In the absence of a sizable anomalous dimension $\eta$ in
the quantum regime of the flow, one has
\begin{equation}
 \Lambda_{cl} \propto T^{1/z} \; ,
\label{Lambda_cl}
\end{equation}
as follows directly the definition of $\tilde T$. It turns out that
$\eta$ is indeed negligible down to the scale $\Lambda_{cl}$, except
in the case $z=2$ in two dimensions (see Sec. \ref{sec:finite}).

\bigskip

Below, we numerically solve the full RG flow equations
(\ref{eq:delta_delta},\ref{eq:delta_u},\ref{eq:etas}) for finite
temperatures. As already announced, the analysis is focused on the
region of the phase diagram where symmetry-breaking occurs. In our
notation this corresponds to sufficiently large values of the
control parameter $\delta$. First we treat the case $z=3$ (in
$d=2,3$) to which
Eqs.~(\ref{eq:delta_delta},\ref{eq:delta_u},\ref{eq:etas}) are
applied directly. In all numerical results we choose an initial
cutoff $\Lambda_0 = 1$, and an initial coupling constant $u = 1$.

\bigskip

We first solve the coupled flow equations
(\ref{eq:delta_delta},\ref{eq:delta_u},\ref{eq:etas})
with the aim of determining the phase boundary $T_c(\delta)$, or,
equivalently $\delta_c(T)$.
To this end, for each given temperature we tune the initial value
of $\delta$ such that at the end of the flow (for $\Lambda \to 0$)
one obtains the critical state with $\delta(\Lambda) \to 0$. The tuned
initial value is then identified as $\delta_c(T)$. Inverting the
function $\delta_c(T)$ yields $T_c(\delta)$.
In the variables $\tilde\delta$, $\tilde{u}$ this corresponds to
seeking for such values of the initial $\delta$,
that both $\tilde\delta(\Lambda)$ and $\tilde{u}(\Lambda)$ reach a
fixed point as the cutoff is removed.

\bigskip

The flow of $\eta$ and $\tilde u$ as a function of the logarithmic
scale variable $s = - \log(\Lambda/\Lambda_0)$ is shown in two
exemplary plots in Fig.~\ref{fig:eta_vs_lambda}, respectively. 
The flow is shown for various temperatures $T$, with $\delta$ tuned
to values very close to $\delta_c(T)$. The plateaus in
Fig.~\ref{fig:eta_vs_lambda} correspond to non-Gaussian fixed point
values of the flowing parameters. From Fig.~\ref{fig:eta_vs_lambda}
where $d=2$, one reads off the value of the anomalous dimension
$\eta\approx 0.5$. The exact value from the Onsager solution to the
Ising model is $\frac{1}{4}$. For the case $d=3$ we find $\eta
\approx 0.1$ within our truncation, which is also about twice as
large as the best estimates for the exact value (Goldenfeld 1992).
To obtain scaling behavior in the range of a few orders of magnitude
one needs to fine-tune the initial conditions with an accuracy of
around 15 digits. The breakdown of scaling behavior observed for
very small $\Lambda$ is due to insufficient accuracy of the initial
value of $\delta$ and numerical errors. The plateaus are more
extended as we go on fine-tuning the initial condition. Only exactly
{\em at} the critical point true scale invariance manifested by
plateaus of infinite size is expected.

\bigskip

\begin{figure}[t]
\begin{center}
\includegraphics[width=71mm]{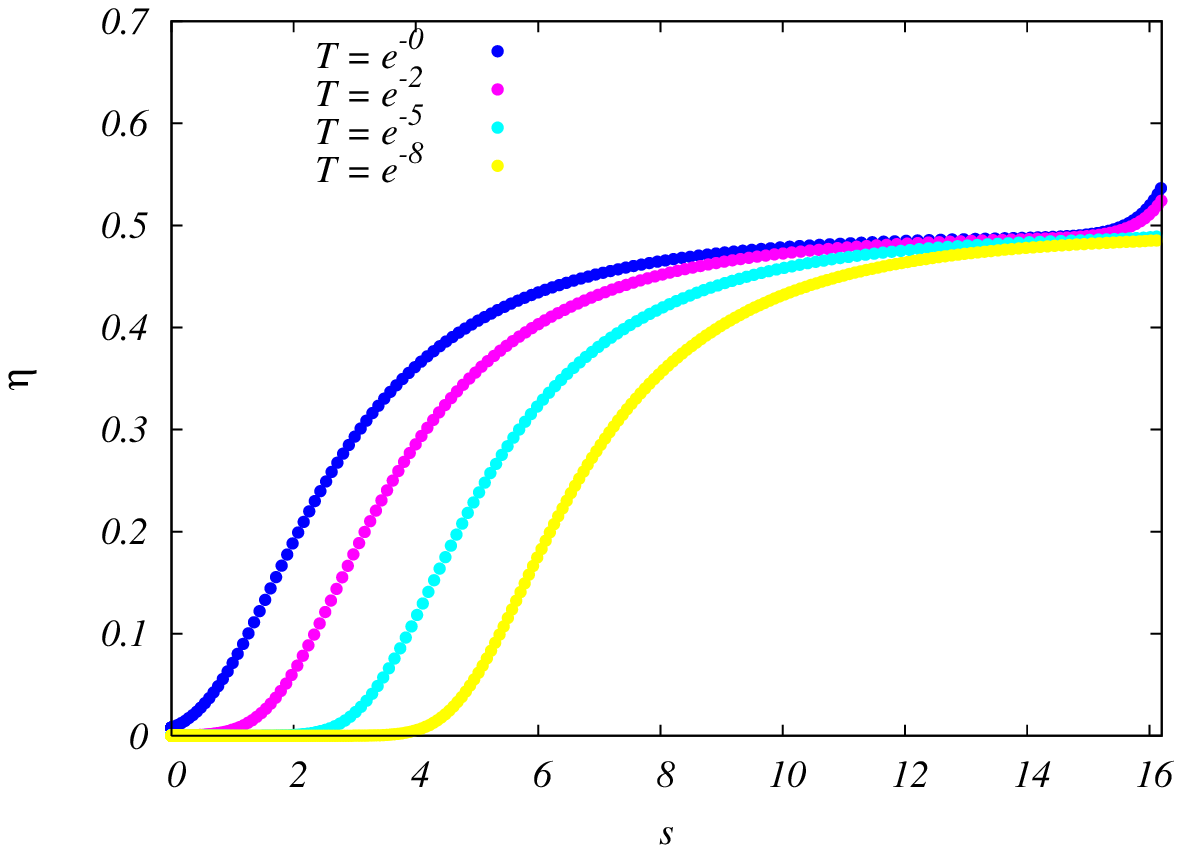}
\includegraphics[width=71mm]{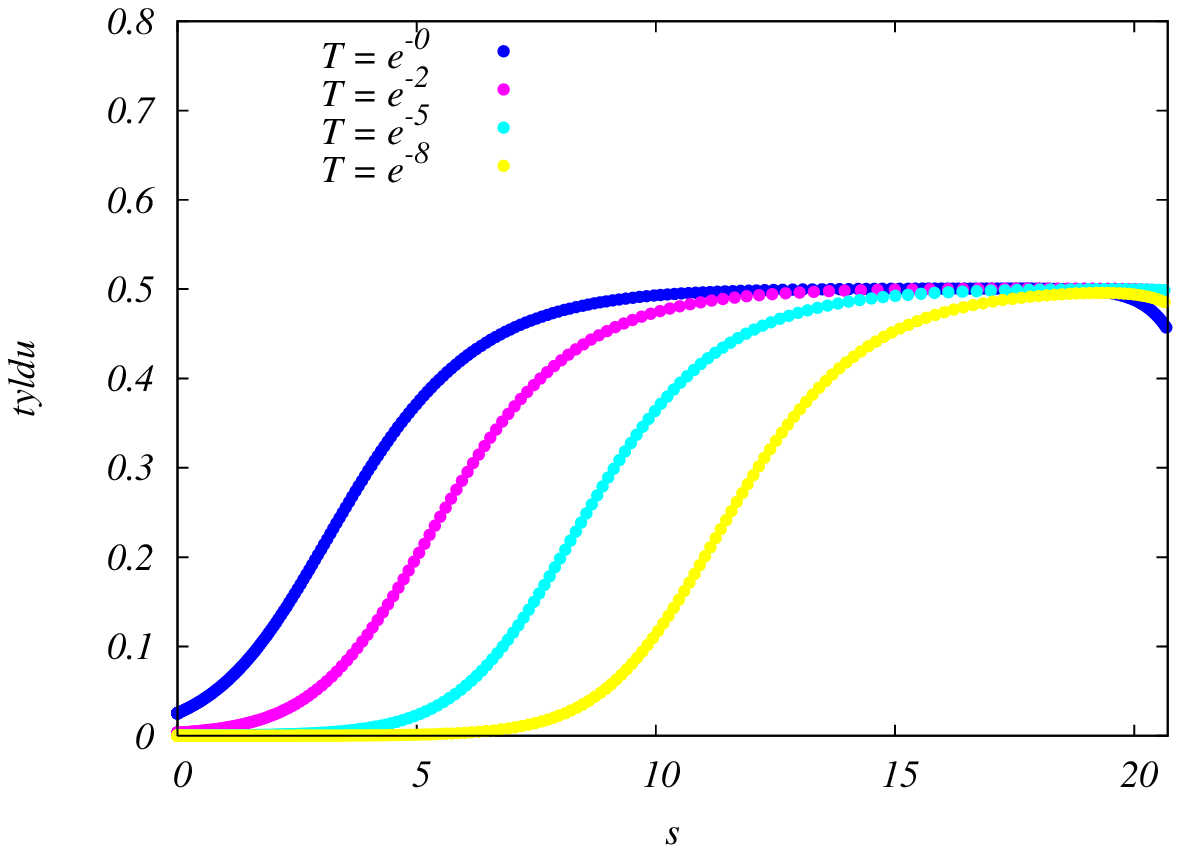}
\caption{Left: Anomalous dimension $\eta(\Lambda)$ plotted as function of
 $s = -\log(\Lambda/\Lambda_0)$ for different values of temperature
 in the case $z=3$ and $d=2$. The function $\eta(s)$ exhibits crossover
 from the mean-field value $\eta=0$, to the non-Gaussian result
 $\eta \approx 0.5$. The crossover scale $\Lambda_G$ is shifted towards smaller values
 (larger $s$) as temperature is reduced.
Right: Quartic coupling $\tilde{u}(\Lambda)$ plotted as function
 of $s = - \log(\Lambda/\Lambda_0)$ for different values of
 temperature in the case $z=3$ and $d=3$.}
\label{fig:eta_vs_lambda}
\end{center}
\end{figure}

The figures also reveal the Ginzburg scale $\Lambda_G$ at which
non-Gaussian fluctuations become dominant, such that the exponent
$\eta$ attains a non-zero value.
By fitting a power-law we observe $\Lambda_G \propto T_{c}$ for $d=3$
and $\Lambda_G \propto \sqrt{T_{c}}$ for $d=2$ with non-universal
proportionality factors.
As expected, $\Lambda_G$ vanishes at the quantum critical point,
because at $T=0$ the effective dimensionality $\mathcal{D}=d+z$
is above the upper critical dimension $d_c = 4$.
At finite temperatures, $\Lambda_G$ is the scale at which $\tilde u$
is promoted from initially small values (of order $T$) to values of
order one. From the linearized flow equations one obtains
\begin{equation}
 \Lambda_G \propto T_c^{\frac{1}{4-d}}
\label{eq:Ginzburg_scale}
\end{equation}
in agreement with the numerical results for $d=2$ and $d=3$.
Note that $\Lambda_G \ll \Lambda_{cl}$, since
$\Lambda_{cl} \propto T_c^{1/3}$ for $z=3$, see Eq.~(\ref{Lambda_cl}).
Hence anomalous scaling (finite $\eta$) is indeed absent in the regime
where quantum fluctuations contribute,
and non-Gaussian fluctuations appear only in the classical regime.
As already mentioned, the quantum contributions influence the flow
only at relatively large $\Lambda$ for $T > 0$.
In particular, they do not alter any fixed point values.
However, at the beginning of the flow
they dominate over the classical part and therefore are crucial
for a correct computation of the initial value of $\delta$ leading
to a scaling solution in the infrared limit.

\bigskip

Results for the transition line $T_c(\delta)$ for $d=2$ and $d=3$ are shown in
Fig.~\ref{fig:T_c_vs_delta_d2}. In both cases we recover the shape of the transition 
line as derived by Millis (1993), who used the Ginzburg temperature in the
symmetric phase as an estimate for $T_c$. Namely, we find
\begin{equation}
 (\delta - \delta_0) \propto T_c \log T_c
\end{equation}
for $d = 2$, and
\begin{equation}
 T_c \propto (\delta - \delta_0)^{0.75}
\end{equation}
\begin{figure}[t]
\begin{center}
\includegraphics[width=71mm]{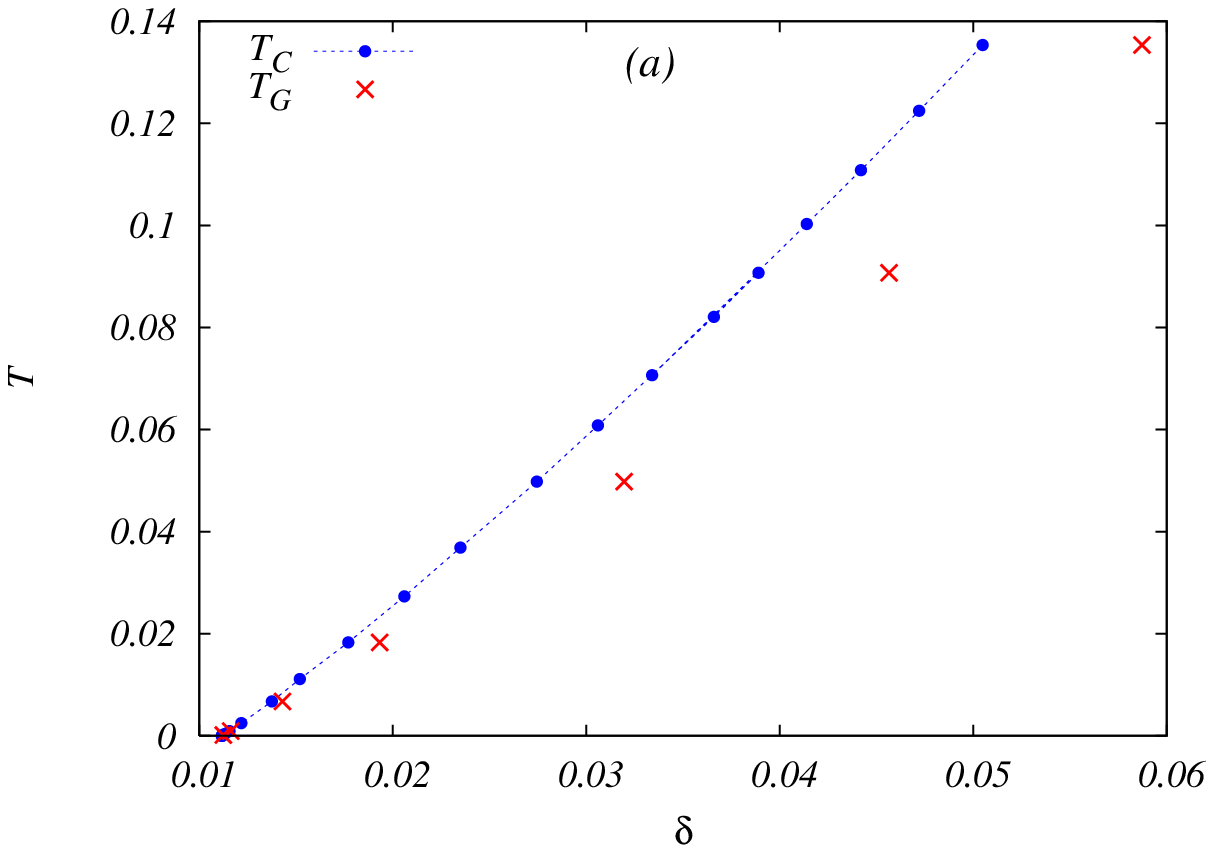}
\includegraphics[width=71mm]{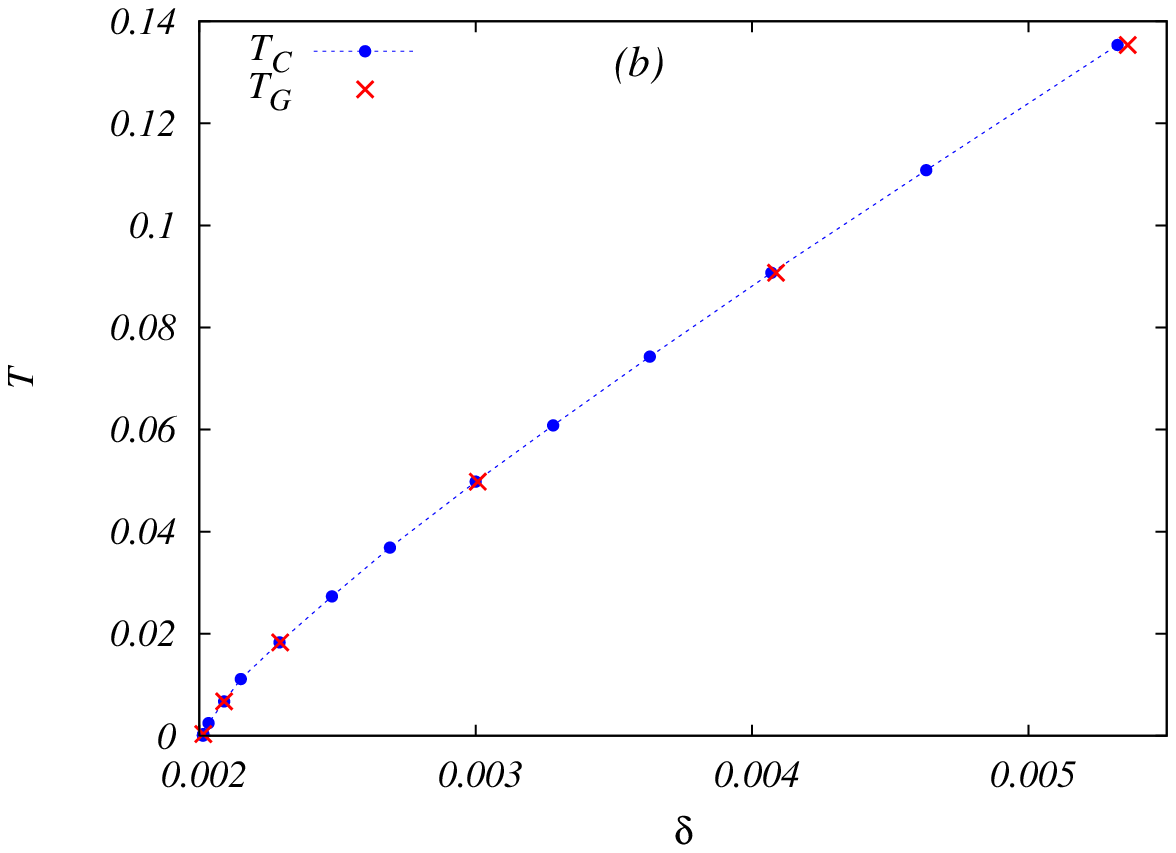}
\caption{The transition lines $T_c(\delta)$ obtained for $z=3$,
 $d=2$ $(a)$ and $z=3$, $d=3$ $(b)$. The phase with broken symmetry is located below the line.
The phase boundary $T_c(\delta)$ obeys
$(\delta-\delta_0) \propto T_c \log T_c$ for $d=2$ and $T_c \propto (\delta - \delta_0)^{0.75}$ in agreement with the result
by Millis (1993). The crosses indicate the Ginzburg temperature $T_G$ for
various choices of $\delta$.}
\label{fig:T_c_vs_delta_d2}
\end{center}
\end{figure}
for $d = 3$. The exponent $3/4$ in the three-dimensional case
matches with the general formula for the shift exponent
\begin{eqnarray}
\psi = \frac{z}{d+z-2}\;
\label{eq:millis_shift}
\end{eqnarray}
for arbitrary $z$ in dimensions $d > 2$ as long as $d+z>4$.
Generally, we see from this formula that \emph{increasing} $z$ has
qualitatively (neglecting possible log-corrections) the same effect
as \emph{decreasing} $d$.
Note that the phase boundary $T_c(\delta)$ approaches the quantum
critical point with vanishing first derivative for $d=2$ and with
singular first derivative in the case $d=3$.

\bigskip

An advantage of the present approach is that one can also follow
the RG flow into the strong coupling regime, where non-Gaussian
critical behavior occurs. This in turn allows an estimate of the
critical region's size as a function of temperature or the control
parameter $\delta$. To evaluate the Ginzburg line in the symmetry-broken phase
one solves the
flow equations
(\ref{eq:delta_delta},\ref{eq:delta_u},\ref{eq:etas})
for fixed $T$ and at different values of
$\delta > \delta_c(T)$, observing the behavior of fixed point
values of the average order parameter $\phi_0$ (or, alternatively,
the correlation length $\xi$) as $\delta$ approaches $\delta_c$.
Typical results are plotted in Fig.~\ref{fig:phi_vs_delta_d2} from which
we read off the value of the exponent $\beta$ describing the decay
of the order parameter upon approaching the transition line $\phi_0
\propto (\delta-\delta_c)^{\beta}$. In the truly critical region
(for $\delta-\delta_c$ small enough) one obtains $\beta\approx 0.11$
for $d=2$ and $\beta\approx 0.30$ for $d=3$. These results come out
close to the correct classical values $0.125$ and $0.31$,
respectively. This is unlike the other critical exponents ($\eta$
and the correlation length exponent $\nu$) which within our
truncation differ by factors close to 2 from their correct values.
%

\begin{figure}[t]
\begin{center}
\includegraphics[width=71mm]{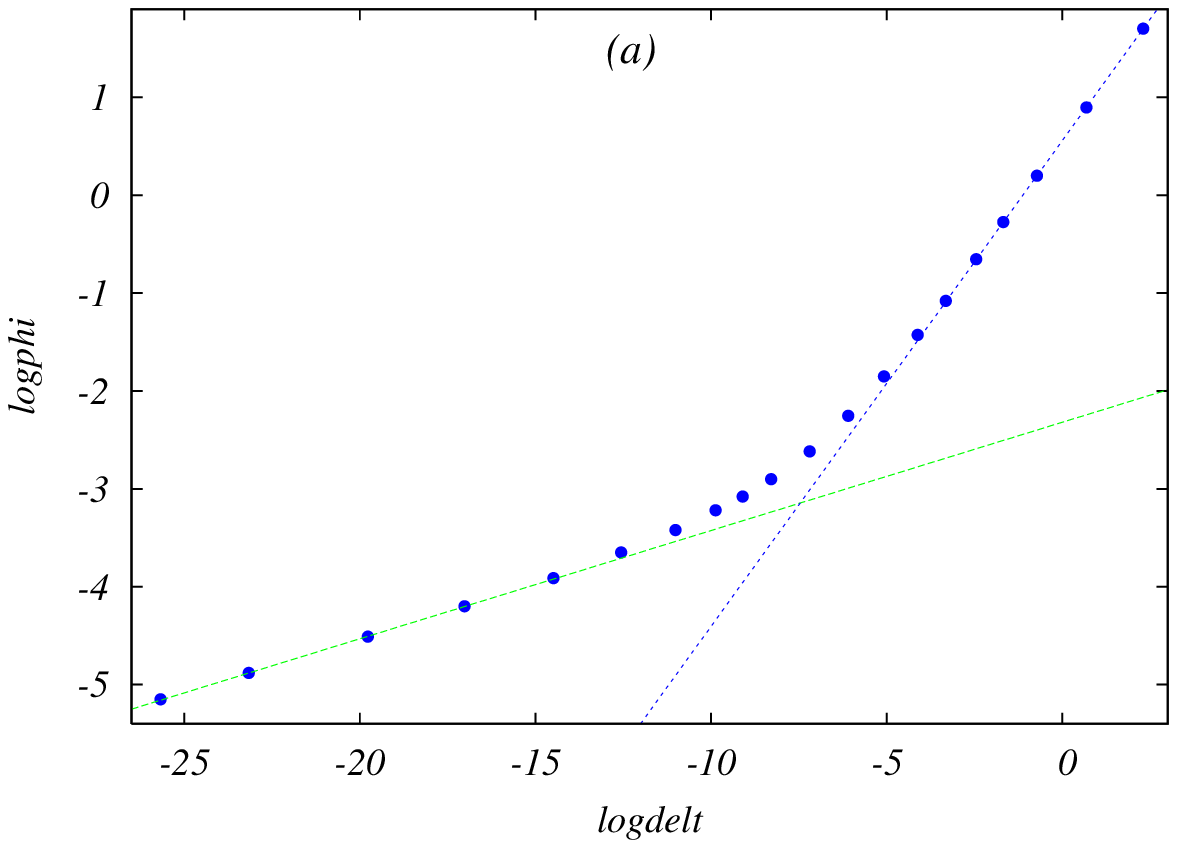}
\includegraphics[width=71mm]{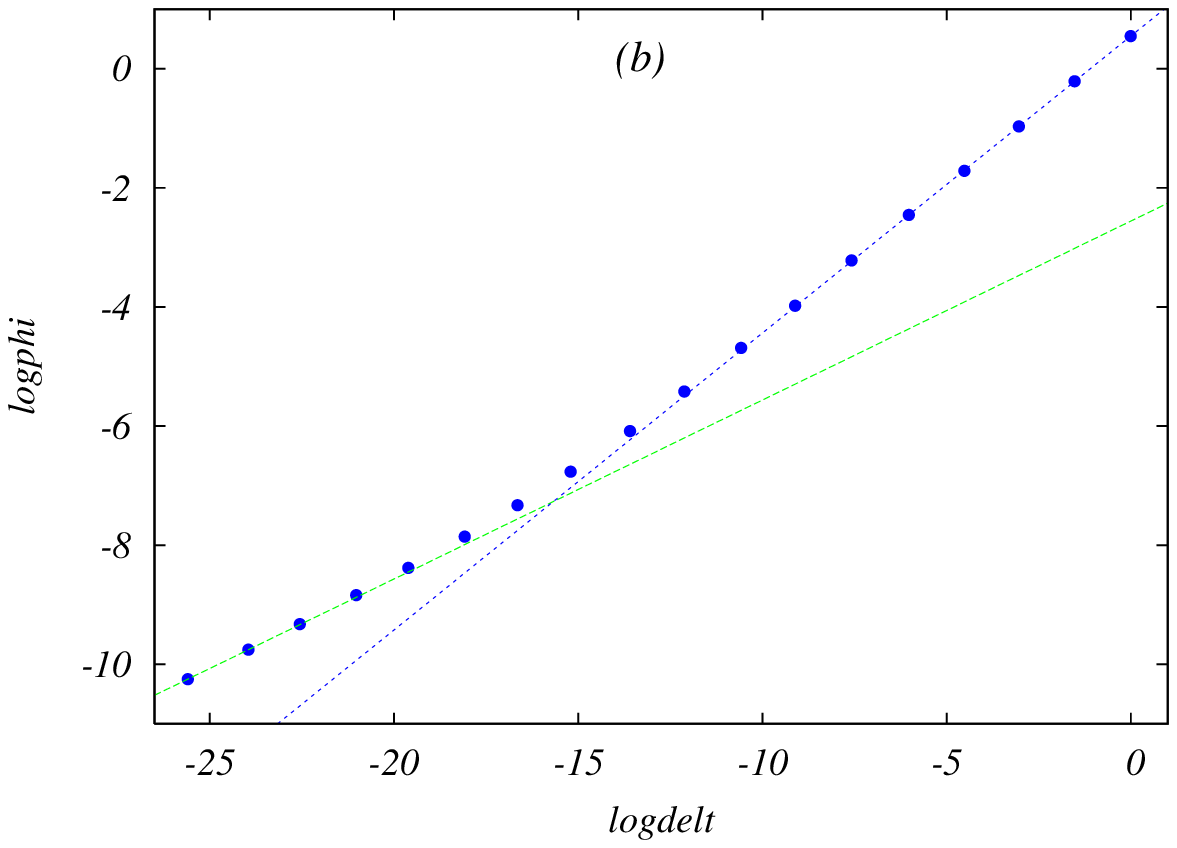}
\caption{Order parameter $\phi_0$ as a function of
$(\delta-\delta_c)$ at $T=e^{-5}$ for $z=3$, $d=2$ $(a)$, and $z=3$, $d=3$ $(b)$.
The exponent $\beta$ governing the decay of $\phi_0$ upon
approaching $\delta_c$ exhibits a crossover from a mean-field
value $\beta = 0.5$ to a non-Gaussian $\beta \approx 0.11$ for $d=2$,
and $\beta \approx 0.30$ for $d=3$.}
\label{fig:phi_vs_delta_d2}
\end{center}
\end{figure}

Indeed, as discussed in (Ballhausen 2004), to obtain accurate values
of the critical exponents in the Ising universality class, and in
particular in $d=2$, one not only needs to consider the full partial
differential equation governing the RG flow of the effective
potential $U[\phi]$, but also the field dependence of the wave
function renormalization and higher orders in the derivative
expansion of the effective action (Berges 2002).

\bigskip

From Fig.~\ref{fig:phi_vs_delta_d2}
we can extract the Ginzburg value $\delta_G$ below which true
critical behavior is found at the chosen temperature $T$.
Around $\delta_G$, the exponent $\beta$ exhibits a crossover
from its mean-field value $\beta = 0.5$ to a non-Gaussian value.
In other words, $\delta_G$ marks the boundary of the non-Gaussian
critical region at a given $T$.
At zero temperature, $\delta_G$ coincides with the quantum
phase transition point $\delta_0$, since there the fluctuations
are effectively $d+z>4$ dimensional, leading to mean-field
behavior. 
Several Ginzburg points in the $\delta-T$ plane are plotted
as $T_G(\delta)$ in Fig.~\ref{fig:T_c_vs_delta_d2},
where they can be compared to the phase transition line.
In three dimensions $T_G$ and $T_c$ almost coincide, such that
$T_G$ provides an accurate estimate for $T_c$.
In two dimensions a sizable region between $T_G$ and $T_c$
appears in the phase diagram. In that region non-Gaussian
classical fluctuations are present.

\bigskip

We stress that accounting for the anomalous exponent $\eta$ is
necessary to describe the classical scaling regime, that is to
obtain the plateaus in Fig.~\ref{fig:eta_vs_lambda}. Upon putting
$\eta=0$ the scaling plateaus do not form. On the other hand, the
shapes of the phase boundaries and the Ginzburg curves become very
similar at $T\to 0$ in the present cases, where $d+z \geq 4$.

%
\subsection{$z = 2$}
\label{subsec:z_2}

\begin{figure}[t]
\begin{center}
\includegraphics[width=71mm]{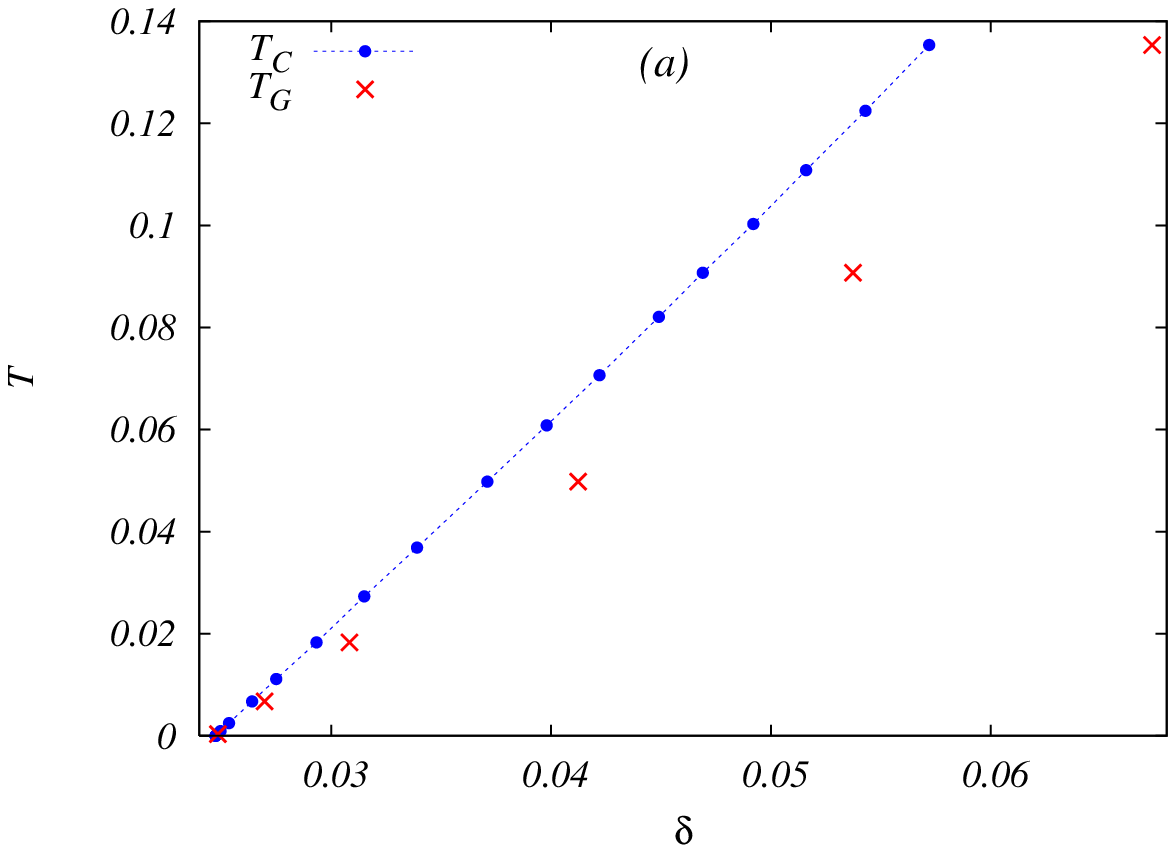}
\includegraphics[width=71mm]{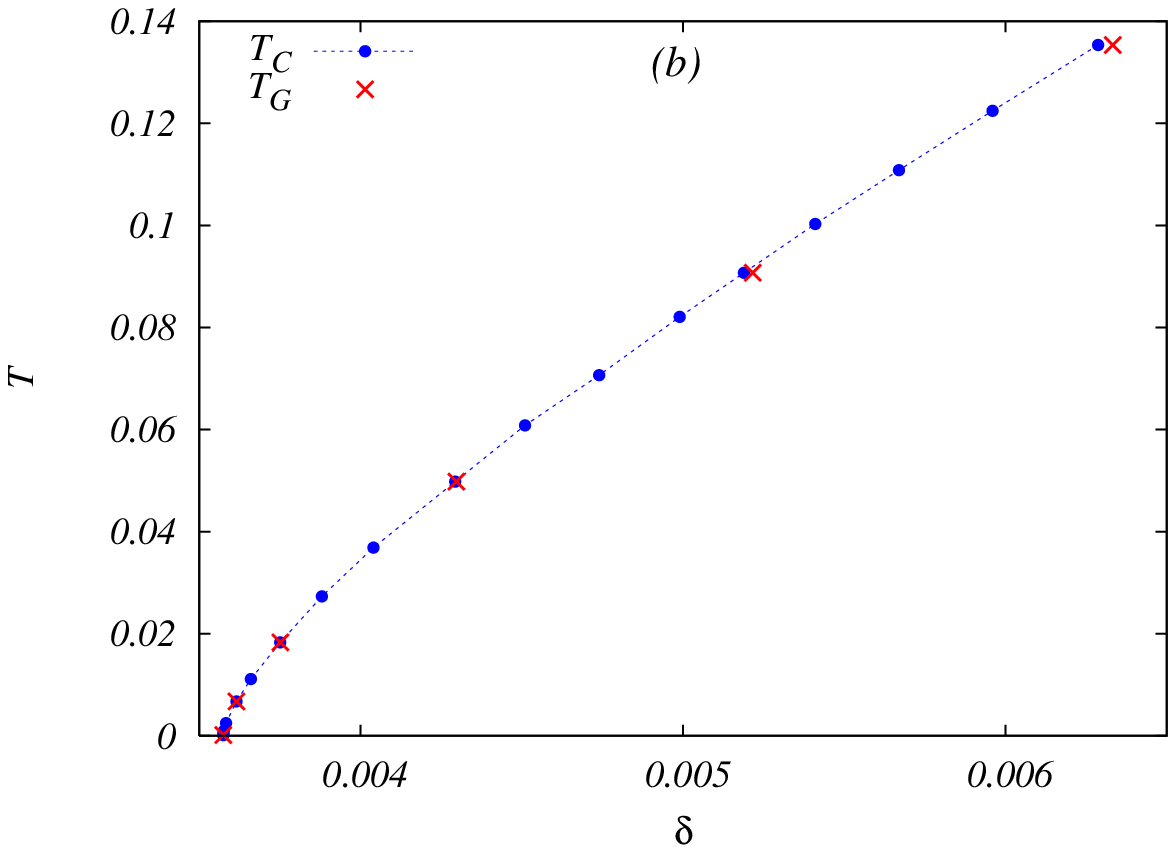}
\caption{The transition lines $T_c(\delta)$ obtained for $z=2$,
 $d=2$ $(a)$ and $z=2$, $d=3$ $(b)$. The phase with broken symmetry is located below the line.
For $d=2$ the phase boundary $T_c(\delta)$ is consistent with the relation
$(\delta-\delta_0) \propto T_c \log\log T_c/\log T_c$
derived by Millis (1993). For $d=3$ it follows the expected power law
$T_c(\delta) \propto (\delta - \delta_0)^{\psi}$ with $\psi = 2/3$. The crosses indicate the Ginzburg temperature $T_G$ for
various choices of $\delta$.}
\label{fig:T_c_vs_delta_d2z2}
\end{center}
\end{figure}

In the case $z=2$ the flow equations
(\ref{eq:delta_delta},\ref{eq:delta_u},\ref{eq:etas}) are
significantly simplified, as all the integrals can be evaluated
analytically. One obtains
\begin{eqnarray}
 \frac{d\tilde\delta}{d \log\Lambda}
 &=& \left(\eta - 2\right)\tilde\delta +
 2\tilde{u} \Bigg[ \frac{2}{d}\frac{1}{(1+\tilde\delta)^2} +
 \frac{4}{d} \tilde{T}^{-2} \Psi_1(y) \Bigg]
 +
 3\tilde{u} \tilde\delta
 \Bigg[ \frac{4}{d} \frac{1}{(1+\tilde\delta)^3} -
 \frac{4}{d} \tilde{T}^{-3}\Psi_2(y) \Bigg]\nonumber
 \\[2mm]
 \frac{d\tilde{u}}{d \log\Lambda}
 &=&\left( d-4+2\eta \right) \tilde{u}
 +
 3\tilde{u}^2\Bigg[\frac{4}{d}\frac{1}{(1+\tilde\delta)^3}
 - \frac{4}{d} \tilde{T}^{-3} \Psi_2(y) \Bigg] \; ,
\end{eqnarray}
and for the anomalous dimension,\label{fig:psi_formula}
\begin{eqnarray}
 \eta &=&
 \frac{6}{d}\frac{\tilde{u}\tilde\delta}{(1+\tilde\delta)^5}
 \Big[2(1+\tilde\delta)-\frac{8}{d+2}\Big]
 +
 \frac{4\tilde{u} \tilde\delta}{d} \tilde{T}^{-5}
 \left[\tilde{T} \Psi_3(y)+\frac{1}{d+2}\Psi_4(y)\right] \; ,
\end{eqnarray}
where the argument of the polygamma functions is given by
$y = 1 + (1+\tilde\delta)/\tilde{T}$.

\bigskip

The procedure to evaluate the phase diagram and the Ginzburg
line is the same as in the previously discussed case $z=3$.
In Fig.~\ref{fig:T_c_vs_delta_d2z2} we show results for the transition
line $T_c(\delta)$ in two and three dimensions.
We also show the Ginzburg temperature $T_G$ for various choices
of $\delta$.

In two dimensions, the transition line is consistent with the
almost linear behavior
$(\delta - \delta_0) \propto T_c \log\log T_c/\log T_c \,$,
derived previously for the Ginzburg temperature (Millis 1993).
However, a sizable region with non-Gaussian fluctuations
opens between $T_c$ and $T_G$.
In three dimensions, $T_c(\delta)$ obeys the expected (Millis 1993)
power law $T_c(\delta) \propto (\delta - \delta_0)^{\psi}$ with
shift exponent $\psi = 2/3$, and the Ginzburg temperature is very
close to $T_c$ for any $\delta$.

\subsection{$z=1$}
\label{subsec:z_1}

We now consider the Quantum Ising model at finite temperatures. 
This case has not been considered before in $d<3$ and also not by Millis (1993). 
The Millis analysis relies on the fact that the QCP is described by a Gaussian fixed point 
which is \emph{not} applicable here. 
The zero-temperature equations (\ref{eq:zero_betas_QI}, \ref{eq:zero_eta_QI}) 
are generalized to finite temperatures. We first define the variables:
\begin{eqnarray}
\tilde{\rho}&=&\frac{A d\rho}{2K_{d}T\Lam^{d-2}}\nonumber\\
\tilde{u}&=&\frac{2 K_{d}T u}{d A^{2}\Lam^{4-d}}\nonumber\\
\tilde{T}&=&\frac{2\pi T}{\Lambda A}\;,
\end{eqnarray}
to write the flow equations as 
\begin{eqnarray}
\frac{d\tilde{\rho}}{d\log \Lambda}&=&\left(2-d-\eta\right)\tilde{\rho}
+\frac{3}{2}
\left[
\frac{1}
{\left(1+\frac{2\tilde{u}\tilde{\rho}}{3}\right)^{2}}
+
2\sum_{n=1}^{\infty}
\frac{1}
{\left(\left(n \tilde{T}\right)^{2} + 1+ \frac{2\tilde{u}\tilde{\rho}}{3}\right)^{2}}
\right]
\nonumber\\
\frac{d\tilde{u}}{d\log \Lambda}&=&\left(d-4+2\eta\right)\tilde{u}
+
3\tilde{u}^{2}
\left[
\frac{1}
{\left(1+\frac{2\tilde{u}\tilde{\rho}}{3}\right)^{3}}
+
2\sum_{n=1}^{\infty}
\frac{1}
{\left(\left(n \tilde{T}\right)^{2} + 1+ \frac{2\tilde{u}\tilde{\rho}}{3}\right)^{3}}
\right]
\;.\nonumber\\
\label{eq:finiteT_u_rho_QI}
\end{eqnarray}
The anomalous dimension is determined by
\begin{eqnarray}
\eta=2\tilde{u}^{2}\tilde{\rho}
\Bigg[
\frac{1}
{\left(1+\frac{2\tilde{u}\tilde{\rho}}{3}\right)^{4}}
-
\frac{2}
{\left(d+2\right)\left(1+\frac{2\tilde{u}\tilde{\rho}}{3}\right)^{5}}
+
2\sum_{n=1}^{\infty}
&&
\frac{1}
{\left(\left(n \tilde{T}\right)^{2}+1+\frac{2\tilde{u}\tilde{\rho}}{3}\right)^{4}}
-
\nonumber\\
&&
\frac{2}
{\left(d+2\right)\left(\left(n \tilde{T}\right)^{2}
+1+\frac{2\tilde{u}\tilde{\rho}}{3}\right)^{5}}
\Bigg]
\;,\nonumber\\
\label{eq:finiteT_eta_QI}
\end{eqnarray}
and the effective temperature scales as 
\begin{eqnarray}
\frac{d \tilde{T}}{d\log\Lam}=\left(\eta-1\right)\tilde{T}\;.
\end{eqnarray}
The Matsubara summation can be performed analytically, which we have done for a 
computationally cheaper numerical 
solution, but they do not deliver analytic insights here.

\begin{wrapfigure}{r}{0.5\textwidth}
  \vspace{-5mm}
  \begin{center}
\includegraphics*[width=80mm,angle=-90]{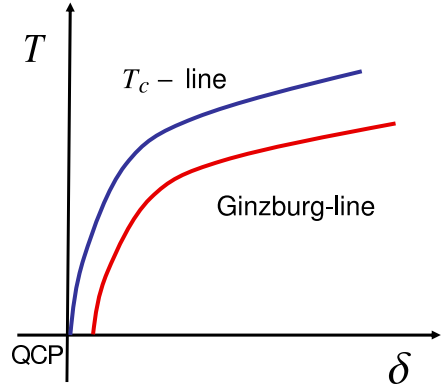}
\end{center}
\vspace{-15mm}
\caption{\textit{Schematic comparison of the Ginzburg line with the $T_{c}$-line 
for $d=2$, $z=1$. The critical region extends down to zero temperature.}}
\label{fig:ginz_vs_Tc}
\vspace*{5mm}
\end{wrapfigure}

We discuss the numerical solution of the finite temperature flow equations 
(\ref{eq:finiteT_u_rho_QI}, \ref{eq:finiteT_eta_QI}) and we compare it to the 
zero-temperature flow equations of subsection \ref{subsec:zero_QI}. Note that now 
$d+z=3<4$, and this case has not been covered previously (Hertz 1976, Millis 1993).
Non-Gaussian fluctuations persist even at zero temperature which is illustrated 
in a qualitative comparison of the Ginzburg line with the $T_{c}$-line in Fig. 
\ref{fig:ginz_vs_Tc}.

In Fig. \ref{fig:u_rho_flows_z_1}, the zero-temperature flow is juxtaposed with finite-temperature 
flows. Clearly, there is no continuous crossover from the finite-temperature behavior to 
the zero-temperature one.  
Hence, taking the limit $T\rightarrow 0$ is a \emph{discontinuous} process. Both, the finite-$T$ 
and the $T=0$ find a description in terms of two distinct non-Gaussian fixed points.
\footnote{\textit{
As a counterexample, consider the merger of the non-Gaussian Wilson-Fisher-type fixed point with the 
Gaussian fixed point when taking the limit $d\rightarrow d_{c}^{+}$, with $d_{c}^{+}$ the upper critical 
dimension, in standard $\phi^{4}$-theory (Goldenfeld 1992).}}
\begin{figure}[t]
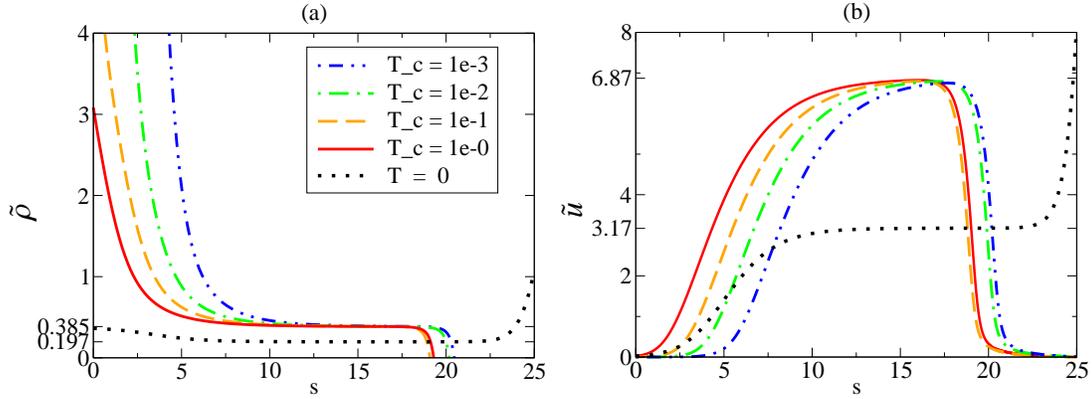

\begin{center}
\includegraphics*[width=71mm]{rho_flows_z_1.eps}
\hspace{1mm}
\includegraphics*[width=69mm]{u_flows_z_1.eps}
\caption{Flows for the Quantum Ising model ($z=1$) in $d=2$. 
(a): Flows of $\tilde{\rho}$ as a function of logarithmic cutoff-scale
$s=-\log\left[\Lam/\Lam_{0}\right]$ for various temperatures and at $T=0$. 
The values of the classical fixed point ($\tilde{\rho}=0.385$) and the QCP ($\tilde{\rho}=0.197$) 
are marked on the vertical axis. We set $\Lambda_{0}=1$. The infrared (ultraviolet)
is to the right (left) of the graphs. (b): Corresponding flows of $\tilde{u}$. The
values of the classical fixed point ($\tilde{u}=6.87$) and the QCP ($\tilde{u}=3.17$) are marked on the vertical axis.}
\label{fig:u_rho_flows_z_1}
\end{center}
\end{figure}
From Fig. \ref{fig:u_rho_flows_z_1} (b), we deduce the Ginzburg-scale to vary with temperature as 
\begin{eqnarray}
\Lambda_{G}\propto T_{c}^{1/2}\;,
\end{eqnarray}
fitting the formula derived for $d+z>4$, $\Lambda_{G}\propto  T_{c}^{1/\left(4-d\right)}$. 
As expected, the QCP-value for $\eta$ shown in 
Fig. \ref{fig:eta_flows_z_1} (a) is smaller than its classical counterpart as 
the QCP is effectively $d+1$-dimensional thus taming critical fluctuations.
The phase boundary in double-logarithmic plot is shown in Fig. \ref{fig:eta_flows_z_1} (b). 
The points do not lie on a straight line, indicating 
log-corrections correcting the power-law behavior. $\psi$ 
attains effective values in the range: $\psi\approx 0.5-0.7$ over six (!) orders 
of magnitude in good agreement with
%
%
%
%
\begin{eqnarray}
T_{c}\sim9.28\frac{\sqrt{\delta-\delta_{c}}}{\left|\ln\left(\delta-\delta_{c}\right)\right|}\; 
\label{eq:logfit}
\end{eqnarray}
which, as anticipated, 
does not fit the Millis formula, Eq. (\ref{eq:millis_shift}), which was derived under the 
assumption $d+z>4$.  However, 
the existence of these log-corrections remains inconclusive. Interestingly, $\psi$ is \emph{reduced} compared to its 
two-dimensional analogs when $z>1$ although the strength of non-Gaussian classical 
fluctuations is \emph{increased} when reducing $z$. This underpins the importance 
of Gaussian quantum fluctuations which on the other hand \emph{decrease} when reducing $z$. 
%
\begin{figure}[b]
\begin{center}
\vspace*{8mm}
\hspace*{-75mm}\includegraphics*[width=65mm]{eta_flows_z_1.eps}\\[-54mm]
\hspace*{70mm}\includegraphics*[width=73mm]{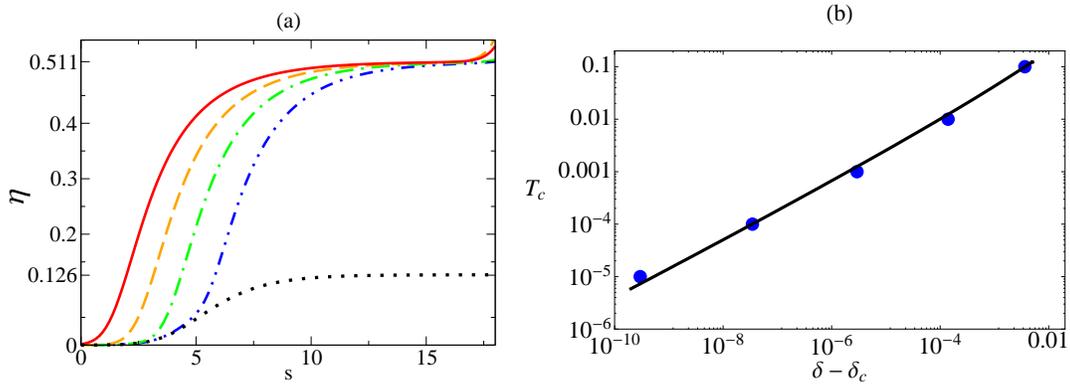}\\[0mm]
\caption{Results for the Quantum Ising model ($z=1$) in $d=2$. 
(a): Cutoff-scale dependences of the anomalous dimension for the same temperatures as
in Fig. \ref{fig:u_rho_flows_z_1}. The value of the anomalous dimension at the classical 
fixed point ($\eta=0.511$) and the QCP ($\eta=0.126$) are marked on the vertical axis. 
(b): Double logarithmic plot of the phase boundary fitted to Eq. (\ref{eq:logfit}).}
\label{fig:eta_flows_z_1}
\end{center}
\end{figure}

\section{Conclusion}
\label{sec:bosonic_qcp_conclusion}

We have analyzed classical and quantum fluctuations in the
symmetry-broken phase near a quantum phase transition in an
itinerant electron system and in the Quantum Ising model. 
The analysis is restricted to the case of
{\em discrete} symmetry breaking, where no Goldstone modes appear.
Following Hertz (1976) and Millis (1993), we use an
effective bosonic action for the order parameter fluctuations as a
starting point. For the Quantum Ising model, 
the effective continuum action is also of the $\phi^{4}$-type but with 
dynamical critical exponent $z=1$. The renormalization of the Hertz 
action and the Quantum Ising model 
by fluctuations is obtained from a system of coupled flow equations,
which are derived as an approximation to the exact flow equation for
one-particle irreducible vertex functions in the functional RG
framework. In addition to the renormalization of the effective mass
and the four-point coupling, we also take the anomalous dimension
$\eta$ of the order parameter fields into account. In the
symmetry-broken phase, contributions to $\eta$ appear already at
one-loop level. Quantum and thermal fluctuations are captured on
equal footing. The flow equations are applicable also in the
immediate vicinity of the transition line at finite temperature,
where fluctuations deviate strongly from Gaussian behavior.

\bigskip

We have computed the transition temperature $T_c$ as a function of
the control parameter $\delta$ near the quantum critical point,
approaching the transition line from the symmetry-broken phase.
Explicit results were presented for dynamical exponents $z=3$ and $z=1$ in two 
and three dimensions corresponding to Pomeranchuck transitions and the Quantum Ising model, respectively. 
For $z=3$, 
$T_c(\delta)$ agrees with the behavior of the Ginzburg
temperature above $T_c$ derived previously by Millis (1993). 
For $z=1$ in two dimensions, even the zero-temperature 
theory is characterized by a non-Gaussian fixed point. This case 
is not covered in the Hertz-Millis theory. We computed the phase boundary and 
found logarithmic corrections to power-law behavior. The effective shift-exponent, 
$\psi\left(d=2,z=1\right)\approx 0.5-0.7$ is smaller than the two-dimensional Hertz-Millis value, 
$\psi\left(d=2,z\geq2\right) =1$. 

We have also computed the Ginzburg temperature $T_G$ below $T_c$,
above which non-Gaussian fluctuations become important. Although $T_G$
and $T_c$ almost coincide in three dimensions, a sizable region
between $T_G$ and $T_c$ opens in two dimensions. While for $z\geq2$ this critical 
region shrinks to zero for $T_{c}\rightarrow0$, a finite difference between the 
Ginzburg and critical coordinates persists even at zero-temperature for $z=1$.

\bigskip

It will be interesting to extend the present approach to the case of
continuous symmetry breaking. Then, symmetry-breaking at finite
temperature is suppressed completely for $d\leq2$ by Goldstone
fluctuations. We consider the fluctuation effects of Goldstone modes on the quantum critical 
behavior at zero and at finite temperature in the next chapter.

\chapter[Quantum critical points with Goldstone modes]
{Quantum critical points with Goldstone modes}
\label{chap:bosonicqcp_goldstone}

\section{Introduction}

A quantum critical point may separate a phase with broken continuous symmetry 
from a normal phase (see  Fig. \ref{fig:af_qcp}) as is the case for example in 
quantum magnets, superfluids (Belitz 2005, Loehneysen 2007), 
or continuum systems with Pomeranchuk  
\begin{wrapfigure}{r}{0.5\textwidth}
  \vspace{-16mm}
  \begin{center}
\hspace*{-7mm}
\includegraphics*[width=80mm,angle=-90]{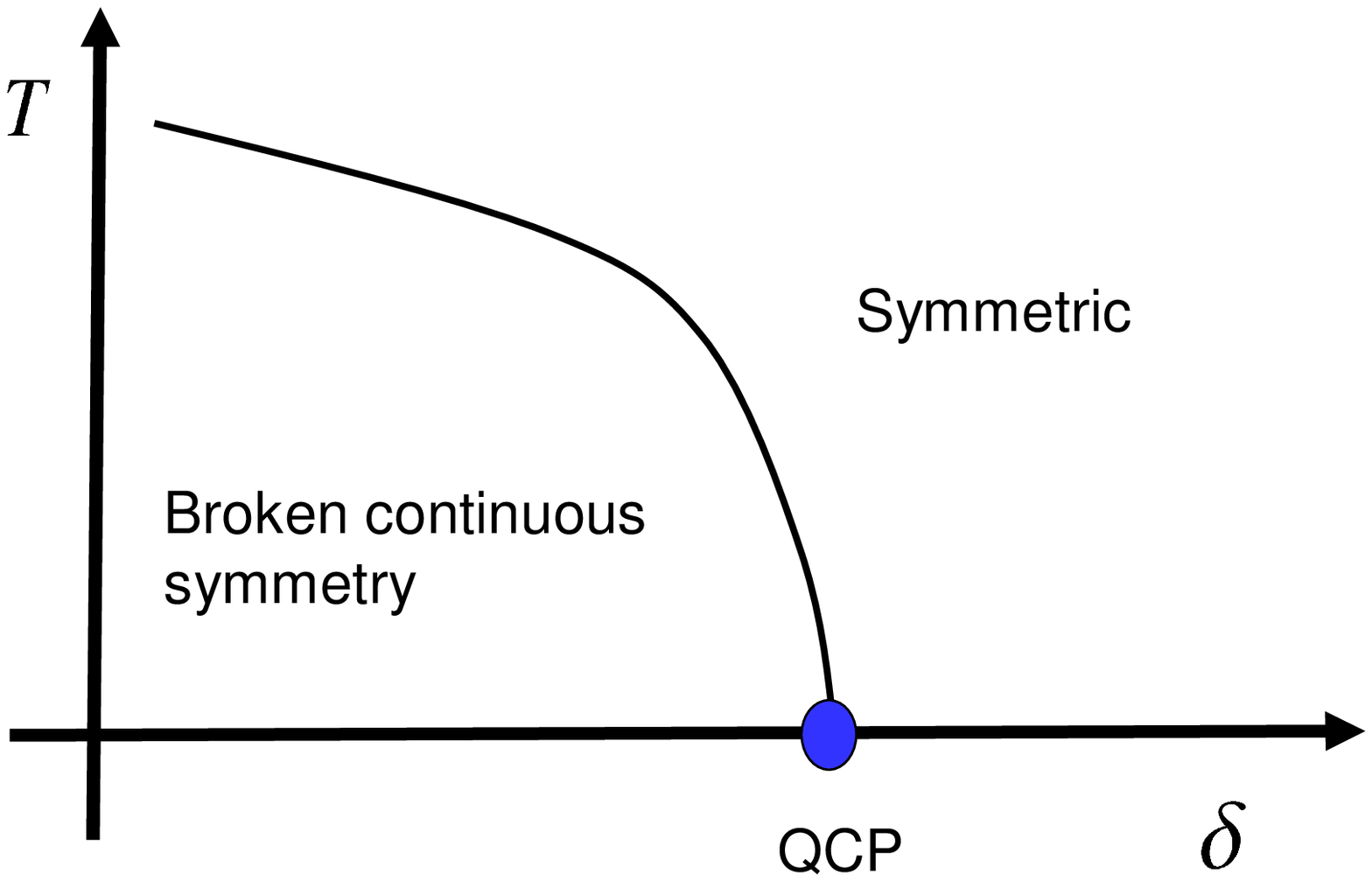}
\end{center}
\vspace{-25mm}
\caption{\textit{Phase diagram of the QCP with Goldstone modes 
for $d>2$. At $d=2$ thermal fluctuations
destroy the order and suppress the $T_{c}$-line to
zero temperature according to the Mermin-Wagner theorem.}} 
\label{fig:af_qcp}
\vspace{-2mm}
\end{wrapfigure}
Fermi surface deformations (Oganesyan 2001, W\"olfle 2007, 
Quintanilla 2008).
The distinct signature of phases with broken continuous symmetry is the occurence 
of Goldstone modes. In a recent experiment on quantum magnets with pressure as non-thermal control parameter (R\"uegg 2008), inelastic neutron scattering 
data reconfirmed the vanishing mass for the Goldstone mode in the magnetic phase and 
provided information about the magnitude of the longitudinal mass in the vicinity of the QCP. Theoretically, it is largely unexplored how Goldstone modes affect the zero and 
finite temperature properties on the ordered side 
of the quantum phase transition (Wetterich 2008).

For classical bosonic $O(N)$-models with Goldstone and longitudinal fluctuations, the functional
RG provided a unifying picture (Berges 2002).
In the present chapter, we compute RG flows in the vicinity of a QCP 
accounting for transversal and 
longitudinal fluctuations of the order parameter restricting 
ourselves to Goldstone modes dispersing linearly in momentum ($\Omega\sim|\mathbf{q}|$)
as is the case in superfluids and antiferromagnets (Belitz 2005).

At finite temperatures, we extend our computation
of the phase boundary for discrete symmetry-breaking
(see chapter \ref{chap:bosonicqcp_discrete}, Jakubczyk 2008) to account for Goldstone modes. We compute the shift-exponent $\psi$ which
characterizes how the $T_{c}$-line varies as a function of control parameter (Loehneysen 2007)
in three dimensions. We also clarify the singularity structure of the theory and establish 
the infrared asymptotics away from the phase boundary. 

\bigskip

We start by presenting the $\sigma$-$\mathbf{\Pi}$ model for phases with broken continuous
symmetry in section \ref{sec:sig_pi_model}. We
clearly distinguish longitudinal fluctuations ($\sigma$) from the Goldstone modes ($\mathbf{\Pi}$) 
both dispersing relativistically with dynamical exponent $z=1$.
In section \ref{sec:qcp_gold_method}, we describe our functional RG set-up.
In section \ref{sec:qcp_gold_finite}, we compute
the shift-exponent characterizing the shape of the phase boundary at finite temperatures
for $d=3$. In section \ref{sec:gold_IR}, we extend our calculation to regions away from the phase boundary and we analytically establish the infrared properties 
of both, the Goldstone and longitudinal propagator, as well as the bosonic self-interaction. 
Summary and conclusions follow in section \ref{sec:qcp_gold_summary}.

\section{$\sigma$-$\mathbf{\Pi}$ Model for continuous symmetry-breaking}
\label{sec:sig_pi_model}

We are interested in the effective action for $N$-component real-valued bosonic fields $\mathbf{\Phi}$ in phases 
with spontaneously broken continuous symmetry ($N\geq2$). The most general 
action to fourth order in the fields that respects the $O(N)$-symmetry and contains 
no more than two 
derivatives reads (Wetterich 1991, Tetradis 1994):
\begin{eqnarray}
 \Gam[\mathbf{\Phi}]&= &\int\left[ \frac{Z}{2} 
\left(\partial\mathbf{\Phi}\right)^{2}
+
\frac{Y}{8}\left(\mathbf{\Phi}
 \partial \,
\mathbf{\Phi}\right)^{2}\right]
+
U^{\text{loc}}[\mathbf{\Phi}]\;,
\label{eq:sig_pi_model}
\end{eqnarray}
where $Z$ and $Y$ are renormalization factors, $\partial$ denotes the 
standard derivative operator, and the integration symbol $\int$ denotes the standard 
space integration. Including the field component 
index $a$ and the dimensionality index $\mu$, 
the products in the derivative terms are defined as: $\left(\partial\mathbf{\Phi}\right)^{2}=\partial_{\mu}\Phi^{a}
\partial^{\mu}\Phi_{a}$, and 
$\left(\mathbf{\Phi}\partial\mathbf{\Phi}\right)^{2}=
\left(\Phi^{a}
\partial_{\mu}\Phi_{a}\right)
\left(\Phi^{b}
\partial^{\mu}\Phi_{b}\right)$ where $a,b=1,...,N$ and $\mu=1,...,d+1$.
Local interaction terms are retained in the effective potential:
\begin{eqnarray}
U^{\text{loc}}[\mathbf{\Phi}]=\frac{u}{8} \int 
\left(\mathbf{\Phi}^{2}-|\alpha|^{2}\right)^{2}\;,
\label{eq:gold_effpot}
\end{eqnarray}
where the effective potential has a
minimum at $\mathbf{\Phi} = (\alf,0,0...)$ with $\alf$ real and positive. 
It is convenient to decompose $\mathbf{\Phi}$ into a 
londitudinal component $\sg$ and $(N-1)$ transverse components $\mathbf{\Pi}$ via
\begin{eqnarray}
 \mathbf{\Phi}  &=& \left(\sg,0,0,...\right) +\left(0,\mathbf{\Pi}\right) \;.
\label{eq:ssb_basis}
\end{eqnarray}
Substituting $\mathbf{\Phi} \to \left(\alf + \sg,\mathbf{\Pi}\right)$ in 
Eq. (\ref{eq:sig_pi_model}) yields various interaction 
terms (see below) and two distinct quadratic parts:
\begin{eqnarray}
 \Gam_{\sg\sg}&= &\frac{1}{2} \int_q \sg_{-q}
 \left(\left(Z+Y|\alpha|^{2}\right)q^2 + u |\alpha|^{2}\right) \,
 \sg_q\nonumber\\
\Gam_{\pi\pi}&= &\frac{1}{2} \int_q \mathbf{\Pi}_{-q}
\left(Z q^2\right)\,\mathbf{\Pi}_q \;,
\end{eqnarray}
where $q=\left(q_{0},\mathbf{q}\right)$ collects the frequency and momenta and $\int_{q}=\int\frac{d q_{0}}{2\pi}\int\frac{d^{d} q}{(2\pi)^{2}}$ comprises
the corresponding integrations. The components of the 
$\mathbf{\Pi}$-field are massless as these are the Goldstone modes. 
The propagators for the $\sg$- and $\mathbf{\Pi}$-field
\begin{wrapfigure}{r}{0.5\textwidth}
  \vspace{-5mm}
  \begin{center}
\begin{fmffile}{legend_ssb_20}
\begin{eqnarray}
\gamma_{\sigma\pi^{2}}&=&
\parbox{20mm}{\unitlength=1mm\fmfframe(2,2)(1,1){
\begin{fmfgraph*}(15,15)\fmfpen{thin}
\fmfleft{l1}
\fmfrightn{r}{2}
\fmf{dashes}{l1,G1}
\fmfpolyn{full,tension=0.8}{G}{3}
\fmf{photon}{G2,r1}
\fmf{photon}{G3,r2}
 \end{fmfgraph*}
}},\hspace{5mm}
\gamma_{\sigma^{3}}\hspace{3mm}=
\parbox{20mm}{\unitlength=1mm\fmfframe(2,2)(1,1){
\begin{fmfgraph*}(15,15)\fmfpen{thin}
\fmfleft{l1}
\fmfrightn{r}{2}
\fmf{dashes}{l1,G1}
\fmfpolyn{full,tension=0.8}{G}{3}
\fmf{dashes}{G2,r1}
\fmf{dashes}{G3,r2}
 \end{fmfgraph*}
}}\nonumber\\[3mm]
\gamma_{\sigma^{4}}&=&
\parbox{20mm}{\unitlength=1mm\fmfframe(2,2)(1,1){
\begin{fmfgraph*}(15,15)\fmfpen{thin}
\fmfleftn{l}{2}
\fmfrightn{r}{2}
\fmf{dashes}{l2,G1}
\fmf{dashes}{l1,G2}
\fmfpolyn{full,tension=1.5}{G}{4}
\fmf{dashes}{r1,G3}
\fmf{dashes}{r2,G4}
 \end{fmfgraph*}
}},\hspace{5mm}
\gamma_{\sigma^{2}\pi^{2}}\hspace{1mm}=
\parbox{20mm}{\unitlength=1mm\fmfframe(2,2)(1,1){
\begin{fmfgraph*}(15,15)\fmfpen{thin}
\fmfleftn{l}{2}
\fmfrightn{r}{2}
\fmf{dashes}{l2,G1}
\fmf{dashes}{l1,G2}
\fmfpolyn{shaded,tension=1.5}{G}{4}
\fmf{photon}{r1,G3}
\fmf{photon}{r2,G4}
 \end{fmfgraph*}
}}\nonumber\\[3mm]
\gamma_{\pi^{4}}\hspace{1mm}&=&
\parbox{20mm}{\unitlength=1mm\fmfframe(2,2)(1,1){
\begin{fmfgraph*}(15,15)\fmfpen{thin}
\fmfleftn{l}{2}
\fmfrightn{r}{2}
\fmf{photon}{l2,G1}
\fmf{photon}{l1,G2}
\fmfpolyn{full,tension=1.5}{G}{4}
\fmf{photon}{r1,G3}
\fmf{photon}{r2,G4}
 \end{fmfgraph*}
}}\nonumber\\[5mm]
G_{\sigma}(k)&=&
\parbox{30mm}{\unitlength=1mm\fmfframe(1,1)(1,1){
\begin{fmfgraph*}(25,5)\fmfpen{thin}
\fmfleft{l1}
 \fmfright{r1}
  \fmf{dashes}{l1,r1}
 \end{fmfgraph*}
}}\nonumber\\[5mm]
G_{\pi}(k)&=&
\parbox{30mm}{\unitlength=1mm\fmfframe(1,1)(1,1){
\begin{fmfgraph*}(25,5)\fmfpen{thin}
\fmfleft{l1}
 \fmfright{r1}
  \fmf{photon}{l1,r1}
 \end{fmfgraph*}
}}\nonumber\\[-2mm]
\nonumber
\end{eqnarray}
\end{fmffile}
  \end{center}
  \vspace{-5mm}
\caption{\textit{Interaction vertices and propagators of the 
$\sg$-$\mathbf{\Pi}$ model for continuous symmetry-breaking.}}
\vspace{-65mm}
\label{fig:ssb_goldstone_constituents}
\end{wrapfigure}
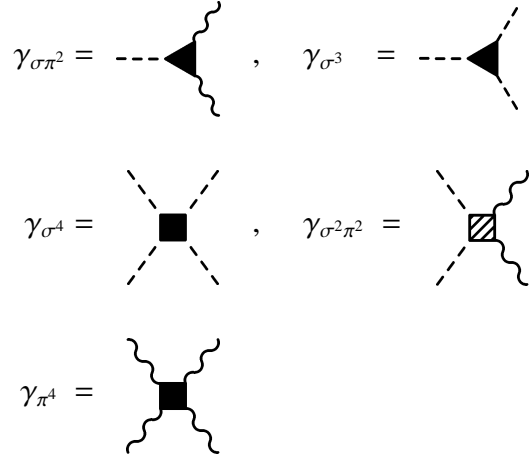
have the form
\begin{eqnarray}
 G_{\sg}(\Omega, \mathbf{q}) &=& -\langle \sg_{q}\sg_{-q}\rangle
=\frac{-1}{Z_{\sg} q^2+m_{\sg}^2}\nonumber\\
 G_{\pi}(\Omega, \mathbf{q})&=& -\langle \mathbf{\Pi}_{q}\mathbf{\Pi}_{-q}\rangle
=\frac{-1}{Z_{\pi} q^2} \;,\nonumber\\
\label{eq:qcp_gold_bose_props}
\end{eqnarray}
with $\rho=|\alpha|^{2}$, $Z_{\pi}=Z$, and
\begin{eqnarray}
Z_{\sg}&=&Z+Y\rho\nonumber\\
m_{\sg}^2&=&u \rho\;.
\label{eq:Y_relations}
\end{eqnarray}
For the interaction terms, we obtain:
\begin{eqnarray}
 \Gam_{\sg^4} &=&  \int_{q,q',p} \gam_{\sg^4}\,
 \sg_{-q-p} \sg_{-q'+p} \sg_{q'} \sg_q \; ,\nonumber \\
 \Gam_{\pi^4} &=&  \int_{q,q',p} \gam_{\pi^4}\,
 \mathbf{\Pi}_{-q-p} \mathbf{\Pi}_{-q'+p}
\mathbf{\Pi}_{q'} \mathbf{\Pi}_q \; ,\nonumber \\
 \Gam_{\sg^2\pi^2} &=&  \int_{q,q',p} \gam_{\sg^2\pi^2}\,
 \sg_{-q-p} \sg_{-q'+p} \mathbf{\Pi}_{q'} \mathbf{\Pi}_q \; , \nonumber\\
 \Gam_{\sg^3} &=& \int_{q,p} \gam_{\sg^3} \,
 \sg_{-q-p} \sg_{p} \sg_q \; ,\nonumber \\
 \Gam_{\sg\pi^2} &=& \int_{q,p} \gam_{\sg\pi^2} \,
 \sg_{-q-p} \mathbf{\Pi}_{p} \mathbf{\Pi}_q \; ,\nonumber\\
\label{eq:gold_ints}
\end{eqnarray}
with $\gam_{\sg^4} = \gam_{\pi^4} = \left(u+Y p^{2}\right)/8$,
$\gam_{\sg^2\pi^2} = \left(u+Y p^{2}\right)/4$, and
$\gam_{\sg^3} = \gam_{\sg\pi^2} = \left(u+Y p^{2}\right)\alf/2$ as 
depicted in Fig. \ref{fig:ssb_goldstone_constituents}. 
In total, the $\sigma-\mathbf{\Pi}$ model contains 
four independent parameters: $\alpha$, $u$, $Z$, and $Y$.

\bigskip

A comment on the choice of field basis in Eq. (\ref{eq:ssb_basis}) is in place.
The rotational $O(N)$-symmetry in internal space is realized only $\emph{linearly}$.
The periodicity of the Goldstone field is deformed
to a linear trough of infinite extension, see. Fig. \ref{fig:trough}. Consequently,
periodic field configurations such as vortices are hard to capture in this
basis.

Alternative choices are the phase-amplitude representation (schematic):
$\Phi\sim\rho_{0}e^{i\mathbf{\Sigma}\mathbf{\Theta}}$, with $\mathbf{\Sigma}$
the $N-1$ group generators, $\mathbf{\Theta}$ the conjugate Goldstone modes, and
$\rho_{0}$ the amplitude of the order-parameter.
This phase-amplitude representation manifestly preserves the rotational symmetry 
and allows, for example, a convenient
description of the non-linear sigma model when the mass of the radial boson is sufficiently
large so that Goldstone bosons remain the only fluctuating fields (Amit 2005). An analytic treatment
of some aspects of the Kosterlitz-Thouless phase for $N=2$ is also easily accessible (Goldenfeld 1992).

At criticality, however, when the effective potential is flat in the center and the order-parameter
is still small, the phase-amplitude basis has a parametric singularity at the origin for $\rho_{0}\gtrsim 0$.
And at higher energies, in the symmetric phase, neither Eq. (\ref{eq:ssb_basis}) nor the phase-amplitude
representation is appropriate: then, the particle representation in terms of the $\Phi$-field is most transparent.

We are therefore confronted with the situation that different choices of basis in field space are useful only in limited regions of the phase diagram. This problem has been the subject of an extensive body of literature (Popov 1987, Nepomnashchy 1992 and references therein, Pistolesi 1997,
Diener 2008, Strack 2008). A derivation of Ward identities for the $\sg$-$\mathbf{\Pi}$ model can be found in Amit (2005).

\begin{figure}[t]
\begin{center}
\includegraphics*[width=41mm,angle=-90]{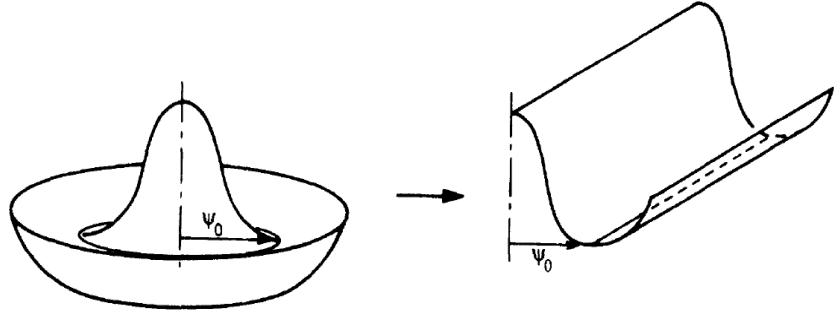}
\vspace{3mm}
\caption{Linearly realized $O(N)$-symmetry. Here, $\psi_{0}=\alpha$ is the
order-parameter and the trough is understood to be $N-1$ dimensional.
From Weichman (1988).}
\label{fig:trough}
\end{center}
\end{figure}

\bigskip

A promising RG approach employing the linear basis in
Eq. (\ref{eq:ssb_basis}) but additionally invoking constraints from Ward identities has recently
been put forward in (Pistolesi 1997, 2004). How one might smoothly connect the linear basis with the phase-amplitude basis via scale-dependent transformations is shown in Diener (2008).

\section{Method}
\label{sec:qcp_gold_method}

In this section, we present the functional RG method and our truncation for the
symmetry-broken phase. 
We have explained in subsection \ref{subsec:ssb_rg} in detail how to account for spontaneous
symmetry-breaking with the exact
flow equation for the effective action
(Wetterich 1993, Berges 2002, Salmhofer 2001, Metzner 2005):
\begin{eqnarray}
\frac{d}{d\Lam} \Gam^{\Lam}[\mathbf{\Phi}] =
 {\rm Tr} \, \frac{\dot{\bR}^{\Lam}}
 {\bGam^{(2) \, \Lam}[\mathbf{\Phi}] + \bR^{\Lam}} \;,
\label{eq:gold_exact_flow}
\end{eqnarray}
where $\bGam^{(2) \, \Lam}=\partial^{2}\Gam^{\Lam}[\mathbf{\Phi}]/\partial \mathbf{\Phi}^{2}$
is the second functional derivative with respect to the order parameter field
and the trace (Tr) traces over all indices (see section \ref{sec:flow_eq} for a detailed
derivation). When expanding this equation around a non-vanishing minimum $\alpha$ of the
effective action, additional terms will arise in the flow equations for the 
n-point vertex involving the scale-derivative of the order-parameter, $\dot{\alpha}$,
and the n+1-point vertex, see Fig. \ref{fig:ssb_hierarchy}.

The objective is now to compute the flow of the independent parameters $\alpha$, $u$, 
$Z$, and $Y$ of the $\sg-\mathbf{\Pi}$ model. A convenient cutoff function $\bR^{\Lam}$ 
that respects the $O(N)$-symmetry is a self-energy independent 
sharp cutoff for space-like momenta:
\begin{equation}
 \bR^{\Lam}(k) = [\bG_{0}(k)]^{-1} -
 [\chi^{\Lam}(\mathbf{k}) \, \bG_{0}(k)]^{-1}
\end{equation}
where $\chi^{\Lam}(\mathbf{k}) = \Theta(|\mathbf{k}| - \Lam)$.
This term replaces the bare propagator $\bG_{0}$ by
$\bG_{0}^{\Lam} = \chi^{\Lam} \bG_{0}$.

For a sharp momentum cutoff the momentum variable running
around the loop is pinned by $\bG'_R(\mathbf{k})$ to $|\mathbf{k}| = \Lam$ 
as the socalled single-scale propagator ${\bG'_R}^{\!\Lam}$ has
support only for momenta at the cutoff, that is, for
$|\mathbf{k}| = \Lam$.
Hence the momentum integral can be performed analytically.
The problem that the integrand contains also step functions
$\chi^{\Lam}(\mathbf{k}) = \Theta(|\mathbf{k}|-\Lam)$ can be treated
by using the identity
\begin{eqnarray}
\int dx \, \delta(x-x_0) \, f[x,\Theta(x-x_0)] =
\int_0^1 du \, f(x_0,u)\;,
\end{eqnarray}
which is valid for any continuous function $f$.
More specifically, in the present case the one-loop diagrams
are evaluated for vanishing external momenta and depend only on the modulus 
$|\mathbf{k}|$, 
such that
all internal propagators carry the same momentum and 
one can use the identity
\begin{equation}
 n \int \frac{d^d k}{(2\pi)^{d}} \, \bG'_{R}(k) \, \bA \,
 [\bG_{R}(k) \, \bA]^{n-1} = \Lam^{d-1} K_{d}
[\bG(k) \, \bA]^n|_{|\mathbf{k}|= \Lam} \, ,
\label{eq:gold_cutoff_identity}
\end{equation}
valid for any matrix $\bA$ and with $K_{d}$ being defined by $\int\frac{d^{d}k}{\left(2\pi\right)^{d}}
=K_{d}\int d|k|\, |k|^{d-1}$. The factor $n$ corresponds to the $n$ possible
choices of positioning $\bG_{R}'$ in a loop with $n$ lines.
For loop integrals in the flow equations, we use the
short-hand notation
\begin{equation}
 \int_{k|\Lam} = \int \frac{d k_{0}}{(2\pi)} 
\int \frac{d^d k}{(2\pi)^{d}}\delta(|\mathbf{k}|-\Lam)\;.
\end{equation}
At finite temperatures, the continuous frequency integration becomes 
a discrete Matsubara sum: $\int \frac{d k_{0}}{2\pi}\rightarrow
T\sum_{\Omega_{n}}$, with $\Omega_{n}=2\pi n T$.

\section{Finite temperature phase boundary in three dimensions}
\label{sec:qcp_gold_finite}

The primary objective of this section is to extend the finite
temperature work presented in chapter \ref{chap:bosonicqcp_discrete}
to systems with Goldstone modes. Most importantly, we
compute the shift-exponent $\psi$ of the finite temperature phase
boundary in $d=3$.

On the phase boundary at $T_{c}$, the effective potential is flat and 
there exists only one bosonic excitation. Following Tetradis and Wetterich (1994), 
we set $Y=0$ in this section and therefore: $Z_{\sg}=Z_{\pi}=Z$. This
truncation gives the best results for critical
exponents and correctly reflects that the phase transition becomes mean-field like 
with vanishing anomalous dimension in the large $N$ limit. 
For an extended discussion we refer to section 9 of Tetradis (1994).

\subsection{Flow equations}

As demanded by the condition of a vanishing bosonic one-point vertex for all
scales, see Eq. (\ref{eq:alpha_dot_graph}), the square of the order parameter 
obeys the flow equation
\begin{wrapfigure}{r}{0.475\textwidth}
\vspace{-3mm}
\begin{fmffile}{20080831_gold_47}
\begin{eqnarray}
\rho &:&\hspace{5mm}
\parbox{30mm}{\unitlength=1mm\fmfframe(2,2)(1,1){
\begin{fmfgraph*}(15,45)\fmfpen{thin} 
\fmfleft{i1}\fmfright{o1}
\fmftop{t1}
\fmfpolyn{empty,tension=0.2}{T}{3}
\fmfpolyn{phantom,tension=100.}{B}{3}
\fmf{phantom,straight,tension=1.2}{i1,B1}
\fmf{phantom,straight,tension=1.2}{o1,B2}
\fmf{dashes,tension=0.6}{B3,T3}
\fmf{dashes,tension=0.2,right=0.8}{T1,t1}\fmf{dashes,tension=0.2,left=0.8}{T2,t1}
\end{fmfgraph*}
}}
\hspace{-6mm}
+
\hspace{5mm}
\parbox{30mm}{\unitlength=1mm\fmfframe(2,2)(1,1){
\begin{fmfgraph*}(15,45)\fmfpen{thin} 
\fmfleft{i1}\fmfright{o1}
\fmftop{t1}
\fmfpolyn{empty,tension=0.2}{T}{3}
\fmfpolyn{phantom,tension=100.}{B}{3}
\fmf{phantom,straight,tension=1.2}{i1,B1}
\fmf{phantom,straight,tension=1.2}{o1,B2}
\fmf{dashes,tension=0.6}{B3,T3}
\fmf{photon,tension=0.2,right=0.8}{T1,t1}\fmf{photon,tension=0.2,right=0.8}{t1,T2}
\end{fmfgraph*}
}}
\hspace{-10mm}
\nonumber\\[-18mm]
u &:&
\hspace{0mm}
\parbox{25mm}{\unitlength=1mm\fmfframe(2,2)(1,1){
\begin{fmfgraph*}(25,17)
\fmfpen{thin} 
\fmfleftn{l}{2}\fmfrightn{r}{2}
\fmfrpolyn{full,tension=0.9}{G}{4}
\fmfpolyn{full,tension=0.9}{K}{4}
\fmf{dashes}{l1,G1}\fmf{dashes}{l2,G2}
\fmf{dashes}{K1,r1}\fmf{dashes}{K2,r2}
\fmf{dashes,left=.5,tension=.2}{G3,K3}
\fmf{dashes,right=.5,tension=.2}{G4,K4}
\end{fmfgraph*}
}}
\hspace{4mm}
+
\parbox{25mm}{\unitlength=1mm\fmfframe(2,2)(1,1){
\begin{fmfgraph*}(25,17)
\fmfpen{thin} 
\fmfleftn{l}{2}\fmfrightn{r}{2}
\fmfrpolyn{shaded,tension=0.9}{G}{4}
\fmfpolyn{shaded,tension=0.9}{K}{4}
\fmf{dashes}{l1,G1}\fmf{dashes}{l2,G2}
\fmf{dashes}{K1,r1}\fmf{dashes}{K2,r2}
\fmf{photon,left=.5,tension=.2}{G3,K3}
\fmf{photon,right=.5,tension=.2}{G4,K4}
\end{fmfgraph*}
}}
\nonumber\\[0mm]
Z &:&\hspace{0mm}
\parbox{25mm}{\unitlength=1mm\fmfframe(2,2)(1,1){
\begin{fmfgraph*}(24,24)\fmfpen{thin}
\fmfleft{l1}
 \fmfright{r1}
 \fmfpolyn{full,tension=0.4}{G}{3}
 \fmfpolyn{full,tension=0.4}{K}{3}
  \fmf{photon}{l1,G1}
   \fmf{photon,straight,tension=0.5,left=0.}{K3,G2}
  \fmf{dashes,tension=0.,left=0.7}{G3,K2}
  \fmf{photon}{K1,r1}
 \end{fmfgraph*}
}}\nonumber\\[-16mm]\nonumber
\end{eqnarray}
\end{fmffile}
\caption{\textit{Feynman diagrams for the flow
 equations (\ref{eq:qcp_gold_rho}-\ref{eq:qcp_gold_Z}).}}
\label{fig:diags_gold1} 
\vspace{-23mm}
\end{wrapfigure}
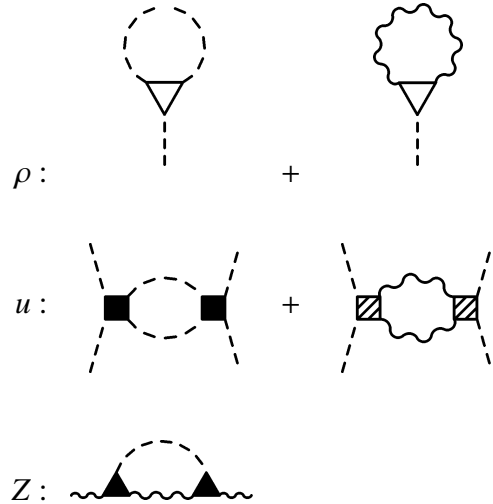
\begin{eqnarray}
\partial_{\Lambda}\rho=
-\int_{q|\Lam}3 G_{\sg}(q) + (N-1)\,G_{\pi}(q)\;,\nonumber\\
\label{eq:qcp_gold_rho}
\end{eqnarray}
The bosonic self-interaction is driven by
\begin{eqnarray}
\partial_{\Lambda}u=
u^{2}\int_{q|\Lam}9 G^{2}_{\sg}(q) + (N-1)\,G^{2}_{\pi}(q)\;.\nonumber\\
\label{eq:qcp_gold_phi4}
\end{eqnarray}
Lastly, the bosonic field renormalization is determined
by the expression
\begin{eqnarray}
\partial_{\Lambda}Z=
u^{2}\rho\int_{q|\Lam}
\frac{\partial^{2}_{\mathbf{k}}}{2}\left[G_{\pi}(q+k)G_{\sg}(q)\right]_{k=0}\;,
\nonumber\\
\label{eq:qcp_gold_Z}
\end{eqnarray}
where $\partial_{\mathbf{k}}^2=\frac{1}{d}\sum_{i=1}^{d}\partial_{k_i}^2$.
The flow equations (\ref{eq:qcp_gold_rho}-\ref{eq:qcp_gold_Z})
are shown in terms of Feynman diagrams in Fig. 
\ref{fig:diags_gold1}.

We split the Matsubara summations into their classical and quantum
parts
\begin{eqnarray}
T\sum_{\Omega_{n}}f(\Omega_{n})=T f(\Omega_{n}=0)
+T\sum_{\Omega_{n}\neq 0}^{(\text{quant})}f(\Omega_{n})\;,
\end{eqnarray}
and utilize the following variables,
\begin{eqnarray}
\tilde{\rho}&=&\frac{Z\,\rho}{\Lambda^{d-2}T K_{d}}\nonumber\\[2mm]
\tilde{u}&=&\frac{ K_{d} T \, u }{Z^{2} \Lambda^{4-d}}\;,
\label{eq:gold_finiteT_variables}
\end{eqnarray}
as well as a rescaled temperature and anomalous dimension
\begin{eqnarray}
\tilde{T}&=&\frac{2\pi T}{\Lam}\nonumber\\[2mm]
\eta&=&-\frac{d \log Z}{d \log \Lam}\;.
\label{eq:gold_eff_temp}
\end{eqnarray}
We therewith obtain for the squared order-parameter and the
quartic coupling:
\begin{eqnarray}
\frac{d\tilde{\rho}}{d\log \Lambda}&=&\left(2-d-\eta\right)\tilde{\rho}
+\frac{3}
{\left(1+\tilde{u}\tilde{\rho}\right)}+ \left(N-1\right)
+ 2\sum_{n=1}^{\infty}
\frac{3}
{\left( \left(n \tilde{T}\right)^{2}+ 1+ \tilde{u}\tilde{\rho}\right)^{2}}+\nonumber\\
&&\hspace{71mm}
\frac{N-1}
{\left(\left(n \tilde{T}\right)^{2}+ 1\right)^{2}}\nonumber\\
\frac{d\tilde{u}}{d\log \Lambda}&=&\left(d-4+2\eta\right)\tilde{u}
+
\tilde{u}^{2}
\Bigg[
\frac{9} {\left(1+\tilde{u}\tilde{\rho}\right)^{2}}+
\left(N-1\right) + 2\sum_{n=1}^{\infty} \frac{9}
{\left(\left( n \tilde{T}\right)^{2}+ 1+
\tilde{u}\tilde{\rho}\right)^{2}} + \nonumber\\
&&\hspace{81mm}
\frac{N-1}
{\left(\left(n \tilde{T}\right)^{2}+ 1\right)^{2}}
\Bigg]
\;.\nonumber\\
\label{eq:gold_finiteT_u_rho}
\end{eqnarray}
Terms proportional to $N-1$ outside the summation stem from classical Goldstone fluctuations. 
But the Goldstone mode also yields new quantum terms proportional to $N-1$ inside the summation.
The anomalous dimension is determined by
\begin{eqnarray}
\eta =\tilde{u}^{2}\tilde{\rho}
&&\Bigg[
\frac{1}
{\left(1+\tilde{u}\tilde{\rho}\right)}
+ 2\sum_{n=1}^{\infty}
\frac{1}
{\left(n \tilde{T}\right)^{2}+ 1+ \tilde{u}\tilde{\rho}}
\frac{1}
{\left(\left(n \tilde{T}\right)^{2}+ 1\right)^{2}}
\Bigg]
\label{eq:gold_finiteT_eta}
\end{eqnarray}
The Matsubara summations can be performed analytically
yielding hyperbolic trigonometric
functions. The rather complex expressions do not
deliver any additional insights.

\subsection{Classical fixed point}
\label{subsec:heisenberg}
In the classical limit, considering only the zeroth Matsubara
frequency/neglecting the summation terms, our equations (\ref{eq:gold_finiteT_u_rho},
\ref{eq:gold_finiteT_eta}) describe the correct classical critical
behavior of the Heisenberg universality class for $N=3$.
Inspection of the first term of $\frac{d \tilde{\rho}}{d \Lam}$ in 
Eq. (\ref{eq:gold_finiteT_u_rho}) yields the lower critical dimension
$d^{-}_{c}=2$ for $N>1$ in agreement with the Mermin-Wagner
theorem. The flow equations therefore correctly account for the 
fact that Goldstone fluctuations suppress $T_{c}$ to zero in two dimensions. 
The upper critical dimension is $d_{c}^{+}=4$
from the first term in $\frac{d \tilde{u}}{d \Lam}$, as expected.
The fixed point value for the anomalous exponent in three dimensions 
comes out as: $\eta=0.058$ to be compared with
the known value $\eta=0.036$ from other methods (Berges 2002).

\subsection{Shift exponent $\psi$}
\label{subsec:gold_shift}

The shift exponent determines how the critical temperature varies with control parameter
in the immediate vicinity of the QCP:
\begin{eqnarray}
T_{c}\propto\left(\delta-\delta_{\text{crit}}\right)^{\psi}\;.
\end{eqnarray}
We compute $\psi$ employing the same procedure as explained 
below Eq. (\ref{Lambda_cl}) in chapter \ref{chap:bosonicqcp_discrete}.
The numerical flows look identical to those of the previous section in Fig. \ref{fig:u_rho_flows_z_1} except that the scaling plateaus form 
at the fixed point values of the Heisenberg universality class for $N=3$. 
We find that the Ginzburg scale decreases with decreasing $T_{c}$ as:
\begin{eqnarray}
\Lam_{G}\propto T_{c}\;,
\label{eq:gold_ginz}
\end{eqnarray}
fitting the formula $\Lam_{G}\propto T_{c}^{1/(4-d)}$ derived in the previous chapter, c.f.
Eq. (\ref{eq:Ginzburg_scale}) for $d=3$. 
For the shift exponent, we find
\begin{eqnarray}
\psi \approx 1/2\;,
\end{eqnarray}
which matches the formula derived by
Millis (1993): $\psi=\frac{z}{d+z-2}$ for $d=3$ and $z=1$. 
We have verified that for any value of $N$, the shift-exponent attains the Millis value 
$1/2$ derived in the symmetric phase without accounting for Goldstone bosons. 

\bigskip

Therefore, the shift-exponent for three dimensional QCPs where a bosonic description is appropriate does not depend on the number of Goldstone bosons. 
This is the main result of this section.

\newpage

\section{Infrared asymptotics in the symmetry-broken phase}
\label{sec:gold_IR}

Away from the QCP and the phase boundary, the system is characterized 
by a finite minimum in the effective potential. 
In this section, we clarify the singularity structure of the $\sg-\mathbf{\Pi}$ model 
in the limit of vanishing 
cutoff $\Lam\rightarrow 0$ in the presence of a finite order parameter $\rho$. 
We will compute at zero temperature but the obtained results for the infrared properties 
may be transferred to finite 
temperature by increasing spatial dimensionality by one. 
We establish the scaling properties of the various 
parameters and show that the Goldstone propagator and longitudinal propagator exhibit 
very different momentum scaling, in stark contrast to the 
phase boundary where both modes become degenerate.

Technically, it is crucial to consider the $Y$-term which distinguishes the 
Goldstone field renormalization $Z_{\pi}$ from that of the longitudinal mode 
$Z_{\sg}$, see Eq. (\ref{eq:Y_relations}). 


\subsection{Flow equations}

The flow equations for $\rho$ and $u$ remain unchanged from Eqs. 
(\ref{eq:qcp_gold_rho},\ref{eq:qcp_gold_phi4}). 
The flow equation for $G_{\sg}^{-1}(k)$ contains terms from the 
four diagrams shown in Fig. \ref{fig:diags_gold2} and has the form:
\begin{eqnarray}
\partial_{\Lambda}G_{\sg}^{-1}(k)&=&
\frac{\rho}{2}
\int_{q|\Lam}
\left(u+Y q^{2}\right)^{2}
\left[9 G_{\sg}(q+k)G_{\sg }(q) +
(N-1)\,G_{\pi}(q+k) G_{\pi}(q)\right]\nonumber\\
&&
+\int_{q|\Lam} 
\frac{Y q^{2}}{2}
\left[(N-1)G_{\pi}(q+k)+
3 G_{\sg}(q+k)\right]\;.
\label{eq:qcp_gold_Zsig}
\end{eqnarray}

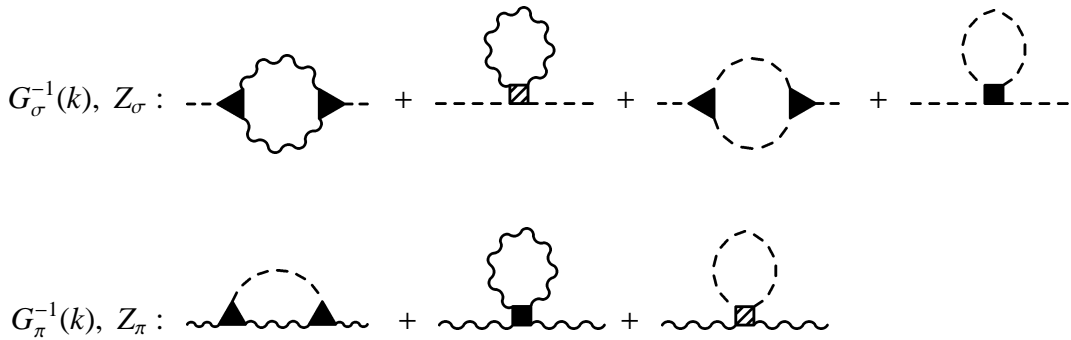
\begin{figure}[b]
\begin{fmffile}{20081227_Z_5}
\begin{eqnarray}
G_{\sg}^{-1}(k),\; Z_{\sigma}
&:&
\parbox{25mm}{\unitlength=1mm\fmfframe(2,2)(1,1){
\begin{fmfgraph*}(25,23)\fmfpen{thin} 
 \fmfleft{l1}
 \fmfright{r1}
 \fmfpolyn{full,tension=0.3}{G}{3}
 \fmfpolyn{full,tension=0.3}{K}{3}
  \fmf{dashes}{l1,G1}
  \fmf{photon,tension=0.2,right=0.8}{G2,K3}
 \fmf{photon,tension=0.2,right=0.8}{K2,G3}
 \fmf{dashes}{K1,r1}
 \end{fmfgraph*}
 }}
\;\;\;
+
\parbox{25mm}{\unitlength=1mm\fmfframe(2,2)(1,1){
\begin{fmfgraph*}(22,25)\fmfpen{thin} 
 \fmfleft{l1}
 \fmfright{r1}
 \fmftop{v1}
 \fmfpolyn{shaded}{G}{4}
 \fmf{dashes,straight}{l1,G4}
 \fmf{dashes,straight}{G1,r1}
 \fmffreeze
\fmf{photon,tension=0.1,right=0.7}{G2,v1}
\fmf{photon,tension=0.1,right=0.7}{v1,G3}
\end{fmfgraph*}
}}
+
\parbox{25mm}{\unitlength=1mm\fmfframe(2,2)(1,1){
\begin{fmfgraph*}(25,23)\fmfpen{thin}
 \fmfleft{l1}
 \fmfright{r1}
 \fmfpolyn{full,tension=0.3}{G}{3}
 \fmfpolyn{full,tension=0.3}{K}{3}
  \fmf{dashes}{l1,G1}
 \fmf{dashes,tension=0.2,right=0.8}{G2,K3}
 \fmf{dashes,tension=0.2,right=0.8}{K2,G3}
 \fmf{dashes}{K1,r1}
 \end{fmfgraph*}
}}\hspace{4mm}
+
\parbox{25mm}{\unitlength=1mm\fmfframe(2,2)(1,1){
\begin{fmfgraph*}(22,25)\fmfpen{thin} 
 \fmfleft{l1}
 \fmfright{r1}
 \fmftop{v1}
 \fmfpolyn{full}{G}{4}
 \fmf{dashes,straight}{l1,G4}
 \fmf{dashes,straight}{G1,r1}
 \fmffreeze
\fmf{dashes,tension=0.1,right=0.7}{G2,v1}
\fmf{dashes,tension=0.1,right=0.7}{v1,G3}
\end{fmfgraph*}
}}
\nonumber\\[0mm]
G_{\pi}^{-1}(k),\;Z_{\pi} &:&\hspace{0mm}
\parbox{25mm}{\unitlength=1mm\fmfframe(2,2)(1,1){
\begin{fmfgraph*}(24,24)\fmfpen{thin}
\fmfleft{l1}
 \fmfright{r1}
 \fmfpolyn{full,tension=0.4}{G}{3}
 \fmfpolyn{full,tension=0.4}{K}{3}
  \fmf{photon}{l1,G1}
   \fmf{photon,straight,tension=0.5,left=0.}{K3,G2}
  \fmf{dashes,tension=0.,left=0.7}{G3,K2}
  \fmf{photon}{K1,r1}
 \end{fmfgraph*}
}}
\hspace*{4mm}
+
\parbox{25mm}{\unitlength=1mm\fmfframe(2,2)(1,1){
\begin{fmfgraph*}(22,25)\fmfpen{thin} 
 \fmfleft{l1}
 \fmfright{r1}
 \fmftop{v1}
 \fmfpolyn{full}{G}{4}
 \fmf{photon,straight}{l1,G4}
 \fmf{photon,straight}{G1,r1}
 \fmffreeze
\fmf{photon,tension=0.1,right=0.7}{G2,v1}
\fmf{photon,tension=0.1,right=0.7}{v1,G3}
\end{fmfgraph*}
}}
+
\parbox{25mm}{\unitlength=1mm\fmfframe(2,2)(1,1){
\begin{fmfgraph*}(22,25)\fmfpen{thin} 
 \fmfleft{l1}
 \fmfright{r1}
 \fmftop{v1}
 \fmfpolyn{shaded}{G}{4}
 \fmf{photon,straight}{l1,G4}
 \fmf{photon,straight}{G1,r1}
 \fmffreeze
\fmf{dashes,tension=0.1,right=0.7}{G2,v1}
\fmf{dashes,tension=0.1,right=0.7}{v1,G3}
\end{fmfgraph*}
}}
\nonumber\\[-16mm]\nonumber
\end{eqnarray}
\end{fmffile}
\caption{\textit{Feynman diagrams for Eqs. (\ref{eq:qcp_gold_Zsig},\ref{eq:qcp_gold_Zgold}).}}
\label{fig:diags_gold2}
\end{figure}

The flow equation for $G_{\pi}^{-1}(k)$ corresponding to the three diagrams in the 
second line of Fig. \ref{fig:diags_gold2} reads:
\begin{eqnarray}
\partial_{\Lambda}G^{-1}_{\pi}(k)&=&
\int_{q|\Lam}
\left(
\frac{\rho}{2}\left(\frac{u+Yq^{2}}{2}\right)^{2}3 G_{\sg}(q)
+
3
\frac{Y q^{2}}{2}
\right)
G_{\pi}(q+k)
+\int_{q|\Lam} 
\frac{Y q^{2}}{2}
G_{\sg}(q+k)\;,\nonumber\\
\label{eq:qcp_gold_Zgold}
\end{eqnarray}
where the factor of three in front of $G_{\sg}(q)$ in the first term arises from 
three different contractions of the first diagram in the second line 
of Fig. \ref{fig:diags_gold2} as drawn in Eq. (4.36) of Pistolesi, et al. (2004).
One obtains the flow for $Z_{\sg}$ and $Z_{\pi}$ by applying the Laplacian $\partial^{2}_{\mathbf{k}}/2$ evaluated at $|\mathbf{k}|=0$ to both sides of 
Eqs. (\ref{eq:qcp_gold_Zsig}, \ref{eq:qcp_gold_Zgold}), respectively.
The flow
of $Y$ follows then from the scale derivative of the relation:
\begin{eqnarray}
Y=\frac{Z_{\sg}-Z_{\pi}}{\rho}\;.
\label{eq:Y}
\end{eqnarray}

\subsection{Analytical results}

We now identify the leading singularities in the flow equations for $u$, $Z_{\sg}$, 
and $Z_{\pi}$ to determine their scaling behavior in the infrared, that is, 
in the limit of vanishing cutoff $\Lam\rightarrow 0$. 
The order parameter $\rho$ is finite and does not scale to zero as we are now in the symmetry-broken phase 
away from the phase boundary and the QCP.

For the quartic coupling $u$, 
the diagram with two Goldstone propagators in Fig. \ref{fig:diags_gold1} is the most singular, leading 
to the expression:
\begin{eqnarray}
\partial_{\Lambda}u=
u^{2}\int_{q|\Lam}(N-1)\,G^{2}_{\pi}(q)\;.
\label{eq:qcp_gold_phi4_IR}
\end{eqnarray}
This equation can be integrated analytically and yields in $d=2$ at zero temperature:
\begin{eqnarray}
u\rightarrow Z_{\pi}^{2} \Lam\;,
\label{eq:u_IR}
\end{eqnarray}
provided $N>1$. As the longitudinal mass is proportional to the quartic coupling, 
see Eq. (\ref{eq:Y_relations}), it also vanishes linearly as a function of scale. 
In three dimensions, the linear behavior becomes logarithmic. 

For $Z_{\sg}$, the first diagram in the first line of Fig. 
\ref{fig:diags_gold2} contains the leading singularity:
\begin{eqnarray}
\partial_{\Lambda}Z_{\sg}&=&
\frac{u^{2}\rho}{2}(N-1)
\int_{q|\Lam}
\frac{\partial_{\mathbf{k}}^2}{2}\left[
\,G_{\pi}(q+k) G_{\pi}(q)\right]_{k=0}\;,
\label{eq:qcp_gold_Zsig_IR}
\end{eqnarray}
Upon inserting Eq. (\ref{eq:u_IR}), we obtain in two dimensions
\begin{eqnarray}
Z_{\sg}\rightarrow \frac{Z_{\pi}^{2}\rho}{\Lam}\;.
 \label{eq:Zsig_IR}
\end{eqnarray}
In three dimensions $Z_{\sg}$ diverges only logarithmically. 
As we will demonstrate below, $Z_{\pi}$ remains finite and therefore 
by virtue of Eq. (\ref{eq:Y}), we obtain for the $Y$-term:
\begin{eqnarray}
Y\sim\frac{Z_{\sg}}{\rho}\rightarrow\frac{Z_{\pi}^{2}}{\Lam}\;,
\label{eq:Y_IR}  
\end{eqnarray}
where we have used Eq. (\ref{eq:Zsig_IR}). 
What remains to be shown is that all singularities for the Goldstone propagator 
cancel such that $Z_{\pi}$ is rendered finite. 
The leading term for the Goldstone propagator from 
Eq. (\ref{eq:qcp_gold_Zgold}) can be written:
\begin{eqnarray}
\partial_{\Lambda}G^{-1}_{\pi}(k)&=&
\int_{q|\Lam}
\gamma_{\pi^{4}}^{\text{eff}}(q)
G_{\pi}(q+k)\;,
\end{eqnarray}
where the effective momentum-dependent 
self-interaction among the Goldstone bosons contains two terms:
\begin{eqnarray}
\gamma_{\pi^{4}}^{\text{eff}}(q)=
\tilde{\gamma}_{\pi^{4}}(q)+\tilde{\gamma}^{2}_{\sigma\pi^{2}}G_{\sigma}(q)\;,
\label{eq:pi_4_eff}
\end{eqnarray}
with $\tilde{\gamma}_{\pi^{4}}(q)=\frac{3}{2}Y q^{2}$ and 
$\tilde{\gamma}_{\sigma\pi^{2}}=\frac{3}{2}\rho \left(\frac{u+Y q^{2}}{2}\right)^{2}$. 
Since $G_{\sigma}(q)$ is negative, the interaction among 
Goldstone bosons from exchange of longitudinal fluctuations has the opposite sign 
of the direct interaction $\tilde{\gamma}_{\pi^{4}}$ (Pistolesi 2004).
By inserting the longitudinal propagator from Eq. 
(\ref{eq:qcp_gold_bose_props}) and the obtained scale 
dependences of Eqs. (\ref{eq:u_IR}, \ref{eq:Zsig_IR}, \ref{eq:Y_IR}) 
into Eq. (\ref{eq:pi_4_eff}), we obtain as the 
central result of this section:
\begin{eqnarray}
\gamma_{\pi^{4}}^{\text{eff}}(q)\rightarrow 0,
\label{eq:effgold_interaction}
\end{eqnarray}
therefore liberating the corrections to the Goldstone propagator from singularities 
and ensuring the finiteness of $Z_{\pi}$ for $d>1$ ($d>2$) at zero (finite) temperature. 
In the important work by 
Pistolesi, Castellani, et al. (2004), this proof has been extended to arbitrary 
loop order for the case of an interacting Bose gas.

We have established 
the following infrared $(\Lam\rightarrow 0)$ behavior for the longitudinal propagator at zero temperature:
\begin{eqnarray}
 G_{\sg}(\Lam) \propto & \frac{-1}{\Lam} & \mbox{for} \; d=2\nonumber\\
 G_{\sg}(\Lam) \propto &\; \log {\Lam}\; & \mbox{for} \; d=3\;,
\label{eq:G_sig}
\end{eqnarray}
while the Goldstone propagator remains quadratic:
\begin{eqnarray}
 G_{\pi}(\Lam) \propto &\; \frac{-1}{\Lam^{2}}\; & \mbox{for} \; d=2,3 \;,
\label{eq:G_pi}
\end{eqnarray}
and the bosonic self-interaction scales as:
\begin{eqnarray}
 u(\Lam) \propto & \Lam  & \mbox{for} \; d=2\nonumber \\
 u(\Lam) \propto &\; -\frac{1}{\log {\Lam}}\; & \mbox{for} \; d=3\;.
\label{eq:self}
\end{eqnarray}
These results may be transferred to finite temperatures by incrementing 
spatial dimensionality $d$ by one.

\section{Conclusion}
\label{sec:qcp_gold_summary}

This chapter was devoted to an analysis of the effects of Goldstone modes in the 
vicinity of quantum critical points where a description in terms of bosonic fields is apppropriate. Within the functional RG framework,
we derived flow equations for the relativistic $\sg-\mathbf{\Pi}$ model with dynamical exponent $z=1$ suitable for the ordered phase at zero and finite temperature.

First, we computed the shift-exponent $\psi$ characterizing the shape of the phase boundary at finite temperature in three dimensions and found that $\psi$ is independent of the number of Goldstone bosons in three dimensions.

We then clarified the infrared properties of the Goldstone and longitudinal propagator as well as the bosonic self-interaction away from the phase boundary. We showed that the longitudinal propagator and the self-interaction are strongly renormalized. The Goldstone propagator, on the other hand, remains quadratic in momenta as the effective interaction among Goldstone bosons flows to zero due to symmetries. 

\bigskip

In the future, it would be worthwile to apply an extended version of the 
flow equations devised in this chapter to quantum criticality in the Kosterlitz-Thouless universality class (see also Outlook \ref{subsec:kt}) and to compute, for example, the shift-exponent for the Kosterlitz-Thouless temperature.

\chapter[Fermi-Bose renormalization group for quantum critical fermion systems]
{Fermi-Bose renormalization group for quantum critical fermion systems}
\label{chap:fermibosetoy}

\section{Introduction}

As noticed already more than 30 years ago by Hertz (1976),
quantum phase transitions in correlated fermion systems
may be described in terms of the order parameter field alone when one integrates
out fermions from the path integral in one stroke and subsequently deals with an
exclusively bosonic theory. The resultant effective action is expanded in powers
of the bosonic field $\phi$ and often truncated after the $\phi^{4}$-term. Although an
analysis in terms of $\phi^{4}$-type theories seems mundane, the presence of \emph{two}
relevant energy scales, one given by temperature and the other given by a non-thermal
control parameter acting as a mass term in the bosonic propagator, gives rise to a rich
finite temperature phase diagram (Millis 1993).

The Hertz-Millis approach relies on integrating out fermions first.
In general, however, the zero temperature fermion propagator is gapless and consequently
may lead to singular coefficients in the effective bosonic action (Belitz 2005, Loehneysen 2007).
In such cases, it is obligatory to keep the gapless fermions in the theory and
consider them on equal footing to the bosons as performed in several works with
(resummed) perturbation theory (Altshuler 1995, Abanov 2000, Abanov 2003, Rech 2006).

Recently, various quantum critical exponents for U(1) gauge theories with Dirac fermions 
and complex-valued bosonic fields in two spatial dimensions were computed in the limit 
of a large number of fermion and boson species $N_{f}$ and $N_{b}$ (Kaul 2008). However,
the complicated interplay of two singular propagators promotes a
controlled perturbative treatment
to a formidable task (Rech 2006) and represents a clear calling for the renormalization group 
(RG).

\bigskip

When computing RG flows for coupled fermion-boson theories, where the propagator 
of the zero temperature fermions becomes unbounded for 
momenta on the Fermi surface and the propagator for massless bosons is singular 
at the origin of momentum space, it is crucial to \emph{synchronize} the
evolution of correlation functions so that both singularities are reached
simultaneously at the end of the flow.

\begin{wrapfigure}{r}{0.5\textwidth}
  \vspace{-10mm}
  \begin{center}
\hspace*{-4mm}
\includegraphics*[width=67mm,angle=-90]{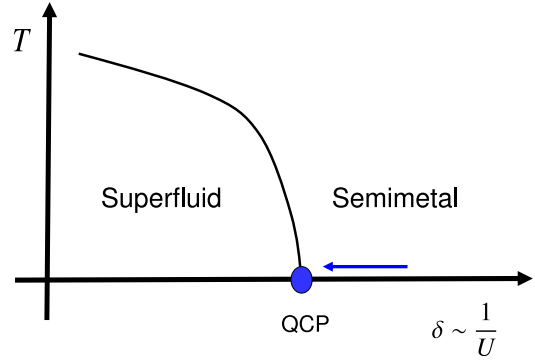}
\end{center}
\vspace{-10mm}
\caption{\textit{Schematic phase diagram of attractive Dirac fermions. 
The QCP at a critical interaction strength $U_{c}$ separates the semimetal from 
the superfluid. In the present chapter, we approach
the QCP at $T=0$ from the semimetallic phase indicated with an arrow}.}
\label{fig:phase_toy}
\vspace{-2mm}
\end{wrapfigure}

Within the functional RG framework formulated for fermionic and bosonic fields
(Berges 2002, Baier 2004, Sch\"utz 2005, Strack 2008), this poses no particular problem as
the regulator functions for fermions and bosons can be chosen to synchronize both
types of fluctuations. The mutual feedback of gapless fermions coupled to massless bosons
has been studied already with functional flow equations in Quantum Electrodynamics 
(Gies 2004), non-abelian gauge theories (Pawlowski 2004), and the Gross-Neveu model 
(Rosa 2001).

\bigskip

In the present chapter, we study a model of attractive Dirac fermions relevant 
for neutral Graphene and cold atoms in the half-filled 
honeycomb lattice (Zhao 2006, Castro Neto 2007, 
Castro Neto 2008) exhibiting a quantum phase transition 
from a semimetal to a superfluid as shown in Fig. \ref{fig:phase_toy}. 
The flow equations are analytically transparent and a simple truncation yields the complete 
set of quantum critical exponents. As a central result, the fermion and order parameter two-point correlation functions develop non-analytic dependences on frequency and momentum 
at the QCP.

\bigskip

In section \ref{sec:fermi_bose_model}, we introduce the \emph{Dirac cone model} and show 
that the mean-field theory of this model leads to a semimetal-to-superfluid quantum phase 
transition at a critical interaction strength $U_{c}$. 
In section \ref{sec:pp_method}, the RG method, truncation and flow equations are
presented. Results for the quantum critical behavior follow in section \ref{sec:sol_qcp}.
We finally summarize and conclude in section \ref{sec:pp_conclusion}.

\section{Dirac cone model}\,\,\,\,
\label{sec:fermi_bose_model}
We consider an attractively interacting
Dirac fermion system with the \emph{bare action} 
\begin{eqnarray}
  &&\Gam_0[\psi,\psib]
   =  - \int_{k\sg}
  \psib_{k\sg} (ik_{0}-\xi_{\bk}) \, \psi_{k\sg} 
+ \int_{k,k',q} U \,
  \psib_{-k+\frac{q}{2}\down} \psib_{k+\frac{q}{2} \up}
  \psi_{k'+\frac{q}{2}\up} \psi_{-k'+\frac{q}{2}\down} \; ,\nonumber\\
  \label{eq:bare_dirac_action}
\end{eqnarray}
where the variables $k = (k_0,\bk)$ and $q = (q_0,\bq)$
collect the zero temperature Matsubara energies and momenta, and we use the short-hand notation
$\int_k = \int_{k_0} \int_{\bk} =
 \int_{-\infty}^{\infty} \frac{d k_{0}}{2\pi}
 \int \frac{d^d \bk}{(2\pi)^d} \,$
for momentum and energy integrals,
and $\int_{k\sg}$ includes also a spin sum.
\begin{wrapfigure}{r}{0.5\textwidth}
  \vspace{2mm}
  \begin{center}
\hspace*{-4mm}
\includegraphics*[width=62mm,angle=0]{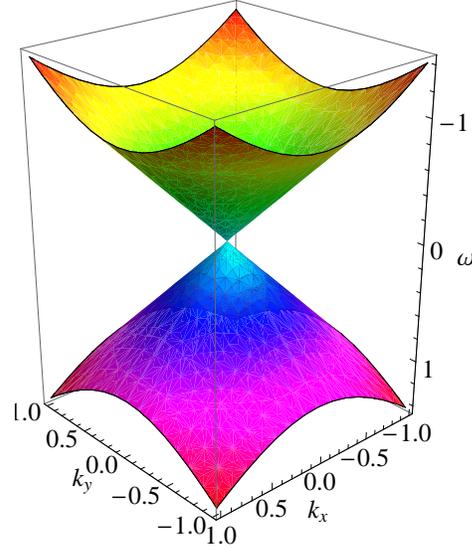}
\end{center}
\vspace{0mm}
\caption{\textit{Energy spectrum of two-dimensional Dirac fermions.}}
\label{fig:cone}
\vspace{5mm}
\end{wrapfigure}
The dispersion for Dirac fermions with the chemical 
potential directly at the Dirac point (see 
Fig. \ref{fig:cone}) is given by:
\begin{eqnarray}
\xi_{\mathbf{k}}=\pm v_{f}|\mathbf{k}|\;,
\end{eqnarray}
with $+v_{f}$ for the empty cone and $-v_{f}$ for the filled cone. 
Energy integrations over $\xi_{\mathbf{k}}$ 
are cut off in the ultraviolet and restricted to the band 
$\left\{\text{min}\,\xi_{\mathbf{k}}=-\Lambda_{0},\;
\text{max}\,\xi_{\mathbf{k}}=+\Lambda_{0}\right\}$. 
For attractive interactions the coupling constant $U$ is negative 
and drives spontaneous breaking of the global $U(1)$ 
gauge symmetry. Therefore, we decouple the Hubbard interaction in the s-wave
spin-singlet pairing channel by introducing a complex bosonic
Hubbard-Stratonovich field $\phi_q$ conjugate to the bilinear
composite of fermionic fields (Popov 1987)
\begin{equation}
 \tilde\phi_{q} = U \int_k \psi_{k+\frac{q}{2}\up}
 \psi_{-k+\frac{q}{2}\down} \; .
\end{equation}
This yields a functional integral over $\psi$, $\psib$ and $\phi$
with the new bare action
\begin{eqnarray}
 \Gam_0[\psi,\psib,\phi]
  = &-& \int_{k\sg} \psib_{k\sg} (ik_{0} - \xi_{\bk}) \, \psi_{k\sg}
  - \int_q \phi^*_q \frac{1}{U} \phi_q
  \nonumber\\
   &+& \int_{k,q} \left(
  \psib_{-k+\frac{q}{2} \down} \psib_{k+\frac{q}{2} \up} \,
  \phi_{q} +
  \psi_{k+\frac{q}{2} \up} \psi_{-k+\frac{q}{2}\down} \,
  \phi^*_q \right) . 
  \label{eq:dirac_finalmodel}
\end{eqnarray}
where $\phi^*$ is the complex conjugate of $\phi$, while $\psi$
and $\psib$ are algebraically independent Grassmann variables. 

\subsection{Mean-field theory}
\label{subsec:dirac_mft}

When neglecting bosonic 
fluctuations by replacing $\phi_{q}$ with its expectation value $\phi_{q=0}$, 
the saddle-point approximation exactly solves the functional integral of Eq. 
(\ref{eq:dirac_finalmodel}) and leads 
to the standard BCS gap equation (Popov 1987):
\begin{eqnarray}
\phi_{0}= -U \int_{k} \frac{\phi_{0}}{k_{0}^{2}+\xi^{2}_{\mathbf{k}}+\phi_{0}^{2}}\;,
\label{eq:bcs_gap}
\end{eqnarray}
with $U$ and $\phi_{0}$ independent of momenta. 
To compute the critical interaction strength $U_{c}$ at which the gap equation 
has a solution with non-zero $\phi_{0}$, we rearrange Eq. (\ref{eq:bcs_gap}):
\begin{eqnarray}
\frac{-1}{U_{c}}= \int \frac{d k_{0}}{2\pi}\int^{\Lambda_{0}}_{-\Lambda_{0}} 
d\xi_{\mathbf{k}}\; N\left(\xi_{\mathbf{k}}\right)\frac{1}{k_{0}^{2}+\xi^{2}_{\mathbf{k}}}\;,
\label{eq:thouless}
\end{eqnarray}
where the density of states has the form:
\begin{eqnarray}
N\left(\xi\right)=\frac{|\xi|^{d-1} K_{d}}{v_{f}^{d}}\;,
\end{eqnarray}
with $K_{d}$ being defined by $\int\frac{d^{d}k}{\left(2\pi\right)^{d}}
=K_{d}\int d|k|\, |k|^{d-1}$. The expression in Eq. (\ref{eq:thouless}) is equivalent 
to the Thouless criterion for superconductivity which involves summing 
the particle-particle ladder in the normal phase to infinite order and determining 
the divergence of the effective interaction. 
Since the density of states vanishes --linearly in two dimensions-- 
at the Dirac point, the zero temperature 
system is stable against formation of a superfluid for any weak 
attraction. Instead, a finite attraction larger than a certain threshold is necessary to cause superfluidity at 
zero temperature. Performing the integrations in Eq. (\ref{eq:thouless}), 
we obtain the mean-field position of the quantum phase transition in 
Fig. \ref{fig:phase_toy}. The control parameter provides a mass term for the boson propagator in 
Eq. (\ref{eq:dirac_finalmodel}) and is therefore inversely related to the four-fermion attraction.
For the physical case of two dimensions, we have:
\begin{eqnarray}
\frac{1}{\delta_{\text{MFT}}}=-U_{c,\,\text{MFT}}= \frac{2\pi v_{f}^{2}}{\Lambda_{0}}\;,
\label{eq:Uc_mft}
\end{eqnarray}
where here $v_{f}$ has units of energy. For electrons in the honeycomb lattice, 
the Fermi velocity in the vicinity of the Dirac points is proportional to the nearest-neighbor hopping matrix element $t$ (Castro Neto 2007). Upon setting $v_{f}=\Lambda_{0}=1$, the numerical value for the mean-field control parameter value is 
$\delta_{\text{MFT}}=0.159$. Note that the position of the QCP is non-universal and 
depends on microscopic parameters such as the ultraviolet band cutoff $\Lambda_{0}$ and 
the Fermi velocity.

\bigskip

In the following, we will conduct a renormalization group study which enables us not only 
to compute the \emph{non-universal} renormalized position of the QCP but 
also yields the complete set of \emph{universal} quantum critical exponents at and in the vicinity of the QCP.

\section{Method}
\label{sec:pp_method}

We derive flow equations for the scale-dependent effective action
$\Gamma_{\Lambda}\left[\psi,\bar{\psi},\phi\right]$
within the functional RG framework for fermionic and
bosonic degrees of freedom in its one-particle irreducible 
representation (Berges 2002, Baier 2004, Sch\"utz 2005, Strack 2008).
Starting from the bare fermion-boson action
$\Gamma_{\Lambda=\Lambda_{0}}\left[\psi,\bar{\psi}, \phi\right]$
in Eq. (\ref{eq:dirac_finalmodel}), fermionic and bosonic fluctuations are integrated along the
continuous flow parameter $\Lambda$ \emph{simultaneously}. In the
infrared limit $\Lambda\rightarrow0$, the renormalized, effective
action $\Gamma_{\Lambda\rightarrow0}\left[\psi,\bar{\psi},
\phi\right]$ is obtained from which physical properties can be
extracted.
Its RG flow is governed by the exact functional flow equation
\begin{eqnarray}
\frac{d}{d\Lam} \Gam^{\Lam}[\mathcal{S},\bar{\mathcal{S}}] =
 {\rm Str} \, \frac{\dot{\bR}^{\Lam}}
 {\bGam^{(2) \, \Lam}[\mathcal{S},\bar{\mathcal{S}}] + \bR^{\Lam}} \;,
\label{eq:pp_exact}
\end{eqnarray}
where we have collected fermionic and bosonic fields in superfields 
$\bar{\mathcal{S}}$, $\mathcal{S}$ defined in subsection \ref{subsec:super} and 
$\bGam^{(2) \, \Lam}=\partial^{2}\Gam^{\Lam}[\mathcal{S},\bar{\mathcal{S}}]/\partial\mathcal{S}
\partial\bar{\mathcal{S}}$ denotes the second functional derivative with respect to the superfields 
and the supertrace (Str) traces over all indices with an additional minus sign for fermionic contractions. We refer to Section \ref{sec:flow_eq} for a detailed derivation and discussion 
of the 1PI functional RG framework for both fermionic and bosonic fields.

\subsection{Truncation}
\label{subsec:toy_trunc}

When evolving $\Gam^{\Lam}$ towards $\Lam\rightarrow 0$, infinitely
many terms involving fermionic and/or bosonic fields with possibly complicated
dependences on frequency and momenta are generated necessitating a truncation of the effective action. The purpose of this subsection is to explain how the effective action for the model
Eq. (\ref{eq:dirac_finalmodel}) is truncated with the objective to capture the most 
relevant quantum critical renormalization effects.

\subsubsection{Fermion propagator}

To account for a renormalization of the single-particle, fermionic 
properties by order parameter fluctuations, 
the quadratic fermionic term in the action is modified
by a field renormalization factor,
\begin{equation}
 \Gam_{\psib\psi} = - \int_{k\sg}
 \psib_{k\sg} Z_{f}(ik_0 - \xi_{\bk}) \,
 \psi_{k\sg} \; ,
\label{eq:toy_Gam_psi_psi}
\end{equation}
yielding the fermion propagator
\begin{equation}
 G_{f}(k) = -\langle\psi_{k}\bar{\psi}_{k}\rangle =
 \frac{Z_{f}^{-1}}{ik_{0}-\xi_{\mathbf{k}}} \; .
\label{eq:pp_green_f}
\end{equation}
A diverging $Z_{f}$ suppresses the quasi-particle weight to zero. If it
diverges as a power law, $Z_{f}\sim 1/\Lambda^{\eta_{f}}$,
the fermion self-energy becomes a non-analytic function of frequency 
($k_{0}=\omega$)
with
\begin{eqnarray}
\Sigma_{f}(\omega)\sim \omega ^{1-\eta_{f}}
\end{eqnarray}
upon identifying the cutoff scale $\Lambda$ with $\omega$. As already mentioned above,
the Fermi velocity is not renormalized separately but kept fixed at unity for the following reasons.
The flow equations for the frequency- and momentum renormalization factors
are obtained by derivatives of expressions corresponding to 1PI-diagrams.
The flow equation for an independently
parametrized $v_{f}$ would come from one derivative of the diagram involving
$G_{f}$ and $G_{b}$, shown in the first line of Fig. \ref{fig:dirac_flow}, with respect to the deviation
of momenta from the Dirac point. In parallel, the flow equation for $Z_{f}$ also comes
from one derivative with respect to frequency of the same expression. Hence,
both expressions are almost identical for a linearized fermion dispersion
and keeping $v_{f}$ does not yield additional information.
The initial condition for $Z_{f}$ is $Z_{f}=1$. 

\subsubsection{Boson propagator}

The bosonic quadratic part of the bare action, Eq. (\ref{eq:dirac_finalmodel}), consists 
only of a local mass term. Upon integrating out fluctuations, the effective four-fermion 
interaction mediated by boson exchange will become momentum- and frequency-dependent. 
To capture this propagating order parameter field, we deploy a 
renormalization factor $Z_{b}$ multiplying the lowest order frequency 
and momentum terms in a derivative expansion of the fermionic particle-particle bubble. 
The real part of the particle-particle bubble spanned by fermion propagators endowed with the regulator $\bR^{\Lam}$ is a quadratic function 
of external frequencies and momenta leading to the bosonic 
quadratic part of the action:
\begin{eqnarray}
\Gam_{\phi^\ast\phi}=
\int_{q}\phi^{\ast}_{q}\,\left(Z_{b}\left(q_{0}^{2}+\mathbf{q}^{2}\right)+\delta\right)\phi_{q}
\;.
\end{eqnarray}
Note that there is no complex linear term in frequency here, as we consider the half-filled 
band and the imaginary linear frequency part of the particle-particle bubble and hence 
the boson propagator vanishes exactly. The control parameter 
term $\delta$ controls
the distance to the continuous quantum phase transition and is 
also renormalized by fluctuations. If the initial value
of $\delta$ is fine-tuned so that $\delta\rightarrow0$ for vanishing cutoff
$\Lam\rightarrow0$, we are in the quantum critical state with infinite susceptibility,
$\chi\sim \frac{1}{\delta}|_{\Lam\rightarrow0}$. 
The boson propagator, parametrized
by two RG parameters, reads
\begin{eqnarray}
G_{b}(q)&=&-\langle\phi_{q}\phi^{\ast}_{q}\rangle=
\frac{-1}{Z_{b}\left(q_{0}^{2}+\mathbf{q}^{2}\right)+\delta}\,\,.
\label{eq:pp_green_b}
\end{eqnarray}
Recall that this initial ansatz for the boson propagator is motivated by the \emph{intermediate}
energy behavior of the particle-particle bubble as the low energy regime is excluded 
from the spectrum by the regulator $\bR^{\Lam}$. The \emph{low energy} regime is controllably 
accessed subsequently in the coupled Fermi-Bose RG flow when 
$\bR^{\Lam}$, $\Lam\rightarrow0$. 
We will see below that, 
although we have started with a differentiable function 
of frequencies and momenta for the boson propagator at intermediate energies, 
at 
criticality in the low energy regime a diverging $Z_{b}$-factor induces strong analyticities in the boson propagator --the order parameter 
field attains an anomalous dimension. 
The initial condition for $Z_{b}$ is $Z_{b}=0$ and for the control parameter 
$\delta=1/U$. 

\subsubsection{Order parameter self interactions}

We now demonstrate how to account for direct self-interactions 
of the order parameter field within our coupled Fermi-Bose scheme. 
Recollect that in the conventional Hertz-Millis theory of quantum criticality (Hertz 1976,
Millis 1993, Loehneysen 2006) one expands the purely bosonic action
in (even) powers of the bosonic fields:
\begin{eqnarray}
\gamma[\phi]&=&\hspace{-3mm}\sum_{n\geq4,\,\text{even}}
\frac{u_{n}}{2\left(\frac{n}{2}\right)!\left(\frac{n}{2}\right)!}
\left(\phi^{\ast}\phi\right)^{n/2}\;,
\end{eqnarray}
where the coefficients $u_{n}$ are generated by fermion loops with $n$ fermion propagators 
and
$n$ external bosonic legs as shown in Fig. \ref{fig:fermion_ring}.
At finite temperatures and when the fermions are gapped, such expressions
are possibly large but finite. 
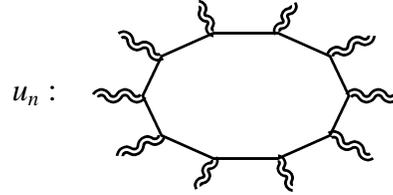
\begin{wrapfigure}{r}{0.5\textwidth}
  \vspace{0mm}
\begin{fmffile}{fermion_ring_8}
\begin{eqnarray}
u_{n}&:&\hspace{2mm}
\parbox{35mm}{\unitlength=1mm\fmfframe(2,2)(1,1){
\begin{fmfgraph*}(40,25)
     \fmfpen{thin}
     \fmfsurroundn{e}{10}
     \begin{fmffor}{n}{1}{1}{10}
       \fmf{dbl_wiggly}{e [n],i [n]}
     \end{fmffor}
     \fmfcyclen{plain,tension=10/8}{i}{10}
   \end{fmfgraph*}}}
\nonumber
\end{eqnarray}
\end{fmffile}
\caption{\textit{Fermion rings which generate the bosonic n-point
vertex (here for $n=10$).}} 
\label{fig:fermion_ring} 
\vspace{-5mm}
\end{wrapfigure}
With the bare, gapless fermion propagator at zero temperature defined
in Eq. (\ref{eq:pp_green_f}), however, these
coefficients are singular for small momenta:
\begin{eqnarray}
u_{n}&\sim& \int d^{d+1}k\,\frac{1}{k^{n}}\sim \frac{1}{k^{n-(d+1)}}\,\,,
\nonumber\\
\end{eqnarray}
leading for instance to a power law singularity for the $\phi^{4}
$ coefficient $u_{4}\equiv u\sim 1/k$ in $d=2$. The presence 
of these singularities completely invalidates the Hertz-Millis approach 
of expanding the effective action in powers of the ordering field alone.
Our coupled Fermi-Bose scheme is capable of renormalizing 
such power law singularities, as during the RG flow, 
fermionic fluctuations will generate order parameter self-interactions 
even if there is no $\phi^{4}$-term present in the bare
action. To capture this, we keep the local fourth-order term in the effective action:
\begin{equation}
 \Gam_{|\phi|^4} = \frac{u}{8} \int_{q,q',p}
 \phi^*_{q+p} \phi^*_{q'-p} \phi_{q'} \phi_q \;.
\end{equation}
The initial condition for $u$ is $u=0$. 
Note that the fermion-boson vertex in Eq. (\ref{eq:dirac_finalmodel}) 
is not renormalized within our truncation. The standard vertex correction one-loop diagram
$\sim g^{3}$ vanishes by particle conservation. 
In the following, we keep the fermion-boson vertex at its bare value $g=1$, 
which can be 
read off from Eq. (\ref{eq:dirac_finalmodel}).

\subsection{Flow equations}

In this subsection, we derive analytic expressions for our flow equations of the
truncated effective action. Both Green functions, Eqs. (\ref{eq:pp_green_f}, \ref{eq:pp_green_b}),
display singularities for certain choices of momenta, the fermion 
propagator everywhere in the phase diagram, and the boson propagator at the QCP when the bosonic mass vanishes. These potential infrared singularities for $|\text{momenta}|<\Lambda$ are
regularized by adding optimized momentum cutoffs (Litim 2001) for
fermions (subscript f) and bosons (subscript b),
\begin{eqnarray}
R_{f\Lambda}(\mathbf{k})&=&Z_{f}\left(-\Lambda\,\text{sgn}\left[\xi_{\mathbf{k}}\right]
+ \xi_{\mathbf{k}}\right)
\theta\left[\Lambda - |\xi_{\mathbf{k}}|\right]\nonumber\\
R_{b\Lambda}(\mathbf{q})&=&Z_{b}\left(\Lambda^{2} -
\mathbf{q}^{2}\right) \theta\left[\Lambda^{2} -
\mathbf{q}^{2}\right]
\label{eq:cutoffs}
\end{eqnarray}
to the inverse of the propagators in Eqs. (\ref{eq:pp_green_f}, \ref{eq:pp_green_b}).
The cutoff-endowed propagators are denoted with by $G_{fR}$ and $G_{bR}$ in the
following. The scale-derivatives of the cutoffs read,
\begin{eqnarray}
\partial_{\Lambda}R_{f\Lambda}=\dot{R}_{f\Lambda}&=&
-Z_{f}\,\text{sgn}\left[\xi_{\mathbf{k}}\right]\,\theta\left[\Lambda
- |\xi_{\mathbf{k}}|\right]\nonumber\\
\dot{R}_{b\Lambda}&=& 2Z_{b}\Lambda\,\theta\left[\Lambda^{2} - \mathbf{q}^{2}\right]\,\,,
\end{eqnarray}
where terms proportional to $\eta_{f}$ and $\eta_{b}$ defined below
in Eq. (\ref{eq:def_etas}) are neglected here. These additional terms 
are of higher order in the vertices and subleading. Further arguments buttressing this
commonly employed procedure are given in (Berges 2002).

A comment on the choice of relative cutoff scale between fermions and bosons is in order.
In principle, fermion fluctuations can be cut off at $\Lambda_{f}$ and their
bosonic counterparts at $\Lambda_{b}$ with both cutoffs being independent functions 
of $\Lam$. Without approximations,
the results do not depend on the concrete choices of $\Lam_{f}(\Lam)$ and $\Lam_{b}(\Lam)$. 
In Eq. (\ref{eq:cutoffs}), we have chosen $\Lambda_{f}=\Lambda_{b}=\Lambda$; the standard choice for critical Fermi-Bose theories (Litim 2001, Gies 2004).

\bigskip

The recipe to obtain the flow equations is now the following: one executes
a cutoff-derivative acting on $R_{f,b\Lambda}$ in the analytic expressions
corresponding to all 1-loop one-particle-irreducible Feynman diagrams for the
parameters $Z_{f}$, $Z_{b}$, $\delta$, $u$, which can be generated
with $G_{fR}$ and $G_{bR}$ as shown in Fig. \ref{fig:dirac_flow}. The subsequent
trace operation is abbreviated by
\begin{eqnarray}
\int_{k,R_{s}} = \int\frac{d k_{0}}{2\pi}\int \frac{d^{d}k}{\left(2\pi\right)^{d}}\sum_{s=f,b}\left(-\dot{R}_{s\Lambda}\right)
\partial_{R_{s\Lambda}}\;.
\end{eqnarray}
The flow equation for the fermion self-energy is obtained from the Fock-type 
diagram in the first line of Fig. \ref{fig:dirac_flow} involving $G_{bR}$ and 
$G_{fR}$. This feedback of bosonic fluctuations on the fermionic propagator is captured via 
the flow equation:
\begin{eqnarray}
\partial_{\Lambda}Z_{f}= g^{2} \int_{q,R_{s}}\partial_{i k_{0}}\,
G_{fR}\left(q-k\right)G_{bR}\left(q\right)|_{k=0}\,\,.
\label{eq:Zf}
\end{eqnarray}
This expression vanishes for a local boson ($Z_{b}=0$) 
from multiple poles in the upper 
complex frequency half-plane. This correctly reflects that 
for a momentum-independent four-fermion interaction, the fermionic $Z_{f}$-factor 
is renormalized only at the two-loop level. When reinserting the particle-particle bubble 
which generates $Z_{b}$ in the third line of Fig. \ref{fig:dirac_flow}  
into the bosonic propagator in the first line of Fig. \ref{fig:dirac_flow}, we 
indeed observe that effectively our flow equations capture two-loop effects by keeping 
the boson propagator as momentum- and frequency dependent ($Z_{b}\neq0$).

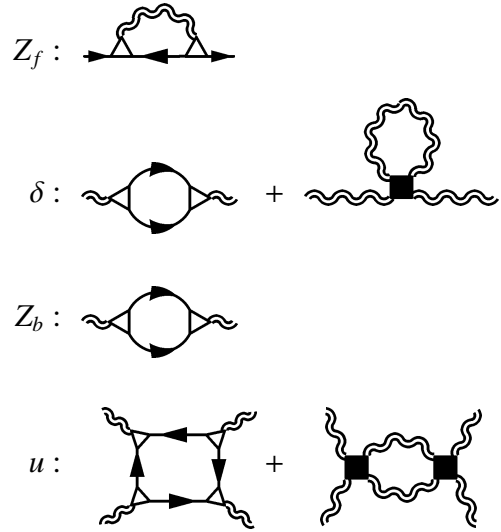
\begin{wrapfigure}{r}{0.5\textwidth}
  \vspace{-5mm}
\begin{fmffile}{20081208_dirac_4}
\begin{eqnarray}
Z_{f}&:&
\parbox{25mm}{\unitlength=1mm\fmfframe(2,2)(1,1){
\begin{fmfgraph*}(20,20)\fmfpen{thin}
\fmfleft{l1}
 \fmfright{r1}
 \fmfpolyn{empty,tension=0.4}{G}{3}
 \fmfpolyn{empty,tension=0.4}{K}{3}
  \fmf{fermion}{l1,G1}
   \fmf{fermion,straight,tension=0.5,left=0.}{K3,G2}
  \fmf{dbl_wiggly,tension=0.,left=0.7}{G3,K2}
  \fmf{fermion}{K1,r1}
 \end{fmfgraph*}
}}\nonumber\\[-8mm]
\delta&:&
\parbox{25mm}{\unitlength=1mm\fmfframe(2,2)(1,1){
\begin{fmfgraph*}(20,20)\fmfpen{thin} 
 \fmfleft{l1}
 \fmfright{r1}
 \fmfpolyn{empty,tension=0.3}{G}{3}
 \fmfpolyn{empty,tension=0.3}{K}{3}
  \fmf{dbl_wiggly}{l1,G1}
 \fmf{fermion,tension=0.2,right=0.6}{G2,K3}
 \fmf{fermion,tension=0.2,left=0.6}{G3,K2}
 \fmf{dbl_wiggly}{K1,r1}
 \end{fmfgraph*}
}}+
\parbox{25mm}{\unitlength=1mm\fmfframe(2,2)(1,1){
\begin{fmfgraph*}(25,25)\fmfpen{thin} 
 \fmfleft{l1}
 \fmfright{r1}
 \fmftop{v1}
 \fmfpolyn{full}{G}{4}
 \fmf{dbl_wiggly,straight}{l1,G4}
 \fmf{dbl_wiggly,straight}{G1,r1}
 \fmffreeze
\fmf{dbl_wiggly,tension=0.1,right=0.7}{G2,v1}
\fmf{dbl_wiggly,tension=0.1,right=0.7}{v1,G3}
\end{fmfgraph*}
}}\nonumber\\[-8mm]
Z_{b}&:&
\parbox{25mm}{\unitlength=1mm\fmfframe(2,2)(1,1){
\begin{fmfgraph*}(20,15)\fmfpen{thin} 
 \fmfleft{l1}
 \fmfright{r1}
 \fmfpolyn{empty,tension=0.3}{G}{3}
 \fmfpolyn{empty,tension=0.3}{K}{3}
  \fmf{dbl_wiggly}{l1,G1}
 \fmf{fermion,tension=0.2,right=0.6}{G2,K3}
 \fmf{fermion,tension=0.2,left=0.6}{G3,K2}
 \fmf{dbl_wiggly}{K1,r1}
 \end{fmfgraph*}
}}\nonumber\\
u&:&
\parbox{25mm}{\unitlength=1mm\fmfframe(2,2)(1,1){
\begin{fmfgraph*}(25,15)
\fmfpen{thin}
\fmfleftn{l}{2}\fmfrightn{r}{2}
 \fmfpolyn{empty,tension=0.8}{OL}{3}
 \fmfpolyn{empty,tension=0.8}{OR}{3}
 \fmfpolyn{empty,tension=0.8}{UR}{3}
 \fmfpolyn{empty,tension=0.8}{UL}{3}
   \fmf{dbl_wiggly}{l1,OL1}
   \fmf{dbl_wiggly}{l2,UL1}
   \fmf{dbl_wiggly}{OR1,r1}
   \fmf{dbl_wiggly}{UR1,r2}
  \fmf{fermion,straight,tension=0.5}{OL2,OR3}
  \fmf{fermion,straight,tension=0.5}{UR3,OR2}
  \fmf{fermion,straight,tension=0.5}{UR2,UL3}
  \fmf{fermion,straight,tension=0.5}{OL3,UL2}
 \end{fmfgraph*}
}}+
\parbox{25mm}{\unitlength=1mm\fmfframe(2,2)(1,1){
\begin{fmfgraph*}(25,15)
\fmfpen{thin} 
\fmfleftn{l}{2}\fmfrightn{r}{2}
\fmfrpolyn{full}{G}{4}
\fmfpolyn{full}{K}{4}
\fmf{dbl_wiggly}{l1,G1}\fmf{dbl_wiggly}{l2,G2}
\fmf{dbl_wiggly}{K1,r1}\fmf{dbl_wiggly}{K2,r2}
\fmf{dbl_wiggly,left=.5,tension=.3}{G3,K3}
\fmf{dbl_wiggly,right=.5,tension=.3}{G4,K4}
\end{fmfgraph*}
}}\nonumber
\end{eqnarray}
\end{fmffile}
\caption{\textit{Feynman diagrams representing the flow equations}.}
\label{fig:dirac_flow}
\vspace{-15mm}
\end{wrapfigure}
For the control parameter, we evaluate the 1PI-diagrams with two external 
bosonic legs, and we obtain the two contributions:
\begin{eqnarray}
\partial_{\Lambda}\delta&=&g^{2}\int_{k,R_{f}} G_{fR}\left(k\right)
G_{fR}\left(-k\right)
\nonumber\\
&+&\frac{u}{2}\int_{q,R_{b}} G_{bR}(q)\,\,,
\label{eq:delta}
\end{eqnarray}
The fermionic contribution on the right-hand-side is positive leading to a reduction of $\delta$ for decreasing $\Lambda$ whereas the bosonic contribution counteracts the fermions and tends 
to increase $\delta$. This is the generic behavior of the bosonic fluctuations as they 
always tend to restore the symmetry.

The flow of the bosonic frequency renormalization is obtained as the second 
frequency derivative of the particle-particle bubble:
\begin{eqnarray}
\partial_{\Lambda}Z_{b}=g^{2} \int_{k,R_{f}}\frac{1}{2}\partial^{2}_{q_{0}} G_{fR}\left(k+q\right)
G_{fR}\left(-k\right)|_{q=0}\,\,.
\label{eq:Zb}
\end{eqnarray}
The bosonic tadpole diagram does not contribute here, as the $\phi^{4}$-vertex $u$ 
is taken as momentum- and frequency-independent.

Finally, the bosonic self-interaction flows according to:
\begin{eqnarray}
 \partial_{\Lambda} u =
 &-& 4 g^4 \int_{k, R_{f}} [G_{fR}(-k)]^2 [G_{fR}(k)]^2 + \frac{5}{4} u^2 \int_{q,R_{b}}
[G_{bR}(q)]^{2}\;,
\label{eq:u}
\end{eqnarray}
where the first terms generates $u$ and the second, bosonic term tends to reduce 
$u$ in the course of the flow.
All frequency and momentum integrations in the above flow equations can be 
performed analytically. We now elegantly reformulate the flow equations by 
employing the following scaling variables:
\begin{eqnarray}
\tilde{\delta}&=&\frac{\delta}{\Lambda^{2}Z_{b}}\nonumber\\
\tilde{g}&=&\frac{g\sqrt{K_{d}}}{\Lambda^{\frac{3-d}{2}}Z_{f}\sqrt{Z_{b}}\sqrt{d}}
\nonumber\\
\tilde{u}&=&\frac{u\,K_{d}}{\Lambda^{3-d}Z_{b}^{2}d}\,\,,
\label{eq:def_variables}
\end{eqnarray}
where $K_{d}$ is defined by $\int\frac{d^{d}k}{\left(2\pi\right)^{d}}
=K_{d}\int d|k|\, |k|^{d-1}$.
By defining the anomalous exponents for the fermionic and bosonic 
$Z$-factor, respectively,
\begin{eqnarray}
\eta_{f}&=&-\frac{d \log Z_{f}}{d \log \Lambda}\nonumber\\
\eta_{b}&=&-\frac{d \log Z_{b}}{d \log \Lambda}\;,
\label{eq:def_etas}
\end{eqnarray}
the explicit dependence on the $Z$-factors disappears from the expressions. 
The flow equations for the control parameter and the bosonic 
self-interaction are obtained as:
\begin{eqnarray}
\frac{d\tilde{\delta}}{d\log \Lambda}&=&\left(\eta_{{b}}-2\right)\tilde{\delta}
+ \tilde{g}^{2} -\frac{\tilde{u}}{4\left(1+\tilde{\delta}\right)^{3/2}}
\nonumber\\[3mm]
\frac{d\tilde{u}}{d\log \Lambda}&=&
\left(d-3+2\eta_{b}\right)\tilde{u}
- 6\, \tilde{g}^{4}
+\frac{15}{16}\, \frac{\tilde{u}^{2}}
{\left(1+\tilde{\delta}\right)^{5/2}}\;.
\label{eq:betas_qcp}
\end{eqnarray}
The rescaled fermion-boson vertex $\tilde{g}$ obeys the equation:
\begin{eqnarray}
\frac{d\tilde{g}}{d\log \Lambda}&=&\left(\eta_{f} +\frac{1}{2} \eta_{b} - \frac{3-d}{2}\right)\tilde{g}\,\,.
\label{eq:beta_g}
\end{eqnarray}
The equations (\ref{eq:betas_qcp}, \ref{eq:beta_g}) have to be considered 
in conjunction with the fermion and boson anomalous exponents:
\begin{eqnarray}
\eta_{b}&=&\frac{3}{4}\tilde{g}^{2}\nonumber\\
\eta_{f}&=&\tilde{g}^{2}
\left(
\frac{1}{\left(1+\tilde{\delta}\right)^{3/2}} +
\frac{2}{\left(1+\tilde{\delta}\right)}\right)
\frac{1}
{2+\tilde{\delta}+2\sqrt{1+\tilde{\delta}}}\;.
\label{eq:dirac_etas}
\end{eqnarray}
We will now investigate the analytic properties of these equations and then 
solve them numerically.
 
\section{Solution at the quantum critical point}
\label{sec:sol_qcp}

At the quantum critical point, the system becomes scale-invariant, that is, the parameters
show no dependence on the scale $\Lambda$. Translated into an algebraic
condition, this becomes
\begin{eqnarray}
\frac{d\tilde{\delta}}{d\log \Lambda}
=\frac{d\tilde{g}}{d\log \Lambda}=\frac{d\tilde{u}}{d\log \Lambda}=0\,\,,
\label{eq:constraint}
\end{eqnarray}
and one has to solve Eqs. (\ref{eq:betas_qcp}, \ref{eq:beta_g}) for 
fixed points together with the anomalous exponents of Eq. (\ref{eq:dirac_etas}).
It is immediately apparent that both, the fermion-boson vertex 
$\tilde{g}$ and the boson self-interaction $\tilde{u}$ are relevant 
couplings below three dimensions. For the $d=2$ case possibly 
relevant for graphene 
and cold atoms in the honeycomb lattice, the RG equations for
$\tilde{g}$ and $\tilde{u}$ have stable non-Gaussian $\left(\tilde{g}\neq 0,\; 
\tilde{u}\neq 0\right)$ solutions with finite 
\begin{wrapfigure}{r}{0.5\textwidth}
\vspace*{3mm}
\begin{center}
\includegraphics*[width=68mm]{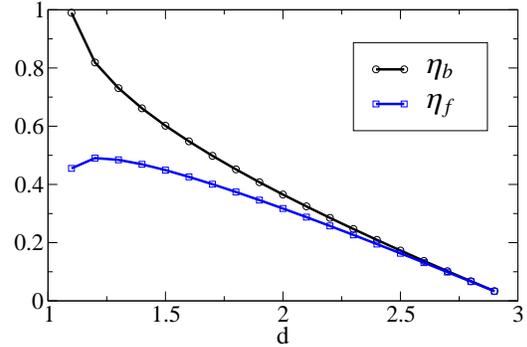}
\caption{\textit{Fermion and boson anomalous exponents at the QCP for $1<d<3$.
}}
\label{fig:fp_line}
\end{center}
\vspace*{-2mm}
\end{wrapfigure}
The anomalous scaling exponents are interrelated at the QCP:
\begin{eqnarray}
\eta_{f}&=&\frac{3-d}{2}-\frac{1}{2}\eta_{b}\;,
\label{eq:inter_etas}
\end{eqnarray}
anomalous exponents $\eta_{f}$ and $\eta_{b}$ 
as follows directly from Eq. (\ref{eq:beta_g}).
The values of the 
anomalous exponents from a numerical solution of Eqs. (\ref{eq:betas_qcp}-\ref{eq:dirac_etas})
under the constraint at the QCP, Eq. (\ref{eq:constraint}), are shown in Fig. \ref{fig:fp_line}.

At quantum criticality, the fermionic properties of the system cannot be described in terms of conventional Fermi liquid theory (Nozi\`eres 1964).
A finite fermion anomalous dimension entails a fermion propagator 
of the form,
\begin{eqnarray}
G_{f}(k_{0}, \mathbf{k}) &\propto&\frac{1}{\left(i k_{0}-\xi_{\mathbf{k}}\right)^{1-\eta_{f}}}
\propto \frac{1}{\Lambda^{1-\eta_{f}}}\,\,,
\label{eq:pp_qcp_etas}
\end{eqnarray}
with a non-analytic frequency-dependence of the fermion self-energy ($k_{0}=\omega$):
\begin{eqnarray}
\Sigma_{f}(\omega)\sim \omega^{1-\eta_{f}}\;,
\label{eq:ferm_self_qcp}
\end{eqnarray}
where $1-\eta_{f}=0.68$ for $d=2$ as can be read off from Fig. \ref{fig:fp_line} (a). 
The quasi-particle picture breaks down. 
A similar quantum criticality induced breakdown of the Fermi liquid has been noted, among others, 
in the context of Fermi surface fluctuations near a Pomeranchuk instability (Metzner 2003, Dell'Anna 2006 and references therein). There, the frequency exponent of the fermion self-energy at the QCP is $\sim \omega^{2/3}$. The fermion propagator also develops a small Luttinger-like 
anomalous dimension in QED$_{3}$ (Franz 2002) and in the Gross-Neveu model 
(Rosa 2001, Herbut 2006).

\bigskip

Concerning the collective properties of the system at quantum criticality, 
a finite boson anomalous dimension entails an order parameter propagator 
of the form,
\begin{eqnarray}
G_{b}(q_{0}, \mathbf{q}) &\propto& \frac{1}
{\left(q_{0}^{2} + \mathbf{q}^{2}\right)^{1-\eta_{b}}}
\propto\frac{1}{\Lambda^{2-\eta_{b}}}\,\,,
\label{eq:pp_qcp_bosonprop}
\end{eqnarray}
with a non-analytic boson self-energy
\begin{eqnarray}
\Sigma_{b}(q)\sim q^{2-\eta_{b}}\;,
\end{eqnarray}
where $2-\eta_{b}=1.63$ in two dimensions, see Fig. \ref{fig:fp_line} (a). 
In a completely different context, at the antiferromagnetic QCP of the
spin-fermion model in two dimensions, Chubukov \textit{et al.} (Abanov 2000, Abanov 2003, Abanov 2004)
estimated for the spin susceptibility the momentum scaling $\sim|\mathbf{q}|^{1.75}$ within
a perturbative $1/N$ calculation where $N$ is the number of hot spots on the Fermi surface. 
The physical origin of the anomalous momentum exponent for 
the spin-fermion model is different from that of the present work. 
There, peculiarities from antiferromagnetic scattering processes cause all $\phi^{n}$-vertices to be marginal 
in $d=2$. Summing these logarithms, Chubukov \textit{et al.} then obtain the power law for the quantum critical 
bosonic self-energy alluded to above. This non-analyticity thus vanishes for $d>2$ as the logarithms are 
special to $d=2$. The present work most closely resembles the Gross-Neveu model 
with two fermion flavors where the boson anomalous dimension is however expected 
to be rather large $\eta_{b, \text{Gross-Neveu}}\approx0.7$ (Rosa 2001, Herbut 2006).

\bigskip

By comparing the exponents for the quantum critical 
fermion and order parameter self-energies of 
attractive Dirac fermions to other physical contexts, we offer that some of the overarching \emph{qualitative} features generic to various QCPs found with various methods are elegantly accessible within our coupled Fermi-Bose RG
framework.

\subsection{Quantum critical flows in two dimensions}
\label{subsec:qcflows}

We now establish a direct continuous link between the microscopic 
bare action in Eq. (\ref{eq:dirac_finalmodel}) and the infrared 
properties of the \emph{effective} \emph{action} at the QCP. For this purpose, we 
solve the flow equations (\ref{eq:betas_qcp}-\ref{eq:dirac_etas}) numerically 
as function of the flow parameter $\Lambda$ in two dimensions. The initial conditions of our parameters 
are chosen to precisely match the bare action in Eq. (\ref{eq:dirac_finalmodel}):
$u=0$, $Z_{b}=0$, $Z_{f}=1$, and $g=1$. As in the mean-field calculation of 
subsection \ref{subsec:dirac_mft}, 
the Fermi velocity and the ultraviolet 
cutoff are set to unity $\Lambda_{0}=v_{f}=1$. To reach the quantum critical state, 
the initial value of $\delta$ is fine-tuned so that at the end of the flow, for 
$\Lambda\rightarrow 0$, $\delta\rightarrow 0$. 

\begin{figure}[t]
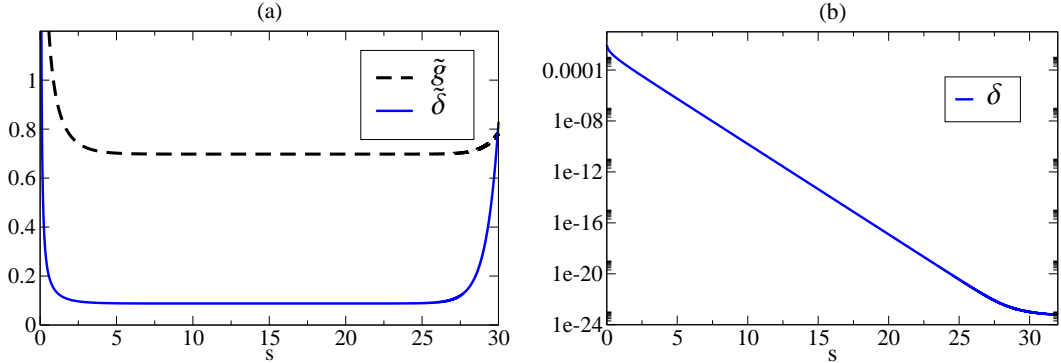

\includegraphics*[width=67mm]{del_g_cone.eps}
\hspace*{2mm}
\includegraphics*[width=69mm]{delta_bare_cone.eps}
\caption{\textit{(a) Flows of the rescaled control parameter and fermion-boson 
vertex versus $s=-\log\left[\Lambda/\Lambda_{0}\right]$. The ultraviolet (infrared) regime 
is on the left (right) side of the plots.
(b): Flow of the control parameter $\delta$ at the QCP.}}
\label{fig:delta_cone_flows}
\vspace{0mm}
\end{figure}

In Fig. \ref{fig:delta_cone_flows} (b), we exhibit such a characteristic flow 
of the control parameter at the QCP versus cutoff scale in a double logarithmic 
plot. We observe scaling behavior over more than 20 (!) orders of magnitude limited 
only by numerical accuracy (note the small values of $\delta$ on the 
vertical axis). The slope is the value of the exponent with which the control 
parameter decreases:
\begin{eqnarray}
\delta\sim\Lambda^{2-\eta_{b}}\;,
\end{eqnarray}
illustrated by the fixed point plateaus of the \emph{rescaled} control parameter $\tilde{\delta}$ and also the rescaled fermion-boson vertex shown 
in Fig. \ref{fig:delta_cone_flows} (a). The point we make here is, that we can verify 
whether the effective action is really \emph{attracted} toward 
the quantum critical fixed point when one starts from the microscopic model 
and whether this behavior is numerically \emph{stable}. 
Beyond doubt, this is indeed the case here.

In Fig. \ref{fig:eta_cone_flows}, we present flows 
for the fermionic and bosonic 
frequency- and momentum renormalization factors $Z_{f}$ and $Z_{b}$ as well 
as the associated anomalous dimensions $\eta_{f}$ and $\eta_{b}$. 
In Fig. \ref{fig:eta_cone_flows} (b), we observe that although 
$Z_{b}$ is zero initially, it is generated for small $s$ (large $\Lambda$) and 
subsequently diverges as a power law with slope $\eta_{b}$: $
Z_{b}\sim 1/\Lambda^{\eta_{b}}$, which is underlined by fixed point plateaus of $\eta_{b}$ shown in Fig. \ref{fig:eta_cone_flows} (a). The fermionic $Z_{f}$ is initially equal to 
unity and then also diverges 
as a power law
\begin{eqnarray}
Z_{f}\sim \frac{1}{\Lambda^{\eta_{f}}}\;, 
\end{eqnarray}
with a slightly smaller slope than $Z_{b}$ to be read off from the fixed point 
plateau of $\eta_{f}$ in Fig. \ref{fig:eta_cone_flows} (a). 
The numerical solution for both Z-factors exactly fulfills 
the interrelation condition Eq. (\ref{eq:inter_etas}) and the fixed point values 
precisely match those of Fig. \ref{fig:fp_line} at $d=2$.
\begin{figure}[t]
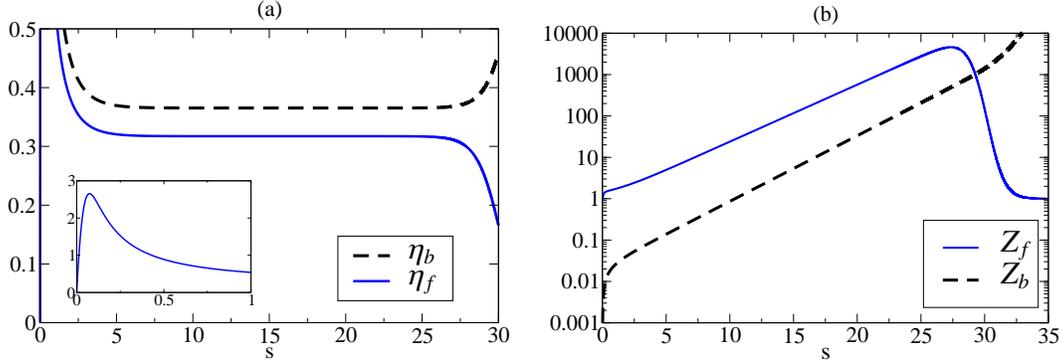

\includegraphics*[width=67mm]{etas_flows_cone.eps}
\hspace*{2mm}
\includegraphics*[width=69mm]{Zb_cone.eps}\\[-25mm]
\hspace*{8mm}\includegraphics*[width=25mm]{etaf_inset.eps}
\vspace*{8mm}
\caption{\textit{(a): Scaling plateaus of the fermion and boson anomalous dimensions at 
quantum criticality versus $s=-\log\left[\Lambda/\Lambda_{0}\right]$. The fermion anomalous dimension starts off at zero (inset) and 
becomes finite as soon as $Z_{b}\neq0$, see discussion below 
Eq. (\ref{eq:Zf}). (b): Flow of the bosonic and fermionic frequency- and momentum factors $Z_{f}$ and $Z_{b}$.}}
\label{fig:eta_cone_flows}
\vspace{0mm}
\end{figure}
We show flows of the $\phi^{4}$-vertex in Fig. \ref{fig:u_cone_flows}. 
Starting from zero initial value, fermion fluctuations quickly generate 
$u$ as is observed in the peak for small $s$ in Fig. \ref{fig:u_cone_flows} (b), and 
then the interplay of fermionic and bosonic fluctuations leads to the power law 
scaling behavior
\begin{eqnarray}
u\sim\Lambda^{3-d-2\eta_{b}}\;,
\end{eqnarray}
again accompanied by fixed point plateaus of $\tilde{u}$ 
depicted in Fig. \ref{fig:u_cone_flows} (a).

\begin{figure}[b]
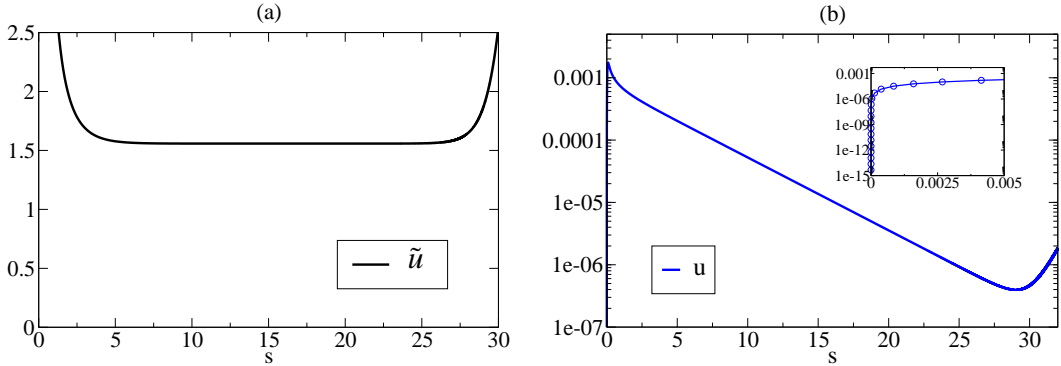

\includegraphics*[width=67mm]{u_cone.eps}
\hspace*{2mm}
\includegraphics*[width=69mm]{u_bare_cone.eps}\\[-40mm]
\hspace*{110mm}\includegraphics*[width=25mm]{u_bare_inset.eps}
\vspace*{25mm}
\caption{\textit{(a): Fixed point plateau of the rescaled $\phi^{4}$-coupling 
versus $s=-\log\left[\Lambda/\Lambda_{0}\right]$. 
(b): Flow of the quartic coupling $u$. Inset: Small-$s$ behavior when fermionic 
fluctuations generate $u$.}}
\label{fig:u_cone_flows}
\vspace{0mm}
\end{figure}

\bigskip 

We emphasize that our RG equations not only yield
various scaling exponents at the QCP, but are also useful in determining 
\emph{non-universal} properties of the system. The characteristic
scale $\Lambda_{\text{QC}}$ at which the quantum critical asymptotics sets in, for example, 
can be determined. From Fig. \ref{fig:eta_cone_flows}, we find that the anomalous dimensions attain their 
fixed point values at $s\approx5$ and therefore 
$\Lambda_{\text{QC}}\approx \Lambda_{0}e^{-5}$.

Additionally, we can compute the renormalized interaction strength $U_{c}$ at which 
the system becomes quantum critical and compare it to the mean-field value derived 
in Eq. (\ref{eq:Uc_mft}) since we have chosen the same values 
for the upper band cutoff ($\Lambda_{0}=1$) and the Fermi velocity ($v_{f}=1$) as in the mean-field calculation. $U_{c}$ is the inverse of the fine-tuned initial 
value of the control parameter ($\delta_{\Lambda=\Lambda_{0}}=0.009032269344423279716$) 
such that $\delta$ vanishes at the end of the flow. 
We find for the ratio between mean-field and RG interaction strengths:
\begin{eqnarray}
\frac{U_{c}}{U_{c,\text{MFT}}}\approx 17.6\;,
\label{eq:ratio}
\end{eqnarray}
therewith renormalizing the mean-field position of the QCP in 
Fig. \ref{fig:phase_toy} to the left. 
Fluctuations drastically reduce the size of the superfluid phase. 

\subsection{Quantum critical exponents}
\label{subsec:suszeb}

When approaching a critical point along the relevant parameter axis, 
the susceptibility and correlation length diverge as a power law. The presence of 
gapless fermions places the quantum phase transition considered in this chapter 
outside the usual Ginzburg-Landau-Wilson paradigm, as distinctively 
signalled by the fermion anomalous dimension $\eta_{f}$ (see Figs. 
\ref{fig:fp_line} and \ref{fig:eta_cone_flows}) which governs the disappearance 
of quasiparticles when approaching the QCP from the semimetallic side. $\eta_{f}$ 
influences all 
\begin{wrapfigure}{r}{0.5\textwidth}
  \vspace{2mm}
  \begin{center}
\includegraphics*[width=63mm]{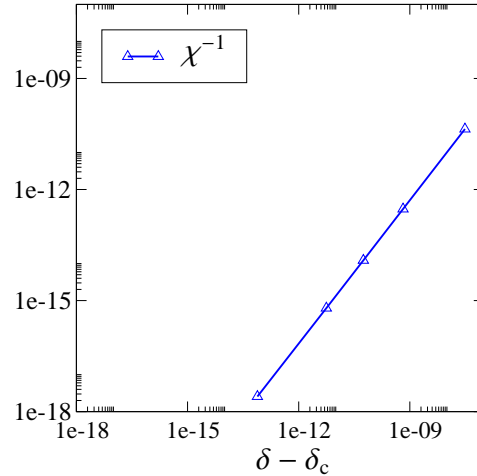}
\end{center}
\vspace*{-4mm}
\caption{\textit{Logarithmic plot of the inverse susceptibility versus 
$\delta$ in $d=2$.}}
\label{fig:pp_nu}
\vspace{0mm}
\end{wrapfigure}
other critical exponents similar to the universality class of the Gross-Neveu model to which the semimetal to antiferromagnetic insulator 
QCP for \emph{repulsively} interacting 
electrons in the 2d-honeycomb lattice has been 
proposed to belong to (Herbut 2006).

We now compute the susceptibility exponent for $d=2$ in the vicinity of the QCP,
\begin{eqnarray}
\chi\sim \frac{1}{\left(\delta -
\delta_{\text{crit}}\right)^{\gamma}}\;,
\end{eqnarray}
where the inverse susceptibility is identified with the (non-rescaled) mass term
$\delta$ of the bosonic propagator, see Eq. (\ref{eq:pp_green_b}). 
In Fig. \ref{fig:pp_nu}, we present a double-logarithmic plot of the susceptibility
at the end of the flow $\chi^{-1}=\delta|_{\Lam\rightarrow 0}$ versus the difference
of initial values of the control parameter, where $\delta_{\text{crit}}$
corresponds to the location of the QCP: 
$\chi^{-1}_{\Lam\rightarrow 0}(\delta_{\text{crit}})\rightarrow 0$. $\gamma$ can 
be read off from the slope in Fig. \ref{fig:pp_nu}:
\begin{eqnarray}
\gamma = 1.3\;.
\end{eqnarray}
The correlation length exponent $\nu$ now follows from the scaling relation 
(Goldenfeld 1992, Kaul 2008): $\gamma=\nu\left(2-\eta_{b}\right)$. 
With $\eta_{b}=0.37$ in $d=2$ from Fig. \ref{fig:eta_cone_flows}, one 
obtains: 
\begin{eqnarray}
\nu=0.8\;.
\end{eqnarray}
The critical exponents of the structurally similar 
Gross-Neveu model were computed with functional RG methods by Rosa et al. (2001). 
In a recent paper on a field theory containing bosons, fermions, 
and a gauge field, various quantum critical exponents were estimated in the limit
of large number of fermion and boson species (Kaul 2008).

\section{Conclusion}
\label{sec:pp_conclusion}
In this chapter, we derived coupled flow equations for fermionic and
bosonic degrees of freedom in quantum critical fermion systems. These
equations capture the mutual feedback of two distinct
types of gapless fluctuations: the first associated with zero temperature fermions
and the second with massless order parameter fluctuations. As a first application, we computed various scaling exponents
for the quantum critical point between the semimetal and superfluid 
phase of attractive Dirac fermions. The fermion and order parameter propagators are non-analytic
functions of frequency and momentum. In two dimensions,
the fermionic self-energy as function of frequency scales $\sim\omega^{0.68}$
leading to a complete breakdown of the Fermi liquid. 
We demonstrated how to compute the susceptibility and correlation length exponents
when approaching the QCP along the control parameter axis and presented first estimates thereof.

\bigskip

Extending the work of this chapter to finite temperatures is interesting. 
For example the correlation length
in the quantum critical regime as a function of temperature, $\xi(T,\delta_{\text{crit}})$,
when approaching the QCP vertically from the top (see Fig. \ref{fig:phase_toy}), would be 
worthwile to investigate within our coupled Fermi-Bose RG.

\bigskip

In the next chapter, we combine the bosonic truncations devised for phases with broken symmetry
in chapters \ref{chap:bosonicqcp_discrete} and \ref{chap:bosonicqcp_goldstone} with the 
normal phase analysis 
in terms of fermionic and bosonic fields in chapter \ref{chap:fermibosetoy} and we compute various physical properties
of fermionic superfluids at $T=0$ with a coupled fermion-boson action.

\cleardoublepage
\thispagestyle{empty}

\chapter[Fermionic superfluids at zero temperature]{Fermionic superfluids at zero temperature}
\label{chap:fermionsuperfluids}

\section{Introduction}

In this chapter, we analyze the attractive Hubbard 
model as a prototype of a Fermi system with
a superfluid low temperature phase (Micnas 1990). 
An experimental realization of the attractive Hubbard model is
conceivable by trapping fermionic atoms in an optical lattice
and tuning the interaction close to a Feshbach resonance
(Hofstetter 2002, Jaksch 2005, Chin 2006).

We here theoretically focus on the superfluid ground state for which 
the importance of quantum fluctuations 
has been emphasized recently in the context of the BCS-BEC
crossover (Diener 2008).
Although the long-range order is not destroyed by fluctuations in
dimensions $d>1$, the order parameter correlations are nevertheless
non-trivial in $d \leq 3$.
The Goldstone mode leads to severe infrared divergences in perturbation
theory. A detailed analysis of the infrared behavior of fermionic
superfluids has appeared earlier in the mathematical literature
(Feldman 1993), where the perturbative renormalizability of the
singularities associated with the Goldstone mode was established
rigorously.
To a large extent divergences of Feynman diagrams cancel due to
Ward identities, while the remaining singularities require a
renormalization group treatment.
Since the fermions are gapped at low energy scales, the infrared
behavior of the collective, bosonic sector in fermionic superfluids
is equivalent to the one of an interacting Bose gas, where
the Goldstone mode of the condensed state strongly affects the
longitudinal correlations, leading to drastic deviations from
mean-field theory in dimensions $d \leq 3$ (Nepomnyashchy 1992, Pistolesi 2004) 
as already demonstrated within the functional RG setting in chapter \ref{chap:bosonicqcp_goldstone}.

\bigskip

Technically, we implement the order parameter via a Hubbard-Stratonovich
field and we compute the renormalized effective action for the coupled theory of bosons and fermions by truncating the exact hierarchy of flow equations for the one-particle irreducible vertex functions (Baier 2004, Sch\"utz 2005, Sch\"utz 2006).
A truncation of this hierarchy has been applied a few years ago to
the antiferromagnetic state of the two-dimensional repulsive Hubbard
model (Baier 2004). Important features of the quantum
antiferromagnet at low temperatures were captured by
the flow.
More recently, various aspects of superfluidity in attractively
interacting Fermi systems have been studied in the fRG framework with
a Hubbard-Stratonovich field for the superfluid order parameter.
Approximate flow equations were discussed previously for the superfluid ground
state (Krippa 2005), for the Kosterlitz-Thouless transition in
two-dimensional superfluids (Krahl 2007) and for the BCS-BEC crossover
in three-dimensional cold atomic Fermi gases (Diehl 2007).

\bigskip

The purpose of this chapter is to construct a relatively simple
truncation of the exact fRG flow which is able to describe the
correct infrared asymptotic behavior, and which yields reasonable
estimates for the magnitude of the order parameter at least for weak and moderate
interaction strength. From a numerical solution of the flow
equations, which we perform in two dimensions, we obtain
information on the importance of Goldstone modes
and other fluctuation effects.

\bigskip

In Sec. \ref{sec:bare_action}, we introduce the bare fermion-boson action obtained from
the attractive Hubbard model by a Hubbard-Stratonovich transformation.
Neglecting bosonic fluctuations, one recovers the standard mean-field
theory for fermionic superfluids, as recapitulated in Sec. \ref{sec:mft}.
By truncating the exact fRG hierarchy, we derive approximate flow
equations involving fermionic and bosonic fluctuations in Sec. \ref{sec:rg_truncation}.
At the end of that section we reconsider mean-field theory from a flow
equation perspective.
Sec. \ref{sec:results} is dedicated to a discussion of results obtained by solving
the flow equations.
We discuss the asymptotic behavior in the infrared limit in two and
three dimensions and then present numerical results for the flow
in two dimensions, where fluctuation effects are most pronounced.
Finally, we summarize our results in Sec. \ref{sec:conclusion}.

\section{Bare action}
\label{sec:bare_action}

As a prototype model for the formation of a superfluid ground state in an
interacting Fermi system we consider the attractive Hubbard model represented
by the Hamiltonian
\begin{equation}
 H = \sum_{\bi,\bj} \sum_{\sg} t_{\bi\bj} \,
 c^{\dag}_{\bi\sg} c^{\pdag}_{\bj\sg} +
 U \sum_{\bi} n_{i\up} n_{i\down} \; ,
\end{equation}
where $c^{\dag}_{\bi\sg}$ and $c^{\pdag}_{\bi\sg}$ are creation and
annihilation operators for spin-$\frac{1}{2}$ fermions with spin orientation
$\sg$ on a lattice site $\bi$. For the hopping matrix we employ
$t_{\bi\bj} = - t$ if $\bi$ and $\bj$
are nearest neighbors on the lattice, and $t_{\bi\bj} = 0$ otherwise.
On a $d$-dimensional simple cubic lattice, this leads to a dispersion
relation $\eps_{\bk} = -2t \sum_{i=1}^d \cos k_i$.
For the attractive Hubbard model the coupling constant $U$ is
negative.

The attractive Hubbard model has a superfluid ground state for any
particle density $n$ in $d \geq 2$ dimensions (Micnas 1990), provided the
lattice is not completely filled ($n=2$) or empty ($n=0$). The Fermi surface
for the fiducial case of a quarter filled band is shown in Fig. \ref{fig:fs}. 
At half filling ($n=1$) the usual U(1) global gauge symmetry becomes a
subgroup of a larger SO(3) symmetry group, and the order parameter for
superfluidity mixes with charge density order (Micnas 1990). 
\begin{wrapfigure}{r}{0.5\textwidth}
  \vspace{2mm}
  \begin{center}
    \includegraphics[width=0.30\textwidth]{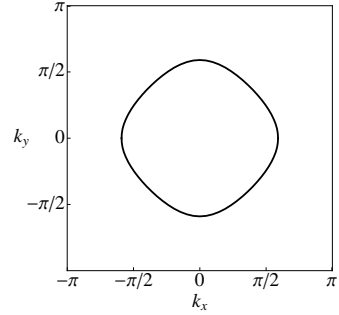}
  \end{center}
  \vspace{-10pt}
  \caption{\textit{Exemplary Fermi surface for a quarter-filled band}.}
  \vspace{-14mm}
\label{fig:fs}
\end{wrapfigure}

Our analysis is based on a functional integral representation of the
effective action, that is, the generating functional of one-particle
irreducible correlation functions.
For the Hubbard model, the starting point is a functional integral
over fermionic fields $\psi$ and $\psib$ with the {\em bare} action
\begin{eqnarray}
  &&\Gam_0[\psi,\psib]
   =  - \int_{k\sg}
  \psib_{k\sg} (ik_{0}-\xi_{\bk}) \, \psi_{k\sg} \nonumber\\
  && + \int_{k,k',q} U \,
  \psib_{-k+\frac{q}{2}\down} \psib_{k+\frac{q}{2} \up}
  \psi_{k'+\frac{q}{2}\up} \psi_{-k'+\frac{q}{2}\down} \; ,\nonumber\\
  \label{eq:bare_action}
\end{eqnarray}
where $\xi_{\bk} = \eps_{\bk} - \mu$ is the single-particle energy
relative to the chemical potential.
The variables $k = (k_0,\bk)$ and $q = (q_0,\bq)$
collect Matsubara energies and momenta.
We use the short-hand notation
$\int_k = \int_{k_0} \int_{\bk} =
 \int_{-\infty}^{\infty} \frac{d k_{0}}{2\pi}
 \int_{-\pi}^{\pi} \frac{d^d \bk}{(2\pi)^d} \,$
for momentum and energy integrals,
and $\int_{k\sg}$ includes also a spin sum.
We consider only {\em ground state} properties, so that
the energy variables are continuous.

The attractive interaction drives spin-singlet pairing with s-wave
symmetry and a spontaneous breaking of the global U(1)
gauge symmetry.
Therefore, we decouple the Hubbard interaction in the s-wave
spin-singlet pairing channel by introducing a complex bosonic
Hubbard-Stratonovich field $\phi_q$ conjugate to the bilinear
composite of fermionic fields (Popov 1987)
\begin{equation}
 \tilde\phi_{q} = U \int_k \psi_{k+\frac{q}{2}\up}
 \psi_{-k+\frac{q}{2}\down} \; .
\end{equation}
This yields a functional integral over $\psi$, $\psib$ and $\phi$
with the new bare action
\begin{eqnarray}
 \Gam_0[\psi,\psib,\phi]
  = &-& \int_{k\sg} \psib_{k\sg} (ik_{0} - \xi_{\bk}) \, \psi_{k\sg}
  - \int_q \phi^*_q \frac{1}{U} \phi_q
  \nonumber\\
   &+& \int_{k,q} \left(
  \psib_{-k+\frac{q}{2} \down} \psib_{k+\frac{q}{2} \up} \,
  \phi_{q} +
  \psi_{k+\frac{q}{2} \up} \psi_{-k+\frac{q}{2}\down} \,
  \phi^*_q \right) . 
  \label{eq:finalmodel}
\end{eqnarray}
where $\phi^*$ is the complex conjugate of $\phi$, while $\psi$
and $\psib$ are algebraically independent Grassmann variables.
Our aim is to compute fermionic and bosonic correlation functions
with focus on the correct description of the low energy (infrared)
behavior. 


\section{Mean-field theory}
\label{sec:mft}

As a warm-up for the renormalization group treatment it is instructive
to recapitulate the mean-field theory for the superfluid phase
in the functional integral formalism (Popov 1987).
In mean-field approximation bosonic fluctuations are neglected,
that is, the bosonic field $\phi$ is fixed instead of being integrated
over all possible configurations.
The fermion fields can be integrated exactly.
The (fixed) bosonic field is determined by minimizing the effective
action as a functional of $\phi$.
For a homogeneous system, the minimizing $\phi_q$ can be be non-zero
only for $q = 0$. We denote the minimum by $\alf$.
Substituting $\phi_0 \to \alf + \phi_0$ yields
\begin{eqnarray}
 \Gam_0[\psi,\psib,\alf+\phi] \! &=&
 - \int_{k\sg} \psib_{k\sg} (ik_{0} - \xi_{\bk}) \, \psi_{k\sg}
 - \alf^* \frac{1}{U} \alf
    + \int_k \left( \psib_{-k\down} \psib_{k\up} \, \alf
  + \psi_{k\up} \psi_{-k\down} \, \alf^* \right)
 \nonumber \\
 && - \frac{1}{U}
 \left( \alf^* \phi_0 + \alf \, \phi_0^* \right)
  + \int_{k,q} \left(
 \psib_{-k+\frac{q}{2} \down} \psib_{k+\frac{q}{2} \up} \,
 \phi_{q} +
 \psi_{k+\frac{q}{2} \up} \psi_{-k+\frac{q}{2}\down} \,
 \phi^*_q \right)
 \nonumber\\
 && - \int_q \phi_q^* \frac{1}{U} \phi_q \; .
\label{eq:mft_Gam_0}
\end{eqnarray}
A necessary condition for a minimum of the effective action is
that its first derivative with respect to $\phi$ (or $\phi^*$),
that is, the bosonic 1-point function $\Gam_b^{(1)}(q)$, vanishes.
In other words, terms linear in $\phi$ (or $\phi^*$) have to vanish
in the effective action.
For $q \neq 0$, $\Gam_b^{(1)}(q)$ vanishes for any choice of $\alf$
in a homogeneous system.
For $q = 0$ and in mean-field approximation, the 1-point function is
given by
\begin{equation}
 \Gam_b^{(1)}(0) = - \frac{1}{U} \alf +
 \int_k \bra \psi_{k\up} \psi_{-k\down} \ket\;,
\label{eq:orderparam_mft}
\end{equation}
where $\bra \dots \ket$ denotes expectation values.
The first term on the right hand side corresponds to the contribution
$- \frac{1}{U} \alf \phi_0^*$ to $\Gam_0$ in the second line of Eq.~(\ref{eq:mft_Gam_0}),
while the second term is generated by contracting the fermions in the
contribution proportional to $\phi_q^*$ in the second line of Eq.~(\ref{eq:mft_Gam_0}).
In the absence of bosonic fluctuations there is no other contribution
to $\Gam_b^{(1)}$.
From the condition $\Gam_b^{(1)}(0) = 0$ one obtains
\begin{equation}
 \alf = U \int_k \bra \psi_{k\up} \psi_{-k\down} \ket \; ,
\end{equation}
which relates $\alf$ to a fermionic expectation value.
We now turn to the fermionic 2-point functions.
The normal fermionic propagator
$G_{f\sg}(k) = - \bra \psi_{k\sg} \psib_{k\sg} \ket$
and the anomalous propagators
$F_f(k) = - \bra \psi_{k\up} \psi_{-k\down} \ket$ and
$\bar F_f(k) = - \bra \psib_{-k\down} \psib_{k\up} \ket$
can be conveniently collected in a Nambu matrix propagator
\begin{equation}
 \bG_f(k) =
 \left( \begin{array}{cc}
 G_{f\up}(k) & F_f(k) \\[1mm]
 \bar F_f(k) & - G_{f\down}(-k)
 \end{array} \right) \; .
\end{equation}
The anomalous propagators satisfy the relations
$\bar F_f(k) = F_f^*(k)$ and $F_f(-k) = F_f(k)$.
In (our) case of spin rotation invariance the normal propagator
does not depend on $\sg$, and therefore,
$G_{f\up}(k) = G_{f\down}(k) = G_f(k)$.

In mean-field theory, the fermionic 2-point vertex function
$\bGam_f^{(2)} = - \bG_f^{-1}$
can be read off directly from the bare action in the form Eq.~(\ref{eq:mft_Gam_0}):
\begin{equation}
 \bGam_f^{(2)}(k) = - \left( \begin{array}{cc}
 ik_0 - \xi_{\bk} & \alf \\
 \alf^* & ik_0 + \xi_{-\bk}
 \end{array} \right) \; .
\end{equation}
The off-diagonal elements are due to the last bracket in the first line
of Eq.~(\ref{eq:mft_Gam_0}). Tadpole contributions which are generated from the terms
in the second and third line of Eq.~(\ref{eq:mft_Gam_0}) cancel exactly by virtue of the condition
$\Gam_b^{(1)} = 0$. In the absence of bosonic fluctuations there are
no other contributions to $\bGam_f^{(2)}$.
Inverting $\bGam_f^{(2)}$ and using $\xi_{-\bk} = \xi_{\bk}$ from
reflection symmetry yields
\begin{eqnarray}
 G_f(k) &=&
 \frac{-ik_0 - \xi_{\bk}}{k_0^2 + E_{\bk}^2}
 \label{eq:Gf}
 \\
 F_f(k) &=&
 \frac{\Delta}{k_0^2 + E_{\bk}^2} \; ,
\label{eq:Ff}
\end{eqnarray}
where $E_{\bk} = (\xi_{\bk}^2 + |\Delta|^2 )^{1/2}$ and
$\Delta = \alf$.
We observe that in mean-field theory the bosonic order parameter $\alf$
is equivalent to the gap $\Delta$ in the fermionic excitation spectrum.
Eq.~(\ref{eq:orderparam_mft}) corresponds to the BCS gap equation
\begin{equation}
 \Delta = - U \int_k F_f(k) \; .
\label{eq:bcs_gap_eqn}
\end{equation}

We finally compute the bosonic 2-point functions in mean-field theory (see Keller (1999) for a more detailed explanation of the diagrammatic ingredients and the notation in the superfluid phase). The bosonic propagators $G_b(q) = - \bra \phi_q \phi^*_q \ket$ and
$F_b(q) = - \bra \phi_q \phi_{-q} \ket$ =
$- \bra \phi^*_{-q} \phi^*_q \ket^*$ form the matrix propagator
\begin{equation}
 \bG_b(q) =
 \left( \begin{array}{cc}
 G_b(q) & F_b(q) \\[1mm]
 F_b^*(q) & G_b(-q)
 \end{array} \right) \; .
\end{equation}
Note that $F_b(-q) = F_b(q)$.
The bosonic 2-point function $\bGam_b^{(2)}$ is equal to
$-\bG_b^{-1}$.
We define a bosonic self-energy $\bSg_b$ via the Dyson equation
$(\bG_b)^{-1} = (\bG_{b0})^{-1} - \bSg_b$, where the bare propagator
corresponding to the bare action $\Gam_0$ is given by
\begin{equation}
 \bG_{b0}(q) =
 \left( \begin{array}{cc}
 U & 0 \\
 0 & U
 \end{array} \right) \; .
\end{equation}
In mean-field theory, only fermionic bubble diagrams contribute to
the bosonic self-energy:
\begin{equation}
 \bSg_b(q) = \left( \begin{array}{cc}
 K(q) & L(q) \\
 L^*(q) & K(-q)
 \end{array} \right) \; ,
\end{equation}
where
\begin{eqnarray}
 K(q) &=& - \int_k G_f(k+q) \, G_f(-k) \\[2mm]
 L(q) &=& {\phantom -} \int_k F_f(k+q) \, F_f(-k)\;,
\end{eqnarray}
see Fig. \ref{fig:sc_KL} for an exemplary plot of $K(q)$
in the normal phase.
\begin{figure}[t]
\includegraphics[width=70mm]{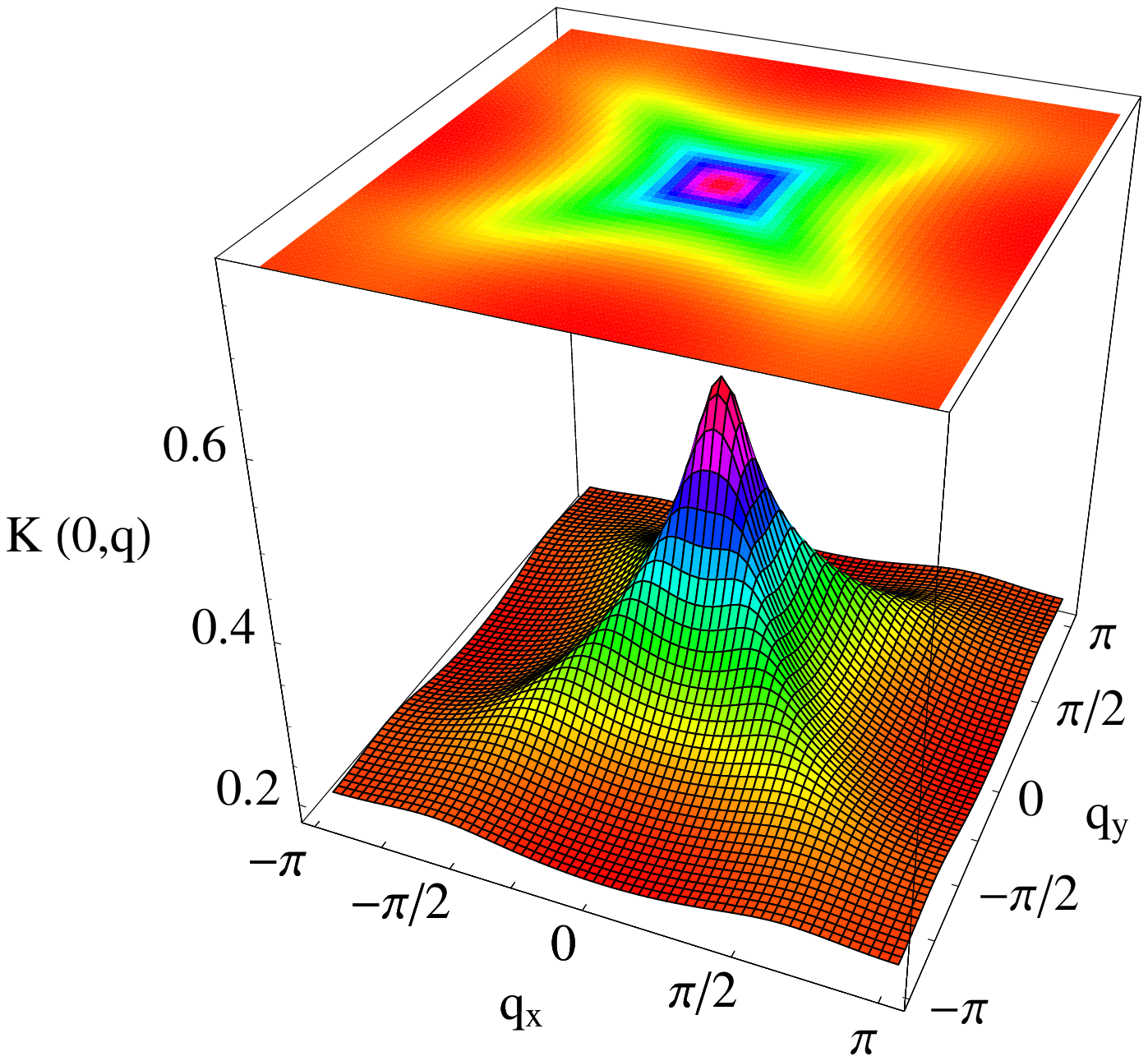}\\[-55mm]
\hspace*{69mm}
\includegraphics[width=76mm]{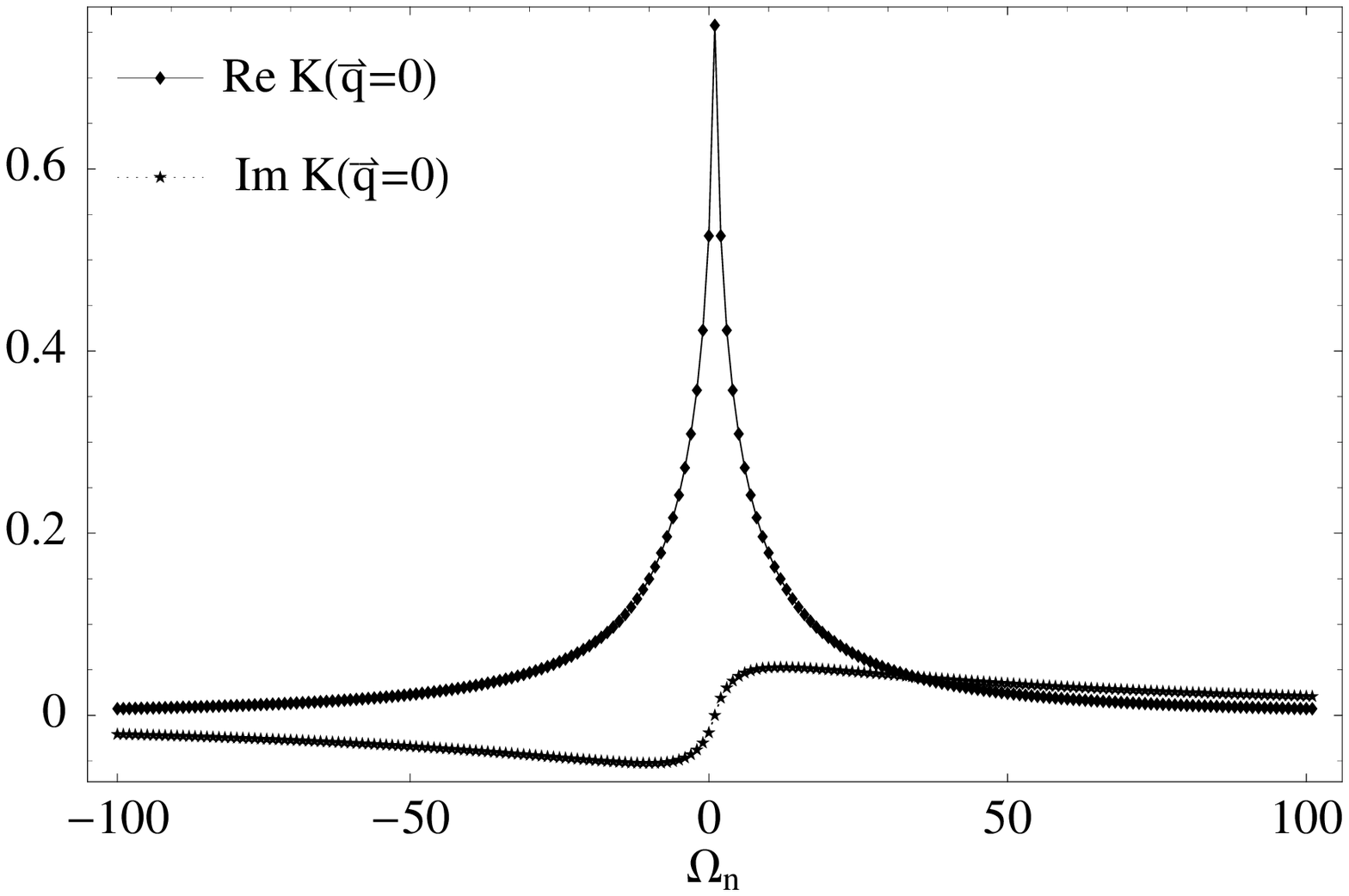}\\[0mm]
\caption{Momentum (left) and frequency (right) dependence of $K(q)$ 
in the normal phase. 
The small linear frequency dependence in the imaginary part vanishes at half-filling and
is discussed in Sec. \ref{sec:rg_truncation}.}
\label{fig:sc_KL}
\vspace*{3mm}
\end{figure}
In the absence of bosonic fluctuations, there are no other contributions
to $\bSg_b$. Tadpole diagrams cancel due to $\Gam_b^{(1)} = 0$.
Note that $K(-q) = K^*(q)$ while $L(-q) = L(q)$.
Inverting the matrix $\bG_{b0} - \bSg_b$ one obtains the bosonic
propagator in mean-field approximation
\begin{equation}
 \bG_b(q) = \frac{1}{d(q)} \left( \begin{array}{cc}
 U^{-1} - K(-q) & L(q) \\
 L^*(q) & U^{-1} - K(q)
 \end{array} \right) \; ,
\end{equation}
with the determinant
$d(q) = |U^{-1} - K(q)|^2 - |L(q)|^2$. %
Using the explicit expressions (\ref{eq:Gf}) and (\ref{eq:Ff})
for $G_f$ and $F_f$, respectively, one can see that
\begin{equation}
 U^{-1} - K(0) + |L(0)| = U^{-1} + \frac{1}{\Delta} \int_k F_f(k) \; ,
\end{equation}
which vanishes if $\Delta$ is non-zero and satisfies the gap equation.
Hence $d(q)$ has a zero and $\bG_b(q)$ a pole in $q=0$.
This pole corresponds to the Goldstone mode associated with the
sponaneous breaking of the U(1) symmetry of the model.
For small finite $\bq$ and $q_0$, the leading $q$-dependences of
$|U^{-1} - K(q)|$ and $L(q)$ are of order $|\bq|^2$ and $q_0^2$. 
Hence the divergence of $\bG_b(q)$ for $q \to 0$ is quadratic in
$\bq$ and $q_0$. Continuing $q_0$ to real frequencies one obtains
a propagating mode with a linear dispersion relation. The second pole
of $d(q)$ is gapped and features a quadratic momentum dispersion.

\bigskip

By appropriately tailoring the fRG-trunction in Sec. \ref{sec:rg_truncation} to the
results of this mean-field calculation in the superfluid phase,
we incorporate the effects of transversal Goldstone fluctuations as well as
longitudinal fluctuations into our computation.

\section{Truncation}
\label{sec:rg_truncation}

At the core of our analysis is the exact functional flow equation for the
scale-dependent effective action $\Gamma^{\Lambda}$, which generates
all 1PI correlation functions (Wetterich 1993, Berges 2002, Salmhofer 2001, Metzner 2005):
\begin{eqnarray}
\frac{d}{d\Lam} \Gam^{\Lam}[\mathcal{S},\bar{\mathcal{S}}] =
 {\rm Str} \, \frac{\dot{\bR}^{\Lam}}
 {\bGam^{(2) \, \Lam}[\mathcal{S},\bar{\mathcal{S}}] + \bR^{\Lam}} \;,
\label{eq:sf_exact_flow}
\end{eqnarray}
where $\bGam^{(2) \, \Lam}=\partial^{2}\Gam^{\Lam}[\mathcal{S},\bar{\mathcal{S}}]/\partial\mathcal{S}\partial\bar{\mathcal{S}}$ denotes the second functional derivative with respect to the superfields and the supertrace (str) traces over all indices
with an additional minus sign for fermionic contractions. We refer to Section \ref{sec:flow_eq} for 
a detailed exposition of the functional RG formalism including symmetry-breaking.

Untruncated, the exact effective action contains an infinite number of
terms of arbitrary order in fermionic and/or bosonic fields.
We now describe which terms are kept and how they are
parametrized.
We keep all terms which are crucial for a qualitatively
correct description of the low-energy behavior of the
system.
We distinguish the symmetric regime, where $\alf = 0$,
from the symmetry broken regime, where $\alf \neq 0$.
The former applies to large $\Lam$, the latter to small
$\Lam$.

\subsection{Symmetric regime}
\label{subsubsec:trunc_sym}
The bare action Eq.~(\ref{eq:finalmodel}) contains quadratic
terms for fermions and bosons, and an interaction term where
bosons couple linearly to a fermion bilinear. In the effective
action we keep these terms with generalized cutoff-dependent
parameters and add a bosonic self-interaction of order
$|\phi|^4$. The latter is generated by the flow and
becomes crucial when the quadratic part of the bosonic
potential changes sign.
Other interactions generated by the flow
are neglected. For our choice of parameters (relatively small $U$), the
fermionic propagator receives only Fermi liquid
renormalizations, leading to a slightly reduced quasi particle
weight and a weakly renormalized dispersion relation.
We neglect these quantitatively small effects and leave
the quadratic
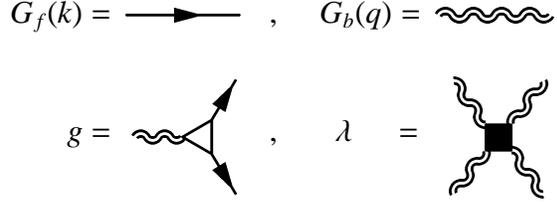
\begin{wrapfigure}{r}{0.5\textwidth}
  \vspace{-5mm}
  \begin{center}
\begin{fmffile}{legend_sym_27}
\begin{eqnarray}
G_{f}(k)&=&
\parbox{20mm}{\unitlength=1mm\fmfframe(1,1)(1,1){
\begin{fmfgraph*}(15,1)\fmfpen{thin}
\fmfleft{l1}
 \fmfright{r1}
  \fmf{fermion}{l1,r1}
 \end{fmfgraph*}
}}\nonumber,\hspace{5mm}
G_{b}(q)=
\parbox{20mm}{\unitlength=1mm\fmfframe(1,1)(1,1){
\begin{fmfgraph*}(15,1)\fmfpen{thin}
\fmfleft{l1}
 \fmfright{r1}
  \fmf{dbl_wiggly}{l1,r1}
 \end{fmfgraph*}
}}\nonumber\\[3mm]
g&=&
\parbox{20mm}{\unitlength=1mm\fmfframe(2,2)(1,1){
\begin{fmfgraph*}(15,15)\fmfpen{thin}
\fmfleft{l1}
\fmfrightn{r}{2}
\fmf{dbl_wiggly}{l1,G1}
\fmfpolyn{empty,tension=0.8}{G}{3}
\fmf{fermion}{G2,r1}
\fmf{fermion}{G3,r2}
 \end{fmfgraph*}
}},\nonumber\hspace{7mm}
\lambda\hspace{5mm}=
\parbox{20mm}{\unitlength=1mm\fmfframe(2,2)(1,1){
\begin{fmfgraph*}(15,15)\fmfpen{thin}
\fmfleftn{l}{2}
\fmfrightn{r}{2}
\fmf{dbl_wiggly}{l2,G1}
\fmf{dbl_wiggly}{l1,G2}
\fmfpolyn{full,tension=1.5}{G}{4}
\fmf{dbl_wiggly}{r1,G3}
\fmf{dbl_wiggly}{r2,G4}
 \end{fmfgraph*}
}}\nonumber\\[-7mm]\nonumber
\end{eqnarray}
\end{fmffile}
  \end{center}
  \vspace{3mm}
\caption{\textit{Diagrammatic constituents of our truncation in the symmetric regime as
described in Sec. \ref{subsubsec:trunc_sym}}.}
  \vspace{3mm}
\label{fig:sym_constituents}
\end{wrapfigure}
fermionic term in the action unrenormalized,
that is,
\begin{equation}
 \Gam_{\psib\psi} = - \int_{k\sg}
 \psib_{k\sg} (ik_0 - \xi_{\bk}) \,
 \psi_{k\sg} \; ,
\label{eq:Gam_psi_psi}
\end{equation}
corresponding to an unrenormalized fermionic propagator
\begin{equation}
 G_{f}(k) = G_{f0}(k) =
 \frac{1}{ik_{0}-\xi_{\mathbf{k}}} \; .
\end{equation}
In the bare action the term quadratic in bosons contains
only a mass term. In the effective action this mass
decreases with decreasing cutoff until it vanishes at
a critical scale $\Lam_c$, which marks the transition to
the symmetry-broken regime. As the mass decreases, the
momentum and frequency dependence of the bosonic 2-point
function becomes important. The latter is generated in
particular by fermionic fluctuations. For small $\bq$,
the leading $\bq$-dependence is of order $|\bq|^2$.
The leading frequency-dependent contribution to the
real part of the bosonic 2-point function is of order
$q_0^2$. The frequency-dependence of the imaginary part
is generally of order $q_0$ but the prefactor is very
small which is related to the fact that it vanishes
completely in case of particle-hole symmetry. Furthermore
this small imaginary part does not have any qualitative
impact on the quantities we compute in the following.
We therefore neglect this term and make the ansatz
\begin{equation}
 \Gam_{\phi^*\phi} = \frac{1}{2} \int_q \phi_q^*
 ( m_b^2 + Z_b q_0^2 + A_b \om_{\bq}^2 ) \, \phi_q \; ,
\end{equation}
where
$\om_{\bq}^2 = 2 \sum_{i=1}^d \left(1-\cos q_i\right)$ is fixed,
while $m_{b}^{2}$, $Z_b$, and $A_b$ are cutoff-dependent numbers.
The function $\om_{\bq}^2$ has been chosen such that the
quadratic momentum dependence for small $\bq$ is
continued to a periodic function defined on the entire
Brillouin zone.
The initial conditions for the parameters in the bosonic
2-point function,
\begin{eqnarray}
G_{b}(q)=-\frac{2}{Z_{b}q_{0}^{2}+A_{b}\omega^{2}_{\mathbf{q}}+m_{b}^{2}}\,\,,
\end{eqnarray}
can be read off from the bare action as
$m^{2}_b = |2/U|$ and $Z_b = A_b = 0$.

The interaction between fermions and bosons remains regular
and finite near $\Lam_c$. It can therefore be parametrized
as
\begin{equation}
 \Gam_{\psi^2\phi^*} =
  g \int_{k,q} \left(
  \psib_{-k+\frac{q}{2} \down} \psib_{k+\frac{q}{2} \up} \,
  \phi_{q} +
  \psi_{k+\frac{q}{2} \up} \psi_{-k+\frac{q}{2}\down} \,
  \phi^*_q \right) \; ,
\label{eq:g_sym}
\end{equation}
where the coupling constant $g$ depends on the cutoff, but
not on momentum and frequency. The initial condition for
$g$ is $g = 1$.

The flow generates a bosonic self-interaction which plays a
crucial role near and in the symmetry-broken regime, that is,
when the bosonic mass term becomes small and finally changes
sign.
The most relevant term is a local $|\phi|^4$-interaction
\begin{equation}
 \Gam_{|\phi|^4} = \frac{\lam}{8} \int_{q,q',p}
 \phi^*_{q+p} \phi^*_{q'-p} \phi_{q'} \phi_q \; ,
\end{equation}
where the coupling constant $\lam$ depends on the cutoff
but not on momentum and frequency.
The initial condition for $\lam$ is $\lam = 0$.

The propagators and vertices are represented
diagrammatically in Fig. \ref{fig:sym_constituents}.

\subsection{Symmetry-broken regime}
\label{subsubsec:trunc_ssb}

For $\Lam < \Lam_c$ the effective action develops a
minimum at $\phi_{q=0} = \alf \neq 0$. Due to the U(1)
symmetry associated with charge conservation the minimum
is degenerate with respect to the phase of $\alf$.
We here employ the $\sg$-$\mathbf{\Pi}$ model presented previously 
in section \ref{sec:sig_pi_model} consisting of the quartic coupling $\lambda$ and the minimum $\alpha$, see Eqs. (\ref{eq:gold_effpot}, \ref{eq:ssb_basis}, \ref{eq:gold_ints}). In this chapter we focus 
on the case $N=2$ degrading the vector field $\mathbf{\Pi}$ to a 
one-component field $\pi$ representing the Goldstone mode. The leading momentum 
and frequency dependence of the bosonic 2-point function is quadratic in $\bq$ and $q_0$, both 
\begin{wrapfigure}{r}{0.5\textwidth}
  \vspace{-4mm}
  \begin{center}
\begin{fmffile}{legend_ssb_red}
\begin{eqnarray}
G_{f}(k)&=&
\parbox{20mm}{\unitlength=1mm\fmfframe(1,1)(1,1){
\begin{fmfgraph*}(15,1)\fmfpen{thin}
\fmfleft{l1}
 \fmfright{r1}
  \fmf{fermion}{l1,r1}
 \end{fmfgraph*}
}},\hspace{5mm}
F_{f}(k)=
\parbox{20mm}{\unitlength=1mm\fmfframe(1,1)(1,1){
\begin{fmfgraph*}(15,1)\fmfpen{thin}
\fmfcmd{%
   style_def gap expr p=
     cdraw p;
     cfill (harrow(reverse p, .5));
     cfill (harrow(p,.5));
   enddef;
   style_def gap_hc expr p=
     cdraw p;
     cfill (tarrow(reverse p, .55));
     cfill (tarrow(p,.55));
   enddef;
     }
\fmfleft{l1}
 \fmfright{r1}
  \fmf{gap}{l1,r1}
 \end{fmfgraph*}
}}\nonumber\\[5mm]
G_{\sigma}(k)&=&
\parbox{20mm}{\unitlength=1mm\fmfframe(1,1)(1,1){
\begin{fmfgraph*}(15,1)\fmfpen{thin}
\fmfleft{l1}
 \fmfright{r1}
  \fmf{dashes}{l1,r1}
 \end{fmfgraph*}
}},\hspace{5mm}
G_{\pi}(k)=
\parbox{20mm}{\unitlength=1mm\fmfframe(1,1)(1,1){
\begin{fmfgraph*}(15,1)\fmfpen{thin}
\fmfleft{l1}
 \fmfright{r1}
  \fmf{photon}{l1,r1}
 \end{fmfgraph*}
}}\nonumber\\[2mm]
g_{\sigma}&=&
\parbox{20mm}{\unitlength=1mm\fmfframe(2,2)(1,1){
\begin{fmfgraph*}(15,15)\fmfpen{thin}
\fmfleft{l1}
\fmfrightn{r}{2}
\fmf{dashes}{l1,G1}
\fmfpolyn{hatched,tension=0.8}{G}{3}
\fmf{fermion}{G2,r1}
\fmf{fermion}{G3,r2}
 \end{fmfgraph*}
}},\hspace{5mm}
g_{\pi}\hspace{4mm}=
\parbox{20mm}{\unitlength=1mm\fmfframe(2,2)(1,1){
\begin{fmfgraph*}(15,15)\fmfpen{thin}
\fmfleft{l1}
\fmfrightn{r}{2}
\fmf{photon}{l1,G1}
\fmfpolyn{filled=30,tension=0.8}{G}{3}
\fmf{fermion}{G2,r1}
\fmf{fermion}{G3,r2}
 \end{fmfgraph*}
}}\nonumber
\end{eqnarray}
\end{fmffile}
  \end{center}
  \vspace{3mm}
\caption{\textit{Propagators and fermion-boson vertex for our truncation in
the symmetry-broken regime as specified in Sec.
\ref{subsubsec:trunc_ssb}. The bosonic self-interactions are exhibited in 
Fig. \ref{fig:ssb_goldstone_constituents}}.}
\label{fig:ssb_constituents}
\vspace*{-20mm}
\end{wrapfigure}
for the
$\sg$- and $\pi$-component.
Hence we make the following ansatz for the quadratic
bosonic contributions to the effective action
\begin{eqnarray}
 \Gam_{\sg\sg} &=& \frac{1}{2} \int_q \sg_{-q}
 ( m_{\sg}^2 + Z_{\sg} q_0^2 + A_{\sg} \om_{\bq}^2 ) \,
 \sg_q\nonumber\\
 \Gam_{\pi\pi} &=& \frac{1}{2} \int_q \pi_{-q}
 ( Z_{\pi} q_0^2 + A_{\pi} \om_{\bq}^2 ) \, \pi_q \; .\nonumber\\
\end{eqnarray}
where $m_{\sg}$, $Z_{\sg}$, $A_{\sg}$, $Z_{\pi}$, and
$A_{\pi}$ are cutoff dependent real numbers. 
The propagators for the $\sg$ and $\pi$ fields thus have
the form
\begin{eqnarray}
 G_{\sg}(q) &=&
 - \frac{1}{m_{\sg}^2 + Z_{\sg} q_0^2 + A_{\sg} \om_{\bq}^2}
\nonumber\\
 G_{\pi}(q)& =&
 - \frac{1}{Z_{\pi} q_0^2 + A_{\pi} \om_{\bq}^2} \; .
\end{eqnarray}
The longitudinal mass is determined by the $|\phi|^4$
coupling $\lam$ and the minimum $\alf$ as
\begin{equation}
 m_{\sg}^2 = \lam \, |\alf|^2 \; .
\label{eq:mSig_condition}
\end{equation}
The small imaginary contribution to $\Gam_{\phi^*\phi}$ generated
by fermionic fluctuations in the absence of particle-hole symmetry,
mentioned already above, gives rise to an off-diagonal quadratic
term $\Gam_{\sg\pi}$ with a contribution linear in $q_0$. Since this
term would complicate the analysis without having any
significant effect, we will neglect it.

We now discuss terms involving fermions in the case of
symmetry breaking.
In addition to the normal quadratic fermionic term $\Gam_{\psib\psi}$
defined as before, see Eq.~(\ref{eq:Gam_psi_psi}), the
anomalous term
\begin{equation}
 \Gam_{\psi\psi} = \int_k \left(
 \Delta \, \psib_{-k\down} \psib_{k\up} +
 \Delta^* \psi_{k\up} \psi_{-k\down} \right)
\end{equation}
is generated in the symmetry-broken regime, where $|\Delta|$ is
a cutoff-dependent energy gap.
The phase of $\Delta$ is inherited from the phase of
$\alf$ while its modulus is generally different,
due to fluctuations.
Since we have chosen $\alf$ real and positive, $\Delta$
is real and positive, too.
The normal and anomalous fermionic propagators $G_f$ and $F_f$
corresponding to $\Gam_{\psib\psi}$ and $\Gam_{\psi\psi}$ have
the standard mean-field form as in Eqs.~(\ref{eq:Gf}) and
(\ref{eq:Ff}), with $E_{\bk} = (\xi_{\bk}^2 + |\Delta|^2 )^{1/2}$,
but now $\Delta$ is not equal to $\alf$.

In addition to the interaction between fermions and bosons
of the form Eq.~(\ref{eq:g_sym}), an anomalous term of the
form
\begin{equation}
 \Gam_{\psi^2\phi} =
  \tilde{g} \int_{k,q} \left(
  \psib_{-k+\frac{q}{2} \down} \psib_{k+\frac{q}{2} \up} \,
  \phi^*_{q} +
  \psi_{k+\frac{q}{2} \up} \psi_{-k+\frac{q}{2}\down} \,
  \phi_q \right)
\end{equation}
is generated in the symmetry-broken regime.
Inserting the decomposition of $\phi$ in longitudinal and
transverse fields into the normal and anomalous interaction
terms, we obtain
\begin{eqnarray}
 \Gam_{\psi^2\sg} &=&
  g_{\sigma} \int_{k,q} \left(
  \psib_{-k+\frac{q}{2} \down} \psib_{k+\frac{q}{2} \up} \,
  \sg_{q} +
  \psi_{k+\frac{q}{2} \up} \psi_{-k+\frac{q}{2}\down} \,
  \sg_{-q} \right) \, , \quad \\
 \Gam_{\psi^2\pi} &=&
  ig_{\pi} \int_{k,q} \left(
  \psib_{-k+\frac{q}{2} \down} \psib_{k+\frac{q}{2} \up} \,
  \pi_{q} -
  \psi_{k+\frac{q}{2} \up} \psi_{-k+\frac{q}{2}\down} \,
  \pi_{-q} \right) \, ,
\end{eqnarray}
where $g_{\sigma}=g+\tilde{g}$ and $g_{\pi}=g-\tilde{g}$.
Fermions couple with different strength to the $\sigma$-
and $\pi$-field, respectively.

A diagrammatic represention of the various progagators and
interaction vertices in the symmetry-broken regime is shown
in Fig. \ref{fig:ssb_constituents}.

We finally note that in a previously reported truncation (Birse 2005)
of the fRG flow in a fermionic superfluid, no distinction between longitudinal and
transverse fields was made for the bosonic Z-factors in
the symmetry-broken regime.

\subsection{Flow equations}
\label{subsec:flow_equations}

Inserting the above ansatz for the truncated effective action
into the exact flow equation and comparing coefficients
yields a set of coupled flow equations for the cutoff
dependent parameters.
The various contributions can be conveniently represented
by Feynman diagrams.
The prefactors and signs in the flow equations could be
extracted from the expansion of the exact flow equation,
Eq.~(\ref{eq:flow_expansion}). However, in practice we determine
them by comparison to a conventional perturbation expansion.

All contributions to our flow equations correspond to one-loop
diagrams with only one momentum and frequency integration,
as dictated by the structure of the exact flow equation in the
form (\ref{eq:flow_expansion}).
One of the propagators in the loop is a bosonic or fermionic
component of the single-scale propagator $\bG'_R$, the others
(if any) are components of $\bG_R$.

In this chapter, we use sharp frequency cutoffs which exclude bosonic
fields with $|\mbox{frequency}| < \Lam_b$ and fermionic fields with
$|\mbox{frequency}| < \Lam_f$ from the functional integral.
Thereby both fermionic and bosonic infrared divergences are
regularized.

Both cutoffs are monotonic functions of the flow parameter,
$\Lam_b(\Lam)$ and $\Lam_f(\Lam)$, which vanish for $\Lam \to 0$ and
tend to infinity for $\Lam \to \infty$. Concretely, we employ
\begin{equation}
 \bR_s^{\Lam}(k) = [\bG_{s0}(k)]^{-1} -
 [\chi_s^{\Lam}(k_0) \, \bG_{s0}(k)]^{-1}
\end{equation}
for $s = b,f$, and $\chi_s^{\Lam}(k_0) = \Theta(|k_0| - \Lam_s)$.
This term replaces the bare propagators $\bG_{s0}$ by
$\bG_{s0}^{\Lam} = \chi_s^{\Lam} \bG_{s0}$.

For a sharp frequency cutoff the frequency variable running
around the loop is pinned by $\bG'_R(k_0)$ to $k_0 = \pm\Lam_b$
or $k_0 = \pm\Lam_f$ as the socalled single-scale propagator ${\bG'_R}^{\!\Lam}$ has
support only for frequencies at the cutoffs, that is, for
$|k_0| = \Lam_s$.
Hence the frequency integral can be performed analytically.
The problem that the integrand contains also step functions
$\chi_s^{\Lam}(k_0) = \Theta(|k_0|-\Lam_s)$ can be treated
by using the identity
\begin{eqnarray}
\int dx \, \delta(x-x_0) \, f[x,\Theta(x-x_0)] =
\int_0^1 du \, f(x_0,u)\;,
\end{eqnarray}
which is valid for any continuous function $f$.

More specifically, in the present case the one-loop diagrams
are evaluated for vanishing external frequencies, such that
all internal propagators carry the same frequency.
In loops involving only either only bosonic or only fermionic
propagators, one can use the identity
\begin{equation}
 n \int dk_0 \, \bG'_{sR}(k_0) \, \bA \,
 [\bG_{sR}(k_0) \, \bA]^{n-1} = \Lam'_s
 \sum_{k_0 = \pm\Lam_s} [\bG_s(k_0) \, \bA]^n \, ,
\label{eq:cutoff_identity}
\end{equation}
valid for any matrix $\bA$,
to replace the frequency integration by a frequency sum
over $\pm\Lam_s$ while replacing all the propagators in the
loop by $\bG_s$. The factor $n$ corresponds to the $n$ possible
choices of positioning $\bG_{sR}'$ in a loop with $n$ lines,
and $\Lam_s' = \partial\Lam_s/\partial\Lam$.
For $\Lam_b = \Lam_f$ the above formula holds also for the
superpropagator $\bG_R$, such that is applies also to loops
with mixed products of bosonic and fermionic propagators.
For $\Lam_b \neq \Lam_f$, mixed loops contribute only if the
single-scale propagator is associated with the larger cutoff.
For example, for $\Lam_b > \Lam_f$, the single-scale propagator
has to be bosonic, since $\bG_b$ vanishes at $|k_0| = \Lam_f$.
On the other hand, for $|k_0| = \Lam_b$ one has
$\bG_{fR}(k_0) = \bG_f(k_0)$, and for the integration of the
bosonic factors in the loop one can again use Eq.~(\ref{eq:cutoff_identity}).

For loop integrals involving the frequency sum over $\pm\Lam_s$
and the momentum integral over the Brillouin zone we use the
short-hand notation
\begin{equation}
 \int_{k|\Lam_s} = \frac{\Lam_s'}{2\pi} \sum_{k_0 = \pm\Lam_s}
 \int \frac{d^dk}{(2\pi)^d} \; ,
\end{equation}
where $\Lam_s' = \partial\Lam_s/\partial\Lam$.

The frequency cutoff is convenient for fermion-boson theories at zero temperature 
as the singularity for both particle species is situated at 
the origin $|q_{0}|=0$ on the frequency axis. This makes mixed diagrams particularly 
easy to treat as alluded to above. The downside, on the other hand, is that 
cutting off frequencies ruins the analyticity properties of Green functions 
in the complex frequency plane. In the present chapter, this issue is not important 
but in other parts of this thesis, when for example calculating a $Z$-factor for 
the fermion self-energy as performed in chapter \ref{chap:fermibosetoy}, we resorted to other cutoffs.

\subsubsection{Symmetric regime}

Here we choose $\Lam_b = \Lam_f = \Lam$. One is in principle free
to choose the fermionic and bosonic cutoff independently. We  have
checked that the concrete choice of $\Lam_b$ and $\Lam_f$ in the symmetric
regime does not change
the final results for $\Lambda\rightarrow 0$ much. The 
diagrams contributing to the flow in the symmetric
regime are shown in Fig.~\ref{fig:flow_sym}.

The flow of the bosonic mass is given by the bosonic
self-energy at vanishing external momentum and frequency,
that is,
\begin{equation}
 \partial_{\Lam} \frac{m_b^2}{2} =
 g^2 \int_{k|\Lam} G_f(k) \, G_f(-k) +
 \frac{\lam}{2} \int_{q|\Lam} G_b(q) \; .
\label{eq:mass_sym}
\end{equation}
The fermionic contribution to $\partial_{\Lam} m_b^2$ is
positive, leading to a reduction of $m_b^2$ upon decreasing
$\Lam$, while the bosonic fluctuation term is negative
(since $G_b(q) < 0$).
The flow of $Z_b$ is obtained from the second frequency
derivative of the bosonic self-energy as
\begin{equation}
 \partial_{\Lam} Z_b =
 g^2 \int_{k|\Lam} \left.
 \partial_{q_0}^2 G_f(k+q) \, G_f(-k) \right|_{q=0}
 \; .
\end{equation}
Similarly, the flow of $A_b$ is obtained from a second
momentum derivative of the bosonic self-energy:
\begin{equation}
 \partial_{\Lam} A_b =
 g^2 \int_{k|\Lam} \left.
 \partial_{\mathbf{q}}^2 G_f(k+q) \, G_f(-k) \right|_{q=0} \; ,
\end{equation}
where $\partial_{\mathbf{q}}^2=\frac{1}{d}\sum_{i=1}^{d}\partial_{q_i}^2$. 
Since the bosonic self-energy is isotropic in $\bq$ to
leading  
\begin{wrapfigure}{r}{0.5\textwidth}
  \vspace{0mm}
  \begin{center}
\begin{fmffile}{20080227_SYM_flow_1}
\begin{eqnarray}
m_{b}^{2}&:&
\parbox{25mm}{\unitlength=1mm\fmfframe(2,2)(1,1){
\begin{fmfgraph*}(20,20)\fmfpen{thin} 
 \fmfleft{l1}
 \fmfright{r1}
 \fmfpolyn{empty,tension=0.3}{G}{3}
 \fmfpolyn{empty,tension=0.3}{K}{3}
  \fmf{dbl_wiggly}{l1,G1}
 \fmf{fermion,tension=0.2,right=0.6}{G2,K3}
 \fmf{fermion,tension=0.2,left=0.6}{G3,K2}
 \fmf{dbl_wiggly}{K1,r1}
 \end{fmfgraph*}
}}+
\parbox{25mm}{\unitlength=1mm\fmfframe(2,2)(1,1){
\begin{fmfgraph*}(25,25)\fmfpen{thin} 
 \fmfleft{l1}
 \fmfright{r1}
 \fmftop{v1}
 \fmfpolyn{full}{G}{4}
 \fmf{dbl_wiggly,straight}{l1,G4}
 \fmf{dbl_wiggly,straight}{G1,r1}
 \fmffreeze
\fmf{dbl_wiggly,tension=0.1,right=0.7}{G2,v1}
\fmf{dbl_wiggly,tension=0.1,right=0.7}{v1,G3}
\end{fmfgraph*}
}}\nonumber\\[-8mm]
Z_{b},\,\,A_{b}&:&
\parbox{25mm}{\unitlength=1mm\fmfframe(2,2)(1,1){
\begin{fmfgraph*}(20,15)\fmfpen{thin} 
 \fmfleft{l1}
 \fmfright{r1}
 \fmfpolyn{empty,tension=0.3}{G}{3}
 \fmfpolyn{empty,tension=0.3}{K}{3}
  \fmf{dbl_wiggly}{l1,G1}
 \fmf{fermion,tension=0.2,right=0.6}{G2,K3}
 \fmf{fermion,tension=0.2,left=0.6}{G3,K2}
 \fmf{dbl_wiggly}{K1,r1}
 \end{fmfgraph*}
}}\nonumber\\
\lambda&:&
\parbox{25mm}{\unitlength=1mm\fmfframe(2,2)(1,1){
\begin{fmfgraph*}(25,15)
\fmfpen{thin} 
\fmfleftn{l}{2}\fmfrightn{r}{2}
 \fmfpolyn{empty,tension=0.8}{OL}{3}
 \fmfpolyn{empty,tension=0.8}{OR}{3}
 \fmfpolyn{empty,tension=0.8}{UR}{3}
 \fmfpolyn{empty,tension=0.8}{UL}{3}
   \fmf{dbl_wiggly}{l1,OL1}
   \fmf{dbl_wiggly}{l2,UL1}
   \fmf{dbl_wiggly}{OR1,r1}
   \fmf{dbl_wiggly}{UR1,r2}
  \fmf{fermion,straight,tension=0.5}{OL2,OR3}
  \fmf{fermion,straight,tension=0.5}{UR3,OR2}
  \fmf{fermion,straight,tension=0.5}{UR2,UL3}
  \fmf{fermion,straight,tension=0.5}{OL3,UL2}
 \end{fmfgraph*}
}}+
\parbox{25mm}{\unitlength=1mm\fmfframe(2,2)(1,1){
\begin{fmfgraph*}(25,15)
\fmfpen{thin} 
\fmfleftn{l}{2}\fmfrightn{r}{2}
\fmfrpolyn{full}{G}{4}
\fmfpolyn{full}{K}{4}
\fmf{dbl_wiggly}{l1,G1}\fmf{dbl_wiggly}{l2,G2}
\fmf{dbl_wiggly}{K1,r1}\fmf{dbl_wiggly}{K2,r2}
\fmf{dbl_wiggly,left=.5,tension=.3}{G3,K3}
\fmf{dbl_wiggly,right=.5,tension=.3}{G4,K4}
\end{fmfgraph*}
}}\nonumber
\end{eqnarray}
\end{fmffile}
  \end{center}
  \vspace{3mm}
\caption{\textit{Feynman diagrams representing the flow equations in the
symmetric regime}.}
\label{fig:flow_sym}
\vspace{-5mm}
\end{wrapfigure}
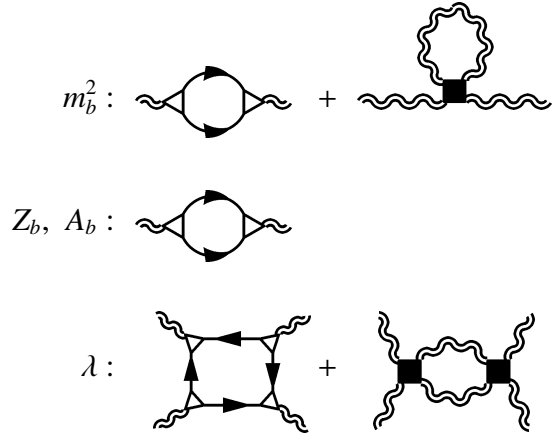
(quadratic) order in $\bq$, the results do not depend on
the direction in which the momentum derivative is taken.
The bosonic tadpole diagram in Fig. \ref{fig:flow_sym} contributes only to
$m_b$, not to $Z_b$ and $A_b$, since it yields a momentum
and frequency independent contribution to the self-energy.

Finally, the flow of the $|\phi|^4$ coupling is given by
\begin{eqnarray}
 \partial_{\Lam} \lam =
 &-&4 g^4 \int_{k|\Lam} [G_f(k)]^2  [G_f(-k)]^2
 \nonumber\\
&+&\frac{5}{4} \lam^2 \int_{q|\Lam} [G_b(q)]^2 \; .
\label{eq:lambda}
\end{eqnarray}
Within the truncation of the effective action described
in Sec. \ref{subsubsec:trunc_sym} there is no contribution to the flow of the
interaction between fermions and bosons in the symmetric
phase. The coupling $g$ remains therefore invariant.

\subsubsection{Symmetry broken regime}
\label{subsubsec:ssb_trunc}

In the limit $\Lam \to 0$ we are forced to choose
$\Lam_f \ll \Lam_b$ to avoid an artificial strong coupling
problem, as will become clear below. We therefore choose
$\Lam_f < \Lam_b$ in the entire symmetry broken regime,
which implies that the frequencies in mixed loops with
bosonic and fermionic propagators are pinned at the
bosonic cutoff. The precise choice of the cutoffs
will be specified later.

\begin{wrapfigure}{r}{0.5\textwidth}
  \vspace{-5mm}
  \begin{center}
\begin{fmffile}{20080227_ferm_self_1}
\begin{eqnarray}
\Gamma_{\sigma}^{(1)}&:&
\parbox{15mm}{\unitlength=1mm\fmfframe(2,2)(1,1){
\begin{fmfgraph*}(15,30)\fmfpen{thin} 
\fmfleft{i1}\fmfright{o1}
\fmftop{t1}
\fmfpolyn{hatched,tension=0.2}{T}{3}
\fmfpolyn{phantom,tension=100.}{B}{3}
\fmf{phantom,straight,tension=1.2}{i1,B1}
\fmf{phantom,straight,tension=1.2}{o1,B2}
\fmf{dashes,tension=0.6}{B3,T3}
\fmf{fermion,tension=0.2,right=0.8}{T1,t1}\fmf{fermion,tension=0.2,left=0.8}{T2,t1}
\end{fmfgraph*}
}}+
\parbox{15mm}{\unitlength=1mm\fmfframe(2,2)(1,1){
\begin{fmfgraph*}(15,30)\fmfpen{thin} 
\fmfleft{i1}\fmfright{o1}
\fmftop{t1}
\fmfpolyn{full,tension=0.2}{T}{3}
\fmfpolyn{phantom,tension=100.}{B}{3}
\fmf{phantom,straight,tension=1.2}{i1,B1}
\fmf{phantom,straight,tension=1.2}{o1,B2}
\fmf{dashes,tension=0.6}{B3,T3}
\fmf{dashes,tension=0.2,right=0.8}{T1,t1}\fmf{dashes,tension=0.2,left=0.8}{T2,t1}
\end{fmfgraph*}
}}
+
\parbox{15mm}{\unitlength=1mm\fmfframe(2,2)(1,1){
\begin{fmfgraph*}(15,30)\fmfpen{thin} 
\fmfleft{i1}\fmfright{o1}
\fmftop{t1}
\fmfpolyn{full,tension=0.2}{T}{3}
\fmfpolyn{phantom,tension=100.}{B}{3}
\fmf{phantom,straight,tension=1.2}{i1,B1}
\fmf{phantom,straight,tension=1.2}{o1,B2}
\fmf{dashes,tension=0.6}{B3,T3}
\fmf{photon,tension=0.2,right=0.8}{T1,t1}\fmf{photon,tension=0.2,right=0.8}{t1,T2}
\end{fmfgraph*}
}}
\nonumber\\[-15mm]
\Delta&:&
\parbox{25mm}{\unitlength=1mm\fmfframe(2,2)(1,1){
\begin{fmfgraph*}(22,15)\fmfpen{thin} 
\fmfcmd{%
   style_def gap expr p=
     cdraw p;
     cfill (harrow(reverse p, .5));
     cfill (harrow(p,.5));
   enddef;
   style_def gap_hc expr p=
     cdraw p;
     cfill (tarrow(reverse p, .55));
     cfill (tarrow(p,.55));
   enddef;
     }
\fmfleft{l1}
 \fmfright{r1}
 \fmfpolyn{hatched,tension=0.4}{G}{3}
 \fmfpolyn{hatched,tension=0.4}{K}{3}
  \fmf{fermion}{l1,G1}
   \fmf{gap_hc,straight,tension=0.5,left=0.}{K3,G2}
  \fmf{dashes,tension=0.,left=0.7}{G3,K2}
  \fmf{fermion}{r1,K1}
 \end{fmfgraph*}
}} -
\parbox{25mm}{\unitlength=1mm\fmfframe(2,2)(1,1){
\begin{fmfgraph*}(22,15)\fmfpen{thin} 
\fmfleft{l1}
 \fmfright{r1}
 \fmfpolyn{filled=30,tension=0.4}{G}{3}
 \fmfpolyn{filled=30,tension=0.4}{K}{3}
  \fmf{fermion}{l1,G1}
   \fmf{gap_hc,straight,tension=0.5,left=0.}{K3,G2}
  \fmf{photon,tension=0.,right=0.7}{K2,G3}
  \fmf{fermion}{r1,K1}
 \end{fmfgraph*}
}}\nonumber\\[-12mm]
\nonumber
\end{eqnarray}
\end{fmffile}
  \end{center}
  \vspace{3mm}
\caption{\textit{Contributions to the bosonic 1-point vertex and fermion gap below $\Lambda_{c}$}.}
\label{fig:fermionic_self}
\vspace{-15mm}
\end{wrapfigure}
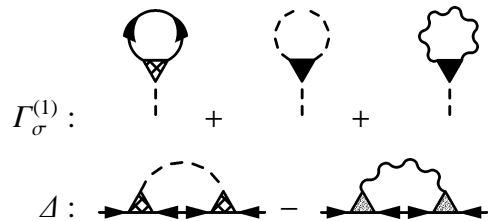

We first derive the flow equation for the minimum of the
bosonic potential $\alf$, which is derived from the
condition that the bosonic 1-point vertex $\Gam_{\sg}^{(1)}$
be zero for all $\Lam$.
The flow equation for $\Gam_{\sg}^{(1)}$ reads
\begin{eqnarray}
 \partial_{\Lam} \Gam_{\sg}^{(1)} &=&
 m_{\sg}^2 \partial_{\Lam} \alf
 + 2g_{\sg} \int_{k|\Lam_f} \! F_f(k)\nonumber\\
 &&+ \frac{\lam\alf}{2} \int_{q|\Lam_b}
 [3G_{\sg}(q) + G_{\pi}(q)] \; .\nonumber\\
\end{eqnarray}
The various contributions are represented diagrammatically
in Fig.~\ref{fig:fermionic_self}.

The first term is due to the cutoff-dependence of the
expansion point around which the effective action is
expanded in powers of the fields.
The condition $\partial_{\Lam} \Gam_{\sg}^{(1)} = 0$ yields
\begin{eqnarray}
 \partial_{\Lam} \alf =
 - \frac{2g_{\sg}}{m_{\sg}^2} \int_{k|\Lam_f} F_f(k)
 - \frac{1}{2\alf} \int_{q|\Lam_b} [3G_{\sg}(q) + G_{\pi}(q)] \; .
\label{eq:alpha_ssb}
\end{eqnarray}
We have used Eq.~(\ref{eq:mSig_condition}) to simplify the last term.
The fermionic contribution to $\partial_{\Lam} \alf$ is
negative, leading to an increase of $\alf$ upon decreasing
$\Lam$, while the bosonic fluctuation term is positive and
therefore reduces $\alf$. The behavior of Eq. (\ref{eq:alpha_ssb}) in the
vicinity of the critical scale, $\Lambda\lesssim\Lambda_{c}$, when $\alpha$
and $m_{\sg}^2$ are small, is
shown below in Sec. \ref{subsec:critflow}.

The flow of $\Delta$ is obtained from the flow of the
anomalous component of the fermionic self-energy as
\begin{equation}
 \partial_{\Lam} \Delta =
 g_{\sg} \, \partial_{\Lam} \alf
 - \! \int_{q|\Lam_b} \! \left. F_f(q-k) \,
 [g^2_{\sg}G_{\sg}(q) - g^2_{\pi}G_{\pi}(q)] \right|_{k=(0,\bk_F)} \; .
\label{eq:gap_ssb}
\end{equation}
The first term, due to the cutoff-dependence of the expansion
point for the effective action, links the flow of the fermionic
gap to the flow of the bosonic order parameter.
The second term captures a correction to the relation
between $\alf$ and $\Delta$ due to bosonic fluctuations as illustrated
in Fig.~\ref{fig:fermionic_self}.

\begin{wrapfigure}{r}{0.5\textwidth}
\vspace{-10mm}
\begin{fmffile}{bos_self_48}
\begin{eqnarray}
m^{2}_{\sigma}&:&
\parbox{25mm}{\unitlength=1mm\fmfframe(2,2)(1,1){
\begin{fmfgraph*}(22,25)\fmfpen{thin} 
 \fmfleft{l1}
 \fmfright{r1}
 \fmfpolyn{hatched,tension=0.3}{G}{3}
 \fmfpolyn{hatched,tension=0.3}{K}{3}
  \fmf{dashes}{l1,G1}
 \fmf{fermion,tension=0.2,right=0.6}{G2,K3}
 \fmf{fermion,tension=0.2,left=0.6}{G3,K2}
 \fmf{dashes}{K1,r1}
 \end{fmfgraph*}
}}
-
\parbox{25mm}{\unitlength=1mm
\fmfframe(2,2)(1,1){
\begin{fmfgraph*}(22,25)
\fmfcmd{%
   style_def gap expr p=
     cdraw p;
     cfill (harrow(reverse p, .5));
     cfill (harrow(p,.5));
   enddef;
   style_def gap_hc expr p=
     cdraw p;
     cfill (tarrow(reverse p, .55));
     cfill (tarrow(p,.55));
   enddef;
     }
 \fmfpen{thin} 
 \fmfleft{l1}
 \fmfright{r1}
 \fmfpolyn{hatched,tension=0.3}{G}{3}
 \fmfpolyn{hatched,tension=0.3}{K}{3}
  \fmf{dashes}{l1,G1}
 \fmf{gap,tension=0.2,right=0.6}{G2,K3}
 \fmf{gap,tension=0.2,left=0.6}{G3,K2}
 \fmf{dashes}{K1,r1}
 \end{fmfgraph*}
}}+
\nonumber\\[-2mm]
&&
\parbox{25mm}{\unitlength=1mm\fmfframe(2,2)(1,1){
\begin{fmfgraph*}(22,25)\fmfpen{thin} 
 \fmfleft{l1}
 \fmfright{r1}
 \fmftop{v1}
 \fmfpolyn{full}{G}{4}
 \fmf{dashes,straight}{l1,G4}
 \fmf{dashes,straight}{G1,r1}
 \fmffreeze
\fmf{dashes,tension=0.1,right=0.7}{G2,v1}
\fmf{dashes,tension=0.1,right=0.7}{v1,G3}
\end{fmfgraph*}
}}
+
\parbox{25mm}{\unitlength=1mm\fmfframe(2,2)(1,1){
\begin{fmfgraph*}(22,25)\fmfpen{thin} 
 \fmfleft{l1}
 \fmfright{r1}
 \fmftop{v1}
 \fmfpolyn{shaded}{G}{4}
 \fmf{dashes,straight}{l1,G4}
 \fmf{dashes,straight}{G1,r1}
 \fmffreeze
\fmf{photon,tension=0.1,right=0.7}{G2,v1}
\fmf{photon,tension=0.1,right=0.7}{v1,G3}
\end{fmfgraph*}
}}+\nonumber\\[-10mm]
&&
\parbox{25mm}{\unitlength=1mm\fmfframe(2,2)(1,1){
\begin{fmfgraph*}(22,25)\fmfpen{thin} 
 \fmfleft{l1}
 \fmfright{r1}
 \fmfpolyn{full,tension=0.4}{G}{3}
 \fmfpolyn{full,tension=0.4}{K}{3}
  \fmf{dashes}{l1,G1}
 \fmf{dashes,tension=0.2,right=0.8}{G2,K3}
 \fmf{dashes,tension=0.2,right=0.8}{K2,G3}
 \fmf{dashes}{K1,r1}
 \end{fmfgraph*}
}}
+
\parbox{25mm}{\unitlength=1mm\fmfframe(2,2)(1,1){
\begin{fmfgraph*}(22,25)\fmfpen{thin} 
 \fmfleft{l1}
 \fmfright{r1}
 \fmfpolyn{full,tension=0.4}{G}{3}
 \fmfpolyn{full,tension=0.4}{K}{3}
  \fmf{dashes}{l1,G1}
 \fmf{photon,tension=0.2,right=0.8}{G2,K3}
 \fmf{photon,tension=0.2,right=0.8}{K2,G3}
 \fmf{dashes}{K1,r1}
 \end{fmfgraph*}
}}
\nonumber\\[-15mm]\nonumber
\end{eqnarray}
\end{fmffile}
\caption{\textit{Diagrammatic representation of the contributions to the
bosonic mass in the symmetry-broken regime}.}
\label{fig:bosonic_self}
\vspace{-50mm}
\end{wrapfigure}
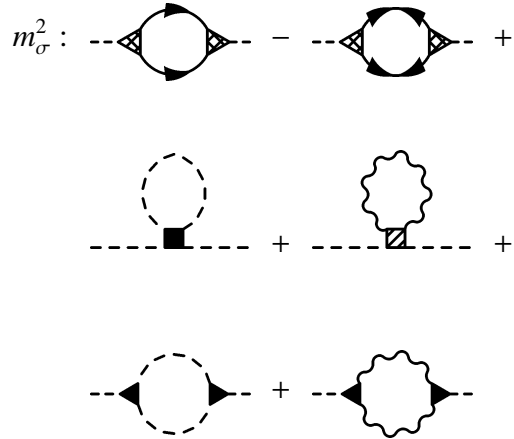
The flow of the mass of the longitudinal order parameter
fluctuations (cf.\ Fig.~\ref{fig:bosonic_self}) is obtained from the
self-energy of the $\sg$ fields at zero momentum and frequency. The flow of $\lam$ can 
then be computed from the flow of
$m_{\sg}^2$ and $\alf$ via the relation Eq.~(\ref{eq:mSig_condition}).
For $m_{\sg}^2$, we have the flow equation
\begin{eqnarray}
 \partial_{\Lam} \frac{m_{\sg}^2}{2} &=&
 g^2_{\sg} \int_{k|\Lam_f}
 \Big[ G_f(k) G_f(-k) \nonumber\\
&-& F_f(k) F_f(-k) \Big]\nonumber\\
 &+& \frac{\lam}{4} \int_{q|\Lam_b}
 \left[ 3G_{\sg}(q) + G_{\pi}(q) \right]\nonumber \\
 &+& \frac{(\lam\alf)^2}{2} \int_{q|\Lam_b}
 \left[ 9G_{\sg}^2(q) + G_{\pi}^2(q) \right]\nonumber\\
 &+& 3 \frac{\lam \alf}{2} \, \partial_{\Lam}\alf \; .
\label{eq:mass_ssb}
\end{eqnarray}
The second term in this equation is due to a product of the 3-point
vertex $\gam_{\sg^3}$ and $\partial_{\Lam}\alf$ arising from the
cutoff dependence of the expansion point for the effective action.
The flow of $Z_{\sg}$ is obtained from the second frequency
derivative of the self-energy of the $\sg$ fields, which yields
\begin{eqnarray}
 \partial_{\Lam} Z_{\sg} &=&
 g^2_{\sg} \! \int_{k|\Lam_f} \!\!\! \partial_{q_0}^2
 [G_f(k+q) \, G_f(-k)
  - F_f(k+q) \, F_f(-k)]
 \Big|_{q=0} \nonumber \\
 + && \hskip -5mm \frac{(\lam\alf)^2}{2} \! \int_{k|\Lam_b} \!\!\!
 \partial_{q_0}^2
 [9G_{\sg}(k+q) \, G_{\sg}(k)
  +  G_{\pi}(k+q) \, G_{\pi}(k)] \Big|_{q=0} \; .
\end{eqnarray}
\begin{figure}[t]
\vspace{-5mm}
\begin{fmffile}{Z_factors_10}
\begin{eqnarray}
Z_{\sigma},\,\,A_{\sigma}&:&
\parbox{25mm}{\unitlength=1mm\fmfframe(2,2)(1,1){
\begin{fmfgraph*}(22,25)\fmfpen{thin}
 \fmfleft{l1}
 \fmfright{r1}
 \fmfpolyn{hatched,tension=0.3}{G}{3}
 \fmfpolyn{hatched,tension=0.3}{K}{3}
  \fmf{dashes}{l1,G1}
 \fmf{fermion,tension=0.2,right=0.6}{G2,K3}
 \fmf{fermion,tension=0.2,left=0.6}{G3,K2}
 \fmf{dashes}{K1,r1}
 \end{fmfgraph*}
}}
-
\parbox{25mm}{\unitlength=1mm
\fmfframe(2,2)(1,1){
\begin{fmfgraph*}(22,25)
\fmfcmd{%
   style_def gap expr p=
     cdraw p;
     cfill (harrow(reverse p, .5));
     cfill (harrow(p,.5));
   enddef;
   style_def gap_hc expr p=
     cdraw p;
     cfill (tarrow(reverse p, .55));
     cfill (tarrow(p,.55));
   enddef;
     }
 \fmfpen{thin} 
 \fmfleft{l1}
 \fmfright{r1}
 \fmfpolyn{hatched,tension=0.3}{G}{3}
 \fmfpolyn{hatched,tension=0.3}{K}{3}
  \fmf{dashes}{l1,G1}
 \fmf{gap,tension=0.2,right=0.6}{G2,K3}
 \fmf{gap,tension=0.2,left=0.6}{G3,K2}
 \fmf{dashes}{K1,r1}
 \end{fmfgraph*}
}}+
%
\parbox{25mm}{\unitlength=1mm\fmfframe(2,2)(1,1){
\begin{fmfgraph*}(22,25)\fmfpen{thin} 
 \fmfleft{l1}
 \fmfright{r1}
 \fmfpolyn{full,tension=0.4}{G}{3}
 \fmfpolyn{full,tension=0.4}{K}{3}
  \fmf{dashes}{l1,G1}
 \fmf{dashes,tension=0.2,right=0.8}{G2,K3}
 \fmf{dashes,tension=0.2,right=0.8}{K2,G3}
 \fmf{dashes}{K1,r1}
 \end{fmfgraph*}
}}
+
\parbox{25mm}{\unitlength=1mm\fmfframe(2,2)(1,1){
\begin{fmfgraph*}(22,25)\fmfpen{thin} 
 \fmfleft{l1}
 \fmfright{r1}
 \fmfpolyn{full,tension=0.4}{G}{3}
 \fmfpolyn{full,tension=0.4}{K}{3}
  \fmf{dashes}{l1,G1}
 \fmf{photon,tension=0.2,right=0.8}{G2,K3}
 \fmf{photon,tension=0.2,right=0.8}{K2,G3}
 \fmf{dashes}{K1,r1}
 \end{fmfgraph*}
}}
\nonumber\\[-10mm]
Z_{\pi},\,\,A_{\pi}&:&
\parbox{25mm}{\unitlength=1mm\fmfframe(2,2)(1,1){
\begin{fmfgraph*}(22,25)\fmfpen{thin}
 \fmfleft{l1}
 \fmfright{r1}
 \fmfpolyn{filled=30,tension=0.3}{G}{3}
 \fmfpolyn{filled=30,tension=0.3}{K}{3}
  \fmf{photon}{l1,G1}
 \fmf{fermion,tension=0.2,right=0.6}{G2,K3}
 \fmf{fermion,tension=0.2,left=0.6}{G3,K2}
 \fmf{photon}{K1,r1}
 \end{fmfgraph*}
}}
+
\parbox{25mm}{\unitlength=1mm
\fmfframe(2,2)(1,1){
\begin{fmfgraph*}(22,25)
\fmfcmd{%
   style_def gap expr p=
     cdraw p;
     cfill (harrow(reverse p, .5));
     cfill (harrow(p,.5));
   enddef;
   style_def gap_hc expr p=
     cdraw p;
     cfill (tarrow(reverse p, .55));
     cfill (tarrow(p,.55));
   enddef;
     }
 \fmfpen{thin} 
 \fmfleft{l1}
 \fmfright{r1}
 \fmfpolyn{filled=30,tension=0.3}{G}{3}
 \fmfpolyn{filled=30,tension=0.3}{K}{3}
  \fmf{photon}{l1,G1}
 \fmf{gap,tension=0.2,right=0.6}{G2,K3}
 \fmf{gap,tension=0.2,left=0.6}{G3,K2}
 \fmf{photon}{K1,r1}
 \end{fmfgraph*}
}}\nonumber\\[-13mm]
\nonumber
\end{eqnarray}
\end{fmffile}
\caption{Diagrams renormalizing the bosonic
Z- and A-factors for $\Lambda<\Lambda_{c}$.}
\label{fig:bosonic_Z}
\end{figure}
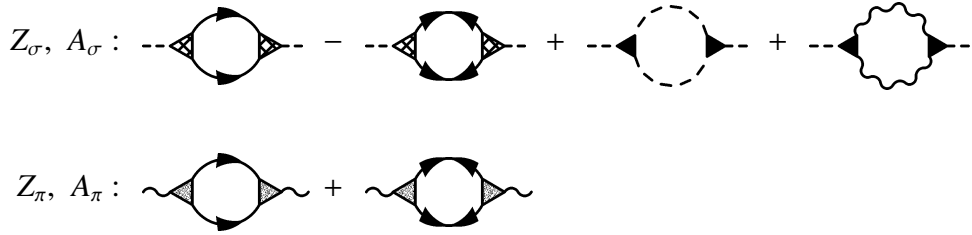
The flow of $A_{\sg}$ is given by the same equation with
$\partial_{q_0}^2$ replaced by $\partial_{\mathbf{q}}^2$.

In the flow of $Z_{\pi}$ there are strong cancellations of
different terms originating from bosonic fluctuations. We have shown 
with a slightly extended truncation in chapter \ref{chap:bosonicqcp_goldstone} that the effective self-interaction among 
Goldstone bosons flows to zero such that $Z_{\pi}$ remains finite and the Goldstone mode is not renormalized substantially (Pistolesi 2004). 
We therefore keep only the fermionic fluctuations, that is,
\begin{equation}
 \partial_{\Lam} Z_{\pi} =
 g^2_{\pi} \! \int_{k|\Lam_f} \!\!\! \left. \partial_{q_0}^2
 [G_f(k+q) \, G_f(-k) + F_f(k+q) \, F_f(-k)]
 \right|_{q=0} \; .
\end{equation}

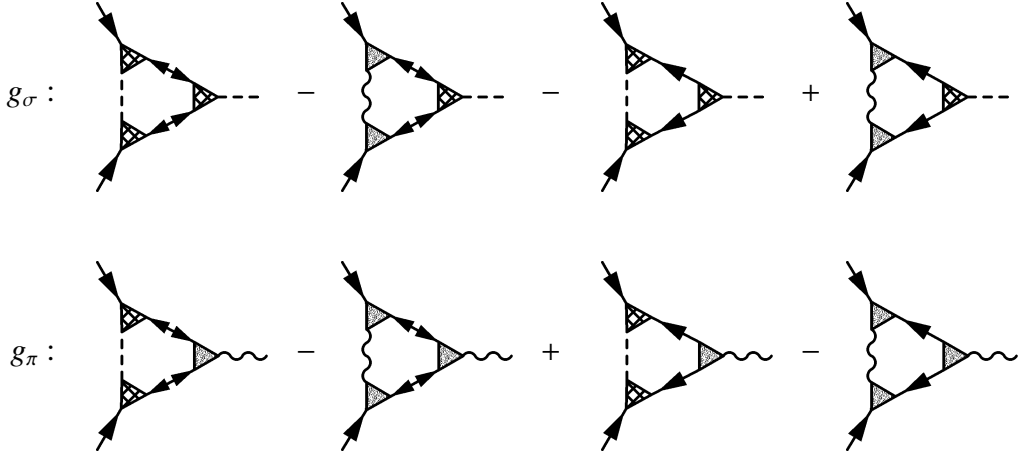
\begin{figure}[t]
\vspace{-5mm}
\begin{fmffile}{20080214_vertex_20}
\begin{eqnarray}
g_{\sigma}&:&
\parbox{30mm}{\unitlength=1mm\fmfframe(2,2)(1,1){
\begin{fmfgraph*}(25,25)
\fmfcmd{%
   style_def gap expr p=
     cdraw p;
     cfill (harrow(reverse p, .5));
     cfill (harrow(p,.5));
   enddef;
   style_def gap_hc expr p=
     cdraw p;
     cfill (tarrow(reverse p, .55));
     cfill (tarrow(p,.55));
   enddef;
     }
\fmfpen{thin}
\fmfleftn{l}{2}\fmfright{r}
\fmfrpolyn{hatched,tension=0.65}{T}{3}
\fmfrpolyn{hatched,tension=0.65}{B}{3}
\fmfpolyn{hatched, tension=0.65}{R}{3}
\fmf{fermion}{l1,T2}\fmf{fermion}{l2,B2}
\fmf{dashes,tension=.6}{T3,B1}
\fmf{gap_hc,left=0.,tension=.5}{T1,R3}
\fmf{gap_hc,right=0.,tension=.5}{B3,R2}
\fmf{dashes}{R1,r}
\end{fmfgraph*}
}}
-
\hspace{-2mm}
\parbox{30mm}{\unitlength=1mm\fmfframe(2,2)(1,1){
\begin{fmfgraph*}(25,25)
\fmfcmd{%
   style_def gap expr p=
     cdraw p;
     cfill (harrow(reverse p, .5));
     cfill (harrow(p,.5));
   enddef;
   style_def gap_hc expr p=
     cdraw p;
     cfill (tarrow(reverse p, .55));
     cfill (tarrow(p,.55));
   enddef;
     }
\fmfpen{thin} 
\fmfleftn{l}{2}\fmfright{r}
\fmfrpolyn{filled=30,tension=0.65}{T}{3}
\fmfrpolyn{filled=30,tension=0.65}{B}{3}
\fmfpolyn{hatched, tension=0.65}{R}{3}
\fmf{fermion}{l1,T2}\fmf{fermion}{l2,B2}
\fmf{photon,tension=.5,left=0.}{B1,T3}
\fmf{gap_hc,left=0.,tension=.5}{T1,R3}
\fmf{gap_hc,right=0.,tension=.5}{B3,R2}
\fmf{dashes}{R1,r}
\end{fmfgraph*}
}}
-
\parbox{30mm}{\unitlength=1mm\fmfframe(2,2)(1,1){
\begin{fmfgraph*}(25,25)
\fmfpen{thin} 
\fmfleftn{l}{2}\fmfright{r}
\fmfrpolyn{hatched,tension=0.65}{T}{3}
\fmfrpolyn{hatched,tension=0.65}{B}{3}
\fmfpolyn{hatched, tension=0.65}{R}{3}
\fmf{fermion}{l1,T2}\fmf{fermion}{l2,B2}
\fmf{dashes,tension=.6}{T3,B1}
\fmf{fermion,left=0.,tension=.5}{R3,T1}
\fmf{fermion,right=0.,tension=.5}{R2,B3}
\fmf{dashes}{R1,r}
\end{fmfgraph*}
}}
+
\hspace{-2mm}
\parbox{30mm}{\unitlength=1mm\fmfframe(2,2)(1,1){
\begin{fmfgraph*}(25,25)
\fmfpen{thin} 
\fmfleftn{l}{2}\fmfright{r}
\fmfrpolyn{filled=30,tension=0.65}{T}{3}
\fmfrpolyn{filled=30,tension=0.65}{B}{3}
\fmfpolyn{hatched, tension=0.65}{R}{3}
\fmf{fermion}{l1,T2}\fmf{fermion}{l2,B2}
\fmf{photon,tension=.5,left=0.}{B1,T3}
\fmf{fermion,left=0.,tension=.5}{R3,T1}
\fmf{fermion,right=0.,tension=.5}{R2,B3}
\fmf{dashes}{R1,r}
\end{fmfgraph*}
}}
\nonumber\\[5mm]
g_{\pi}&:&
\parbox{30mm}{\unitlength=1mm\fmfframe(2,2)(1,1){
\begin{fmfgraph*}(25,25)
\fmfcmd{%
   style_def gap expr p=
     cdraw p;
     cfill (harrow(reverse p, .5));
     cfill (harrow(p,.5));
   enddef;
   style_def gap_hc expr p=
     cdraw p;
     cfill (tarrow(reverse p, .55));
     cfill (tarrow(p,.55));
   enddef;
     }
\fmfpen{thin} 
\fmfleftn{l}{2}\fmfright{r}
\fmfrpolyn{hatched,tension=0.65}{T}{3}
\fmfrpolyn{hatched,tension=0.65}{B}{3}
\fmfpolyn{filled=30, tension=0.65}{R}{3}
\fmf{fermion}{l1,T2}\fmf{fermion}{l2,B2}
\fmf{dashes,tension=.6}{T3,B1}
\fmf{gap_hc,left=0.,tension=.5}{T1,R3}
\fmf{gap_hc,right=0.,tension=.5}{B3,R2}
\fmf{photon}{R1,r}
\end{fmfgraph*}
}}
-
\hspace{-2mm}
\parbox{30mm}{\unitlength=1mm\fmfframe(2,2)(1,1){
\begin{fmfgraph*}(25,25)
\fmfcmd{%
   style_def gap expr p=
     cdraw p;
     cfill (harrow(reverse p, .5));
     cfill (harrow(p,.5));
   enddef;
   style_def gap_hc expr p=
     cdraw p;
     cfill (tarrow(reverse p, .55));
     cfill (tarrow(p,.55));
   enddef;
     }
\fmfpen{thin} 
\fmfleftn{l}{2}\fmfright{r}
\fmfrpolyn{filled=30,tension=0.65}{T}{3}
\fmfrpolyn{filled=30,tension=0.65}{B}{3}
\fmfpolyn{filled=30, tension=0.65}{R}{3}
\fmf{fermion}{l1,T2}\fmf{fermion}{l2,B2}
\fmf{photon,tension=.5,left=0.}{B1,T3}
\fmf{gap_hc,left=0.,tension=.5}{T1,R3}
\fmf{gap_hc,right=0.,tension=.5}{B3,R2}
\fmf{photon}{R1,r}
\end{fmfgraph*}
}}
+
\parbox{30mm}{\unitlength=1mm\fmfframe(2,2)(1,1){
\begin{fmfgraph*}(25,25)
\fmfpen{thin} 
\fmfleftn{l}{2}\fmfright{r}
\fmfrpolyn{hatched,tension=0.65}{T}{3}
\fmfrpolyn{hatched,tension=0.65}{B}{3}
\fmfpolyn{filled=30, tension=0.65}{R}{3}
\fmf{fermion}{l1,T2}\fmf{fermion}{l2,B2}
\fmf{dashes,tension=.6}{T3,B1}
\fmf{fermion,left=0.,tension=.5}{R3,T1}
\fmf{fermion,right=0.,tension=.5}{R2,B3}
\fmf{photon}{R1,r}
\end{fmfgraph*}
}}
-
\hspace{-2mm}
\parbox{30mm}{\unitlength=1mm\fmfframe(2,2)(1,1){
\begin{fmfgraph*}(25,25)
\fmfpen{thin} 
\fmfleftn{l}{2}\fmfright{r}
\fmfrpolyn{filled=30,tension=0.65}{T}{3}
\fmfrpolyn{filled=30,tension=0.65}{B}{3}
\fmfpolyn{filled=30, tension=0.65}{R}{3}
\fmf{fermion}{l1,T2}\fmf{fermion}{l2,B2}
\fmf{photon,tension=.5,left=0.}{B1,T3}
\fmf{fermion,left=0.,tension=.5}{R3,T1}
\fmf{fermion,right=0.,tension=.5}{R2,B3}
\fmf{photon}{R1,r}
\end{fmfgraph*}
}}\nonumber\\[-5mm]
\nonumber
\end{eqnarray}
\end{fmffile}
\caption{Fermion-boson vertex corrections below $\Lambda_{c}$.}
\label{fig:vertex_ssb}
\end{figure}

For the flow of $A_{\pi}$ we obtain the same equation with
$\partial_{q_0}^2$ replaced by $\partial_{\mathbf{q}}^2$.
The terms contributing to the flow of the $Z$- and $A$-factors
are illustrated in Fig.~\ref{fig:bosonic_Z}.

In the symmetry broken regime, there are also contributions
to the flow of the interaction between fermions and bosons
due to vertex corrections with bosonic fluctuations
(see Fig.~\ref{fig:vertex_ssb}), yielding
\begin{eqnarray}
 \partial_{\Lam} g_{\sigma} &=& g_{\sigma} \int_{q|\Lam_b}
 \left[F^{2}_f(k-q) - \left|G_f(k-q)\right|^2\right]_{k=(0,\bk_F)}
 \left[g_{\sigma}^{2}G_{\sg}(q) - g_{\pi}^{2}G_{\pi}(q)\right]
 \nonumber\\
 \partial_{\Lam} g_{\pi} &=& g_{\pi} \int_{q|\Lam_b}
 \left[F^{2}_f(k-q) + \left|G_f(k-q)\right|^2\right]_{k=(0,\bk_F)}
 \left[g_{\sigma}^{2}G_{\sg}(q) - g_{\pi}^{2}G_{\pi}(q)\right]
 \; .\nonumber\\
\label{eq:g_pi_ssb}
\end{eqnarray}
The right hand sides are dominated by the contribution from
the $\pi$ propagator, which tends to reduce $g_{\sigma}$, $g_{\pi}$  for
decreasing $\Lam$.

\subsection{Relation to mean-field theory}
\label{subsec:rel_mft}

Before solving the flow equations derived above, we first
analyze what happens when contributions due to bosonic
fluctuations are neglected, and relate the reduced set of
equations to the usual mean-field theory (Sec. \ref{sec:mft}).

In the absence of bosonic fluctuations, $g=g_{\sg}=g_{\pi}=1$.
Furthermore, the bosonic order parameter $\alf$ and the
fermionic gap $\Delta$ are identical: $\alf = \Delta$.
Since the bosonic cutoff is irrelevant here, we can
choose $\Lam_f = \Lam$. 
The flow equation for the bosonic mass in the symmetric
regime, Eq.~(\ref{eq:mass_sym}), simplifies to
\begin{equation}
 \partial_{\Lam} \frac{m_b^2}{2} =
 \int_{k|\Lam} G_f(k) \, G_f(-k) \; ,
\end{equation}
where $G_f(k) = G_{f0}(k) = (ik_0 - \xi_{\bk})^{-1}$.
This equation can be easily integrated, yielding
\begin{equation}
 \frac{m_b^2}{2} = \frac{1}{|U|} -
 \int_{|k_0|> \Lam} \int_{\bk}
 \frac{1}{k_0^2 + \xi_{\bk}^2} \; .
\end{equation}
$m_b$ vanishes at a critical scale $\Lam_c > 0$.
The flow equation for $\Delta$ (= $\alf$) in the symmetry
broken regime $\Lam < \Lam_c \,$, Eq.~(\ref{eq:gap_ssb}), is reduced to
\begin{eqnarray}
 \partial_{\Lam} \Delta =
 - \frac{2}{m_{\sg}^2} \int_{k|\Lam} F_f(k)  \; .
\label{eq:gap_mftflow}
\end{eqnarray}
It is complemented by the flow equation for the mass of
the $\sg$ field, Eq.~(\ref{eq:mass_ssb}), which becomes
\begin{eqnarray}
 \partial_{\Lam} \frac{m_{\sg}^2}{2} =
 \int_{k|\Lam}
 \left[ |G_f(k)|^2 - F_f^2(k) \right]
 + 3 \gam_{\sg^3} \, \partial_{\Lam}\Delta
\label{eq:mass_mftflow}
\end{eqnarray}
in the absence of bosonic fluctuations,
with $\gam_{\sg^3} = \lam\Delta/2 = m_{\sg}^2/(2\Delta)$.
The propagators $G_f$ and $F_f$ have the usual BCS form,
as in Eqs.~(\ref{eq:Gf}) and (\ref{eq:Ff}).

A numerical solution of the coupled flow equations (\ref{eq:gap_mftflow})
and (\ref{eq:mass_mftflow}) yields a gap $\Delta$ which is a bit smaller than
the BCS result obtained from the gap equation (\ref{eq:bcs_gap_eqn}).
The reason for this discrepancy is the relatively simple
quartic ansatz (\ref{eq:gold_effpot}) for the bosonic potential.
The complete bosonic potential is non-polynomial in
$|\phi|^2$ even in mean-field theory.
Restricted to the zero momentum and frequency component
of $\phi$ it has the form (Popov 1987)
\begin{equation}
 U^{\rm MF}(\phi) = \frac{|\phi|^2}{|U|} - \int_k \ln
 \frac{k_0^2 + \xi_{\bk}^2 + |\phi|^2}{k_0^2 + \xi_{\bk}^2}
 \; .
\end{equation}
The kernel of the 3-point vertex $\gam_{\sg^3}$ obtained
from an expansion of this mean-field potential around a
finite order parameter $\Delta$ reads
\begin{equation}
 \gam_{\sg^3} = - 2 \int_k \; \Big[
 \frac{1}{3} F_f^3(k) - F_f(k) \, |G_f(k)|^2 \Big]
\end{equation}
at zero frequencies and momenta.
Inserting this into (\ref{eq:mass_mftflow}), the flow of $m_{\sg}^2$ can
be written as a total derivative
\begin{equation}
 \partial_{\Lam} \frac{m_{\sg}^2}{2} =
 - \partial_{\Lam} \int_{|k_0| > \Lam} \int_{\bk}
 \left[ |G_f(k)|^2 - F_f^2(k) \right] \; ,
\label{eq:mass_mftflow_correct}
\end{equation}
where the $\Lam$-derivative on the right hand side acts
also on $\Delta$, generating the term proportional to
$\gam_{\sg^3}$.
Integrating this equation with the initial condition
$m_{\sg} = 0$ at $\Lam=\Lam_c \,$, yields
\begin{equation}
 \frac{m_{\sg}^2}{2} = K(0) + L(0) - U^{-1} \; ,
\label{eq:mass_mft_correct}
\end{equation}
which is the correct mean-field result.
With $m_{\sg}$ given by (\ref{eq:mass_mft_correct}), the
flow equation (\ref{eq:gap_mftflow}) yields the correct mean-field gap.
The easiest way to see this, is to write the BCS gap
equation in the presence of a cutoff in the form
$1 = - U \int_{|k_0| > \Lam} \int_{\bk} \Delta^{-1} \,
 F_f(k)$, and take a derivative with respect to $\Lam$.

It is instructive to relate the above flow equations
for $\Delta$ and $m_{\sg}$ to the flow equations for
the BCS mean-field model obtained in a purely fermionic
RG (Salmhofer 2004).
For a sharp frequency cutoff, those flow equations
have the form
\begin{eqnarray}
 \partial_{\Lam} \Delta &=&
 - (V+W) \int_k' F_f(k)  \; , \\
 \partial_{\Lam} (V \!+\! W) &=&
 (V \!+\! W)^2 \, \partial_{\Lam} \!\!
 \int_{|k_0| > \Lam} \int_{\bk}
 \left[ |G_f(k)|^2 - F_f^2(k) \right] \, , \hskip 5mm
\end{eqnarray}
where $V$ is a normal two-fermion interaction in the Cooper
channel, while $W$ is an anomalous interaction
corresponding to annihilation (or creation) of four
particles. With the identification $2/m_{\sg}^2 = V+W$
these equations are obviously equivalent to (\ref{eq:gap_mftflow})
and (\ref{eq:mass_mftflow_correct}).
The above flow equation for $V+W$ is obtained from a one-loop
truncation complemented by additional self-energy
insertions drawn from higher order diagrams with tadpoles
(Salmhofer 2004, Katanin 2004).
These additional terms correspond to the contractions with
$\gam_{\sg^3}$ in the present bosonized RG.

\section{Results}
\label{sec:results}

In subsections \ref{subsec:critflow} and \ref{subsec:asymptotics}, 
analytic properties of our flow equations are discussed. Numerical 
results for two-dimensional systems are exhibited and interpreted in 
subsection \ref{subsec:num_results}.

\subsection{Flow for $\Lam \lesssim \Lam_c$}
\label{subsec:critflow}

For $\Lam$ slightly below $\Lam_c$, the flow equations can
be expanded in the order parameter. To leading order, the
order parameter $\alf$ and the gap $\Delta$ are identical,
$\alf = \Delta \,$.
The fluctuation term in the flow equation (\ref{eq:gap_ssb}) for $\Delta$
is quadratic in $\alf$, and also the flow of $g_{\sigma}$ yields
only corrections beyond linear order to the relation between
$\alf$ and $\Delta$.

Near $\Lam_c$, the flow equation (\ref{eq:alpha_ssb}) for $\alf$ can be
written as
\begin{equation}
 \partial_{\Lam} \alf^2 =
 - \frac{4}{\lam} \, I
 - \int_{q|\Lam_b} [3G_{\sg}(q) + G_{\pi}(q)] \; ,
\end{equation}
where $I = \int_{k|\Lam_f} (k_0^2 + \tilde\xi_{\bk}^2)^{-1}$,
evaluated for $\Lam = \Lam_c$. Note that we have replaced
the ratio $m_{\sg}^2/\alf^2$ by $\lam$ in the first term
on the right hand side of the flow equation.
Integrating the flow equation one obtains
\begin{equation}
 \alf^2 =
 \left[ \frac{4}{\lam} \, I +
 \int_{q|\Lam_b} [3G_{\sg}(q) + G_{\pi}(q)] \right]_{\Lam=\Lam_c}
 (\Lam_c - \Lam)
\label{eq:alpha_L_c}
\end{equation}
for $\Lam \lesssim \Lam_c$. The order parameter $\alf$ and
the fermionic gap are thus proportional to
$(\Lam_c - \Lam)^{1/2}$ for $\Lam \lesssim \Lam_c$.

Inserting Eq.~(\ref{eq:alpha_L_c}) into the flow equation
(\ref{eq:mass_ssb})
for $m_{\sg}^2$ and neglecting the last fluctuation term,
which is of higher order in $\alf$, one obtains
\begin{equation}
 \partial_{\Lam} m_{\sg}^2 =
 - 4 \, I
 - \lam \int_{q|\Lam_b} [3G_{\sg}(q) + G_{\pi}(q)] \; .
\end{equation}
This shows that the flow of $\alf$ and $m_{\sg}^2$ is
indeed consistent with the relation
$m_{\sg}^2 = \lam \alf^2$ following from the ansatz for
the bosonic potential.

\subsection{Infrared asymptotics}
\label{subsec:asymptotics}

In the infrared limit ($\Lam \to 0$), the key properties of the
flow can be extracted from the flow equations analytically as already 
demonstrated in chapter \ref{chap:bosonicqcp_goldstone} for the general $\sigma-\mathbf{\Pi}$ model with $N-1$ Goldstone bosons.
The behavior of the bosonic sector depends strongly on the
dimensionality of the system. We consider dimensions $d \geq 2$,
focusing in particular on the two- and three-dimensional case.
The bosonic order parameter and the fermionic gap saturate at
finite values in the limit $\Lam \to 0$. The fluctuation
corrections to $\partial_{\Lam} \alf$ and $\partial_{\Lam} \Delta$
involve the singular Goldstone propagator $G_{\pi}$ only linearly
and are therefore integrable in $d > 1$. The fermion-boson interactions $g_{\sg}$ and $g_{\pi}$ also saturate.
The finiteness of $Z_{\pi}$ and $A_{\pi}$ is guaranteed by the correct implementation of symmetries (see chapter \ref{chap:bosonicqcp_goldstone}, Pistolesi 2004). 
We choose $\Lam_b = \Lam$ in the following. The choice of $\Lam_f$
(as a function of $\Lam$) will be discussed and specified below.
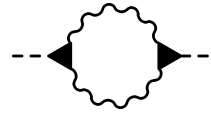
\begin{wrapfigure}{r}{0.5\textwidth}
\vspace{-15mm}
\begin{fmffile}{goldstone_3}
\begin{eqnarray}
\parbox{30mm}{\unitlength=1mm\fmfframe(2,2)(1,1){
\begin{fmfgraph*}(27,30)\fmfpen{thin} 
 \fmfleft{l1}
 \fmfright{r1}
 \fmfpolyn{full,tension=0.4}{G}{3}
 \fmfpolyn{full,tension=0.4}{K}{3}
  \fmf{dashes}{l1,G1}
 \fmf{photon,tension=0.2,right=0.8}{G2,K3}
 \fmf{photon,tension=0.2,right=0.8}{K2,G3}
 \fmf{dashes}{K1,r1}
 \end{fmfgraph*}
}}
\nonumber\\[-17mm]
\nonumber
\end{eqnarray}
\end{fmffile}
\caption{\textit{Goldstone fluctuations determining the infrared asymptotics
for $\Lambda\rightarrow0$}.}
\label{fig:goldstone}
\end{wrapfigure}

The flows of $m_{\sg}^2$, $\lam$, $Z_{\sg}$, and $A_{\sg}$ are
dominated by terms quadratic in $G_{\pi}$ for $\Lam \to 0$, see
Fig.~\ref{fig:goldstone}. Using $m_{\sg}^2 = \lam \alf^2$, we obtain the asymptotic flow
equation for $\lam$ from Eq.~(\ref{eq:mass_ssb}) in the simple
form
\begin{equation}
 \partial_{\Lam} \lam = \lam^2 \int_{q|\Lam} G_{\pi}^2(q) \; .
\end{equation}
The above integral over $G_{\pi}^2$ is proportional to $\Lam^{d-4}$
for small $\Lam$ in dimensions $d < 4$, implying that $\lam$
scales to zero in $d \leq 3$.
In two dimensions one obtains $\int_{q|\Lam} G_{\pi}^2(q) =
\frac{1}{4\pi^2 A_{\pi} Z_{\pi}} \Lam^{-2}$ for small $\Lam$,
such that the rescaled variable $\tilde\lam = \lam/\Lam$ obeys
the flow equation
\begin{equation}
 \frac{d\tilde\lam}{d\log\Lam} = - \tilde\lam +
 \frac{\tilde\lam^2}{4\pi^2 A_{\pi} Z_{\pi}} \; ,
\end{equation}
which has a stable fixed point at
$\tilde\lam^* = 4\pi^2 A_{\pi} Z_{\pi}$. Hence, the bosonic
self-interaction vanishes as
\begin{equation}
 \lam \to 4\pi^2 A_{\pi} Z_{\pi} \, \Lam \quad
 \mbox{for} \quad \Lam \to 0
\end{equation}
in two dimensions in agreement with Eq. (\ref{eq:self}). Consequently, also the radial mass $m_{\sg}^2$
of the Bose fields vanishes linearly in $\Lam$.
In three dimensions one has
$\int_{q|\Lam} G_{\pi}^2(q) \propto \Lam^{-1}$
for small $\Lam$ such that $\lam$ and $m_{\sg}^2$ scale to zero
logarithmically for $\Lam \to 0$.
Since $m_{\sg}^2$ is the dominant contribution to the denominator
of $G_{\sg}$ at small momenta and frequencies, the scaling of
$m_{\sg}^2$ to zero as a function of $\Lam$ implies that $G_{\sg}$
(at $\Lam = 0$) diverges as
\begin{eqnarray}
 G_{\sg}(sq) \propto & s^{-1} & \mbox{for} \; d=2 \\
 G_{\sg}(sq) \propto & \log s & \mbox{for} \; d=3
\end{eqnarray}
in the limit $s \to 0$ matching Eq. (\ref{eq:G_sig}). 
Although derived from an approximate truncation of the functional
flow equation, this result is {\em exact} even in two dimensions,
where the renormalization of $m_{\sg}^2$ is very strong.
This is due to the fact that the scaling dimension of $m_{\sg}^2$
is fully determined by the scaling dimension of the Goldstone
propagator and the existence of a fixed point for $\tilde\lam$,
but does not depend on the position of the fixed point (Pistolesi 2004).

The flow of $Z_{\sg}$ is given by
\begin{equation}
 \partial_{\Lam} Z_{\sg} = \frac{(\lam\alf)^2}{2}
 \int_{q|\Lam} \left.
 \partial_{p_0}^2 G_{\pi}(p+q) G_{\pi}(q) \right|_{p=0}
\end{equation}
for small $\Lam$. The integral over the second derivative of
$G_{\pi} G_{\pi}$ is of order $\Lam^{d-6}$.
In two dimensions the coupling $\lam$ vanishes linearly in
$\Lam$, such that $\partial_{\Lam} Z_{\sg} \propto \Lam^{-2}$,
implying that $Z_{\sg}$ diverges as $\Lam^{-1}$ for $\Lam \to 0$.
Hence, the term $Z_{\sg} q_0^2$ with $|q_0|=\Lam$ in the
denominator of $G_{\sg}$ scales linearly in $\Lam$, as $m_{\sg}^2$.
In three dimensions
$\int_{q|\Lam} \partial_{p_0}^2 G_{\pi}(p+q) G_{\pi}(q) |_{p=0}$
diverges as $\Lam^{-3}$, while $\lam$ vanishes only logarithmically.
Hence $\partial_{\Lam} Z_{\sg} \propto (\log\Lam)^{-2} \Lam^{-3}$,
which is larger than in two dimensions. Integrating over $\Lam$
one finds $Z_{\sg} \propto (\Lam\log\Lam)^{-2}$, which means that
$Z_{\sg} q_0^2$ vanishes as $(\log\Lam)^{-2}$ in the infrared
limit. This yields a subleading logarithmic correction to the
mass term $m_{\sg}^2$ in the denominator of $G_{\sg}$. An analogous
analysis with the same results as just obtained for $Z_{\sigma}$ also holds
for the momentum renormalization factor $A_{\sigma}$.
A strong renormalization of longitudinal correlation functions
due to Goldstone fluctuations appears in various physical contexts
(Weichman 1988, Zwerger 2004).

Recently, a singular effect of Goldstone fluctuations on the
fermionic excitations in a superfluid was found in
Gaussian approximation (Lerch 2008). This singularity appears
only after analytic continuation to real frequencies, and its fate
beyond Gaussian approximation remains to be clarified.

\bigskip

Since $m_{\sg}^2$ and $\lam$ scale to zero in the infrared limit
in $d \leq 3$, all purely bosonic contributions to the effective
action scale to zero. On the other hand, the fermion-boson coupling
remains finite.
One is thus running into a strong coupling problem, indicating a
failure of our truncation, if fermionic fields are
integrated too slowly, compared to the bosons.
The problem manifests itself particularly strikingly in the flow
equation for the order parameter, Eq.~(\ref{eq:alpha_ssb}), in
two dimensions.
Since $m_{\sg}^2 \propto \Lam_b$ for small $\Lam_b$, the fermionic
contribution to $\partial_{\Lam}\alf$ is of order $\Lam^{-1}$
if one chooses $\Lam_f = \Lam_b = \Lam$, leading to a spurious
divergence of $\alf$ for $\Lam \to 0$.

The problem can be easily avoided by integrating the fermions
fast enough, choosing $\Lam_f \ll \Lam_b$ in the infrared limit.
In our numerical solution of the flow equations in the following
section we will choose $\Lam_b = \Lam$ and $\Lam_f = \Lam^2/\Lam_c$
for $\Lam < \Lam_c$, which matches continuously with the equal
choice of cutoffs for $\Lam > \Lam_c$.
The fermionic contribution to $\partial_{\Lam}\alf$ in
Eq.~(\ref{eq:alpha_ssb}) is then finite for small $\Lam$, since
the factor $\Lam'_f = 2\Lam$ in $\int_{k|\Lam_f}$ compensates
the divergence of $m_{\sg}^{-2}$ in front of the integral.
Since the fermions are gapped below $\Lam_c$, one could also
integrate them completely (set $\Lam_f$ to zero),
and then compute the flow driven by $\Lam_b$ only.

The freedom to choose fermionic and bosonic cutoffs independently
was exploited also in a recent fRG-based computation of the
fermion-dimer scattering amplitude in vacuum (Diehl 2007).

\subsection{Numerical results in two dimensions}
\label{subsec:num_results}

In this section, we present a numerical solution of our flow
equations from Sec.~\ref{subsec:flow_equations} in two dimensions.
Technically, we employ a fifth-order Runge-Kutta integration
routine to solve coupled, ordinary differential equations.
At each increment of the Runge-Kutta routine, two-dimensional
integrations over the whole Brillouin zone have to be executed.
For this purpose, we employ an integrator for singular functions 
with relative error of less then $1\%$.
In particular, for $\Lam\approx\Lam_c$ it is imperative to
operate with sufficiently accurate routines as the integrands
are large and small deviations result in a significant spread in
the final values for $\Lambda\rightarrow0$. For further details 
on the numerical procedure, we refer to Appendix \ref{app:num_proc}.

We fix our energy units by setting the hopping amplitude $t=1$.
We choose a chemical potential $\mu=-1.44$ corresponding to an
average electron density of $1/2$ (quarter-filled band).
This choice represents the generic case of a convex Fermi surface
remote from van Hove singularities.
The only varying input parameter is the Hubbard U, which determines
the initial value of the bosonic mass via $m^{2}_b = |2/U|$.
Initially, the flow starts in the symmetric regime with
$\Lam = \Lam_0 = 100$, where Eqs.~(\ref{eq:mass_sym} -
\ref{eq:lambda}) determine the evolution.
The critical scale is determined by the condition
$m_{b}^{2}(\Lambda_c)=0$.
In the symmetry-broken regime ($\Lambda<\Lambda_{c}$),
Eqs.~(\ref{eq:alpha_ssb} - \ref{eq:g_pi_ssb}) determine the
evolution.

\bigskip

\begin{figure}[b]
\begin{center}
\includegraphics*[width=56mm,angle=-90]{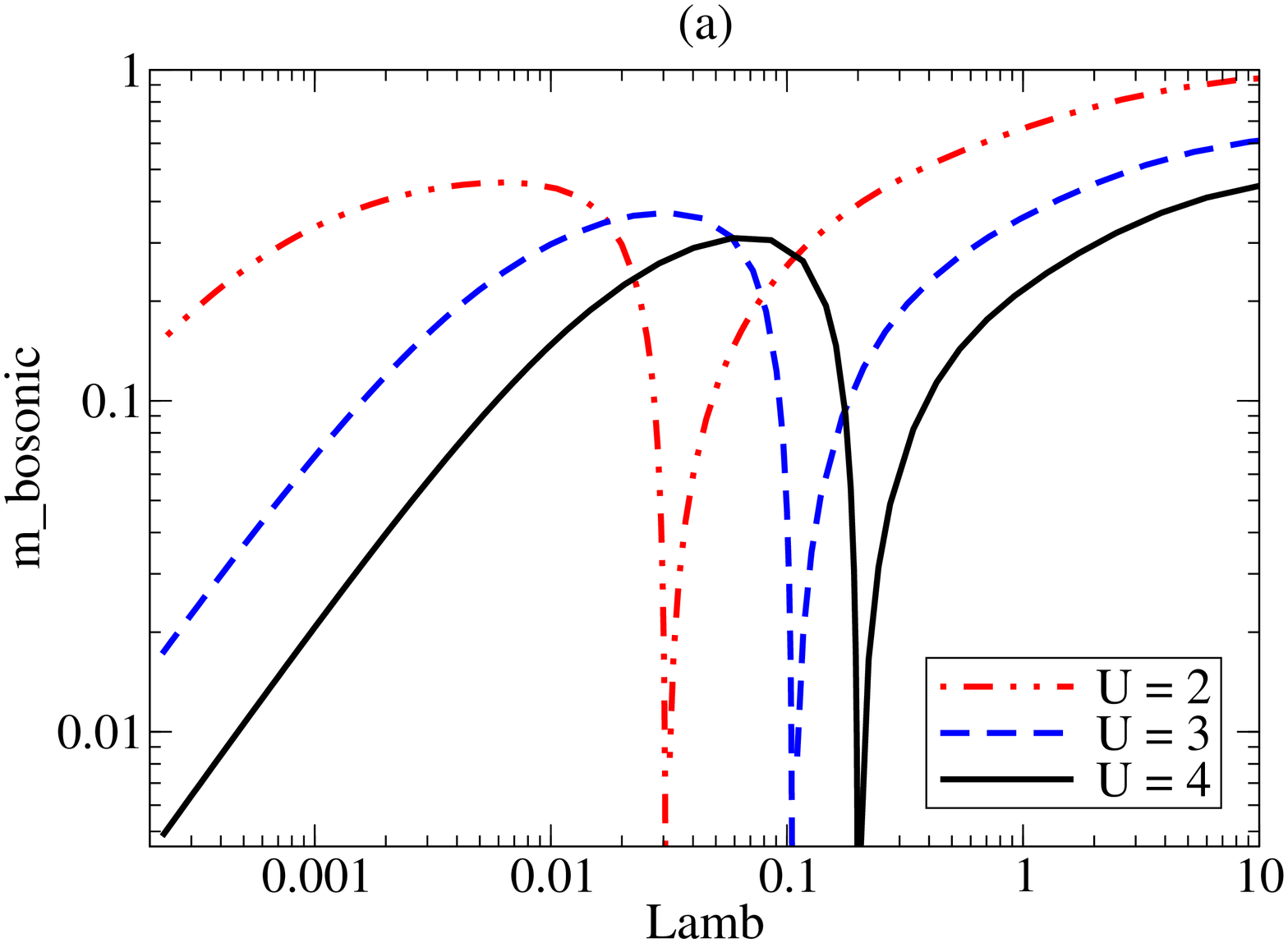}
\hspace*{-4mm}
\includegraphics*[width=56mm,angle=-90]{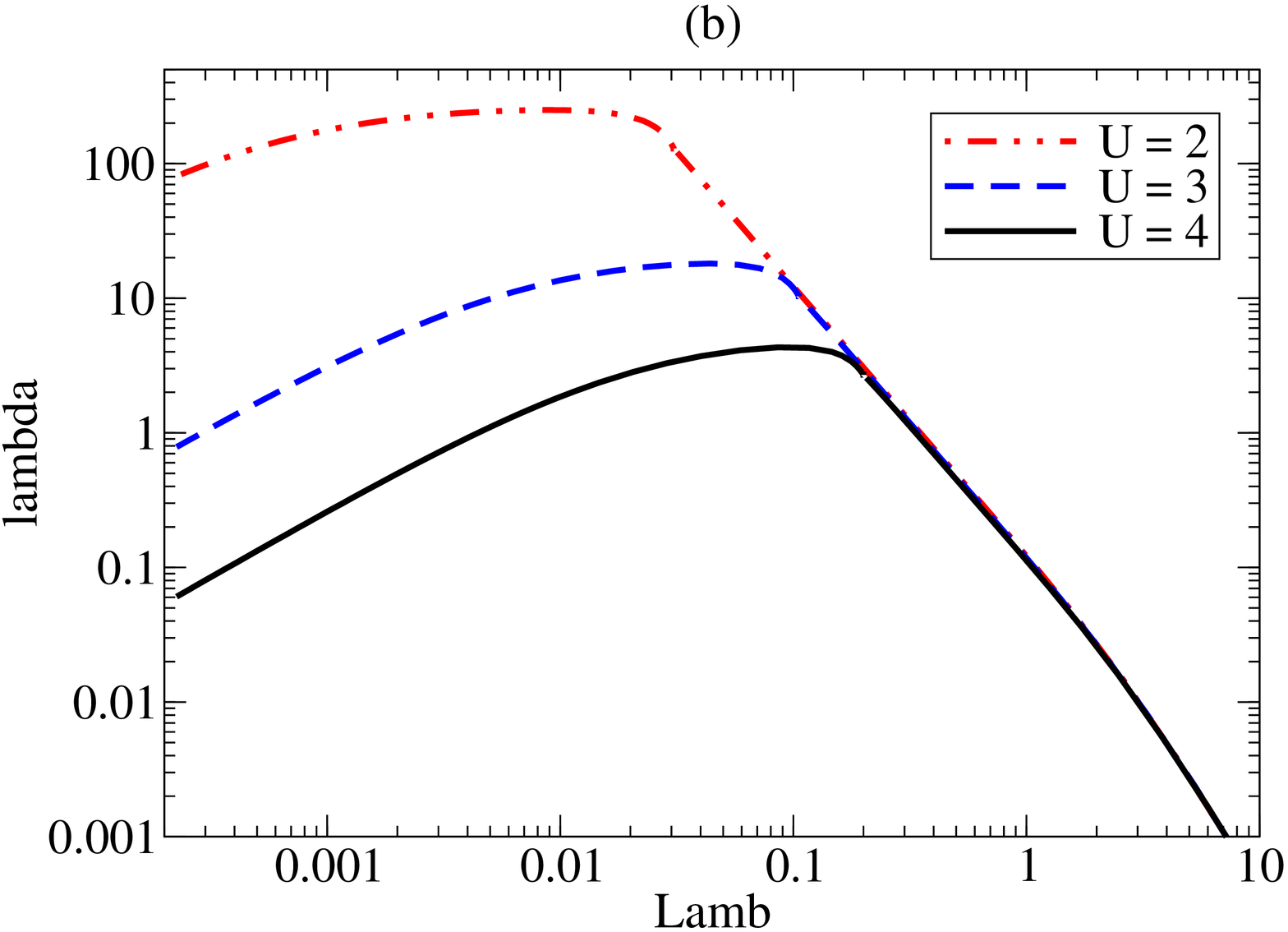}
\caption{(a): Flows of the bosonic mass, $m^{2}_{b}$ for
 $\Lambda>\Lambda_{c}$ and $m^{2}_{\sigma}$ for
 $\Lambda<\Lambda_{c}$.
 (b): Quartic coupling $\lambda$.}
\label{fig:mSig}
\end{center}
\end{figure}

In Fig.~\ref{fig:mSig} characteristic flows of the bosonic mass
and the quartic coupling are shown in double-logarithmic plots for
different choices of the Hubbard U. The sharp de- and increase of
the bosonic mass marks the region around $\Lam_c$.
For small $\Lam$, the flow reaches the infrared asymptotic regime
(see Sec.~\ref{subsec:asymptotics}).
The scale $\Lambda_{\text{IR}}$ at which this scaling sets in
decreases for decreasing $U$. The numerically obtained scaling
$m^{2}_{\sigma},\,\lambda \propto \Lam$ is consistent with the
analytical result of Sec.~\ref{subsec:asymptotics}.

\begin{figure}[t]
\begin{center}
\includegraphics*[width=56mm,angle=-90]{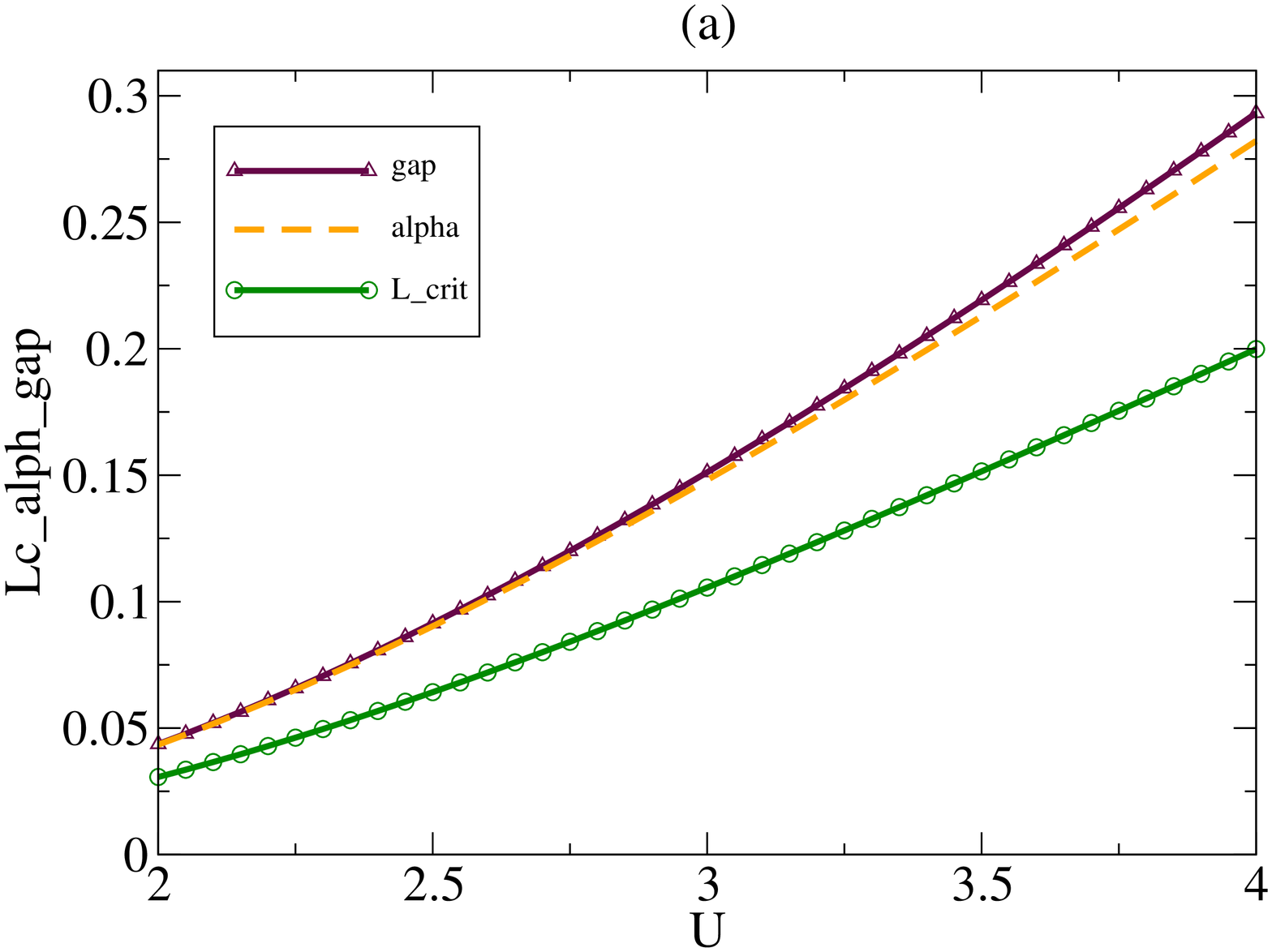}
\hspace*{-4mm}
\includegraphics*[width=56mm,angle=-90]{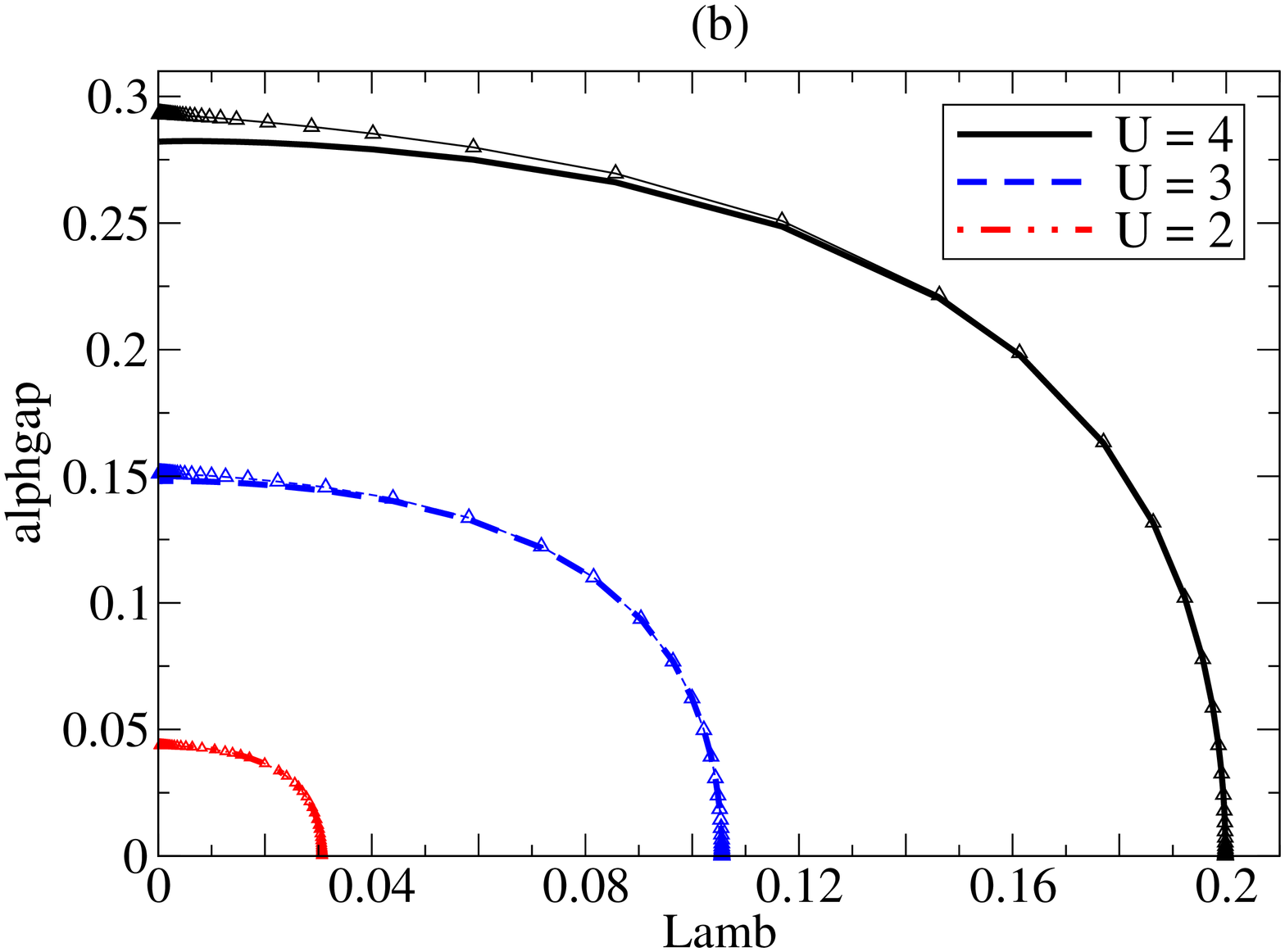}
\caption{(a): Fermion gap $\Delta$, order parameter $\alpha$, and the
 critical scale $\Lambda_{c}$ versus Hubbard $U$.
 (b): Exemplary flows for $\Delta$ (triangles) and $\alpha$ (lines)
 each corresponding to one point in (a).}
\label{fig:gap}
\end{center}
\end{figure}

\bigskip

In Fig.~\ref{fig:gap} (a) we compare the fermion gap with the
bosonic order parameter (final values at $\Lam = 0$) and the
critical scale as a function of $U$.
In Fig.~\ref{fig:gap} (b) the flow of $\Delta$ and $\alf$ as
a function of $\Lam$ is shown for various choices of $U$.
We observe $\Delta,\,\alpha \propto (\Lam_c - \Lam)^{1/2}$ for
$\Lambda\lesssim\Lambda_{c}$ as derived below Eq.~(\ref{eq:alpha_L_c}).
The ratio $\Delta/\Lam_c$, where $\Delta$ is the final gap for
$\Lam \to 0$, is approximately 1.4 for the values of $U$ studied
here.
As a result of fluctuations, the gap is reduced considerably
compared to the mean-field result
\begin{equation}
 \frac{\Delta}{\Delta_{\rm BCS}} \approx 0.25
 \;
\end{equation}
for $2 \leq U \leq 4$.
The main reduction here stems from the bosonic self-interactions
in the symmetric regime leading to a substantial decrease of
$\Lam_{c}$ via the second term of Eq.~(\ref{eq:mass_sym}).
A reduction of the gap compared to the mean-field value is
generally present even in the weak coupling limit $U \to 0$.
In fermionic perturbation theory second order corrections reduce
the prefactor of the BCS gap formula even for $U \to 0$ (Martin 1992).
The reduction obtained here is slightly stronger than what is
expected from a fermionic renormalization group calculation (Gersch 2008).

\begin{figure}[t]
\begin{center}
\includegraphics*[width=56mm,angle=-90]{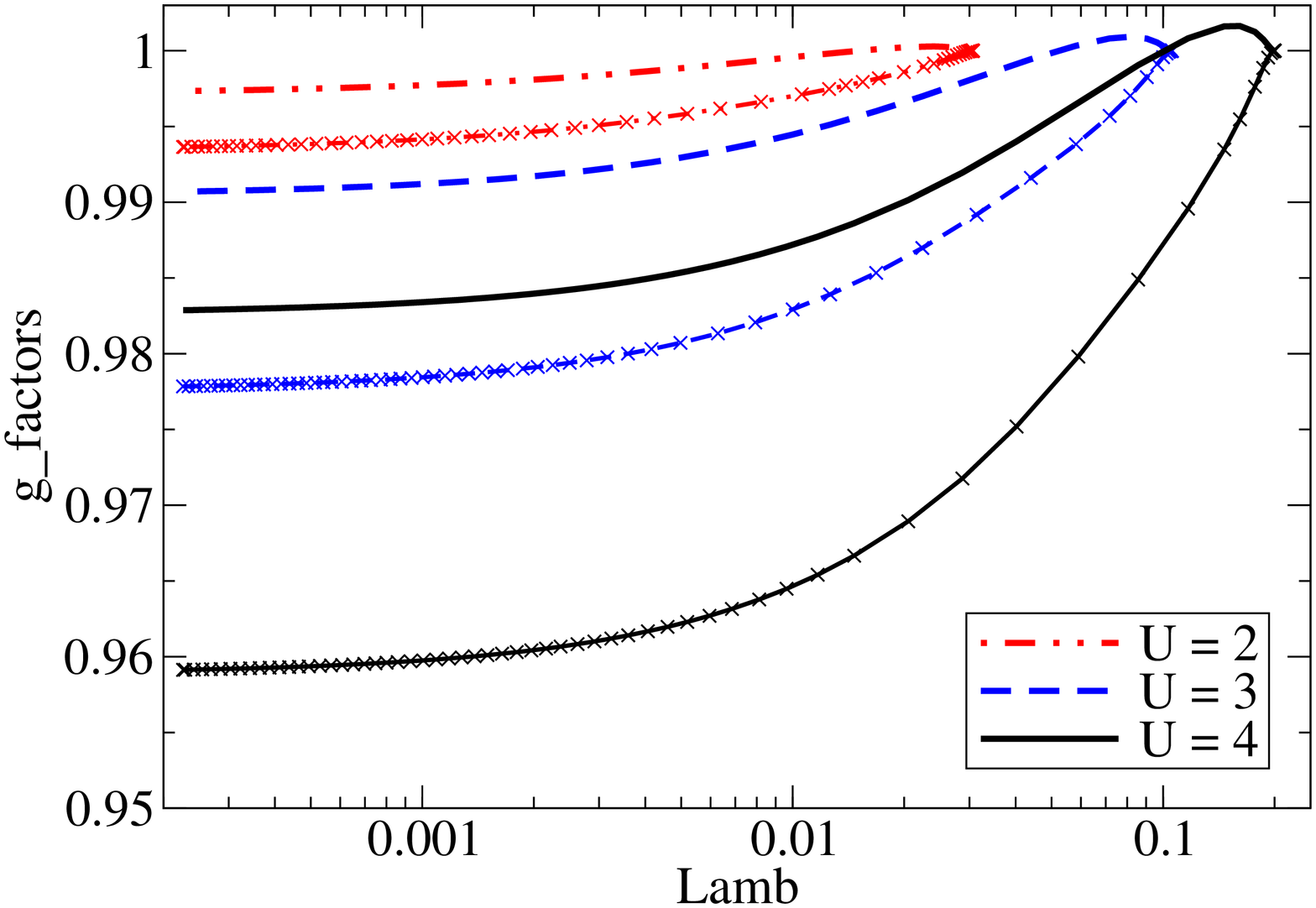}
\hspace*{-4mm}
\includegraphics*[width=56mm,angle=-90]{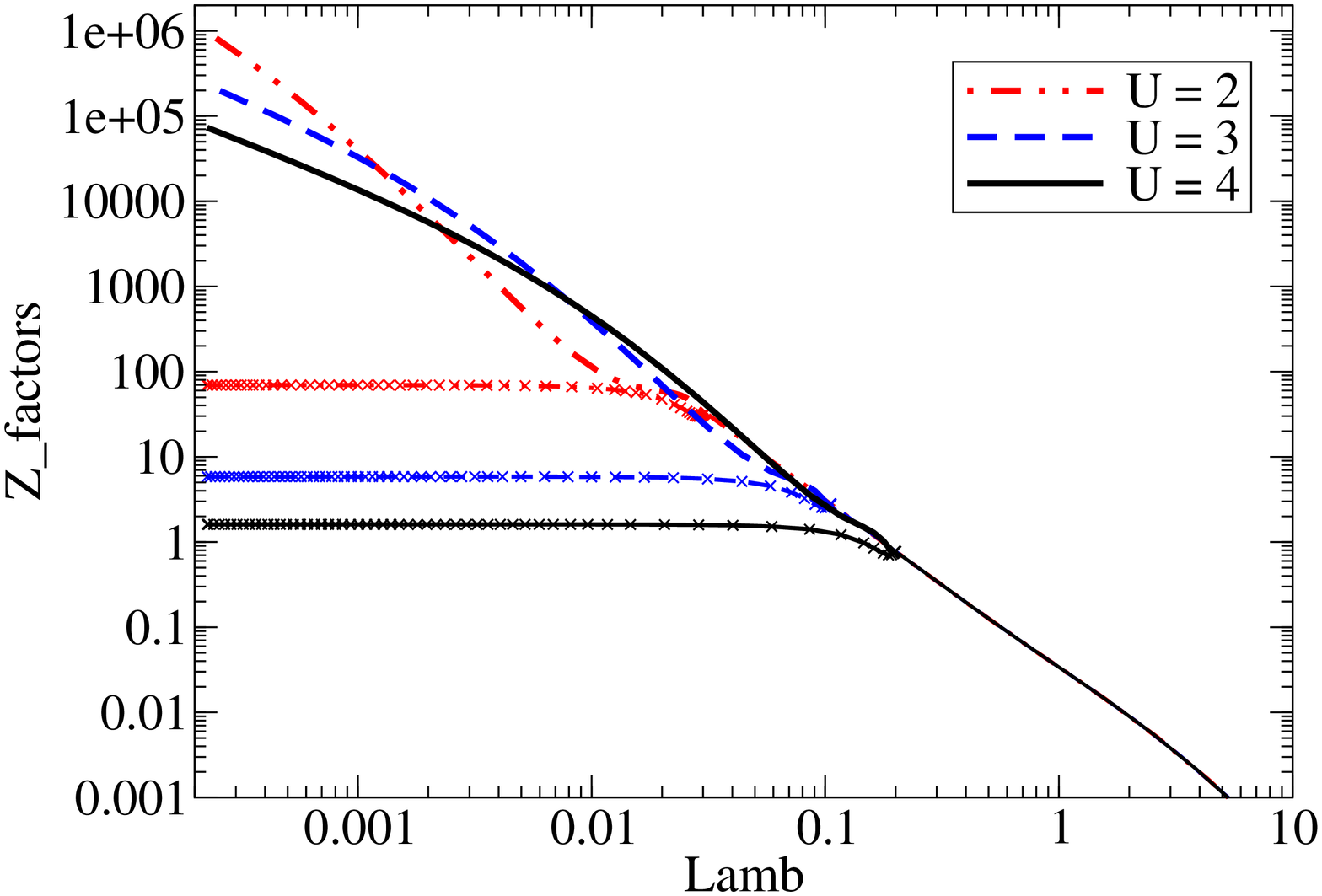}
\caption{Left: Flows of fermion-boson vertices, $g_{\sigma}$ (lines),
 and $g_{\pi}$ (lines with crosses), for $\Lam < \Lam_c$.
Right: Flows of $Z_{\sigma}$ (lines) and $Z_{\pi}$ (crosses). }
\label{fig:vertex_Z}
\end{center}
\end{figure}

For $\Lam < \Lam_c$, Goldstone fluctuations slightly reduce $\alf$
via the term involving $G_{\pi}(q)$ in Eq.~(\ref{eq:alpha_ssb}).
On the other hand, the term due to the Goldstone mode
in Eq.~(\ref{eq:gap_ssb}) enhances the fermionic gap $\Delta$
relative to $\alf$, such that $\Delta$ is generally slightly
larger than $\alf$. This difference will become larger upon increasing
the interaction strength as one enters the regime of a Bose gas made of tightly bound fermions.
Here, however, for relatively weak interactions, the impact of Goldstone fluctuations
on both $\alf$ and $\Delta$ is very modest.
By contrast, the impact of Goldstone fluctuations is known to be
dramatic at finite temperatures (not treated here) in two dimensions,
since they drive the order parameter to zero (Goldenfeld 1992).

\bigskip

In the right plot of Fig.~\ref{fig:vertex_Z}, we show flows of the Z-factors
of the $\sigma$- and $\pi$-field, respectively.
In the symmetric regime, the evolution is independent of $U$.
At $\Lam_{c}$, the $\phi$-field splits into the $\sigma$- and
$\pi$-modes with $Z_{\sigma}$ diverging in the limit $\Lam \to 0$
as $Z_{\sigma} \propto \Lam^{-1}$
(cf.\ Sec.~\ref{subsec:asymptotics}).
The Z-factor of the Goldstone field saturates for
$\Lam \ll \Delta$.
The flows for the A-factors (not shown) parametrizing the
momentum dependence of the $\sigma$- and $\pi$-propagators
exhibit very similar behavior.

Finally, in the left plot of Fig.~\ref{fig:vertex_Z} we show flows of the
fermion-boson vertices $g_{\sigma}$ and $g_{\pi}$ for $\Lam < \Lam_c$.
Their relative changes are only of the order of a few percent
(note the scale of the vertical axis) with $g_{\sigma}$ being a
bit larger than $g_{\pi}$.

\section{Conclusion}

\label{sec:conclusion}

Truncating the exact fRG flow, we have derived approximate flow
equations which capture the non-trivial order parameter fluctuations
in the superfluid ground state of the attractive Hubbard model,
which has been chosen as a prototype model for attractively
interacting fermions.
The superfluid order parameter is associated with a bosonic field
which is introduced via a Hubbard-Stratonovich decoupling of the
fermionic interaction.
Below a critical scale $\Lam_c$, the bosonic effective potential
assumes a mexican hat shape leading to spontaneous symmetry
breaking and a Goldstone mode. The bosonic order parameter is
linked to but not equivalent to a fermionic gap.
The fermionic gap is significantly smaller than the mean-field gap,
mostly due to fluctuations above the scale $\Lam_c$.
Transverse order parameter fluctuations (Goldstone mode) below
$\Lam_c$ lead to a strong renormalization of radial fluctuations.
The radial mass and the bosonic self-interaction vanish linearly
as a function of the scale in two dimensions, and logarithmically
in three dimensions, in agreement with the exact behavior of an
interacting Bose gas (Pistolesi 2004).
On the other hand, the average order parameter, the fermionic gap,
and the interaction between fermions and bosons are affected only
very weakly by the Goldstone mode.

\bigskip

Supplementing the flow equations derived above by a shift of the
chemical potential, to keep the density fixed, one may also try
to deal with larger values of $U$ (see also Outlook \ref{subsec:crossover}).
Eagles (1969) and Leggett (1980) have shown
that already the BCS mean-field theory captures many features of
the condensed Bose gas ground state made from strongly bound
fermion pairs in the limit of strong attraction.
Beyond mean-field theory, the difference between the fermionic
gap and the order parameter $\alpha$ increases at larger $U$.

\bigskip

It will also be interesting to extend the present analysis to $T>0$,
in particular in view of the possibility of a finite fermionic gap
in the absence of long-range order in a Kosterlitz-Thouless phase
at low finite temperatures (see also Outlook \ref{subsec:kt}).


\part{Summary}
\label{part:summary}

\chapter[Conclusions]{Conclusions}
\label{chap:conclusions}

The topic of this thesis has been the computation of the effects of
order parameter fluctuations in two- and three-dimensional
interacting Fermi systems with emphasis on symmetry-broken phases 
and quantum critical behavior. 
For this purpose, we have solved flow equations within the 
functional RG framework for fermionic (single-particle) and bosonic
(collective) degrees of freedom at zero and finite temperature.

\section{Key results}
\label{sec:key}

Three important methodological advancements were achieved during this work.

\subsection{Non-Gaussian fixed points}

We demonstrated in chapters \ref{chap:bosonicqcp_discrete}, 
\ref{chap:bosonicqcp_goldstone}, and \ref{chap:fermibosetoy} how to compute RG flows of
quantum critical fermion systems at zero and finite temperature
where the interaction vertex is attracted towards a 
\emph{non-Gaussian fixed point} with nontrivial critical
exponents.

In chapters \ref{chap:bosonicqcp_discrete} and \ref{chap:bosonicqcp_goldstone}, 
we extended the Hertz-Millis theory formulated entirely in terms
of a bosonic order parameter field to phases with discrete and continuous symmetry-breaking. 
Extending the previous work by Millis (1993), we computed the 
shape of the $T_{c}$-line in the vicinity of the QCP for the case 
where even the zero temperature theory is described by a non-Gaussian fixed point. 
We fully captured the interplay and relative importance
of quantum and thermal fluctuations in the
vicinity of the QCP.

In chapter \ref{chap:fermibosetoy}, we set up a renormalization group framework for 
coupled theories of fermions interacting with their own collective order parameter fluctuations at quantum criticality. As a first application, we considered the semi-metal to superfluid QCP 
of attractive Dirac fermions in two dimensions possibly relevant for cold atoms on the honeycomb lattice and graphene. At the QCP, the fermion-boson coupling becomes relevant 
and induces anomalous exponents for both the fermion and boson propagator, respectively.

\subsection{Goldstone modes}

Systems whose spectrum contains massless \emph{Goldstone
bosons} as a consequence of spontaneously breaking a continuous
symmetry were analyzed in chapters \ref{chap:bosonicqcp_goldstone}
and \ref{chap:fermionsuperfluids}. 

In chapter \ref{chap:bosonicqcp_goldstone}, we devised a truncation of 
the effective action for zero temperature phases with continuous symmetry-breaking 
($\sigma$-$\mathbf{\Pi}$ model) that yields the \emph{exact} momentum behavior 
of the longitudinal and transversal (Goldstone) mode, respectively. 
The Goldstone mode strongly renormalizes the longitudinal propagator while 
all singularities in the Goldstone channel cancel as a consequence 
of symmetries --both modes therefore exhibit strongly different dependences on momenta. 
An even simpler truncation also yields the expected critical exponents 
directly at the critical point where both modes become degenerate. Additionally, 
the $\sigma$-$\mathbf{\Pi}$ model permits investigation of the finite temperature effects in particular correctly reproducing the Mermin-Wagner theorem.

In chapter \ref{chap:fermionsuperfluids}, we thoroughly studied the impact of the Goldstone mode 
in fermionic superfluids at zero temperature. 
The Goldstone mode only weakly affects the fermion-boson vertex, 
the fermionic gap, and the order parameter. In the infrared regime, the fermions 
decouple from the flow and the collective bosonic sector described by the $\sigma$-$\mathbf{\Pi}$ model yields the exact momentum scaling of the transversal and longitudinal mode. 

\subsection{Non-universal and universal quantities}

In chapters \ref{chap:fermibosetoy} and \ref{chap:fermionsuperfluids}, 
the functional RG
was implemented as a versatile framework for coupled fermion-boson systems 
to compute \emph{universal}
as well as \emph{non-universal} quantities simultaneously. The emergence 
of --bosonic and universal-- effective theories from microscopic --fermionic and non-universal-- models upon lowering the energy scale is a strength of the Fermi-Bose RG approach. In addition to the singular 
terms also collected in other RG methods, our flow equations also 
capture finite but possibly large contributions which affect 
non-universal quantities such as the position 
of a QCP, or the size of a fermionic gap. Furthermore, the energy scale 
at which the universal asymptotics sets in can also be computed conveniently within 
this framework enabling a controlled comparison of the size of the various couplings 
with their residual low energy phase-space volume.

\section{Criticism}
\label{sec:criticism}

Despite the methodological advancements of this thesis,
some sore spots, subject to valid criticism, remain. Three central
shortcomings of this work are as follows.

\bigskip

As outlined in section \ref{sec:fermi_lickit}, in many microscopic
models such as the repulsive Hubbard model the situation of
competing ordering tendencies arises. For an accurate computation of
non-universal properties and to obtain certain qualitative phenomena
such as the generation of an attractive d-wave coupling from
antiferromagnetic spin fluctuations, it is not sufficient to take
into account only one ordering channel.

In chapter \ref{chap:fermionsuperfluids}, 
the attractive Hubbard model was considered in a
parameter regime where the Cooper channel is undoubtedly the
most dominant channel and it was therefore justified to utilize
only one Hubbard-Stratonovich field for superfluid order. In the
future, one would want to either (i) decouple the microscopic
four-fermion interaction into multiple bosonic channels and a
posteriori check how much the results depend on the precise
decoupling procedure (Baier 2004), or (ii) perform a combined
fermionic plus bosonic RG study, that is, first integrate out
the high energy modes with a fermionic RG and subsequently
perform the Hubbard-Stratonovich decoupling at a scale
$\Lam_{b}$ when only a few dominant bosonic susceptibilities are
left over. Then, the results have to be checked on their
dependence on the switching scale $\Lambda_{b}$. A similar
strategy has been adopted within a fermionic RG plus mean-field
theory approach in Reiss (2007).

\bigskip

The number of couplings employed in the truncations of this thesis
has been kept to a minimum and often involved merely multiplicative
Z-factors for the frequency and momentum dependencies of the various
propagators. Although it is impressive how much interesting physics
one can extract already on this level, to ascertain the results and
improve on quantitative accuracy, it is necessary to parameterize
the self-energies and vertices on a frequency and/or momentum grid resulting 
in a more elaborate numerical effort. In some circumstances, as for example 
in the superfluid Kosterlitz-Thouless phase (see subsection \ref{subsec:kt}), 
this is necessary to obtain even the correct qualitative behavior.

\bigskip

It is under debate how well the functional RG copes with strongly
coupled systems (Salmhofer 2007). On the one hand, non-Gaussian
fixed points in the infrared regime are captured at least qualitatively. 
The system is strongly coupled in only a limited region of phase-space and the
small phase-space volume when the cutoff goes to zero is helpful in
controlling the vertex expansion when truncating the exact effective action. 
On the other hand, truly strongly coupled
systems when the local interaction exceeds the bandwidth are strongly
coupled everywhere in phase-space and cannot be treated in
manageable truncations yet (Metzner 2005).

\section{Outlook}
\label{sec:outlook}

There exist numerous interesting directions for future research based on the work presented in this thesis.
We here list three of them and describe the strategy for an RG treatment as well as the operational risks associated with each.

\subsection{QED$_{3}$ (extending chapter \ref{chap:fermibosetoy})}
\label{subsec:qed_3}

We have shown in chapter \ref{chap:fermibosetoy} how to compute RG flows for 
coupled fermion-boson systems at quantum criticality. An interesting 
extension concerns Quantum Electrodynamics in $2+1$ dimensions which has been 
advocated as a
model for the zero temperature behavior in
underdoped high $T_{c}$ superconductors
\mbox{(Franz 2002, Herbut 2002)}. 

\bigskip

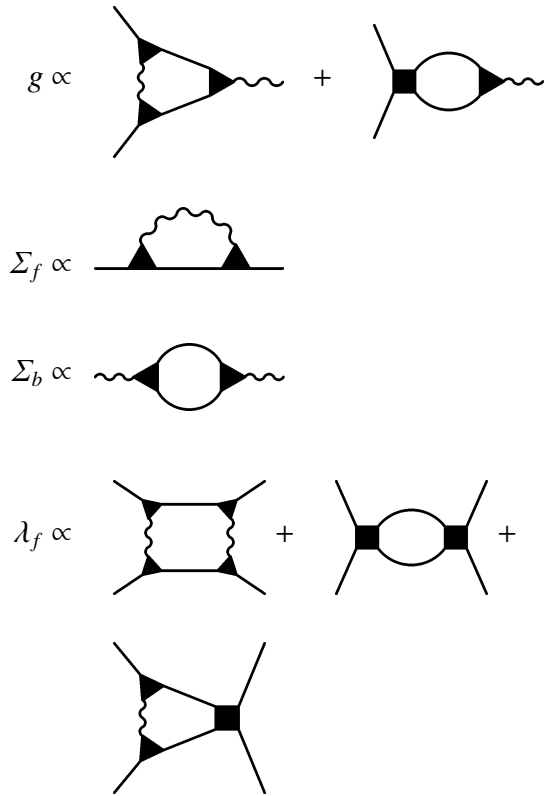
\begin{wrapfigure}{r}{0.5\textwidth}
\vspace*{-5mm}
\begin{fmffile}{20081024_qed_20}
\begin{eqnarray}
g&\propto&
\parbox{30mm}{\unitlength=1mm\fmfframe(2,2)(1,1){
\begin{fmfgraph*}(25,20)
\fmfpen{thin} 
\fmfleftn{l}{2}\fmfright{r}
\fmfrpolyn{full,tension=0.65}{T}{3}
\fmfrpolyn{full,tension=0.65}{B}{3}
\fmfpolyn{full, tension=0.65}{R}{3}
\fmf{plain}{l1,T2}\fmf{plain}{l2,B2}
\fmf{photon,tension=.6}{T3,B1}
\fmf{plain,left=0.,tension=.5}{R3,T1}
\fmf{plain,right=0.,tension=.5}{R2,B3}
\fmf{photon}{R1,r}
\end{fmfgraph*}
}}
+
\parbox{25mm}{\unitlength=1mm\fmfframe(2,2)(1,1){
\begin{fmfgraph*}(25,15)
\fmfpen{thin} 
\fmfleftn{l}{2}\fmfright{r}
\fmfrpolyn{full,tension=1}{G}{4}
\fmfpolyn{full,tension=0.4}{K}{3}
\fmf{plain}{l1,G1}\fmf{plain}{l2,G2}
\fmf{photon}{r,K2}
\fmf{plain,left=.5,tension=.3}{G3,K3}
\fmf{plain,right=.5,tension=.3}{G4,K1}
\end{fmfgraph*}
}}
\nonumber\\[3mm]
\Sigma_{f}&\propto&
\parbox{15mm}{\unitlength=1mm\fmfframe(2,2)(1,1){
\begin{fmfgraph*}(25,15)\fmfpen{thin} 
\fmfleft{l1}
 \fmfright{r1}
 \fmfpolyn{full,tension=0.4}{G}{3}
 \fmfpolyn{full,tension=0.4}{K}{3}
  \fmf{plain}{l1,G1}
   \fmf{plain,straight,tension=0.5,left=0.}{K3,G2}
  \fmf{wiggly,tension=0.,left=0.7}{G3,K2}
  \fmf{plain}{K1,r1}
 \end{fmfgraph*}
}}\nonumber\\[-10mm]
\Sigma_{b}&\propto&
\parbox{25mm}{\unitlength=1mm\fmfframe(2,2)(1,1){
\begin{fmfgraph*}(25,25)\fmfpen{thin} 
 \fmfleft{l1}
 \fmfright{r1}
 \fmfpolyn{full,tension=0.4}{G}{3}
 \fmfpolyn{full,tension=0.4}{K}{3}
  \fmf{wiggly}{l1,G1}
 \fmf{plain,tension=0.3,right=0.6}{G2,K3}
 \fmf{plain,tension=0.3,left=0.6}{G3,K2}
 \fmf{wiggly}{K1,r1}
 \end{fmfgraph*}
}}
\nonumber\\[-3mm]
\lambda_{f}&\propto&
\parbox{25mm}{\unitlength=1mm\fmfframe(2,2)(1,1){
\begin{fmfgraph*}(25,15)
\fmfpen{thin} 
\fmfleftn{l}{2}\fmfrightn{r}{2}
 \fmfpolyn{full,tension=0.8}{OL}{3}
 \fmfpolyn{full,tension=0.8}{OR}{3}
 \fmfpolyn{full,tension=0.8}{UR}{3}
 \fmfpolyn{full,tension=0.8}{UL}{3}
   \fmf{plain}{l1,OL1}
   \fmf{plain}{UL1,l2}
   \fmf{plain}{OR1,r1}
   \fmf{plain}{r2,UR1}
  \fmf{plain,straight,tension=0.5}{OR3,OL2}
  \fmf{wiggly,straight,tension=0.5}{UR3,OR2}
  \fmf{plain,straight,tension=0.5}{UL3,UR2}
  \fmf{wiggly,straight,tension=0.5}{OL3,UL2}
 \end{fmfgraph*}
}}+
\parbox{25mm}{\unitlength=1mm\fmfframe(2,2)(1,1){
\begin{fmfgraph*}(25,15)
\fmfpen{thin} 
\fmfleftn{l}{2}\fmfrightn{r}{2}
\fmfrpolyn{full}{G}{4}
\fmfpolyn{full}{K}{4}
\fmf{plain}{l1,G1}\fmf{plain}{G2,l2}
\fmf{plain}{K1,r1}\fmf{plain}{r2,K2}
\fmf{plain,left=.5,tension=.3}{G3,K3}
\fmf{plain,left=.5,tension=.3}{K4,G4}
\end{fmfgraph*}
}}+\nonumber\\[2mm]
&&
\parbox{30mm}{\unitlength=1mm\fmfframe(2,2)(1,1){
\begin{fmfgraph*}(25,20)
\fmfpen{thin}
\fmfleftn{l}{2}\fmfrightn{r}{2}
\fmfrpolyn{full,tension=0.65}{T}{3}
\fmfrpolyn{full,tension=0.65}{B}{3}
\fmfpolyn{full, tension=1.4}{R}{4}
\fmf{plain}{l1,T2}\fmf{plain}{l2,B2}
\fmf{photon,tension=.6}{T3,B1}
\fmf{plain,left=0.,tension=.5}{R3,T1}
\fmf{plain,right=0.,tension=.5}{R2,B3}
\fmf{plain}{R1,r2}
\fmf{plain}{R4,r1}
\end{fmfgraph*}
}}\nonumber
\end{eqnarray}
\end{fmffile}
\caption{\textit{Diagrammatic representation of the set of flow
equations for QED$_3$. Wiggly lines stand for gauge bosons and
straight lines for fermions.}}
\label{fig:qed_3}
\vspace*{0mm}
\end{wrapfigure}
%
%
%
Massless $\text{QED}_{3}$ can be
tuned to a QCP by varying the number of fermion flavors to a
critical $N_{f,\text{crit}}$. For fewer flavors than
$N_{f,\text{crit}}$ the ground state spontaneously breaks the chiral
symmetry while for more flavors than $N_{f,\text{crit}}$ the gauge
interaction is weakened and the ground state is chirally invariant.
The properties of $\text{QED}_{3}$, in particular 
the momentum dependence of the fermion and gauge boson 
propagator (\emph{universal}) at the QCP 
and $N_{f,\text{crit}}$ (\emph{non-universal}), are presently only poorly 
known (Kaveh 2005, Strouthos 2008 and references therein).
%
%
To investigate this in a functional RG setting, one would replace the effective action written
in Eq. (\ref{eq:dirac_finalmodel}) with the fermion-boson action of QED$_{3}$.
The minimal truncation includes a fermion and (gauge) boson self-energy, possibly
in the form of $Z$-factors. The flow of the renormalized gauge coupling ($g$)
generates four-fermion interactions ($\lambda_{f}$) in the chiral sector
necessitating two interaction channels for the truncation. Diverging
four-fermion interactions signal the onset of chiral symmetry-breaking and thus permit the
determination of $N_{f,\text{crit}}$ (Gies 2006). The set of flow equations is depicted
graphically in Fig. \ref{fig:qed_3}.

\bigskip

Operational risks here involve 
the fulfillment of gauge symmetry and 
the dependence on the choice of gauge-fixing, typically Landau gauge. 
To get warmed up, one may want to start with an
effective model for QED$_{3}$ without gauge bosons --the Thirring model in $2+1$ dimension (Hands 1995, Christofi 2007, Gies 2008).

\subsection{Superfluid Kosterlitz-Thouless phase (ext. chapter \ref{chap:fermionsuperfluids})}
\label{subsec:kt}

The correct finite temperature extension of the truncation employed in the ground-state study for
two-dimensional fermionic superfluids in chapter \ref{chap:fermionsuperfluids} must yield the qualitative features peculiar to the Kosterlitz-Thouless phase 
(Kosterlitz 1973, Kosterlitz 1974, Goldenfeld 1992).
Although there is \emph{no spontaneous symmetry-breaking}, this phase is characterized by a line of fixed points and a temperature dependent anomalous dimension
for the order parameter field. The order parameter correlation function
decays algebraically due to the presence of a massless Goldstone mode. At the transition temperature, the superfluid density jumps to a universal value due to a conspiracy of vortex and Goldstone excitations (Nelson 1977). The single-particle spectrum displays a gap around the Fermi level reminiscent of pseudo-gap phenomena in the cuprates which also
fall into the Kosterlitz-Thouless universality class (Rohe 2001, Eckl 2002).

\bigskip

To date, no satisfactory study of the Kosterlitz-Thouless phase starting from a microscopic
fermionic model and capturing the above features has been performed and would clearly be desirable.
First attempts within the functional RG framework for
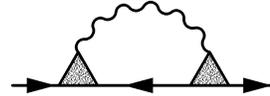
\begin{wrapfigure}{r}{0.5\textwidth}
\vspace*{-5mm}
\begin{fmffile}{20080909_ferm_normal_7}
\begin{eqnarray}
\parbox{35mm}{\unitlength=1mm\fmfframe(2,2)(1,1){
\begin{fmfgraph*}(35,28)\fmfpen{thin} 
\fmfleft{l1}
 \fmfright{r1}
 \fmfpolyn{filled=30,tension=0.4}{G}{3}
 \fmfpolyn{filled=30,tension=0.4}{K}{3}
  \fmf{fermion}{l1,G1}
   \fmf{fermion,straight,tension=0.5,left=0.}{K3,G2}
  \fmf{photon,tension=0.,right=0.7}{K2,G3}
  \fmf{fermion}{K1,r1}
 \end{fmfgraph*}
}}\nonumber\\[-20mm]
\nonumber
\end{eqnarray}
\end{fmffile}
\caption{\textit{Contribution to the normal fermion self-energy
in the superfluid phase. The wiggly line stands for the boson propagator and
the straight line for fermions.}}
\label{fig:kt}
\end{wrapfigure}
bosonic systems
have been put forward in (Gr\"ater 1995, v. Gersdorff 2001). However, in both studies the flow runs back into the symmetric phase, though very slowly, therefore not capturing the vanishing of the $\beta$-function of the superfluid density in the Kosterlitz-Thouless phase. 
Notwithstanding the notorious difficulties of the bosonic sector, capturing the gap in the single-particle
spectrum without symmetry-breaking seems in reach.
Since the anomalous part of the fermionic self-energy, that normally gaps the spectrum, vanishes, one
has to parametrize the frequency- and momentum dependence of the \emph{normal} self-energy of the fermions rather accurately.

The crucial diagram is shown in Fig. \ref{fig:kt} where the order parameter field 
renormalizes the
fermion self-energy. The gap should be visible in the temperature dependence of the specific heat deviating strongly from the linear temperature behavior expected
in a (gapless) Fermi liquid. 
%
%
The scale-dependent specific heat can be computed via temperature derivatives of the thermodynamic potential in Eq. (\ref{eq:eff_pot}):
\begin{eqnarray}
C^{\Lam}(T)=-T\frac{\partial^{2} U^{\Lam}\left(T\right)}{\partial T^{2}}\;.
\end{eqnarray}
Operational challenges lie in the numerical implementation of the normal fermion self-energy
on a momentum- and frequency grid.

\subsection{BCS-Bose crossover (ext. chapter \ref{chap:fermionsuperfluids})}
\label{subsec:crossover}

Upon increasing the attractive interaction in a superfluid Fermi gas, the nature of the fermion pairs
evolves from BCS-type Cooper pairs to tightly bound bosons which undergo Bose-Einstein condensation
(Leggett 1980). Nozi\`eres and Schmitt-Rink first considered this BCS-Bose crossover for lattice fermions
and they extended Leggett's approach to finite temperatures (Nozi\`eres 1985). 

\bigskip

An interesting extension of chapter \ref{chap:fermionsuperfluids} concerns 
the properties of the attractive Hubbard model when increasing the 
attraction $U$ to larger values (Singer 1996, Keller 1999, Keller 2001, Burovski 2006). 
The main addition to the RG truncation already included in chapter
\ref{chap:fermionsuperfluids} is a constraint to keep the density fixed during
the RG flow. Implementing a flow equation
for the density can be achieved by differentiating the flow equation
of the thermodynamic potential, Eq. (\ref{eq:eff_pot}), with respect
to the chemical potential:
\begin{eqnarray}
\partial_{\Lambda}n^{\Lam}=
-\frac{\partial}{\partial \mu}\, \partial_{\Lam}U^{\Lam}
\;.
\end{eqnarray}
Including the fermionic, bosonic and
fluctuation contributions (Ohashi 2002, Diehl 2007) to the density is necessary
to self-consistently adjust the chemical potential to keep the
density constant during the RG flow. As a first step, one would try
to understand and reproduce the Hartree mean-field theory in the
zero sound channel within the RG equations analogously to the
mean-field BCS model (Salmhofer 2004). Subsequently, a Hartree plus
BCS-model with fixed density would have to be, at least
qualitatively, reproduced by the RG flow. Finally, the
order parameter fluctuations are added in and
the effects thereof studied.

\chapter[Deutsche Zusammenfassung]{Deutsche Zusammenfassung}


Die vorliegende Dissertation entwickelt Renormierungsgruppenstrategien zur 
Berechnung makroskopischer Eigenschaften 
von wechselwirkenden Fermi-Systemen. Besonders das Verhalten bei niederen Energien ist 
interessant, da hier wechselwirkungsinduzierte, 
kollektive Ph\"anomene auftreten. Durch Herleitung und L\"osung von Re-normierungsgruppengleichungen analysieren wir in dieser Arbeit den Einfluss von Ordnungsparameterfluktuationen auf thermodynamische Observable und Korrelationsfunktionen in Fermi-Systemen mit 
spontaner Symmetriebrechung und quantenkritischem \mbox{Verhalten}.

\bigskip

Die Dissertation ist in zwei Hauptteile gegliedert. Teil \ref{part:one} stellt das theoretische R\"ustzeug der Arbeit bereit; in Teil \ref{part:apple} wird dieses dann auf verschiedene Problemstellungen angewendet. 

In Kapitel \ref{chap:theo_concepts} werden fundamentale Konzepte der Physik korrelierter Fermionen und der Renormierungsgruppe, die 
zum Verst\"andnis der in Teil \ref{part:apple} folgenden Anwendungen notwendig sind, vorgestellt.
Das Herzst\"uck des ersten Teils stellt Kapitel \ref{chap:functional} dar, in dem 
die einteilchenirreduzible Darstellung der funktionalen Renormierungsgruppe hergeleitet wird.

In Abschnitt \ref{subsec:super} wird erl\"autert wie man die in Superfelder geb\"undelten fermionischen und bosonischen Freiheitsgrade in den Formalismus einbringt, so dass man simultan zu den fermionischen auch die kollektiven, bosonischen Fluktuationen ausintegrieren kann; dies wird in den Kapiteln \ref{chap:fermibosetoy} und \ref{chap:fermionsuperfluids} in die Praxis umgesetzt.

Im Hinblick auf Anwendung in den Kapiteln \ref{chap:bosonicqcp_discrete}, \ref{chap:bosonicqcp_goldstone} und \ref{chap:fermionsuperfluids}, erweitern wir in 
Abschnitt \ref{subsec:ssb_rg} den Formalismus auf symmetriegebrochene Systeme. 
In Abb. \ref{fig:ssb_hierarchy} zeigen wir die durch einen nichtverschwindenden Erwartungswerts einer bosonischen Feldkomponente im Minimum des effektiven Potentials modifizierte Hierarchie der Vertexfunktionen. 

\bigskip

Der zweite Teil der Dissertation ist vier verschiedenen Anwendungen der in Teil \ref{part:one} 
vorgestellten Konzepte und Methoden gewidmet. 

In Kapitel \ref{chap:bosonicqcp_discrete} verallgemeinern wir die urspr\"unglich von Hertz und Millis entwickelte Renormierungsgruppentheorie, die Fermionen auf einen Schlag ausintegriert und die daraus resultierende effektive, bosonische Theorie mit Hilfe von Flussgleichungen analysiert, 
auf Phasen mit diskreter Symmetriebrechung. Im Gegensatz zu Millis k\"onnen wir durch das L\"osen unserer Flussgleichungen bei endlichen Temperaturen auch stark wechselwirkende, nicht-Gau\ss sche Fluktuationen behandeln, die das Verhalten entlang der im quantenkritischen Punkt m\"undenden Phasengrenze dominieren. 

Im ersten Schritt studieren wir bosonische Propagatoren mit 
dynamischen Exponenten $z\geq2$ in zwei und drei Raumdimensionen.
%
Obwohl das Infrarotverhalten bei endlichen Temperaturen durch einen nicht-Gau\ss schen Fixpunkt beschrieben wird, best\"atigen wir das von Millis in Gau\ss scher Approximation von der symmetrischen Phase kommend abgeleitete Potenzgesetz, das die Phasengrenze als Funktion des Kontrollparameters in der N\"ahe des quantenkritischen Punktes beschreibt. Dies ist eine Konsequenz des geringen Phasenraumvolumens, das den Effekt der starken Kopplung im Infraroten unterdr\"uckt.
Da wir die Phasengrenze durch Ausloten der Position des Fixpunktes bestimmen, sind wir in der Lage, die kritische Region \mbox{zwischen} der von Millis als Phasengrenze angenommenen Ginzburg-Linie und der \emph{wahren} Phasengrenze numerisch auszumessen: W\"ahrend beide Linien in drei Raumdimensionen kaum voneinander abweichen, 
\"offnet sich in zwei Raumdimensionen eine ausgepr\"agte 
L\"ucke zwischen Ginzburg-Linie und Phasengrenze bei endlichen Temperaturen.

Im zweiten Schritt nehmen wir uns im Abschnitt \ref{subsec:z_1} das bis dato nicht untersuchte Quanten-Ising-Modell in zwei Raumdimensionen mit dynamischem Exponent $z=1$ vor. Nun liegt 
auch die Grundzustandstheorie unterhalb der oberen kritischen Dimension, was die von Millis gemachte N\"aherung --um den nicht-wechselwirkenden Gau\ss schen Fixpunkt zu entwickeln-- vollst\"andig invalidiert. Wie in Abb. \ref{fig:ginz_vs_Tc} illustriert, weicht in diesem Fall die Ginzburg-Linie auch bei Temperatur null von der Phasengrenze ab. Resultate der numerischen L\"osung der Flussgleichungen sind in Abb. \ref{fig:u_rho_flows_z_1} und \ref{fig:eta_flows_z_1} abgebildet. F\"ur die Phasengrenze als Funktion des Kontrollparameters finden wir Wurzelverhalten mit einer logarithmischen Korrektur, siehe Gl. (\ref{eq:logfit}).

\bigskip

In Kapitel \ref{chap:bosonicqcp_goldstone} analysieren wir F\"alle, wo der quantenkritische Punkt einer Universalit\"atsklasse zugeh\"orig ist, in der eine kontinuierliche Symmetriegruppe gebrochen wird. Dann treten masselose Goldstone-Moden auf, die das kritische Verhalten radikal beeinflussen --besonders in niedrigen Dimensionen.

Basierend auf dem in Abschnitt \ref{sec:sig_pi_model} definierten $\sigma-\mathbf{\Pi}$ Modell, dass den Goldstone-Anregungen mit einem separaten Propagator Rechnung tr\"agt, f\"uhren wir analog zu Kapitel \ref{chap:bosonicqcp_discrete} eine Flussgleichungsstudie in der N\"ahe eines quantenkritischen Punktes durch. Wir fokussieren uns dabei auf Systeme mit quadratischer Frequenzabh\"angigkeit --also dynamischer Exponent $z=1$-- in den Propagatoren beider, radialer und transversaler, Anregungen des Systems. Die Flussgleichungen spiegeln das Mermin-Wagner-Theorem richtig wider: Der Ordnungsparameter, und damit die kritische Temperatur, wird in Raumdimensionen kleiner gleich \emph{zwei} von thermisch angeregten Goldstone-Moden zu null gedr\"uckt.

Als ein unerwartetes Hauptresultat finden wir in Abschnitt \ref{subsec:gold_shift}, 
dass die Form der Phasengrenze bei endlicher Temperatur in der N\"ahe des quantenkritischen Punktes in \emph{drei} Raumdimensionen nicht von der Anzahl der Goldstone-Moden abh\"angt.

Abseits der Phasengrenze, in der symmetriegebrochenen Phase, extrahieren wir in 
Abschnitt \ref{sec:gold_IR} das exakte Infrarotverhalten beider Progagatoren und der Wechselwirkungen analytisch aus den Flussgleichungen. Die effektive Selbstwechselwirkung der Goldstone-Moden flie\ss t aus Symmetriegr\"unden zu Null 
(siehe Gl. (\ref{eq:effgold_interaction})), und der Goldstone-Propagator beh\"alt seine unrenormierte, quadratische Frequenz- und Impulsabh\"angigkeit bei. 
Die radiale Mode und deren Wechselwirkungen hingegen werden, wie in 
Gl. (\ref{eq:G_sig}}-\ref{eq:self}) zusammengefasst, stark durch die Goldstone-Moden renormiert.

\bigskip

In Kapitel \ref{chap:fermibosetoy} berechnen wir Renormierungsgruppenfl\"usse gekoppelter Fermi-Bose Theorien, die sich durch ihr kompliziertes, und ob der beiden masselosen Propagatoren am quantenkritischen Punkt daher mit anderen Methoden bisher nur unzu-reichend erforschbares, Infrarotverhalten auszeichnen.

Wir analysieren attraktiv wechselwirkende Dirac-Fermionen 
(siehe Abb. \ref{fig:cone}); hier treibt die Wechselwirkungst\"arke bereits in Molekularfeldtheorie einen kontinuierlichen Quantenphasen\"ubergang von einem Halbmetall zu einer Superfl\"ussigkeit (siehe Abb. \ref{fig:phase_toy} und Abschnitt \ref{subsec:dirac_mft}). Physikalisch treten Dirac-Fermionen in Kohlenstoffschichten und in optischen Honigwabengittern in kalten, atomaren Quantengasen auf. 

Unsere Trunkierung und Flussgleichungen der Fermion-Boson Wirkung, in Abb. \ref{fig:dirac_flow} diagrammatisch dargestellt, erm\"oglichen die Berechnung aller quantenkritischen Exponenten in unmittelbarer N\"ahe des quantenkritischen Punktes. 
Abb. \ref{fig:fp_line} zeigt die Werte der anomalen Dimensionen der Fermionen und Boson direkt am quantenkritschen Punkt f\"ur verschieden Raumdimensionen. Wie unter Gl. (\ref{eq:ferm_self_qcp}) 
erl\"autert, impliziert eine anomale Dimension des Fermionfeldes den Zusammenbruch des Fermifl\"ussigkeitsverhalten: Die lineare Frequenzabh\"angigkeit der Selbstenergie wird zu einem fraktionellen Potenzgesetz, das die hohe Zerfallsrate der niederenergetischen Quasiteilchen in der N\"ahe des Dirac-Punktes beschreibt.

In Abschnitt \ref{subsec:qcflows} werden numerische Fl\"usse der in der Trunkierung vorkommenden Parameter pr\"asentiert. Durch kontrolliertes Verkn\"upfen der mikroskopischen Anfangsbedingungen, die auch der Molekularfeldrechung von Abschnitt \ref{subsec:dirac_mft} zugrundeliegen, ist es uns m\"oglich, zus\"atzlich zu den universellen Exponenten auch die nicht-universelle, renormierte Position des quantenkritischen Punktes entlang der Kontrollparameterachse mit dem Wert der Molekularfeldrechnung zu vergleichen; Gl. (\ref{eq:ratio}) veranschaulicht, dass Quantenfluktuationen die Ausma\ss e des Superfl\"ussigkeitsbereichs im Phasendiagram Abb. \ref{fig:phase_toy} drastisch reduzieren.

Die Berechnung des Satzes quantenkritischer Exponenten wird durch die Bestimmung der Potenzgesetze f\"ur die Korrelationsl\"ange und die Suszeptibilit\"at als Funktion des Kontrollparameters in Abschnitt \ref{subsec:suszeb} abgerundet.

\bigskip

In Kapitel \ref{chap:fermionsuperfluids} wird ein Gros der in den Kapiteln  \ref{chap:bosonicqcp_discrete} bis \ref{chap:fermibosetoy} aufgebauten Neuerungen synthetisiert, und wir pr\"asentieren eine umfassende Analyse des attraktiven Hubbard-Modells bei Viertelbandf\"ullung, als Prototyp eines fermionischen Systems mit suprafluidem Grundzustand. 
In einem gekoppelten Fermion-Boson Fluss, der sowohl die Selbstwechselwirkungen des Ordnungsparameters, zwei verschiedene Fermion-Boson Kopplungen, als auch die anomalen Komponenten der \mbox{fermionischen} Selbstenergie ber\"ucksichtigt, werden simultan zu diversen nicht-universellen Gr\"o\ss en, wie der Energiel\"ucke des fermionischen Energiespektrums, auch universelle Infrarot-eigenschaften des kollektiven, bosonischen Sektors bestimmt.

Strukturell zeichnet sich unsere Analyse durch Unterscheidung des Minimums des bosonischen, effektiven Potentials und der fermionisches Energiel\"ucke --beide fallen nur in Molekularfeldtheorie zusammen, siehe Abschnitt \ref{subsubsec:ssb_trunc}-- und Trennung der radialen und transversalen Moden (siehe Abschnitt \ref{subsubsec:ssb_trunc}), aus. Letzteres erm\"oglicht in Anlehnung an Abschnitt \ref{sec:gold_IR} die exakte Behandlung des Infrarotsektors (siehe Abschnitt \ref{subsec:asymptotics}).

In Abschnitt \ref{subsec:num_results} werden numerische L\"osungen der gekoppelten Renormierungsgruppengleichungen in zwei Raumdimensionen pr\"asentiert. Ein wesentliches Resultat ist der in Abb. \ref{fig:gap} dargestellte Vergleich zwischen fermionischer Energiel\"ucke, Minimum des effektiven Potentials und kritischer Skala. Die Werte f\"ur die fermionische Energiel\"ucke sind um etwa einen Faktor vier gegen\"uber der Molekularfeldtheorie reduziert; haupts\"achlich verursacht durch Ordnungsparameterfluktuationen, die bei relativ hohen Energieskalen bereits im symmetrischen Teil des Flusses der Formierung suprafluider Ordnung entgegenwirken.

Die linke Graphik in Abb. \ref{fig:vertex_Z} zeigt, dass beide Fermion-Boson Vertizes, obschon diese durch Goldstone-Fluktuationen renormiert werden (siehe Abb. \ref{fig:vertex_ssb}), quantitativ nur um wenige Prozentpunkte von ihren nicht-renormierten Ausgangswerten abweichen. Dies ist eine Konsequenz des geringen den Goldstone-Moden zur Verf\"ugung stehenden Phasenraumvolumens.

Der unterschiedliche Verlauf der bosonischen Z-Faktoren ist in der rechten Abbildung von Abb. \ref{fig:vertex_Z} sichtbar: W\"ahrend der Z-Faktor der radialen Mode als Funktion der Skala divergiert, saturiert der Z-Faktor der Goldstone-Mode, so dass diese ihre quadratische Frequenz- und Impulsabh\"angigkeit beh\"alt --wie bereits in Abschnitt \ref{sec:gold_IR} analytisch demonstriert.

\bigskip

Die Dissertation schliesst mit einer kritischen Diskussion in Kapitel \ref{chap:conclusions}, 
die sowohl methodische Errungenschaften als auch Optimierungspotenziale der Arbeit und 
interessante zuk\"unftige Forschungsvorhaben umrei\ss t. Im Anhang \ref{app:num_proc} wird die zur numerischen L\"osung der Flussgleichungen entwickelte numerische Prozedur erkl\"art.

\backmatter

\appendix
\part{Appendices}


\chapter[Numerical procedure]{Numerical procedure}
\label{app:num_proc}

In this Appendix, we present the programs which were assembled and
subsequently used to solve the RG flow equations in this thesis. The
exact flow equation, Eq. (\ref{eq:exact_flow_eqn}), projected on a
finite number of parameters represents a system of coupled ordinary
differential equations. The right-hand-sides of these equations involve
numerical integrations. In chapter \ref{chap:bosonicqcp_discrete},
the right-hand-sides involve one-dimensional integrations and in
chapter \ref{chap:fermionsuperfluids} two-dimensional momentum
integrations over the whole Brillouin zone have to be executed.

\bigskip

\begin{figure}[t]
\hspace*{-15mm}
\includegraphics[width=165mm,angle=-90]{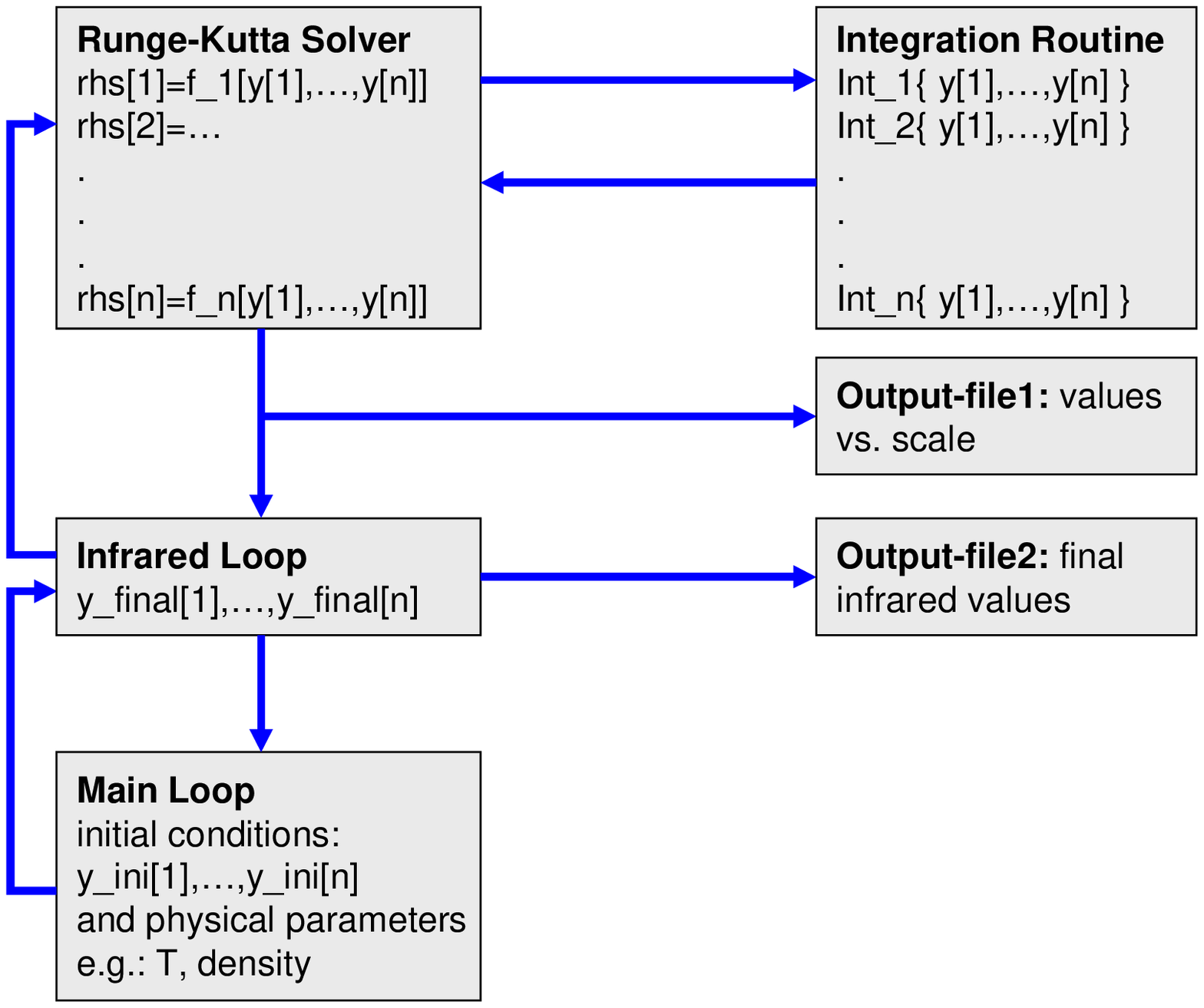}
\vspace*{-20mm}
\caption{\textit{Schematic data flow of the program employed to solve the flow equations 
in this thesis. The maximum number of coupled equations solved simultaneously 
in chapter \ref{chap:fermionsuperfluids} was $n\lesssim 15$. Typical computation times 
for forty consecutive flows as shown in Fig. \ref{fig:gap} were of the order of several 
days up to one week on a single modern processor.}}
\label{fig:data_flow}
\end{figure}

Hence, a routine for coupled differential equations was
combined with a numerical integration routine which provides the
values of the right-hand-sides at each increment of the differential
equation solver. A data flow picture is exhibited in Fig.
\ref{fig:data_flow}.

The routines employed were taken from the GSL-GNU scientific library
\footnote{\textit{http://www.gnu.org/software/gsl/}}
and then assembled and modified in a C/C++
code. Specifically, we used the Runge-Kutta-Fehlberg (4, 5)
stepping function, gsl\textunderscore odeiv\textunderscore step\_rkf45, 
for the differential
equation solver. This routine adaptively adjusts the step-size for
given error margins and proved to be computationally efficient as
well as accurate even at very low temperatures. We used
relative error margins of $1e$-$4$ to $1e$-$8$.

\bigskip

For the numerical integrations, we employed the routine 
gsl\textunderscore integration\textunderscore qags. This routine performs well for singular integrands 
as we encounter at the critical scale or toward the end of the flow. It adaptively adjusts 
the stepsize within each integration interval to satisfy the error margins. 
We employed relative error margins of $1e$-$2$ to $1e$-$4$. 
Two-dimensional integrations were executed by linearly combining 
two one-dimensional routines.

\chapter[Academic curriculum vitae]{Academic curriculum vitae}
\label{app:cv}

\vspace*{-13mm}

\begin{center}
\textit{Born on April 24th, 1981 in Frankfurt am Main}\\[5mm]
\end{center}

\begin{tabbing}
\textsc{\textbf{Education}}\\[0mm]
\hspace{43mm}\=\\[-2.5mm]

\it{10.2005 - 02.2009} \> \textit{Doctorate - Theoretical Physics (grade: with distinction)}\\[1mm]
$\;\;\;\;\,$\footnotesize{10.2005 - 02.2009}   \> $\;\;\;\;\bullet
$ PhD-Student of W. Metzner, Max-Planck-Institute\\
\> $\;\;\;\;$\hspace{2mm} for Solid State Research, Stuttgart\\[1mm]
$\;\;\;\;\,$\footnotesize{07.2007 - 09.2007}   \> $\;\;\;\;\bullet$
Guest researcher with G. Lonzarich, Trinity College,\\
\> $\;\;\;\;$ \hspace{1mm} University of Cambridge, UK\\[1mm]
$\;\;\;\;\,$\footnotesize{10.2005 - 07.2007}   \> $\;\;\;\;\bullet$
Regular visitor to the Group of C. Wetterich,\\
\> $\;\;\;\;$\hspace{2mm} University of Heidelberg\\[1mm]
$\;\;\;\;\,$\footnotesize{04.2006 - 08.2006}   \> $\;\;\;\;\bullet$
Tutor in Statistical Physics and Electrodynamics,\\
\> $\;\;\;\;$\hspace{2mm} University of Stuttgart\\[3mm]
\it{10.2001 - 05.2005}   \>\textit{Master of Science - Physics (grade: superior)}\\[1mm]
$\;\;\;\;\,$\footnotesize{08.2004 - 05.2005}   \> $\;\;\;\;\bullet$
M. Sc. in Theoretical Astrophysics\\
\> $\;\;\;\;$\hspace{2mm} with A. Burrows, University of Arizona, Tucson, USA\\[1mm]
$\;\;\;\;\,$\footnotesize{04.2003 - 07.2004}    \> $\;\;\;\;\bullet$
Diploma exams, University of Heidelberg\\[1mm]
$\;\;\;\;\,$\footnotesize{09.2001 - 03.2003}    \> $\;\;\;\;\bullet$
Prediploma, Technical University Dresden\\[3mm]
\it{06.2000}           \> \textit{Abitur, Justus-Liebig-School, Darmstadt}\\[1mm]
$\;\;\;\;\,$\footnotesize{06.1997 - 08.1998}  \> $\;\;\;\;\bullet$
Queen Elizabeth High School, Edmonton, Canada\\[5mm]

\textsc{\textbf{Scholarships}}\\[-5mm]

 \> Max Planck Contract for Doctoral Research\\[1mm]
 \> Full scholarship -- The University of Arizona\\[1mm]
 \> Fulbright Commission Travel Grant\\[1mm]
 \> Baden-W\"urttemberg Stipendium
\end{tabbing}
\vspace{7mm}
Stuttgart, February 2009\\[-13mm]

\cleardoublepage \thispagestyle{empty}

\noindent
Abanov, A., and Chubukov, A.V., 
Phys. Rev. Lett. {\bf 84}, 5608 (2000).\\[4mm]
Abanov, A., Chubukov, A.V., and Schmalian, J.,
Adv. in Phys. {\bf 52}, 119-218 (2003).\\[4mm]
Abanov, A., and Chubukov, A.V., 
Phys. Rev. Lett. {\bf 93}, 255702 (2004).\\[4mm]
Abrikosov, A.A., Gorkov, L.P., and Dzyaloshinski, I.E., 
{\em Methods of Quantum Field Theory in Statistical Physics} 
(Dover Publications, New York, 2005).\\[4mm]
Altshuler, B.L., Ioffe, L.B., and Millis, A.J., 
Phys. Rev. B {\bf 50}, 14048 (1994); \emph{ibid.} {\bf 52}, 5563 (1995).\\[4mm]
Amit, D.J., and Martin-Mayor, V., {\em Field Theory, the Renormalization Group and Critical Phenomena} 
(World Scientific, 2005).\\[4mm]
Baeriswyl, D., {\it et al.}, {\em The Hubbard Model} (NATO ASI Series, Plenum Press, 1993).\\[4mm]
Baier, T., Bick, E., and Wetterich, C., Phys. Rev. B {\bf 70}, 125111 (2004).\\[4mm]
Ballhausen, H., Berges, J., and Wetterich, C., 
Phys. Lett. B {\bf 582}, 144 (2004).\\[4mm]
Bedaque, P.F., Caldas, H., and Rupak, G., 
Phys. Rev. Lett. {\bf 91}, 247002 (2003).\\[4mm]
%
%
%
Belitz, D., Kirkpatrick, T.R., and Vojta, T., 
Rev. Mod. Phys. {\bf 70}, 580 (2005).\\[4mm]
%
%
Berges, J., Tetradis, N., and Wetterich, C., Phys. Rep. {\bf 363}, 223 (2002).\\[4mm]
Birse, M.C., Krippa, B., McGovern, J.A., Walet, N.R., Phys. Lett. B {\bf 605}, 287 (2005).\\[4mm]
Burovski, E., Prokofev, N., Svistunov, B., and Troyer, M., 
Phys. Rev. Lett. {\bf 96}, 160402 (2006); 
New Journal of Physics {\bf 8}, 153 (2006) .\\[4mm]
%
%
%
Castro Neto, A.H., Guinea, F., Peres, N.M.R., Novoselov, K.S., 
and Geim, A.K., Rev. Mod. Phys. {\bf 81}, 109 (2009).\\[4mm]
Chin, J.K., Miller, D.E., Liu, Y., Stan, C., Setiawan, W., Sanner, C., Xu, K., 
and Ketterle, W., Nature {\bf 446}, 961 (2006).\\[4mm]
Chitov, G.Y., and Millis, A.J., Phys. Rev. Lett. {\bf 86}, 5337 (2001); 
Phys. Rev. B {\bf 64}, 054414 (2001).\\[4mm]
Christofi, S., Hands, S., and Strouthos, C., 
Phys. Rev. D  {\bf 75}, 101701(R) (2007).\\[4mm]
Chubukov, A.V., and Maslov, D.L., 
Phys. Rev. B {\bf 68}, 155113 (2003).\\[4mm]
Chubukov, A.V., Maslov, D.L., Gangadharaiah, S., and Glazman, L.I., 
Phys. Rev. B {\bf 71}, 205112 (2005).\\[4mm]
Chubukov, A.V., and Khveshchenko, D.V., 
Phys. Rev. Lett. {\bf 97}, 226403 (2006).\\[4mm]
Delamotte, B., Mouhanna, D., and Tissier, M., 
Phys.Rev. B {\bf 69} 134413 (2004).\\[4mm]
Dell'Anna, L., and Metzner, W., 
Phys. Rev. B {\bf 73}, 045127 (2006).\\[4mm]
Diehl, S., Gies, H., Pawlowski, J.M., and Wetterich, C., 
Phys. Rev. A {\bf 76}, 021602(R) (2007).\\[4mm]
Diehl, S., Krahl, H.C., and Scherer, M., Phys. Rev. C {\bf 78}, 034001 (2008).\\[4mm]
Diener, R.B., Sensarma, R., and Randeria, M., 
Phys. Rev. A {\bf 77}, 023626 (2008).\\[4mm]
%
%
Eagles, D.M., Phys. Rev. {\bf 186}, 456 (1969).\\[4mm]
Eckl, T., Scalapino, D.J., Arrigoni, E., and Hanke, W., 
Phys. Rev. B {\bf 66}, 140510(R) (2002).\\[4mm]
%
%
Enss, T., PhD Thesis, Max Planck Institute for Solid State Research, 
arXiv:cond-mat/0504703 (2005).\\[4mm] 
Feldman, J., Magnen, J., Rivasseau, V., 
and Trubowitz, E., Helv. Phys. Acta {\bf 66}, 497 (1993).\\[4mm]
%
%
Franz, M., Tesanovic, Z., and Vafek, O., 
Phys. Rev. B {\bf 66}, 054535 (2002).\\[4mm]
Gersch, R., Honerkamp, C., Rohe, D., and Metzner W., Eur. Phys. J. B {\bf 48}, 349 (2005).\\[4mm]
Gersch, R., Reiss, J., and Honerkamp, C., New J. Phys. {\bf 8}, 320 (2006).\\[4mm]
Gersch, R., Honerkamp, C., and Metzner, W., 
New J. Phys. {\bf 10}, 045003 (2008).\\[4mm]
v. Gersdorff, G., and Wetterich, C., Phys. Rev. B {\bf 64}, 054513 (2001).\\[4mm]
Gies, H., and Jaeckel, J., Phys. Rev. Lett. {\bf 93}, 110405 (2004).\\[4mm]
Gies, H., and Jaeckel, J., Eur. Phys. J. C  {\bf 46}, 433 (2006).\\[4mm]
Gies, H., arXiv:hep-ph/0611146 (2006).\\[4mm]
Gies, H., private communication (2008).\\[4mm]
GSL-GNU scientific library, 
http://www.gnu.org/software/gsl/.\\[4mm]
Goldenfeld, N., {\em Lectures on Phase Transitions and the Renormalization Group} 
(Perseus Publishing, Oxford, 1992).\\[4mm]
%
%
Gr\"ater, M., and Wetterich, C., Phys. Rev. Lett. {\bf 75}, 3 (1995).\\[4mm]
Gross, D.J., and Wilczek, F., Phys. Rev. Lett. {\bf 30}, 1343 (1973); 
Phys. Rev. D {\bf 8}, 3633 (1973).\\[4mm]   
Halboth, C.J., and Metzner, W., Phys. Rev. Lett. {\bf 85}, 5162 (2000); 
Phys. Rev. B {\bf 61}, 7364 (2000).\\[4mm]  
Hands, S., Kocic, A., and Kogut, J.B., Annals of Physics, {\bf 224}, 
29 (1993).\\[4mm]
Hands, S., Phys. Rev. D {\bf 51}, 5816 (1995).\\[4mm] 
%
%
Herbut, I.F., Phys. Rev. B {\bf 66}, 094504 (2002).\\[4mm]
Herbut, I.F., Phys. Rev. Lett. {\bf 97}, 146401 (2006).\\[4mm]
Hertz, J.A., Phys. Rev. B {\bf 14}, 1165 (1976).\\[4mm]
Hofstetter, W., \emph{et al.}, 
Phys. Rev. Lett. {\bf 89}, 220407 (2002).\\[4mm]
Honerkamp, C., and Salmhofer, M., Phys. Rev. Lett. {\bf 87}, 187004 (2001), 
Phys. Rev. B {\bf 64}, 184516 (2001).\\[4mm]
Jaksch, D., and Zoller, P., Ann. Phys. {\bf 315}, 52 (2005).\\[4mm]
Jakubczyk, P., Strack, P., Katanin, A.A., and Metzner, W., 
Phys. Rev. B {\bf 77}, 195120 (2008).\\[4mm]
Katanin, A.A., Phys. Rev. B {\bf 70}, 115109 (2004).\\[4mm]
Kaul, R.K, and Sachdev, S., Phys. Rev. B {\bf 77}, 155105 (2008).\\[4mm]
Kaveh, K., and Herbut, I.F., Phys. Rev. B {\bf 71}, 184519 (2005).\\[4mm]
Keller, M., Metzner, W., and Schollw\"ock, U., Phys. Rev. B {\bf 60}, 3499 (1999); 
Phys. Rev. Lett. {\bf 86}, 4612 (2001).\\[4mm]
Kopper, C., and Magnen, J., 
Ann. Henri Poincar\'e {\bf 2}, 513 (2001).\\[4mm]
Kosterlitz, J.M., and Thouless, D.J., J. Phys. C {\bf 6}, 1181 (1973).\\[4mm]
Kosterlitz, J.M., J. Phys. C {\bf 7}, 1046 (1974).\\[4mm]
Krahl, H.C., and Wetterich, C., Phys. Lett. A {\bf 367}, 263 (2007).\\[4mm]
Leggett, A.J., in {\em Modern Trends in the Theory of Condensed Matter}, 
edited by A. Pekalski and R. Przystawa (Springer, Berlin, 1980).\\[4mm]
Lerch, N., Bartosch, L., and Kopietz, P., 
Phys. Rev. Lett. {\bf 100}, 050403 (2008).\\[4mm]
Litim, D.F., Phys. Rev. D {\bf 64}, 105007 (2001).\\[4mm]
%
%
Lobo, C., Recati, A., Giorgini, S., and Stringari, S.,
Phys. Rev. Lett. {\bf 97}, 200403 (2006).\\[4mm]
L\"ohneysen, H. v., Rosch, A., Vojta, M., and W\"olfle, P., 
Rev. Mod. Phys. {\bf 79}, 1015 (2007).\\[4mm]
Mart\'in-Rodero A., and Flores, F., 
Phys. Rev. B {\bf 45}, 13008 (1992).\\[4mm]
Metzner, W., Castellani, C., and Di Castro, C., 
Adv. Phys. {\bf 47}, 317-445 (1998).\\[4mm]
Metzner, W., Rohe, D., and Andergassen, S., 
Phys. Rev. Lett. {\bf 91}, 066402 (2003).\\[4mm] 
Metzner, W., Prog. Theor. Phys. Suppl. {\bf 160}, 58 (2005).\\[4mm] 
Micnas, R., Ranninger, J., and Robaszkiewicz, S., 
Rev. Mod. Phys. {\bf 62}, 113 (1990).\\[4mm]
Millis, A.J., Phys. Rev. B {\bf 48}, 7183 (1993).\\[4mm]
Nambu, Y., http://nobelprize.org/nobel\_prizes/physics/laureates/2008/index.html (2008).\\[4mm]
\newpage
\noindent
Negele, J.W., and Orland, H., 
{\em Quantum Many-Particle Systems} (Addison-Wesley, Reading, MA, 1987).\\[4mm]
Nelson, D.R., and Kosterlitz, J.M., 
Phys. Rev. Lett. {\bf 39}, 19 (1977).\\[4mm]
Nepomnashchy, Y.A., Phys. Rev. B {\bf 46}, 6611 (1992), and references therein.\\[4mm]
Neumayr, A., and Metzner, W., 
Phys. Rev. B {\bf 58}, 15449 (1998).\\[4mm]
Nozi\`eres, P., {\em Theory of Interacting Fermi Systems} 
(Westview Press, 1964).\\[4mm]
Nozi\`eres, P., and Schmitt-Rink, S., 
Journal of Low Temp. Phys. {\bf 59}, 195 (1985).\\[4mm] 
Oganesyan, V., Kivelson, S.A., and Fradkin, E.,
Phys. Rev. B {\bf 64}, 195109 (2001).\\[4mm]
Ohashi, Y., and Griffin, A., 
Phys. Rev. Lett. {\bf 89}, 130402 (2002).\\[4mm]
Pawlowski, J.M., Litim, D.F., Nedelko, S., 
and L. v. Smekal, Phys. Rev. Lett. {\bf 93}, 152002 (2004).\\[4mm]
Peskin, M.E., and Schroeder, D.V.,
{\em An Introduction to Quantum Field Theory} (Westview Press, 1995).\\[4mm]
Pistolesi, F., Castellani, C., Di Castro, C., 
and Strinati, G.C., Phys. Rev. B {\bf 69}, 024513 (2004); 
Phys. Rev. Lett. {\bf 78}, 1612 (1997).\\[4mm]
Polchinski, J., Nucl. Phys. B {\bf 231}, 269 (1984).\\[4mm]
Politzer, H.D., Phys. Rev. Lett. {\bf 30}, 1346 (1973).\\[4mm]
%
%
Popov, V.N., 
{\em Functional integrals and collective excitations} 
 (Cambridge University Press, Cambridge, 1987).\\[4mm]
Quintanilla, J., Haque, M., and Schofield, A.J., 
Phys. Rev. B {\bf 78}, 035131 (2008).\\[4mm] 
Randeria, M., in
 {\em Bose-Einstein Condensation}, 
 edited by A. Griffin, D. Snoke, and S. Stringari
 (Cambridge University Press, Cambridge, England, 1995),
 pp.~355-392.\\[4mm]
Rech, J., Pepin, C., and Chubukov, A.V., 
Phys. Rev. B {\bf 74}, 195126 (2006).\\[4mm]
Reiss, J., Rohe, D., and Metzner, W., 
Phys. Rev. B {\bf 75}, 075110 (2007).\\[4mm]
Rohe, D., and Metzner, W., 
Phys. Rev. B {\bf 63}, 224509 (2001).\\[4mm]
Rosa, L., Vitale, P., and Wetterich, C., 
Phys. Rev. Lett. {\bf 86}, 958 (2001).\\[4mm]
Rosch, A., Phys. Rev. B {\bf 64}, 174407 (2001).\\[4mm]
R\"uegg, Ch., Normand, B., Matsumoto, M., Furrer, A., McMorrow, D.F., 
Kr\"amer, K.W., G\"udel, H.-U., Gvasaliya, S.N., Mutka, H., 
and Boehm, M., 
Phys. Rev. Lett. {\bf 100}, 205701 (2008).\\[4mm] 
Sachdev, S., {\em Quantum Phase Transitions}
 (CUP, Cambridge, U.K., 1999).\\[4mm]
Salmhofer, M., {\em Renormalization An Introduction}
 (Springer, Berlin Heidelberg, 1999).\\[4mm]
Salmhofer, M., Honerkamp, C., Metzner, W., 
and Lauscher, O., Prog. Theor. Phys. {\bf 112}, 943 (2004).\\[4mm]
Salmhofer, M., Ann. Phys. (Leipzig) {\bf 16}, No. 3, 171 (2007).\\[4mm]
Sch\"utz, F., Bartosch, L., and Kopietz, P., 
Phys. Rev. B {\bf 72}, 035107 (2005).\\[4mm]
Sch\"utz, F., and Kopietz, P., J. Phys. A {\bf 39}, 8205 (2006).\\[4mm]
%
%
Sebastian, S.E., Harrison, N., Batista, C.D., Balicas, L., 
Jaime, M., Sharma, P.A., Kawashima, N., and Fisher, I.R., 
Nature {\bf 441}, 617 (2006).\\[4mm]
Shankar, R., Rev. Mod. Phys. {\bf 66}, 129 (1994).\\[4mm]
Shin, Y., Schunck, C.H., Schirotzek, A., and Ketterle, W., 
Nature {\bf 451}, 689 (2008).\\[4mm]
Singer, J.M., Pedersen, M.H., Schneider, T., Beck, H., and 
Matuttis, H.G., Phys. Rev. B {\bf 54}, 2 (1996).\\[4mm] 
%
%
Solyom, J., Adv. Phys. {\bf 28}, 201 (1979).\\[4mm]
Stewart, G.R., Rev. Mod. Phys. {\bf 73}, 797 (2001).\\[4mm]
Strack, P., Gersch, R., and Metzner, W., Phys. Rev. B {\bf 78}, 014522 (2008).\\[4mm]
Strouthos, C., and Kogut, J.B., arXiv:0804.0300 (2008).\\[4mm]
Tetradis, N, and Wetterich, C., 
Nucl. Phys. B {\bf 422}, 541 (1994).\\[4mm]
Uchoa, B., and Castro Neto, A.H., Phys. Rev. Lett. {\bf 98}, 146801 (2007).\\[4mm]
Voit, J., Rep. Prog. Phys. {\bf 57}, 977 (1994).\\[4mm]
Weichman, P.B., Phys. Rev. B {\bf 38}, 8739 (1988).\\[4mm]
Weinberg, S., {\em The Quantum Theory of Fields Vol. I}, 
(CUP, Cambridge, 2005).\\[4mm] 
Wetterich, C., Z. Phys. C {\bf 57}, 451 (1991).\\[4mm]
Wetterich, C., Phys. Lett. B {\bf 301}, 90 (1993).\\[4mm]
Wetterich, C., Phys. Rev. B {\bf 77}, 064504 (2008).\\[4mm]
Wilson, K.G., and Kogut, J., Phys. Rep. C {\bf 12}, 75 (1974).\\[4mm]
W\"olfle, P., and Rosch, A., 
J. Low Temp. Phys. {\bf 147}, 165 (2007).\\[4mm]
Zanchi, D., and Schulz, H.J., 
Phys. Rev. B {\bf 61}, 13609 (2000).\\[4mm]
Zhao, E., and Paramekanti, A., Phys. Rev. Lett. {\bf 97}, 230404 (2006).\\[4mm]
Zwerger, W., Phys. Rev. Lett. {\bf 92}, 027203 (2004).\\[4mm]
%

%



\end{document}